\documentclass[11pt]{article} 

\usepackage{textcomp, amssymb}  
\usepackage{anysize}            
\usepackage{amsfonts}
\usepackage{amsmath}
\usepackage{dsfont}
\usepackage{MnSymbol}
\usepackage{mathtools}
\usepackage{graphicx}
\usepackage{epsfig}
\usepackage{epstopdf}
\usepackage{lmerge}
\usepackage{tikz}
\usepackage{amsthm}
\usepackage{ifpdf}
\usepackage{thm-restate}
\usepackage{stmaryrd}
\usepackage{cleveref}
\usepackage{subcaption}

\usetikzlibrary{calc,decorations.pathmorphing,shapes,graphs,positioning,automata,angles}

\newcounter{sarrow}

\newcounter{sarrow1}
\newcommand\xnrsquigarrow[1]{%
\stepcounter{sarrow1}%
\mathrel{\begin{tikzpicture}[baseline= {( $ (current bounding box.south) + (0,-0.5ex) $ )}]
\node[inner sep=.5ex] (\thesarrow) {$\scriptstyle #1$};
\path[draw,<-,decorate,
  decoration={zigzag,amplitude=0.7pt,segment length=1.2mm,pre=lineto,pre length=4pt}]
    (\thesarrow1.south east) -- (\thesarrow1.south west);
    $\slashedarrowfill@\relbar\relbar/$
    \end{tikzpicture}}%
}

\makeatletter
\def\slashedarrowfill@#1#2#3#4#5{%
  $\m@th\thickmuskip0mu\medmuskip\thickmuskip\thinmuskip\thickmuskip
   \relax#5#1\mkern-7mu%
   \cleaders\hbox{$#5\mkern-2mu#2\mkern-2mu$}\hfill
   \mathclap{#3}\mathclap{#2}%
   \cleaders\hbox{$#5\mkern-2mu#2\mkern-2mu$}\hfill
   \mkern-7mu#4$%
}
\def\rightslashedarrowfillb@{%
  \slashedarrowfill@\relbar\relbar/\rightarrow}
\newcommand\xnrightarrow[2][]{%
  \ext@arrow 0055{\rightslashedarrowfillb@}{#1}{#2}}

\def\rightslashedarrowfille@{%
  \slashedarrowfill@\relbar\relbar/\twoheadrightarrow}
\newcommand\xntworightarrow[2][]{%
  \ext@arrow 0055{\rightslashedarrowfille@}{#1}{#2}}

\def\rightslashedarrowfillg@{%
  \slashedarrowfill@\relbar\relbar{\raisebox{.12em}{}}\twoheadrightarrow}
\newcommand\xtworightarrow[2][]{%
  \ext@arrow 0055{\rightslashedarrowfillg@}{#1}{#2}}

\def\rightslashedarrowfillx@{%
  \slashedarrowfill@\Relbar\Relbar/\rightrightarrows}
\newcommand\xnTworightarrow[2][]{%
  \ext@arrow 0055{\rightslashedarrowfillx@}{#1}{#2}}

\def\rightslashedarrowfilly@{%
  \slashedarrowfill@\Relbar\Relbar{\raisebox{.12em}{}}\rightrightarrows}
\newcommand\xTworightarrow[2][]{%
  \ext@arrow 0055{\rightslashedarrowfilly@}{#1}{#2}}

\pgfdeclareshape{slash underlined}
{
  \inheritsavedanchors[from=rectangle] 
  \inheritanchorborder[from=rectangle]
  \inheritanchor[from=rectangle]{north}
  \inheritanchor[from=rectangle]{north west}
  \inheritanchor[from=rectangle]{north east}
  \inheritanchor[from=rectangle]{center}
  \inheritanchor[from=rectangle]{west}
  \inheritanchor[from=rectangle]{east}
  \inheritanchor[from=rectangle]{mid}
  \inheritanchor[from=rectangle]{mid west}
  \inheritanchor[from=rectangle]{mid east}
  \inheritanchor[from=rectangle]{base}
  \inheritanchor[from=rectangle]{base west}
  \inheritanchor[from=rectangle]{base east}
  \inheritanchor[from=rectangle]{south}
  \inheritanchor[from=rectangle]{south west}
  \inheritanchor[from=rectangle]{south east}
  \inheritanchorborder[from=rectangle]
  \foregroundpath{
    \southwest \pgf@xa=\pgf@x \pgf@ya=\pgf@y
    \northeast \pgf@xb=\pgf@x \pgf@yb=\pgf@y
    \pgf@xc=\pgf@xa
    \advance\pgf@xc by .5\pgf@xb
    \pgf@yc=\pgf@ya
    \advance\pgf@xc by -1.3pt
    \advance\pgf@yc by -1.8pt
    \pgfpathmoveto{\pgfqpoint{\pgf@xc}{\pgf@yc}}
    \advance\pgf@xc by  2.6pt
    \advance\pgf@yc by  3.6pt
    \pgfpathlineto{\pgfqpoint{\pgf@xc}{\pgf@yc}}
    \pgfpathmoveto{\pgfqpoint{\pgf@xa}{\pgf@ya}}
    \pgfpathlineto{\pgfqpoint{\pgf@xb}{\pgf@ya}}
 }
}
\tikzset{nomorepostaction/.code=\let\tikz@postactions\pgfutil@empty}

\newcommand{\semangle}[1]{\langle\!|#1|\!\rangle}
\newcommand{\sembrack}[1]{\llbracket #1\rrbracket}
\newcommand*{\rmbrace}{|\mskip-4mu\}}
\newcommand*{\lmbrace}{\{\mskip-4mu|}
\newcommand*{\mset}[1]{\lmbrace#1\rmbrace}

\newtheorem{theorem}{Theorem}[section]
\newtheorem{definition}[theorem]{Definition}

\newtheorem{lemma}[theorem]{Lemma}

\newtheorem{corollary}[theorem]{Corollary}

\newcommand{\subsubsubsection}[1]{\paragraph{#1}\mbox{}\\} 
\begin{document}

\begin{titlepage}
\thispagestyle{empty}

\hrule
\begin{center}
{\bf\LARGE Computation and Concurrency\\}
%
\vspace{0.5cm}
--- Yong Wang ---

\vspace{2cm}

\end{center}
\end{titlepage}

\newpage 

\setcounter{page}{1}\pagenumbering{roman}

\tableofcontents

\newpage
\setcounter{page}{1}\pagenumbering{arabic}

        \section{Introduction}\label{intro}

There are mainly two kinds of models of concurrency \cite{MOC}: the models of interleaving concurrency and the models of true concurrency. Among the models of interleaving concurrency, the representatives are process algebras based on bisimilarity semantics, such as CCS \cite{CC} \cite{CCS} and ACP \cite{ACP}. And among the models of true concurrency, the representatives are event structure \cite{PED} \cite{ES} \cite{IES}, Petri net \cite{PN00} \cite{PN01} \cite{PN02} \cite{PN1} \cite{PN2} \cite{PN3}, and also automata and concurrent Kleene algebra \cite{CKA1} \cite{CKA2} \cite{CKA3} \cite{CKA4} \cite{CKA5} \cite{CKA6} \cite{CKA7}. The relationship between interleaving concurrency vs. true concurrency is not clarified, the main work on the relationship is giving interleaving concurrency a semantics of true concurrency \cite{NISCCS} \cite{POSCCS} \cite{ESSCCS}.

Event structure \cite{PED} \cite{ES} \cite{IES} is a model of true concurrency. In an event structure, there are a set of atomic events and arbitrary causalities and conflictions among them, and concurrency is implicitly defined. Based on the definition of an event structure, truly concurrent behaviours such as pomset bisimulation, step bisimulation, history-preserving (hp-) bisimulation and the finest hereditary history-preserving (hhp-) bisimulation \cite{HHPS1} \cite{HHPS2} can be introduced. Since the relationship between process algebra vs. event structure (interleaving concurrency vs. true concurrency in nature) is not clarified before the introduction of truly concurrent process algebra \cite{APTC} \cite{APTC2}, the work on the relationship between process algebra vs. event structure usually gives traditional process algebra an event structure-based semantics, such as giving CCS a event structure-based semantics \cite{ESSCCS}. Petri net \cite{PN00} \cite{PN01} \cite{PN02} \cite{PN1} \cite{PN2} \cite{PN3} is also a model of true concurrency. In a Petri net, there are two kinds of nodes: places (conditions) and transitions (actions), and causalities among them. On the relationship between process algebra and Petri net, one side is giving process algebra a Petri net semantics, the other side is giving Petri net a process algebra foundation \cite{BOXA1} \cite{BOXA2} \cite{PNA} \cite{PAPN}, among them, Petri net algebra \cite{PNA} gives Petri net a CCS-like foundation. 

Kleene algebra (KA) \cite{KA00} \cite{KA0} \cite{KA1} \cite{KA2} \cite{KA3} \cite{KA4} \cite{KA5} \cite{KA6} is an important algebraic structure with operators $+$, $\cdot$, $^*$, $0$ and $1$ to model computational properties of regular expressions. Kleene algebra can be used widely in computational areas, such as relational algebra, automata, algorithms, program logic and semantics, etc. A Kleene algebra of the family of regular set over a finite alphabet $\Sigma$ is called the algebra of regular events denoted $\mathbf{Reg}_{\Sigma}$, which was firstly studied as an open problem by Kleene \cite{KA00}. Then, Kleene algebra was widely studied and there existed several definitions on Kleene algebra \cite{KA0} \cite{KA1} \cite{KA2} \cite{KA3} \cite{KA4} \cite{KA5} for the almost same purpose of modelling regular expressions, and Kozen \cite{KA6} established the relationship among these definitions. Then Kleene algebra has been extended in many ways to capture more computational properties, such as hypotheses \cite{EKA1} \cite{EKA2}, tests \cite{EKA3} \cite{EKA4} \cite{EKA5}, observations \cite{EKA6}, probabilistic KA \cite{EKA7}, etc. Among these extensions, a significant one is concurrent KA (CKA) \cite{CKA1} \cite{CKA2} \cite{CKA3} \cite{CKA4} \cite{CKA5} \cite{CKA6} \cite{CKA7} and its extensions \cite{CKA70} \cite{CKA8} \cite{CKA9} \cite{CKA10} \cite{CKA11} \cite{CKA12} to capture the concurrent and parallel computations, which is a combination of computation and concurrency.

It is well-known that process algebras are theories to capture concurrent and parallel computations, for CCS \cite{CC} \cite{CCS} and ACP \cite{ACP} are with bisimilarity semantics. A natural question is that how automata theory is related to process algebra and how (concurrent) KA is related to process algebra? J. C. M. Baeten et al have done a lot of work on the relationship between automata theory and process algebra \cite{AP1} \cite{AP2} \cite{AP3} \cite{AP4} \cite{AP5} \cite{AP6}. It is essential of the work on introducing Kleene star into the process algebra based on bisimilarity semantics to answer this question, firstly initialized by Milner's proof system for regular expressions modulo bisimilarity (Mil) \cite{MF1}. Since the completeness of Milner's proof system remained open, some efforts were done, such as Redko's incompleteness proof for Klneene star modulo trace semantics \cite{MF10n}, completeness for BPA (basic process algebra) with Kleene star \cite{MF2} \cite{MF20}, work on ACP with iteration \cite{MF15} \cite{MF16} \cite{MF17}, completeness for prefix iteration \cite{MF10} \cite{MF11} \cite{MF12}, multi-exit iteration \cite{MF13}, flat iteration \cite{MF14}, 1-free regular expressions \cite{MF3} modulo bisimilarity. But these are not the full sense of regular expressions, most recently, Grabmayer \cite{MF4} \cite{MF5} \cite{MF6} \cite{MF7} has prepared to prove that Mil is complete with respect to a specific kind of process graphs called LLEE-1-charts which is equal to regular expressions, modulo the corresponding kind of bisimilarity called 1-bisimilarity.

But for computation, concurrency and parallelism, the relationship between CKA and process algebra has remained open from Hoare \cite{CKA1} \cite{CKA2} to the recent work of CKA \cite{CKA7} \cite{CKA70}. Since most CKAs are based on the so-called true concurrency, we can draw the conclusion that the concurrency of CKA includes the interleaving one which the bisimilarity based process algebra captures, as the extended Milner's expansion law $a\parallel b=a\cdot b+b\cdot a+a\parallel b$ says, where $a,b$ are primitives (atomic actions), $\parallel$ is the parallel composition, $+$ is the alternative composition and $\cdot$ is the sequential composition with the background of computation. In contrast, Milner's expansion law is that $a\parallel b=a\cdot b+b\cdot a$ in bisimilarity based process algebras CCS and ACP.

As Chinese, we love "big" unification, i.e., the unification of interleaving concurrency vs. true concurrency. In concurrency theory, we refer to parallelism, denoted $a\parallel b$ for $a,b$ are atomic actions, which means that there are two parallel branches $a$ and $b$, they executed independently (without causality and confliction) and is captured exactly by the concurrency relation. But the whole thing, we prefer to use the word \emph{concurrency}, denoted $a\between b$, is that the actions in the two parallel branches may exist causalities or conflictions. In the background of computation and concurrency, we only consider the structurization of unstructured causalities. The causalities between two parallel branches are usually not the sequence relation, but communications (the sending/receiving or writing/reading pairs). Concurrency is made up of several parallel branches, in each branch which can be a model of concurrency, there exists communications among these branches. This is well supported by computational systems in reality from the smaller ones to bigger ones: threads, cores, CPUs, processes, and communications among them inner one computer system; distributed applications, communications via computer networks and distributed locks among them, constitute small or big scale distributed systems and the whole Internet. Base on the above assumptions, we have done some work on the so-called truly concurrent process algebra CTC and APTC \cite{APTC} \cite{APTC2}, which are generalizations of CCS and ACP from interleaving concurrency to true concurrency. 

In this small book, we deep the relationship between computation and concurrency, especially, base on the so-called pomsetc automata, we introduce communication and more operators, and establish the algebras modulo language equivalence and truly concurrent bisimilarities. Based on the work of truly concurrent process algebra APTC \cite{APTC} which is process algebra based on truly concurrent semantics, we can introduce Kleene star (and also parallel star) into APTC. Both for CKA with communications and APTC with Kleene star and parallel star, the extended Milner's expansion law $a\between b=a\cdot b+b\cdot a+a\parallel b +a\mid b$ with the concurrency operator $\between$ and communication merge $\mid$ holds. CKA and APTC are all the truly concurrent computation models, can have the same syntax (primitives and operators), the similar axiomatizations, and maybe have the same or different semantics. That's all. 

Note that, we write some conclusions without any proof.
\newpage\section{Preliminaries}\label{pre} 

For self-satisfactory, in this chapter, we introduce the preliminaries on set, language, and rational language and automata in \cref{sl}, and also algebras for rational expressions modulo language equivalence and bisimilarity in \cref{rla}.

\subsection{Set and Language}\label{sl}

\begin{definition}[Set]
A set contains some objects, and let $\{-\}$ denote the contents of a set. For instance, $\mathbb{N}=\{1,2,3,\cdots\}$. Let $a\in A$ denote that $a$ is an element of the set $A$ and $a\notin A$ denote that $a$ is not an element of the set $A$. For all $a\in A$, if we can get $a\in B$, then we say that $A$ is a subset of $B$ denoted $A\subseteq B$. If $A\subseteq B$ and $B\subseteq A$, then $A=B$. We can define a new set by use of predicates on the existing sets, such that $\{n\in\mathbb{N}|\exists k\in\mathbb{N},n=2k\}$ for the set of even numbers. We can also specify a set to be the smallest set satisfy some inductive inference rules, for instance, we specify the set of even numbers $A$ satisfying the following rules:

$$\frac{}{0\in A}\quad\quad\frac{n\in A}{n+2\in A}$$
\end{definition}

\begin{definition}[Set composition]
The union of two sets $A$ and $B$, is denoted by $A\cup B=\{a|a\in A\textrm{ or }a\in B\}$, and the intersection of $A$ and $B$ by $A\cap B=\{a|a\in A\textrm{ and }a\in B\}$, the difference of $A$ and $B$ by $A\setminus B=\{a|a\in A\textrm{ and }a\notin B\}$. The empty set $\emptyset$ contains nothing. The set of all subsets of a set $A$ is called the powerset of $A$ denoted $2^A$.
\end{definition}

\begin{definition}[Tuple]
A tuple is a finite and ordered list of objects and denoted $\langle -\rangle$. For sets $A$ and $B$, the Cartesian product of $A$ and $B$ is denoted by $A\times B=\{\langle a,b\rangle|a\in A,b\in B\}$. $A^n$ is the $n$-fold Cartesian product of set $A$, for instance, $A^2=A\times A$. Tuples can be flattened as $A\times (B\times C)=(A\times B)\times C=A\times B\times C$ for sets $A$, $B$ and $C$.
\end{definition}

\begin{definition}[Relation]
A relation $R$ between sets $A$ and $B$ is a subset of $A\times B$, i.e., $R\subseteq A\times B$. We say that $R$ is a relation on set $A$ if $R$ is a relation between $A$ and itself, and,

\begin{itemize}
  \item $R$ is reflexive if for all $a\in A$, $aRa$ holds; it is irreflexive if for all $a\in A$, $aRa$ does not hold.
  \item $R$ is symmetric if for all $a,a'\in A$ with $aRa'$, then $a'Ra$ holds; it is antisymmetric if for all $a,a'\in A$ with $aRa'$ and $a'Ra$, then $a=a'$.
  \item $R$ is transitive if for all $a,a',a''\in A$ with $aRa'$ and $a'Ra''$, then $aRa''$ holds.
\end{itemize}
\end{definition}

\begin{definition}[Preorder, partial order, strict order]
If a relation $R$ is reflexive and transitive, we call that it is a preorder; When it is a preorder and antisymmetric, it is called a partial order, and a partially ordered set (poset) is a pair $\langle A,R\rangle$ with a set $A$ and a partial order $R$ on $A$; When it is irreflexive and transitive, it is called a strict order.
\end{definition}

\begin{definition}[Equivalence]
A relation $R$ is called an equivalence, if it is reflexive, symmetric and transitive. For an equivalent relation $R$ and a set $A$, $[a]_R=\{a'\in A|aRa'\}$ is called the equivalence class of $a\in A$.
\end{definition}

\begin{definition}[Relation composition]
For sets $A$, $B$ and $C$, and relations $R\subseteq A\times B$ and $R'\subseteq B\times C$, the relational composition denoted $R\circ R'$, is defined as the smallest relation $a(R\circ R')c$ satisfying $aRb$ and $bR'c$ with $a\in A$, $b\in B$ and $c\in C$. For a relation $R$ on set $A$, we denote $R^*$ for the reflexive and transitive closure of $R$, which is the least reflexive and transitive relation on $A$ that contains $R$.
\end{definition}

\begin{definition}[Function]
A function $f:A\rightarrow B$ from sets $A$ to $B$ is a relation between $A$ and $B$, i.e., for every $a\in A$, there exists one $b=f(a)\in B$, where $A$ is called the domain of $f$ and $B$ the codomain of $f$. $\sembrack{-}$ is also used as a function with $-$ a placeholder, i.e., $\sembrack{x}$ is the value of $\sembrack{-}$ for input $x$. A function $f$ is a bijection if for every $b\in B$, there exists exactly one $a\in A$ such that $b=f(a)$. For functions $f:A\rightarrow B$ and $g:B\rightarrow C$, the functional composition of $f$ and $g$ denoted $g\circ f$ such that $(g\circ f)(a)=g(f(a))$ for $a\in A$.
\end{definition}

\begin{definition}[Poset morphism]
For posets $\langle A,\leq\rangle$ and $\langle A',\leq'\rangle$ and function $f:A\rightarrow A'$, $f$ is called a poset morphism if for $a_0,a_1\in A$ with $a_0\leq a_1$, then $f(a_0)\leq' f(a_1)$ holds.
\end{definition}

\begin{definition}[Multiset]
A multiset is a kind of set of objects which may be repetitive denoted $\mset{-}$, such that $\mset{0,1,1}$ is significantly distinguishable from $\mset{0,1}$.
\end{definition}

\begin{definition}[Alphabet, word, language]
An alphabet $\Sigma$ is a (maybe infinite) set of symbols. A word over some alphabet $\Sigma$ is a finite sequence of symbols from $\Sigma$. Words can be concatenated horizontally and the horizontal concatenation operator is denoted by $\cdot$, for instance $ab\cdot c=abc$. The empty word is denoted $1$ with $1\cdot w=w=w\cdot 1$ for word $w$. For $n\in\mathbb{N}$ and $a\in\Sigma$, $a^n$ is the $n$-fold concatenation of $a$ with $a^0=1$ and $a^{n+1}=a\cdot a^n$. Words can be concatenated vertically and the vertical concatenation operator is denoted by $\parallel$, for instance $a\parallel b\parallel c$. $1\parallel w=w=w\parallel 1$ for word $w$. For $n\in\mathbb{N}$ and $a\in\Sigma$, $a^{\langle n\rangle}$ is the $n$-fold vertical concatenation of $a$ with $a^{\langle 0\rangle}=1$ and $a^{\langle n+1\rangle}=a\parallel a^{\langle n\rangle}$. A language is a set of words, and the language of all words over an alphabet $\Sigma$ is denoted $\Sigma^{\langle *\rangle^*}$.
\end{definition}

\begin{definition}[Expressions]
Expressions are builded by function symbols and constants over a fixed alphabet inductively. For instance, the set of numerical expressions over some fixed set of variables $V$ are defined as the smallest set $\mathcal{T}$ satisfying the following inference rules:

$$\frac{n\in\mathbb{N}}{n\in \mathcal{T}}\quad \frac{v\in V}{v\in \mathcal{T}}\quad \frac{x,y\in \mathcal{T}}{x+y\in \mathcal{T}}\quad \frac{x,y\in \mathcal{T}}{x\times y\in \mathcal{T}}\quad \frac{x\in \mathcal{T}}{-x\in \mathcal{T}}$$

The above inference rules are equal to the following Backus-Naur Form (BNF) grammer.

$$\mathcal{T}\ni x,y::=n\in\mathbb{N}|v\in V|x+y|x\times y|-x$$
\end{definition}

\begin{definition}[Congruence, precongruence]
A relation $R$ on a set of expressions is a congruence if it is an equivalence compatible with the operators; and a relation $R$ on a set of expressions is a precongruence if it is a preorder compatible with the operators.
\end{definition}

\subsection{Rational Language and Automata}\label{rla}

\subsubsection{Automata}

\begin{definition}[Automaton]
An automaton is a tuple $A=(Q,F,\delta)$ where $Q$ is a finite set of states, $F\subseteq Q$ is the set of final states, and $\delta$ is the finite set of the transitions of $A$ and $\delta\subseteq Q\times \Sigma\times Q$.
\end{definition}

It is well-known that automata recognize rational languages.

\begin{definition}[Transition relation]
Let $p,q\in Q$. We define the transition relation $\xrightarrow[A]{}\subseteq Q\times \Sigma\times Q$ on $A$ as the smallest relation satisfying:

\begin{enumerate}
  \item $p\xrightarrow[A]{1}p$ for all $p\in Q$;
  \item $p\xrightarrow[A]{a}q$ if and only if $(p,a,q)\in \delta$.
\end{enumerate}
\end{definition}

\begin{definition}[Bisimulation based on automata]
Let $A=(Q,F,\delta)$ and $A'=(Q',F',\delta')$ be two automata with the same alphabet, and $p,q\in Q$ and $p',q'\in Q'$. The automata $A$ and $A'$ are bisimilar, $A\sim_{HM}A'$, if and only if there is a relation $R$ between their reachable states that preserves transitions and termination:

\begin{enumerate}
  \item $R$ relate reachable states, i.e., every reachable state of $A$ is related to a reachable state of $A'$ and every reachable state of $A'$ is related to a reachable state of $A$;
  \item whenever $p$ is related to $p'$, $pRp'$ and $p\xrightarrow[A]{a}q$, then there is state $q'$ in $A'$ with $p'\xrightarrow[A']{a}q'$ and $qRq'$;
  \item whenever $p$ is related to $p'$, $pRp'$ and $p'\xrightarrow[A']{a}q'$, then there is state $q$ in $A$ with $p\xrightarrow[A]{a}q$ and $qRq'$;
  \item whenever $pRp'$, then $p\in F$ if and only if $p'\in F'$.
\end{enumerate}
\end{definition}

\begin{definition}[Simulation based on automata]
Let $A=(Q,F,\delta)$ and $A'=(Q',F',\delta')$ be two automata with the same alphabet, and $p,q\in Q$ and $p',q'\in Q'$. The automata $A$ and $A'$ are similar, $A\lesssim_{HM}A'$, if and only if there is a relation $R$ between their reachable states that preserves transitions and termination:

\begin{enumerate}
  \item $R$ relate reachable states, i.e., every reachable state of $A$ is related to a reachable state of $A'$;
  \item whenever $p$ is related to $p'$, $pRp'$ and $p\xrightarrow[A]{a}q$, then there is state $q'$ in $A'$ with $p'\xrightarrow[A']{a}q'$ and $qRq'$;
  \item whenever $pRp'$, if $p\in F$ then $p'\in F'$.
\end{enumerate}
\end{definition}

\subsubsection{Algebra Modulo Language Equivalence}

Traditionally, we note Kleene algebra in the context of rational language and expressions. We fix a finite alphabet $\Sigma$ and a word formed over $\Sigma$ is a finite sequence of symbols from $\Sigma$, and the empty word is denoted $1$. Let $\Sigma^*$ denote the set of all words over $\Sigma$ and a language is a set of words. For words $u,v\in\Sigma^*$, we define $u\cdot v$ as the concatenation of $u$ and $v$, $u\cdot v=uv$. Then for $U,V\subseteq\Sigma^*$, we define $U\cdot V=\{uv|u\in U,v\in V\}$, $U+V=U\cup V$, $U^*=\bigcup_{n\in\mathbb{N}}U^n$ where $U^0=\{1\}$ and $U^{n+1}=U\cdot U^n$.

\begin{restatable}[Syntax of rational expressions]{definition}{rationalexpressions}\label{rationalexpressions}
We define the set of rational expressions $\mathcal{T}_{R}$ as follows.

$$\mathcal{T}_{R}\ni x,y::=0|1|a\in\Sigma|x+y|x\cdot y|x^*$$
\end{restatable}

\begin{definition}[Language semantics of rational expressions]
We define the interpretation of rational expressions $\sembrack{-}:\mathcal{T}_{R}\rightarrow\mathcal{P}(\Sigma^*)$ inductively as Table \ref{SRE} shows.
\end{definition}

\begin{center}
    \begin{table}
        $$\sembrack{0}_{R}=\emptyset \quad \sembrack{a}_{R}=\{a\} \quad \sembrack{x\cdot y}_{R}=\sembrack{x}_{R}\cdot \sembrack{y}_{R}$$
        $$\sembrack{1}_{R}=\{1\} \quad \sembrack{x+y}_{R}=\sembrack{x}_{R}+\sembrack{y}_{R} \quad\sembrack{x^*}_{R}=\sembrack{x}^*_{R}$$
        \caption{Language semantics of rational expressions}
        \label{SRE}
    \end{table}
\end{center}

\begin{definition}[Axiomatization]
An axiomatization over an alphabet $\Sigma$ is a finite set of equations, called axioms, of the form $x=y$ with $x,y\in\mathcal{T}_{R}$.
\end{definition}

An axiomatization gives rise to an equality relation $=$ over $\mathcal{T}_{R}$.

\begin{definition}[Equality relation]
The binary relation equality $=$ is defined as follows, $x,y\in \mathcal{T}_{R}$:

\begin{itemize}
  \item (Substitution) If $x = y$ and $\sigma$ a substitution, then $\sigma(x) = \sigma(y)$.
  \item (Equivalence) The relation $=$ is an equivalence, i.e., closed under reflexivity, symmetry and transitivity.
  \item (Context) The relation $=$ is closed under contexts, i.e., if $x=y$ and $f$ is an operator and $ar(f)$ is the arity of $f$, then
  
  $$f(x_1,\cdots, x_{i-1}, x, x_{i+1}, \cdots, x_{ar(f)})=f(x_1,\cdots, x_{i-1}, y, x_{i+1}, \cdots, x_{ar(f)})$$
\end{itemize}
\end{definition}

We define a Kleene algebra as a tuple $(\Sigma,+,\cdot,^*,0,1)$, where $\Sigma$ is a set, $^*$ is a unary operator, $+$ and $\cdot$ are binary operators, and $0$ and $1$ are constants, which satisfies the axioms in Table \ref{AxiomsForKA} for all $x,y,z\in \mathcal{T}_{R}$, where $x\leqq y$ means $x+y=y$.

\begin{center}
    \begin{table}
        \begin{tabular}{@{}ll@{}}
            \hline No. &Axiom\\
            $A1$ & $x+y=y+z$\\
            $A2$ & $x+(y+z)=(x+y)+z$\\
            $A3$ & $x+x=x$\\
            $A4$ & $(x+y)\cdot z=x\cdot z+y\cdot z$\\
            $A5$ & $x\cdot(y+z)=x\cdot y+x\cdot z$\\
            $A6$ & $x\cdot(y\cdot z)=(x\cdot y)\cdot z$\\
            $A7$ & $x+0=x$\\
            $A8$ & $0\cdot x=0$\\
            $A9$ & $x\cdot 0=0$\\
            $A10$ & $x\cdot 1=x$\\
            $A11$ & $1\cdot x=x$\\
            $A12$ & $1+x\cdot x^*=x^*$\\
            $A13$ & $1+x^*\cdot x=x^*$\\
            $A14$ & $x+y\cdot z\leqq z\Rightarrow y^*\cdot x\leqq z$\\
            $A15$ & $x+y\cdot z\leqq y\Rightarrow x\cdot z^*\leqq y$\\
        \end{tabular}
        \caption{Axioms of Kleene algebra modulo language equivalence}
        \label{AxiomsForKA}
    \end{table}
\end{center}

Since language equivalence is a congruence w.r.t. the operators of KA, we can only check the soundness of each axiom according to the definition of semantics of rational expressions. Then we can get the following soundness and completeness theorem, which is proven by Kozen \cite{KA3}.

\begin{theorem}[Soundness and completeness of Kleene algebra]
For all $x,y\in\mathcal{T}_{R}$, $x=y$ if and only if $\sembrack{x}_{R}=\sembrack{y}_{R}$.
\end{theorem}

\subsubsection{Milner's Proof System for Rational Expressions Modulo Bisimilarity}

Process algebras CCS \cite{CC} \cite{CCS} and ACP \cite{ACP} have a bisimilarity-based operational semantics.

Milner wanted to give rational expressions a bisimilarity-based semantic foundation and designed a proof system \cite{MF1} denoted Mil. Similarly to Kleene algebra, the signature of Mil as a tuple $(\Sigma,+,\cdot,^*,0,1)$ includes a set of atomic actions $\Sigma$ and $a,b,c,\cdots\in \Sigma$, two special constants with inaction or deadlock denoted $0$ and empty action denoted $1$, two binary functions with sequential composition denoted $\cdot$ and alternative composition denoted $+$, and also a unary function iteration denoted $^*$. 

Note that Kleene algebra KA and Mil have almost the same grammar structures to express rational language and expressions, but different backgrounds for the former usually initialized to axiomatize the rational expressions and the latter came from process algebra to capture computation.

\begin{definition}[Operational semantics of Mil]
Let the symbol $\downarrow$ denote the successful termination predicate. Then we give the TSS (Transition System Specification) of Mil as Table \ref{SMil} shows, where $a,b,c,\cdots\in \Sigma$, $x,y,x',y'\in\mathcal{T}_{R}$.
\end{definition}

\begin{center}
    \begin{table}
        $$\frac{}{1\downarrow}\quad\frac{}{a\xrightarrow{a}1}$$
        $$\frac{x\downarrow}{(x+y)\downarrow}\quad\frac{y\downarrow}{(x+y)\downarrow}\quad\frac{x\xrightarrow{a}x'}{x+y\xrightarrow{a}x'}\quad\frac{y\xrightarrow{b}y'}{x+y\xrightarrow{b}y'}$$
        $$\frac{x\downarrow\quad y\downarrow}{(x\cdot y)\downarrow} \quad\frac{x\xrightarrow{a}x'}{x\cdot y\xrightarrow{a}x'\cdot y} \quad\frac{x\downarrow\quad y\xrightarrow{b}y'}{x\cdot y\xrightarrow{b}y'}$$
        $$\frac{x\downarrow}{(x^*)\downarrow} \quad\frac{x\xrightarrow{a}x'}{x^*\xrightarrow{a}x'\cdot x^*}$$
        \caption{Operational semantics of Mil}
        \label{SMil}
    \end{table}
\end{center}

Note that there is no any transition rules related to the constant $0$. 

\begin{definition}[Bisimulation based on expressions]
A bisimulation relation $R$ is a binary relation on expressions $\mathcal{T}_{R}$ with $x,y\in\mathcal{T}_{R}$such that: (1) if $x R y$ and $x\xrightarrow{a}x'$ then $y\xrightarrow{a}y'$ with $x' R y'$; (2) if $x R y$ and $y\xrightarrow{a}y'$ then $x\xrightarrow{a}x'$ with $x' R y'$; (3) if $x R y$ and $xP$, then $yP$; (4) if $x R y$ and $yP$, then $xP$. Two expressions $x$ and $y$ are bisimilar, denoted by $x\sim_{HM} y$, if there is a bisimulation relation $R$ such that $x R y$. Note that $x,x',y,y'$ are expressions, $a$ is a primitive, and $P$ is a predicate.
\end{definition}

\begin{definition}[Simulation based on expressions]
A simulation relation $R$ is a binary relation on expressions $\mathcal{T}_{R}$ with $x,y,x',y'\in\mathcal{T}_{R}$such that: (1) if $x R y$ and $x\xrightarrow{a}x'$ then $y\xrightarrow{a}y'$ with $x' R y'$; (2) if $x R y$ and $xP$, then $yP$. Two expressions $x$ and $y$ are similar, denoted by $x\lesssim_{HM} y$, if there is a simulation relation $R$ such that $x R y$. Note that $a$ is a primitive, and $P$ is a predicate.
\end{definition}

Then the axiomatic system of Mil is shown in Table \ref{AxiomsForMil}.

\begin{center}
    \begin{table}
        \begin{tabular}{@{}ll@{}}
            \hline No. &Axiom\\
            $A1$ & $x+y=y+z$\\
            $A2$ & $x+(y+z)=(x+y)+z$\\
            $A3$ & $x+x=x$\\
            $A4$ & $(x+y)\cdot z=x\cdot z+y\cdot z$\\
            $A5$ & $x\cdot(y\cdot z)=(x\cdot y)\cdot z$\\
            $A6$ & $x+0=x$\\
            $A7$ & $0\cdot x=0$\\
            $A8$ & $x\cdot 1=x$\\
            $A9$ & $1\cdot x=x$\\
            $A10$ & $1+x\cdot x^*=x^*$\\
            $A11$ & $(1+x)^*=x^*$\\
            $A12$ & $x+y\cdot z\leqq z\Rightarrow y^*\cdot x\leqq z$\\
            $A13$ & $x+y\cdot z\leqq y\Rightarrow x\cdot z^*\leqq y$\\
        \end{tabular}
        \caption{Axioms of Mil modulo bisimilarity}
        \label{AxiomsForMil}
    \end{table}
\end{center}

Note that there are two significant differences between the axiomatic systems of Mil and KA, the axioms $x\cdot 0=0$ and $x\cdot(y+z)=x\cdot y+x\cdot z$ of KA do not hold in Mil.

Since bisimilarity is a congruences w.r.t. the operators $\cdot$, $+$ and $^*$, and similarity is a precongruences w.r.t. the operators $\cdot$, $+$ and $^*$, we can only check the soundness of each axiom according to the definition of TSS of rational expressions in Table \ref{SMil}. As mentioned in section \ref{intro}, Milner proved the soundness of Mil and remained the completeness open. Just very recently, Grabmayer \cite{MF6} claimed to have proven that Mil is complete with respect to a specific kind of process graphs called LLEE-1-charts which is equal to rational expressions, modulo the corresponding kind of bisimilarity called 1-bisimilarity.

\begin{theorem}[Soundness and completeness of Mil]
For all $x,y\in\mathcal{T}_{R}$, $x=y$ if and only if $x\sim_{HM}y$.
\end{theorem} 
\newpage\section{Concurrency and Pomsetcs}\label{cp}

In this chapter, we analyze concurrency in section \ref{conc}, introduce the concept of Series-Communication-Parallelism in section \ref{scp}, introduce Pomsetc language in section \ref{pl}. Then we introduce truly concurrent bisimilarities based on expressions in section \ref{tcbbe}. Finally, in section \ref{scre}, we discuss the so-called series-communication rational expressions.

\subsection{Concurrency}\label{conc}

There were always two ways of concurrency from 1970's: the interleaving concurrency and its representative process algebras vs. the true concurrency and its representatives Petri net, event structure, and directed graph, etc. Through the work on truly concurrent process algebras \cite{APTC} \cite{APTC2}, we showed that truly concurrent process algebras are generalizations to the corresponding traditional process algebras. 

It is well-known that process algebras are based on a structured way by atomic actions and operators manipulated on the actions, and true concurrency is based on graph-like models by atomic actions and unstructured causalities and conflictions among the actions. The key challenge of truly concurrent process algebra is how to structurize the unstructured causalities and conflictions in true concurrency. In the background of computation and concurrency, we only consider the structurization of unstructured causalities.

Before we give the basic model of true concurrency in the form of labelled partially ordered sets (labelled posets), firstly, we fix an alphabet $\Sigma$ of symbols usually called actions or events.

\begin{definition}[Labelled poset]
A labelled poset is a tuple $\mathbf{u}=\langle S, \leq, \lambda\rangle$, where $S$ is the carrier set, $\leq$ is a partial order on $S$ and $\lambda$ is a labelling function $\lambda:S\rightarrow\Sigma$.

For a labelled poset $\mathbf{u}$, $S_{\mathbf{u}}$, $\leq_{\mathbf{u}}$ and $\lambda_{\mathbf{u}}$ denote the carrier, the partial order and the labelling of $\mathbf{u}$ respectively. The set of labelled posets is denoted $\mathsf{LP}$ and the empty labelled poset is $\mathbf{1}$.
\end{definition}

\begin{definition}[Labelled poset isomorphism]
Let $\mathbf{u}=\langle S_1,\leq_1,\lambda_1\rangle$ and $\mathbf{v}=\langle S_2,\leq_2,\lambda_2\rangle$ be labelled posets. A labelled poset morphism $h$ from $\mathbf{u}=\langle S_1,\leq_1,\lambda_1\rangle$ to $\mathbf{v}=\langle S_2,\leq_2,\lambda_2\rangle$ is a poset morphism from $\langle S_1,\leq_1\rangle$ and $\langle S_2,\leq_2\rangle$ with $\lambda_2\circ h=\lambda_1$. Moreover, $h$ is a labelled poset isomorphism if it is a bijection with $h^{-1}$ is a poset isomorphism from $\langle S_2,\leq_2,\lambda_2\rangle$ to $\langle S_1,\leq_1,\lambda_1\rangle$. We say that $\mathbf{u}=\langle S_1,\leq_1,\lambda_1\rangle$ is isomorphic to $\mathbf{v}=\langle S_2,\leq_2,\lambda_2\rangle$ denoted $\langle S_1,\leq_1,\lambda_1\rangle\sim\langle S_2,\leq_2,\lambda_2\rangle$, if there exists a poset isomorphism $h$ between $\langle S_1,\leq_1,\lambda_1\rangle$ and $\langle S_2,\leq_2,\lambda_2\rangle$.
\end{definition}

It is easy to see that $\sim$ is an equivalence and can be used to abstract from the carriers.

\begin{definition}[Pomset]
A partially ordered multiset, pomset, is a $\sim$-equivalence class of labelled posets. The $\sim$-equivalence class of $\mathbf{u}\in\mathsf{LP}$ is denoted $[\mathbf{u}]$; the set of pomsets is denoted $\mathsf{Pom}$; the empty labelled poset is denoted $\mathbf{1}$ and the $\sim$-equivalence class of $\mathbf{1}$ is denoted by $1$; the pomset containing exactly one action $a\in\Sigma$ is called primitive.
\end{definition}

We assume that the partial order $\leq$ can be divided into two kinds: execution order $\leq^{e}$ and communication $\leq^{c}$. In the same parallel branch, the partial orders usually execution orders and communication usually exists among different parallel branches. Of course, parallel branches can be nested. Then, we can get the following definitions naturally.

\begin{definition}[Labelled poset with communications]
A labelled poset with communications is a tuple $\mathbf{u}=\langle S, \leq^{e}, \leq^{c}, \lambda\rangle$, where $S$ is the carrier set, $\leq^{e}$ is an execution order on $S$, $\leq^{c}$ is a communication on $S$, and $\lambda$ is a labelling function $\lambda:S\rightarrow\Sigma$. We usually use $\mathbf{u},\mathbf{v}$ to denote labelled posets with communications. And the set of labelled posets with communications is denoted $\mathsf{LPC}$, and the empty labelled poset with communications is $\mathbf{1}$.
\end{definition}

\begin{definition}[Labelled poset isomorphism]
Let $\mathbf{u}=\langle S_1,\leq^{e}_1,\leq^{c}_1,\lambda_1\rangle$ and $\mathbf{v}=\langle S_2,\leq^{e}_2, \leq^{c}_2,\lambda_2\rangle$ be labelled posets. A labelled poset morphism $h$ from $\langle S_1,\leq^{e}_1,\leq^{c}_1,\lambda_1\rangle$ to $\langle S_2,\leq^{e}_2,\leq^{c}_2,\lambda_2\rangle$ is a poset morphism from $\langle S_1,\leq^{e}_1,\leq^{c}_1\rangle$ and $\langle S_2,\leq^{e}_2,\leq^{c}_2\rangle$ with $\lambda_2\circ h=\lambda_1$. Moreover, $h$ is a labelled poset isomorphism if it is a bijection with $h^{-1}$ is a poset isomorphism from $\langle S_2,\leq^{e}_2,\leq^{c}_2,\lambda_2\rangle$ to $\langle S_1,\leq^{e}_1,\leq^{c}_1,\lambda_1\rangle$. We say that $\mathbf{u}=\langle S_1,\leq^{e}_1,\leq^{c}_1,\lambda_1\rangle$ is isomorphic to $\mathbf{v}=\langle S_2,\leq^{e}_2,\leq^{c}_2,\lambda_2\rangle$ denoted $\langle S_1,\leq^{e}_1,\leq^{c}_1,\lambda_1\rangle\sim\langle S_2,\leq^{e}_2,\leq^{c}_2,\lambda_2\rangle$, if there exists a poset isomorphism $h$ between $\langle S_1,\leq^{e}_1,\leq^{c}_1,\lambda_1\rangle$ and $\langle S_2,\leq^{e}_2,\leq^{c}_2,\lambda_2\rangle$.
\end{definition}

It is easy to see that $\sim$ is an equivalence and can be used to abstract from the carriers.

\begin{definition}[Pomset with communications]
A partially ordered multiset with communications, pomsetc, is a $\sim$-equivalence class of labelled posets with communications $\mathbf{u}$, written as $[\mathbf{u}]$, i.e., $[\mathbf{u}]=\{\mathbf{v}\in \mathsf{LPC}:\mathbf{u}\sim\mathbf{v}\}$. The set of pomsetcs is denoted $\mathsf{Pomc}$; the empty labelled poset with communications is denoted $\mathbf{1}$ and the $\sim$-equivalence class of $\mathbf{1}$ is denoted by $1$; the pomsetc containing exactly one action $a\in\Sigma$ is called primitive.
\end{definition}

Concurrency includes parallelism and communication, then, we can get the following definitions of Pomsetc compositions.

\begin{definition}[Pomsetc composition in parallel]
Let $U,V\in\mathsf{Pomc}$ with $U=[\mathbf{u}]$ and $V=[\mathbf{v}]$. We write $U\parallel V$ for the parallel composition of $U$ and $V$, which is the pomsetc represented by $\mathbf{u}\parallel\mathbf{v}$, where

$$S_{\mathbf{u}\parallel\mathbf{v}}=S_{\mathbf{u}}\cup S_{\mathbf{v}}
\quad\quad\leq^{e}_{\mathbf{u}\parallel\mathbf{v}}=\leq^{e}_{\mathbf{u}}\cup\leq^{e}_{\mathbf{v}}
\quad\quad\leq^{c}_{\mathbf{u}\parallel\mathbf{v}}=\leq^{c}_{\mathbf{u}}\cup\leq^{c}_{\mathbf{v}}
\quad\quad\lambda_{\mathbf{u}\parallel\mathbf{v}}(x)=\begin{cases}
\lambda_{\mathbf{u}}\quad x\in S_{\mathbf{u}}\\
\lambda_{\mathbf{v}}\quad x\in S_{\mathbf{v}}
\end{cases}$$
\end{definition}

\begin{definition}[Pomsetc composition in communication]
Let $U,V\in\mathsf{Pomc}$ with $U=[\mathbf{u}]$ and $V=[\mathbf{v}]$. We write $U\mid V$ for the communicative composition of $U$ and $V$, which is the pomsetc represented by $\mathbf{u}\mid\mathbf{v}$, where

$$S_{\mathbf{u}\mid\mathbf{v}}=S_{\mathbf{u}}\cup S_{\mathbf{v}}
\quad\quad\leq^{e}_{\mathbf{u}\mid\mathbf{v}}=\leq^{e}_{\mathbf{u}}\cup\leq^{e}_{\mathbf{v}}
\quad\quad\leq^{c}_{\mathbf{u}\mid\mathbf{v}}=\leq^{c}_{\mathbf{u}}\cup\leq^{c}_{\mathbf{v}}\cup(S_{\mathbf{u}}\times S_{\mathbf{v}})
\quad\quad\lambda_{\mathbf{u}\mid\mathbf{v}}(x)=\begin{cases}
\lambda_{\mathbf{u}}\quad x\in S_{\mathbf{u}}\\
\lambda_{\mathbf{v}}\quad x\in S_{\mathbf{v}}
\end{cases}$$
\end{definition}

\begin{definition}[Pomsetc composition in concurrency]
Let $U,V\in\mathsf{Pomc}$ with $U=[\mathbf{u}]$ and $V=[\mathbf{v}]$. We write $U\between V$ for the concurrent composition of $U$ and $V$, which is the pomsetc represented by $\mathbf{u}\between\mathbf{v}$, where

$$S_{\mathbf{u}\between\mathbf{v}}=S_{\mathbf{u}}\cup S_{\mathbf{v}}
\quad\leq^{e}_{\mathbf{u}\between\mathbf{v}}=\leq^{e}_{\mathbf{u}}\cup\leq^{e}_{\mathbf{v}}
\quad\leq^{c}_{\mathbf{u}\between\mathbf{v}}=\leq^{c}_{\mathbf{u}}\cup\leq^{c}_{\mathbf{v}}
\textrm{ or }\leq^{c}_{\mathbf{u}\between\mathbf{v}}=\leq^{c}_{\mathbf{u}}\cup\leq^{c}_{\mathbf{v}}\cup(S_{\mathbf{u}}\times S_{\mathbf{v}})
\quad\lambda_{\mathbf{u}\between\mathbf{v}}(x)=\begin{cases}
\lambda_{\mathbf{u}}\quad x\in S_{\mathbf{u}}\\
\lambda_{\mathbf{v}}\quad x\in S_{\mathbf{v}}
\end{cases}$$
\end{definition}

\begin{definition}[Pomsetc composition in sequence]
Let $U,V\in\mathsf{Pomc}$ with $U=[\mathbf{u}]$ and $V=[\mathbf{v}]$. We write $U\cdot V$ for the sequential composition of $U$ and $V$, which is the pomsetc represented by $\mathbf{u}\cdot\mathbf{v}$, where

$$S_{\mathbf{u}\cdot\mathbf{v}}=S_{\mathbf{u}}\cup S_{\mathbf{v}}
\quad\quad\leq^{e}_{\mathbf{u}\cdot\mathbf{v}}=\leq^{e}_{\mathbf{u}}\cup\leq^{e}_{\mathbf{v}}\cup(S_{\mathbf{u}}\times S_{\mathbf{v}})
\quad\quad\leq^{c}_{\mathbf{u}\cdot\mathbf{v}}=\leq^{c}_{\mathbf{u}}\cup\leq^{c}_{\mathbf{v}}
\quad\quad\lambda_{\mathbf{u}\cdot\mathbf{v}}(x)=\begin{cases}
\lambda_{\mathbf{u}}\quad x\in S_{\mathbf{u}}\\
\lambda_{\mathbf{v}}\quad x\in S_{\mathbf{v}}
\end{cases}$$
\end{definition}

The following definitions and conclusions are coming from \cite{CKA7}, we retype them.

\begin{definition}[Pomset types]
Let $U\in\mathsf{Pom}$, $U$ is sequential (resp. parallel) if there exist non-empty pomsets $U_1$ and $U_2$ such that $U=U_1\cdot U_2$ (resp. $U=U_1\parallel U_2$).
\end{definition}

\begin{definition}[Factorization]
Let $U\in\mathsf{Pom}$. (1) When $U=U_1\cdot\cdots \cdot U_i\cdot\cdots \cdot U_n$ with each $U_i$ non-sequential and non-empty, the sequence $U_1,\cdots,U_i,\cdots,U_n$ is called a sequential factorization of $U$. (2) When $U=U_1\parallel\cdots \parallel U_i\parallel\cdots \parallel U_n$ with each $U_i$ non-parallel and non-empty, the multiset $\mset{U_1,\cdots,U_i,\cdots,U_n}$ is called a parallel factorization of $U$.
\end{definition}

\begin{lemma}[Factorization]\label{LemmaFactorization}
Sequential and parallel factorizations exist uniquely.
\end{lemma}

On the proof of Lemma \ref{LemmaFactorization}, please refer to \cite{CKA7} for details.

\begin{lemma}
For $U\in\mathsf{Pomc}$, then the following two conclusions hold:

\begin{enumerate}
  \item $U$ is either sequential or parallel, and there are not other types in $U$.
  \item Sequential and parallel factorizations exist in $U$ uniquely.
\end{enumerate}
\end{lemma}

\subsection{Series-Communication-Parallelism}\label{scp}

\begin{definition}[Series-parallel pomset]
The set of series-parallel pomset, or sp-pomsets denoted $\mathsf{SP}$, is the smallest set satisfying the following rules:

$$\frac{}{1\in\mathsf{SP}}
\quad\frac{a\in\Sigma}{a\in\mathsf{SP}}
\quad\frac{U,V\in\mathsf{SP}}{U\cdot V\in\mathsf{SP}}
\quad\frac{U,V\in\mathsf{SP}}{U\parallel V\in\mathsf{SP}}$$
\end{definition}

\begin{definition}[Series-communication-parallel pomsetc]
The set of series-communication-parallel pomsetcs, or scp-pomsetcs denoted $\mathsf{SCP}$, is the smallest set satisfying the following rules:

$$\frac{}{1\in\mathsf{SCP}}
\quad\frac{a\in\Sigma}{a\in\mathsf{SCP}}
\quad\frac{U,V\in\mathsf{SCP}}{U\cdot V\in\mathsf{SCP}}
\quad\frac{U,V\in\mathsf{SCP}}{U\parallel V\in\mathsf{SCP}}
\quad\frac{U,V\in\mathsf{SCP}}{U\mid V\in\mathsf{SCP}}
\quad\frac{U,V\in\mathsf{SCP}}{U\between V\in\mathsf{SCP}}$$
\end{definition}

\begin{definition}[N-shape1]\label{nshape1s}
Let $U=[\mathbf{u}]$ be a pomset. An N-shape1 in $U$ is a quadruple $u_0,u_1,u_2,u_3\in S_{\mathbf{u}}$ of distinct points such that $u_0\leq_{\mathbf{u}}u_1$, $u_2\leq_{\mathbf{u}}u_3$ and $u_0\leq_{\mathbf{u}}u_3$ and their exists no other relations among them. A pomset $U$ is N-free if it has no N-shape1s.
\end{definition}

\begin{definition}[N-shape2]\label{nshape2s}
Let $U=[\mathbf{u}]$ be a pomsetc. An N-shape2 in $U$ is a quadruple $u_0,u_1,u_2,u_3\in S_{\mathbf{u}}$ of distinct points such that $u_0\leq^{e}_{\mathbf{u}}u_1$, $u_2\leq^{e}_{\mathbf{u}}u_3$ and $u_0\leq^{c}_{\mathbf{u}}u_3$ and their exists no other relations among them.
\end{definition}

The definition of N-shape2 in Definition \ref{nshape2s} is based on the assumption that partial orders (causalities) among different parallel branches are all communications.

\begin{theorem}[N-shape1]
A pomset is series-parallel if and only if it is N-shape1-free in Definition \ref{nshape1s}.
\end{theorem}

\begin{theorem}[N-shape2]
A pomsetc is series-communication-parallel if and only if it only contains N-shape2s in Definition \ref{nshape2s}.
\end{theorem}

\begin{theorem}[Series-communication-parallelism to series-parallelism]
A series-communication-parallel pomsetc $U$ can be translated into a series-parallel pomset $U'$ if all the communications are all synchronous, i.e., for all $u_i\leq^{c} u_j$ in $U$, $u_i,u_j$ can merge into a single $\rho(u_i,u_j)$ in $U'$, where $\rho(u_i,u_j)$ is the communication function between $u_i$ and $u_j$.
\end{theorem}

\begin{definition}[Subsumption]
Let $U=[\mathbf{u}]\in\mathsf{Pomc}$ and $V=[\mathbf{v}]\in\mathsf{Pomc}$, $U$ is subsumed by $V$, denoted $U\sqsubseteq V$, if there exists a labelled poset isomorphism from $\mathbf{v}$ to $\mathbf{u}$ that is also a bijection.
\end{definition}

The following conclusions are natural extensions from $\mathsf{Pom}$ to $\mathsf{Pomc}$, the new cases are concurrency composition $\between$ and communication composition $\mid$ and can be proven similarly to the case of parallel composition $\parallel$.

\begin{lemma}
Let $U,V\in\mathsf{Pomc}$ with $U\sqsubseteq V$ or $V\sqsubseteq U$. If $U$ is empty, then $U=V$; if $U=a$ for some $a\in\Sigma$, then $V=a$.
\end{lemma}

\begin{lemma}[Separation]
Let $U,V\in\mathsf{Pomc}$ with $U\sqsubseteq V$.

\begin{enumerate}
  \item If $V=V_0\cdot V_1$, then $U=U_0\cdot U_1$ such that $U_0\sqsubseteq V_0$ and $U_1\sqsubseteq V_1$.
  \item If $U=U_0\parallel U_1$, then $V=V_0\parallel V_1$ such that $U_0\sqsubseteq V_0$ and $U_1\sqsubseteq V_1$.
  \item If $U=U_0\mid U_1$, then $V=V_0\mid V_1$ such that $U_0\sqsubseteq V_0$ and $U_1\sqsubseteq V_1$.
  \item If $U=U_0\between U_1$, then $V=V_0\between V_1$ such that $U_0\sqsubseteq V_0$ and $U_1\sqsubseteq V_1$.
\end{enumerate}
\end{lemma}

\begin{lemma}[Interpolation]
Let $U,V,W,X\in\mathsf{Pomc}$ such that $U\cdot V\sqsubseteq W\parallel X$, $U\cdot V\sqsubseteq W\mid X$ and $U\cdot V\sqsubseteq W\between X$, then, there exist pomsetcs $W_0,W_1,X_0,X_1$ such that the following hold:

$$W_0\cdot W_1\sqsubseteq W\quad X_0\cdot X_1\sqsubseteq X\quad U\sqsubseteq W_0\parallel X_0\quad V\sqsubseteq W_1\parallel X_1$$
$$U\sqsubseteq W_0\mid X_0\quad V\sqsubseteq W_1\mid X_1\quad U\sqsubseteq W_0\between X_0\quad V\sqsubseteq W_1\between X_1$$
\end{lemma}

\subsection{Pomsetc Language}\label{pl}

\begin{definition}[Pomsetc language]
A pomsetc language is a set of pomsetcs. A pomsetc language made up of scp-pomsetcs is referred to as series-communication-parallel language, or scp-language for short.
\end{definition}

\begin{definition}[Pomsetc language composition]
Let $L,K\subseteq\mathsf{Pomc}$. Then we define the following compositions.

$$L+K=L\cup K\quad L\cdot K=\{U\cdot V:U\in L,V\in K\}\quad L\parallel K=\{U\parallel V:U\in L,V\in K\}$$
$$L\mid K=\{U\mid V:U\in L,V\in K\}\quad L\between K=\{U\between V:U\in L,V\in K\}$$
$$L^*=\bigcup_{n\in\mathbb{N}}L^n\textrm{ where }L^0=\{1\}\textrm{ and }L^{n+1}=L^n\cdot L$$
$$L^{\langle *\rangle}=\bigcup_{n\in\mathbb{N}}L^{\langle n\rangle}\textrm{ where }L^{\langle 0\rangle}=\{1\}\textrm{ and }L^{\langle{n+1}\rangle}=L^{\langle n\rangle}\parallel L$$
\end{definition}

\begin{definition}[Pomsetc language substitution]
Let $\Delta$ be an alphabet. A substitution is a function $\zeta:\Sigma\rightarrow 2^{\mathsf{Pomc}(\Delta)}$ and lift to $\zeta:\mathsf{Pomc}(\Delta)\rightarrow 2^{\mathsf{Pomc}(\Delta)}$:

$$\zeta(1)=\{1\}\quad \zeta(U\cdot V)=\zeta(U)\cdot\zeta(V)\quad\zeta(U\parallel V)=\zeta(U)\parallel\zeta(V)$$
$$\zeta(U\mid V)=\zeta(U)\mid\zeta(V)\quad \zeta(U\between V)=\zeta(U)\between\zeta(V)$$
\end{definition} 

\subsection{Truly Concurrent Bisimilarities Based on Expressions}\label{tcbbe}

\begin{definition}[Configuration]
Let $x\in\mathcal{T}$ be an expression. A (finite) configuration in $x$ is a (finite) sub-pomset of $x$, $\mathbf{C}\subseteq x$. The set of finite configurations of $x$ is denoted by $\mathcal{C}(x)$.
\end{definition}

\begin{definition}[Pomset transitions and step]
Let $x\in\mathcal{T}$ be an expression and let $\mathbf{C}\in\mathcal{C}(x)$, and $\emptyset\neq X\subseteq \Sigma^*$, if $\mathbf{C}\cap X=\emptyset$ and $\mathbf{C}'=\mathbf{C}\cup X\in\mathcal{C}(x)$, then $\mathbf{C}\xrightarrow{X} \mathbf{C}'$ is called a pomset transition from $\mathbf{C}$ to $\mathbf{C}'$. When the events in $X$ are pairwise concurrent, we say that $\mathbf{C}\xrightarrow{X}\mathbf{C}'$ is a step.
\end{definition}

\begin{definition}[Pomset, step bisimulation]\label{PSB}
Let $x,y\in\mathcal{T}$ be expressions. A pomset bisimulation is a relation $R\subseteq\mathcal{C}(x)\times\mathcal{C}(y)$, such that if $(\mathbf{C}_1,\mathbf{C}_2)\in R$, and $\mathbf{C}_1\xrightarrow{X_1}\mathbf{C}_1'$ then $\mathbf{C}_2\xrightarrow{X_2}\mathbf{C}_2'$, with $X_1\subseteq \Sigma^*$, $X_2\subseteq \Sigma^*$, $X_1\sim X_2$ and $(\mathbf{C}_1',\mathbf{C}_2')\in R$, and vice-versa. We say that $x$, $y$ are pomset bisimilar, written $x\sim_p y$, if there exists a pomset bisimulation $R$, such that $(\emptyset,\emptyset)\in R$. By replacing pomset transitions with steps, we can get the definition of step bisimulation. When $x$ and $y$ are step bisimilar, we write $x\sim_s y$.
\end{definition}

\begin{definition}[Pomset, step simulation]\label{PSS}
Let $x,y\in\mathcal{T}$ be expressions. A pomset simulation is a relation $R\subseteq\mathcal{C}(x)\times\mathcal{C}(y)$, such that if $(\mathbf{C}_1,\mathbf{C}_2)\in R$, and $\mathbf{C}_1\xrightarrow{X_1}\mathbf{C}_1'$ then $\mathbf{C}_2\xrightarrow{X_2}\mathbf{C}_2'$, with $X_1\subseteq \Sigma^*$, $X_2\subseteq \Sigma^*$, $X_1\sim X_2$ and $(\mathbf{C}_1',\mathbf{C}_2')\in R$. We say that $x$, $y$ are pomset similar, written $x\lesssim_p y$, if there exists a pomset simulation $R$, such that $(\emptyset,\emptyset)\in R$. By replacing pomset transitions with steps, we can get the definition of step simulation. When $x$ and $y$ are step similar, we write $x\lesssim_s y$.
\end{definition}

\begin{definition}[Posetal product]
Given two expressions $x,y\in\mathcal{T}$, the posetal product of their configurations, denoted $\mathcal{C}(x)\overline{\times}\mathcal{C}(y)$, is defined as

$$\{(\mathbf{C}_1,f,\mathbf{C}_2)|\mathbf{C}_1\in\mathcal{C}(x),\mathbf{C}_2\in\mathcal{C}(y),f:\mathbf{C}_1\rightarrow \mathbf{C}_2 \textrm{ isomorphism}\}$$

A subset $R\subseteq\mathcal{C}(x)\overline{\times}\mathcal{C}(y)$ is called a posetal relation. We say that $R$ is downward closed when for any $(\mathbf{C}_1,f,\mathbf{C}_2),(\mathbf{C}_1',f',\mathbf{C}_2')\in \mathcal{C}(x)\overline{\times}\mathcal{C}(y)$, if $(\mathbf{C}_1,f,\mathbf{C}_2)\subseteq (\mathbf{C}_1',f',\mathbf{C}_2')$ pointwise and $(\mathbf{C}_1',f',\mathbf{C}_2')\in R$, then $(\mathbf{C}_1,f,\mathbf{C}_2)\in R$.

For $f:X_1\rightarrow X_2$, we define $f[a_1\mapsto a_2]:X_1\cup\{a_1\}\rightarrow X_2\cup\{a_2\}$, $z\in X_1\cup\{a_1\}$,(1)$f[a_1\mapsto a_2](z)=
a_2$,if $z=a_1$;(2)$f[a_1\mapsto a_2](z)=f(z)$, otherwise. Where $X_1\subseteq x$, $X_2\subseteq y$, $a_1\in x$, $a_2\in y$.
\end{definition}

\begin{definition}[(Hereditary) history-preserving bisimulation]\label{HHPB}
A history-preserving (hp-) bisimulation is a posetal relation $R\subseteq\mathcal{C}(x)\overline{\times}\mathcal{C}(y)$ such that if $(\mathbf{C}_1,f,\mathbf{C}_2)\in R$, and $\mathbf{C}_1\xrightarrow{a_1} \mathbf{C}_1'$, then $\mathbf{C}_2\xrightarrow{a_2} \mathbf{C}_2'$, with $(\mathbf{C}_1',f[a_1\mapsto a_2],\mathbf{C}_2')\in R$, and vice-versa. $x,y$ are history-preserving (hp-)bisimilar and are written $x\sim_{hp}y$ if there exists a hp-bisimulation $R$ such that $(\emptyset,\emptyset,\emptyset)\in R$.

A hereditary history-preserving (hhp-)bisimulation is a downward closed hp-bisimulation. $x,y$ are hereditary history-preserving (hhp-)bisimilar and are written $x\sim_{hhp}y$.
\end{definition}

\begin{definition}[(Hereditary) history-preserving simulation]\label{HHPS}
A history-preserving (hp-) simulation is a posetal relation $R\subseteq\mathcal{C}(x)\overline{\times}\mathcal{C}(y)$ such that if $(\mathbf{C}_1,f,\mathbf{C}_2)\in R$, and $\mathbf{C}_1\xrightarrow{a_1} \mathbf{C}_1'$, then $\mathbf{C}_2\xrightarrow{a_2} \mathbf{C}_2'$, with $(\mathbf{C}_1',f[a_1\mapsto a_2],\mathbf{C}_2')\in R$. $x,y$ are history-preserving (hp-)similar and are written $x\lesssim_{hp}y$ if there exists a hp-simulation $R$ such that $(\emptyset,\emptyset,\emptyset)\in R$.

A hereditary history-preserving (hhp-)simulation is a downward closed hp-simulation. $x,y$ are hereditary history-preserving (hhp-)similar and are written $x\lesssim_{hhp}y$.
\end{definition}

\subsection{Series-Communication Rational Expressions}\label{scre}

We define the syntax and semantics of the series-communication rational (scr-) expressions.

\begin{definition}[Syntax of scr-expressions]
We define the set of scr-expressions $\mathcal{T}_{SCR}$ as follows.

$$\mathcal{T}_{SCR}\ni x,y::=0|1|a,b\in\Sigma|\rho(a,b)|x+y|x\cdot y|x^*|x\parallel y|x\mid y|x\between y$$
\end{definition}

In the definition of scr-expressions, the atomic actions include actions in $a,b\in\Sigma$, the constant $0$ denoted inaction without any behaviour, the constant $1$ denoted empty action which terminates immediately and successfully, and also the communication action $\rho(a,b)$. The operator $+$ is the alternative composition, i.e., the program $x+y$ either executes $x$ or $y$ alternatively. The operator $\cdot$ is the sequential composition, i.e., the program $x\cdot y$ firstly executes $x$ followed $y$. The Kleene star $x^*$ can execute $x$ for some number of times sequentially (maybe zero). The operator $\parallel$ is the parallel composition, i.e., the program $x\parallel y$ executes $x$ and $y$ in parallel. The program $x\mid y$ executes with synchronous communications. The program $x\between y$ means $x$ and $y$ execute concurrently, i.e., in parallel but may be with unstructured communications.

\subsubsection{Algebra modulo Language Equivalence}

\begin{definition}[Language semantics of scr-expressions]
Then we define the interpretation of scr-expressions $\sembrack{-}_{SCR}:\mathcal{T}_{SCR}\rightarrow 2^{\mathsf{SCP}}$ inductively as Table \ref{LSSCRE} shows.
\end{definition}

\begin{center}
    \begin{table}
        $$\sembrack{0}_{SCR}=\emptyset \quad \sembrack{a}_{SCR}=\{a\} \quad \sembrack{x\cdot y}_{SCR}=\sembrack{x}_{SCR}\cdot \sembrack{y}_{SCR}$$
        $$\sembrack{1}_{SCR}=\{1\} \quad \sembrack{x+y}_{SCR}=\sembrack{x}_{SCR}+\sembrack{y}_{SCR} \quad\sembrack{x^*}_{SCR}=\sembrack{x}^*_{SCR}$$
        $$\sembrack{x\parallel y}_{SCR}=\sembrack{x}_{SCR}\parallel\sembrack{y}_{SCR} \quad\sembrack{x\mid y}_{SCR}=\sembrack{x}_{SCR}\mid\sembrack{y}_{SCR}$$
        $$\sembrack{x\between y}_{SCR}=\sembrack{x}_{SCR}\between\sembrack{y}_{SCR}$$
        \caption{Language semantics of scr-expressions}
        \label{LSSCRE}
    \end{table}
\end{center}


We define a Bi-Kleene algebra with communication (BKAC) as a tuple $(\Sigma,+,\cdot,^*,\parallel,\between,\mid,0,1)$, where $\Sigma$ is a set, $^*$ is unary, $+$, $\cdot$, $\parallel$, $\between$ and $\mid$ are binary operators, and $0$ and $1$ are constants, which satisfies the axioms in Table \ref{AxiomsForBKACL} for all $x,y,z,h\in \mathcal{T}_{SCR}$ and $a,b,a_0,a_1,a_2,a_3\in\Sigma$, where $x\leqq y$ means $x+y=y$.

\begin{center}
    \begin{table}
        \begin{tabular}{@{}ll@{}}
            \hline No. &Axiom\\
            $A1$ & $x+y=y+z$\\
            $A2$ & $x+(y+z)=(x+y)+z$\\
            $A3$ & $x+x=x$\\
            $A4$ & $(x+y)\cdot z=x\cdot z+y\cdot z$\\
            $A5$ & $x\cdot(y+z)=x\cdot y+x\cdot z$\\
            $A6$ & $x\cdot(y\cdot z)=(x\cdot y)\cdot z$\\
            $A7$ & $x+0=x$\\
            $A8$ & $0\cdot x=0$\\
            $A9$ & $x\cdot 0=0$\\
            $A10$ & $x\cdot 1=x$\\
            $A11$ & $1\cdot x=x$\\
            $P1$ & $x\between y=x\parallel y+x\mid y$\\
            $P2$ & $x\parallel y=y\parallel x$\\
            $P3$ & $x\parallel(y\parallel z)=(x\parallel y)\parallel z$\\
            $P4$ & $(x+y)\parallel z=x\parallel z+y\parallel z$\\
            $P5$ & $x\parallel(y+z)=x\parallel y+x\parallel z$\\
            $P6$ & $x\parallel 0=0$\\
            $P7$ & $0\parallel x=0$\\
            $P8$ & $x\parallel 1=x$\\
            $P9$ & $1\parallel x=x$\\
            $C1$ & $x\mid y=y\mid x$\\
            $C2$ & $(x+y)\mid z=x\mid z+y\mid z$\\
            $C3$ & $x\mid(y+z)=x\mid y+x\mid z$\\
            $C4$ & $x\mid 0=0$\\
            $C5$ & $0\mid x=0$\\
            $C6$ & $x\mid 1=0$\\
            $C7$ & $1\mid x=0$\\
            $A12$ & $1+x\cdot x^*=x^*$\\
            $A13$ & $1+x^*\cdot x=x^*$\\
            $A14$ & $x+y\cdot z\leqq z\Rightarrow y^*\cdot x\leqq z$\\
            $A15$ & $x+y\cdot z\leqq y\Rightarrow x\cdot z^*\leqq y$\\
        \end{tabular}
        \caption{Axioms of BKAC modulo language equivalence}
        \label{AxiomsForBKACL}
    \end{table}
\end{center}

Since language equivalence is a congruence w.r.t. the operators of BKAC, we can only check the soundness of each axiom according to the definition of semantics of scr-expressions. And also by use of communication merge, the scr-expressions are been transformed into the so-called series-parallel ones \cite{CKA7} \cite{CKA8} free of N-shapes. Then we can get the following soundness and completeness theorem with reference to \cite{CKA8}, the new cases are the operators $\between$ and $\mid$, which can be added to the proof similarly to the operator $\parallel$.

\begin{theorem}[Soundness and completeness of BKAC modulo language equivalence]
For all $x,y\in\mathcal{T}_{SCR}$, $x= y$ if and only if $\sembrack{x}_{SCR}=\sembrack{y}_{SCR}$.
\end{theorem}

\begin{definition}
We define $\mathcal{F}_{SCR}$ as smallest subset of $\mathcal{T}_{SCR}$ satisfying the following rules:

$$\frac{}{1\in\mathcal{F}_{SCR}}\quad \frac{x\in\mathcal{F}_{SCR}\quad y\in\mathcal{T}_{SCR}}{x+y\in\mathcal{F}_{SCR}\quad y+x\in\mathcal{F}_{SCR}} \quad\frac{x\in\mathcal{T}_{SCR}}{x^*\in\mathcal{F}_{SCR}}$$
$$\frac{x\in\mathcal{F}_{SCR}\quad y\in\mathcal{F}_{SCR}}{x\cdot y\in\mathcal{F}_{SCR}\quad x\between y\in\mathcal{F}_{SCR}\quad x\parallel y\in\mathcal{F}_{SCR}\quad x\mid y\in\mathcal{F}_{SCR}}$$
\end{definition}

\begin{theorem}
Let $x, y\in\mathcal{T}_{SCR}$. It is decidable whether $\sembrack{x}_{SCR}=\sembrack{y}_{SCR}$.
\end{theorem}

\begin{lemma}
Let $x\in\mathcal{T}_{SCR}$, $x\in\mathcal{F}_{SCR}$ if and only if $1\in\sembrack{x}$, which holds precisely when $1\leqq x$.
\end{lemma}

\subsubsection{Algebra modulo Bisimilarities}

\begin{definition}[Operational semantics of scr-expressions]
Let the symbol $\downarrow$ denote the successful termination predicate. Then we give the TSS of scr-expressions as Table \ref{OSSCRE}, where $a,b,c,\cdots\in \Sigma$, $x,y,x',y'\in\mathcal{T}_{SCR}$.
\end{definition}

\begin{center}
    \begin{table}
        $$\frac{}{1\downarrow}\quad\frac{}{a\xrightarrow{a}1}$$
        $$\frac{x\downarrow}{(x+y)\downarrow}\quad\frac{y\downarrow}{(x+y)\downarrow}\quad\frac{x\xrightarrow{a}x'}{x+y\xrightarrow{a}x'}\quad\frac{y\xrightarrow{b}y'}{x+y\xrightarrow{b}y'}$$
        $$\frac{x\downarrow\quad y\downarrow}{(x\cdot y)\downarrow} \quad\frac{x\xrightarrow{a}x'}{x\cdot y\xrightarrow{a}x'\cdot y} \quad\frac{x\downarrow\quad y\xrightarrow{b}y'}{x\cdot y\xrightarrow{b}y'}$$
        $$\frac{x\downarrow\quad y\downarrow}{(x\between y)\downarrow} \quad\frac{x\xrightarrow{a}x'\quad y\xrightarrow{b}y'}{x\between y\xrightarrow{\mset{a,b}}x'\between y'}
        \quad\frac{x\xrightarrow{a}x'\quad y\xrightarrow{b}y'}{x\between y\xrightarrow{\rho(a,b)}x'\between y'}$$
        $$\frac{x\downarrow\quad y\downarrow}{(x\parallel y)\downarrow} \quad\frac{x\xrightarrow{a}x'\quad y\xrightarrow{b}y'}{x\parallel y\xrightarrow{\mset{a,b}}x'\between y'}$$
        $$\frac{x\downarrow\quad y\downarrow}{(x\mid y)\downarrow} \quad\frac{x\xrightarrow{a}x'\quad y\xrightarrow{b}y'}{x\mid y\xrightarrow{\rho(a,b)}x'\between y'}$$
        $$\frac{x\downarrow}{(x^*)\downarrow} \quad\frac{x\xrightarrow{a}x'}{x^*\xrightarrow{a}x'\cdot x^*}$$
        \caption{Operational semantics of scr-expressions}
        \label{OSSCRE}
    \end{table}
\end{center}

Note that there is no any transition rules related to the constant $0$. Then the axiomatic system of BKAC modulo pomset, step and hp-bisimilarities is shown in Table \ref{AxiomsForBKACB1}.

\begin{center}
    \begin{table}
        \begin{tabular}{@{}ll@{}}
            \hline No. &Axiom\\
            $A1$ & $x+y=y+z$\\
            $A2$ & $x+(y+z)=(x+y)+z$\\
            $A3$ & $x+x=x$\\
            $A4$ & $(x+y)\cdot z=x\cdot z+y\cdot z$\\
            $A5$ & $x\cdot(y\cdot z)=(x\cdot y)\cdot z$\\
            $A6$ & $x+0=x$\\
            $A7$ & $0\cdot x=0$\\
            $A8$ & $x\cdot 1=x$\\
            $A9$ & $1\cdot x=x$\\
            $P1$ & $x\between y=x\parallel y+x\mid y$\\
            $P2$ & $x\parallel y=y\parallel x$\\
            $P3$ & $x\parallel(y\parallel z)=(x\parallel y)\parallel z$\\
            $P4$ & $(x+y)\parallel z=x\parallel z+y\parallel z$\\
            $P5$ & $x\parallel(y+z)=x\parallel y+x\parallel z$\\
            $P6$ & $x\parallel 0=0$\\
            $P7$ & $0\parallel x=0$\\
            $P8$ & $x\parallel 1=x$\\
            $P9$ & $1\parallel x=x$\\
            $C1$ & $x\mid y=y\mid x$\\
            $C2$ & $(x+y)\mid z=x\mid z+y\mid z$\\
            $C3$ & $x\mid(y+z)=x\mid y+x\mid z$\\
            $C4$ & $x\mid 0=0$\\
            $C5$ & $0\mid x=0$\\
            $C6$ & $x\mid 1=0$\\
            $C7$ & $1\mid x=0$\\
            $A10$ & $1+x\cdot x^*=x^*$\\
            $A11$ & $(1+x)^*=x^*$\\
            $A12$ & $x+y\cdot z\leqq z\Rightarrow y^*\cdot x\leqq z$\\
            $A13$ & $x+y\cdot z\leqq y\Rightarrow x\cdot z^*\leqq y$\\
        \end{tabular}
        \caption{Axioms of BKAC modulo pomset, step and hp-bisimilarities}
        \label{AxiomsForBKACB1}
    \end{table}
\end{center}

Note that there are two significant differences between the axiomatic systems of BKAC modulo language equivalence and bisimilaties, $x\cdot 0=0$ and $x\cdot(y+z)=x\cdot y+x\cdot z$ of BKAC do not hold modulo bisimilarities.

Since pomset, step and hp-bisimilarities are all congruences w.r.t. the operators $\cdot$, $+$, $^*$, $\between$, $\parallel$ and $\mid$, and pomset, step and hp-similarities are all precongruences w.r.t. the operators $\cdot$, $+$, $^*$, $\between$, $\parallel$ and $\mid$, we can only check the soundness of each axiom according to the definition of TSS of scr-expressions in Table \ref{OSSCRE}.

\begin{theorem}[Soundness of BKAC modulo pomset (bi)similarity]
BKAC is sound modulo pomset (bi)similarity w.r.t. scr-expressions.
\end{theorem}

\begin{theorem}[Soundness of BKAC modulo step (bi)similarity]
BKAC is sound modulo step (bi)similarity w.r.t. scr-expressions.
\end{theorem}

\begin{theorem}[Soundness of BKAC modulo hp-(bi)similarity]
BKAC is sound modulo hp-(bi)similarity w.r.t. scr-expressions.
\end{theorem}

For hhp-bisimilarity, an auxiliary binary operator called left-parallelism denoted $\leftmerge$ would be added into the syntax of $\mathcal{T}_{SCR}$. The following transition rules of $\leftmerge$ should be added into the operational semantics of scr-expressions.

$$\frac{x\downarrow\quad y\downarrow}{(x\leftmerge y)\downarrow} \quad\frac{x\xrightarrow{a}x'\quad y\xrightarrow{b}y'\quad a\leq b}{x\leftmerge y\xrightarrow{\mset{a,b}}x'\between y'}$$

Then the axiomatic system of BKAC modulo hhp-bisimilarity is shown in Table \ref{AxiomsForBKACB2}.

Note that, the left-parallelism operator $\leftmerge$ is unnecessary to be added into the language semantics, pomset bisimilarity, step bisimilarity and hp-bisimilarity semantics.

\begin{center}
    \begin{table}
        \begin{tabular}{@{}ll@{}}
            \hline No. &Axiom\\
            $A1$ & $x+y=y+z$\\
            $A2$ & $x+(y+z)=(x+y)+z$\\
            $A3$ & $x+x=x$\\
            $A4$ & $(x+y)\cdot z=x\cdot z+y\cdot z$\\
            $A5$ & $x\cdot(y\cdot z)=(x\cdot y)\cdot z$\\
            $A6$ & $x+0=x$\\
            $A7$ & $0\cdot x=0$\\
            $A8$ & $x\cdot 1=x$\\
            $A9$ & $1\cdot x=x$\\
            $P1$ & $x\between y=x\parallel y+x\mid y$\\
            $P2$ & $x\parallel y=y\parallel x$\\
            $P3$ & $x\parallel(y\parallel z)=(x\parallel y)\parallel z$\\
            $P4$ & $x\parallel y = x\leftmerge y+y\leftmerge x$\\
            $P5$ & $(x+y)\leftmerge z=x\leftmerge z+y\leftmerge z$\\
            $P6$ & $0\leftmerge x=0$\\
            $P7$ & $x\leftmerge 1=x$\\
            $P8$ & $1\leftmerge x=x$\\
            $C1$ & $x\mid y=y\mid x$\\
            $C2$ & $(x+y)\mid z=x\mid z+y\mid z$\\
            $C3$ & $x\mid(y+z)=x\mid y+x\mid z$\\
            $C4$ & $x\mid 0=0$\\
            $C5$ & $0\mid x=0$\\
            $C6$ & $x\mid 1=0$\\
            $C7$ & $1\mid x=0$\\
            $A10$ & $1+x\cdot x^*=x^*$\\
            $A11$ & $(1+x)^*=x^*$\\
            $A12$ & $x+y\cdot z\leqq z\Rightarrow y^*\cdot x\leqq z$\\
            $A13$ & $x+y\cdot z\leqq y\Rightarrow x\cdot z^*\leqq y$\\
        \end{tabular}
        \caption{Axioms of BKAC modulo hhp-bisimilarity}
        \label{AxiomsForBKACB2}
    \end{table}
\end{center}

Since hhp-bisimilarity is a congruences w.r.t. the operators $\cdot$, $+$, $^*$, $\between$, $\parallel$, $\leftmerge$ and $\mid$, and hhp-similarity is a precongruences w.r.t. the operators $\cdot$, $+$, $^*$, $\between$, $\parallel$, $\leftmerge$ and $\mid$, we can only check the soundness of each axiom according to the definition of TSS of scr-expressions in Table \ref{OSSCRE} and the additional transition rules of $\leftmerge$.

\begin{theorem}[Soundness of BKAC modulo hhp-(bi)similarity]
BKAC is sound modulo hhp-(bi)similarity w.r.t. scr-expressions.
\end{theorem}

\begin{lemma}
Let $x\in\mathcal{T}_{SCR}$, $x\in\mathcal{F}_{SCR}$ if and only if $1\in\sembrack{x}$, which holds precisely when $1\leqq x$ modulo pomset, step, hp- and hhp-similarities.
\end{lemma}

Then there are two questions: (R) the problem of recognizing whether a given process graph is bisimilar to one in the image of the process interpretation of a $\mathcal{T}_{SCR}$ expression, and (A) whether a natural adaptation of Salomaa’s complete proof system for language equivalence of $\mathcal{T}_{SCR}$ expressions is complete for bisimilarities of the process interpretation of $\mathcal{T}_{SCR}$ expressions. While (R) is decidable in principle, it is just a pomset extension to the problem of recognizing whether a given process graph is bisimilar to one in the image of the process interpretation of a star expression \cite{AP8}.

As mentioned in the section \ref{intro}, just very recently, Grabmayer \cite{MF6} claimed to have proven that Mil is complete w.r.t. a specific kind of process graphs called LLEE-1-charts which is equal to regular expressions, modulo the corresponding kind of bisimilarity called 1-bisimilarity. Based on this work, we believe that we can get the completeness conclusions based on the corresponding truly concurrent bisimilarities and let the proof of the completeness be open.

\begin{theorem}[Completeness of BKAC modulo pomset (bi)similarity]
BKAC is complete modulo pomset (bi)similarity w.r.t. scr-expressions.
\end{theorem}

\begin{theorem}[Completeness of BKAC modulo step (bi)similarity]
BKAC is complete modulo step (bi)similarity w.r.t. scr-expressions.
\end{theorem}

\begin{theorem}[Completeness of BKAC modulo hp-(bi)similarity]
BKAC is complete modulo hp-(bi)similarity w.r.t. scr-expressions.
\end{theorem}

\begin{theorem}[Completeness of BKAC modulo hhp-(bi)similarity]
BKAC is complete modulo hhp-(bi)similarity w.r.t. scr-expressions.
\end{theorem}

\begin{theorem}
Let $x, y\in\mathcal{T}_{SCR}$. It is decidable whether $x\sim_{p}y$.
\end{theorem}

\begin{theorem}
Let $x, y\in\mathcal{T}_{SCR}$. It is decidable whether $x\sim_{s}y$.
\end{theorem}

\begin{theorem}
Let $x, y\in\mathcal{T}_{SCR}$. It is decidable whether $x\sim_{hp}y$.
\end{theorem}

\begin{theorem}
Let $x, y\in\mathcal{T}_{SCR}$. It is decidable whether $x\sim_{hhp}y$.
\end{theorem}

\subsubsection{Series-Communication Rational Systems}

We have already defined five kinds of $=$ relations of BKAC modulo language equivalence, pomset bisimilarity, step bisimilarity, hp-bisimilarity, and hhp-bisimilarity and the corresponding preorders $\leqq$ in Tables \ref{AxiomsForBKACL}, \ref{AxiomsForBKACB1} and \ref{AxiomsForBKACB2}, we denote the corresponding $=$ and $\leqq$ as $=_1$ and $\leqq_1$, $=_2$ and $\leqq_2$, $=_3$ and $\leqq_3$, $=_4$ and $\leqq_4$, and $=_5$ and $\leqq_5$ respectively.

\begin{definition}[Series-communication rational system modulo language equivalence]
Let $Q$ be a finite set. A series-communication rational system modulo language equivalence on $Q$, or called scr-system modulo language equivalence, is a pair $\mathcal{S}=\langle M,b\rangle$, where $M:Q^2\rightarrow\mathcal{T}_{SCR}$ and $b:Q\rightarrow\mathcal{T}_{SCR}$. Let $=_1$ be a BKAC language equivalence on $\mathcal{T}_{SCR}(\Delta)$ with $\Sigma\subseteq\Delta$ and $x\in\mathcal{T}_{SCR}$. We call $s:Q\rightarrow\mathcal{T}_{SCR}(\Delta)$ a $\langle =_1,x\rangle$-solution to $\mathcal{S}$ if for $q\in Q$:

$$b(q)\cdot x+\sum_{q'\in Q}M(q,q')\cdot s(q')\leqq_1 s(q)$$

Lastly, $s$ is the least $\langle =_1,x\rangle$-solution, if for every such solution $s'$ and every $q\in Q$, we have $s(q)\leqq_1 s'(q)$.
\end{definition}

\begin{definition}[Series-communication rational system modulo pomset bisimilarity]
Let $Q$ be a finite set. A series-communication rational system modulo pomset bisimilarity on $Q$, or called scr-system modulo posmet bisimilarity, is a pair $\mathcal{S}=\langle M,b\rangle$, where $M:Q^2\rightarrow\mathcal{T}_{SCR}$ and $b:Q\rightarrow\mathcal{T}_{SCR}$. Let $=_2$ be a BKAC pomset bisimilarity on $\mathcal{T}_{SCR}(\Delta)$ with $\Sigma\subseteq\Delta$ and $x\in\mathcal{T}_{SCR}$. We call $s:Q\rightarrow\mathcal{T}_{SCR}(\Delta)$ a $\langle =_2,x\rangle$-solution to $\mathcal{S}$ if for $q\in Q$:

$$b(q)\cdot x+\sum_{q'\in Q}M(q,q')\cdot s(q')\leqq_2 s(q)$$

Lastly, $s$ is the least $\langle =_2,x\rangle$-solution, if for every such solution $s'$ and every $q\in Q$, we have $s(q)\leqq_2 s'(q)$.
\end{definition}

\begin{definition}[Series-communication rational system modulo step bisimilarity]
Let $Q$ be a finite set. A series-communication rational system modulo step bisimilarity on $Q$, or called scr-system modulo step bisimilarity, is a pair $\mathcal{S}=\langle M,b\rangle$, where $M:Q^2\rightarrow\mathcal{T}_{SCR}$ and $b:Q\rightarrow\mathcal{T}_{SCR}$. Let $=_3$ be a BKAC step bisimilarity on $\mathcal{T}_{SCR}(\Delta)$ with $\Sigma\subseteq\Delta$ and $x\in\mathcal{T}_{SCR}$. We call $s:Q\rightarrow\mathcal{T}_{SCR}(\Delta)$ a $\langle =_3,x\rangle$-solution to $\mathcal{S}$ if for $q\in Q$:

$$b(q)\cdot x+\sum_{q'\in Q}M(q,q')\cdot s(q')\leqq_3 s(q)$$

Lastly, $s$ is the least $\langle =_3,x\rangle$-solution, if for every such solution $s'$ and every $q\in Q$, we have $s(q)\leqq_3 s'(q)$.
\end{definition}

\begin{definition}[Series-communication rational system modulo hp-bisimilarity]
Let $Q$ be a finite set. A series-communication rational system modulo hp-bisimilarity on $Q$, or called scr-system modulo hp-bisimilarity, is a pair $\mathcal{S}=\langle M,b\rangle$, where $M:Q^2\rightarrow\mathcal{T}_{SCR}$ and $b:Q\rightarrow\mathcal{T}_{SCR}$. Let $=_4$ be a BKAC hp-bisimilarity on $\mathcal{T}_{SCR}(\Delta)$ with $\Sigma\subseteq\Delta$ and $x\in\mathcal{T}_{SCR}$. We call $s:Q\rightarrow\mathcal{T}_{SCR}(\Delta)$ a $\langle =_4,x\rangle$-solution to $\mathcal{S}$ if for $q\in Q$:

$$b(q)\cdot x+\sum_{q'\in Q}M(q,q')\cdot s(q')\leqq_4 s(q)$$

Lastly, $s$ is the least $\langle =_4,x\rangle$-solution, if for every such solution $s'$ and every $q\in Q$, we have $s(q)\leqq_4 s'(q)$.
\end{definition}

\begin{definition}[Series-communication rational system modulo hhp-bisimilarity]
Let $Q$ be a finite set. A series-communication rational system modulo hhp-bisimilarity on $Q$, or called scr-system modulo hhp-bisimilarity, is a pair $\mathcal{S}=\langle M,b\rangle$, where $M:Q^2\rightarrow\mathcal{T}_{SCR}$ and $b:Q\rightarrow\mathcal{T}_{SCR}$. Let $=_5$ be a BKAC hhp-bisimilarity on $\mathcal{T}_{SCR}(\Delta)$ with $\Sigma\subseteq\Delta$ and $x\in\mathcal{T}_{SCR}$. We call $s:Q\rightarrow\mathcal{T}_{SCR}(\Delta)$ a $\langle =_5,x\rangle$-solution to $\mathcal{S}$ if for $q\in Q$:

$$b(q)\cdot x+\sum_{q'\in Q}M(q,q')\cdot s(q')\leqq_5 s(q)$$

Lastly, $s$ is the least $\langle =_5,x\rangle$-solution, if for every such solution $s'$ and every $q\in Q$, we have $s(q)\leqq_5 s'(q)$.
\end{definition}

For $z:Q\rightarrow\mathcal{T}_{SCR}$ and $z\in\mathcal{T}_{SCR}$, we write $z^x$ for the vector given by $z^x(q)=x(q)\cdot x$. $M$ can be regarded as a $Q$-indexed matrix, and $b$ and $s$ as $Q$-indexed vectors, i.e., for a scr-system on $Q$, the elements of $Q$ can be deemed as states of an operational description, and $M$ as the transitions relation, while $b$ as the halting behavior (the behavior that occurs when a state decides to halt execution).

\begin{lemma}
Let $\mathcal{S}=\langle M,b\rangle$ be an scr-system on $Q$ modulo language equivalence, and $=_1$ be a BKAC equivalence on $\mathcal{T}_{SCR}(\Delta)$ with $\Sigma\subseteq\Delta$ and $x\in\mathcal{T}_{SCR}$. For the least $\langle =_1,x\rangle$-solution $s$ to $\mathcal{S}$, we have for $q\in Q$:

$$b^x(q)+\sum_{q'\in Q}M(q,q')\cdot s(q')=_1 s(q)$$
\end{lemma}

\begin{lemma}
Let $\mathcal{S}=\langle M,b\rangle$ be an scr-system on $Q$ modulo pomset bisimilarity, and $=_2$ be a BKAC equivalence on $\mathcal{T}_{SCR}(\Delta)$ with $\Sigma\subseteq\Delta$ and $x\in\mathcal{T}_{SCR}$. For the least $\langle =_2,x\rangle$-solution $s$ to $\mathcal{S}$, we have for $q\in Q$:

$$b^x(q)+\sum_{q'\in Q}M(q,q')\cdot s(q')=_2 s(q)$$
\end{lemma}

\begin{lemma}
Let $\mathcal{S}=\langle M,b\rangle$ be an scr-system on $Q$ modulo step bisimilarity, and $=_3$ be a BKAC equivalence on $\mathcal{T}_{SCR}(\Delta)$ with $\Sigma\subseteq\Delta$ and $x\in\mathcal{T}_{SCR}$. For the least $\langle =_3,x\rangle$-solution $s$ to $\mathcal{S}$, we have for $q\in Q$:

$$b^x(q)+\sum_{q'\in Q}M(q,q')\cdot s(q')=_3 s(q)$$
\end{lemma}

\begin{lemma}
Let $\mathcal{S}=\langle M,b\rangle$ be an scr-system on $Q$ modulo hp-bisimilarity, and $=_4$ be a BKAC equivalence on $\mathcal{T}_{SCR}(\Delta)$ with $\Sigma\subseteq\Delta$ and $x\in\mathcal{T}_{SCR}$. For the least $\langle =_4,x\rangle$-solution $s$ to $\mathcal{S}$, we have for $q\in Q$:

$$b^x(q)+\sum_{q'\in Q}M(q,q')\cdot s(q')=_4 s(q)$$
\end{lemma}

\begin{lemma}
Let $\mathcal{S}=\langle M,b\rangle$ be an scr-system on $Q$ modulo hhp-bisimilarity, and $=_5$ be a BKAC equivalence on $\mathcal{T}_{SCR}(\Delta)$ with $\Sigma\subseteq\Delta$ and $x\in\mathcal{T}_{SCR}$. For the least $\langle =_5,x\rangle$-solution $s$ to $\mathcal{S}$, we have for $q\in Q$:

$$b^x(q)+\sum_{q'\in Q}M(q,q')\cdot s(q')=_5 s(q)$$
\end{lemma}

\begin{theorem}\label{ls1}
Let $\mathcal{S}=\langle M,b\rangle$ be an scr-system on $Q$ modulo language equivalence. We can construct an $s:Q\rightarrow \mathcal{T}_{SCR}$ such that, for any BKAC equivalence $=_1$ on $\mathcal{T}_{SCR}(\Delta)$ with $\Sigma\subseteq\Delta$ and any $x\in\mathcal{T}_{SCR}$, the $Q$-vector $s^x:Q\rightarrow\mathcal{T}_{SCR}$ is the least $\langle =_1,x\rangle$-solution to $\mathcal{S}$; we call such an $s$ the least solution to $\mathcal{S}$.
\end{theorem}

\begin{theorem}\label{ls2}
Let $\mathcal{S}=\langle M,b\rangle$ be an scr-system on $Q$ modulo pomset bisimilarity. We can construct an $s:Q\rightarrow \mathcal{T}_{SCR}$ such that, for any BKAC equivalence $=_2$ on $\mathcal{T}_{SCR}(\Delta)$ with $\Sigma\subseteq\Delta$ and any $x\in\mathcal{T}_{SCR}$, the $Q$-vector $s^x:Q\rightarrow\mathcal{T}_{SCR}$ is the least $\langle =_2,x\rangle$-solution to $\mathcal{S}$; we call such an $s$ the least solution to $\mathcal{S}$.
\end{theorem}

\begin{theorem}\label{ls3}
Let $\mathcal{S}=\langle M,b\rangle$ be an scr-system on $Q$ modulo step bisimilarity. We can construct an $s:Q\rightarrow \mathcal{T}_{SCR}$ such that, for any BKAC equivalence $=_3$ on $\mathcal{T}_{SCR}(\Delta)$ with $\Sigma\subseteq\Delta$ and any $x\in\mathcal{T}_{SCR}$, the $Q$-vector $s^x:Q\rightarrow\mathcal{T}_{SCR}$ is the least $\langle =_3,x\rangle$-solution to $\mathcal{S}$; we call such an $s$ the least solution to $\mathcal{S}$.
\end{theorem}

\begin{theorem}\label{ls4}
Let $\mathcal{S}=\langle M,b\rangle$ be an scr-system on $Q$ modulo hp-bisimilarity. We can construct an $s:Q\rightarrow \mathcal{T}_{SCR}$ such that, for any BKAC equivalence $=_4$ on $\mathcal{T}_{SCR}(\Delta)$ with $\Sigma\subseteq\Delta$ and any $x\in\mathcal{T}_{SCR}$, the $Q$-vector $s^x:Q\rightarrow\mathcal{T}_{SCR}$ is the least $\langle =_4,x\rangle$-solution to $\mathcal{S}$; we call such an $s$ the least solution to $\mathcal{S}$.
\end{theorem}

\begin{theorem}\label{ls5}
Let $\mathcal{S}=\langle M,b\rangle$ be an scr-system on $Q$ modulo hhp-bisimilarity. We can construct an $s:Q\rightarrow \mathcal{T}_{SCR}$ such that, for any BKAC equivalence $=_5$ on $\mathcal{T}_{SCR}(\Delta)$ with $\Sigma\subseteq\Delta$ and any $x\in\mathcal{T}_{SCR}$, the $Q$-vector $s^x:Q\rightarrow\mathcal{T}_{SCR}$ is the least $\langle =_5,x\rangle$-solution to $\mathcal{S}$; we call such an $s$ the least solution to $\mathcal{S}$.
\end{theorem}

\subsubsection{More Operators}

We introduce more operators to the algebras modulo language equivalence and bisimilarities, including prefix, recursion, encapsulation, silent step and abstraction.

\subsubsubsection{Prefix}

\begin{definition}[Syntax of prefix-expressions]
We define the set of prefix-expressions $\mathcal{T}_{PRE}$ as follows.

$$\mathcal{T}_{PRE}\ni x,y::=0|1|a,b\in\Sigma|\rho(a,b)|x+y|a.x|(a\parallel b).x|x\parallel y|x\mid y|x\between y$$
\end{definition}

The definitions of $0,1,a,b,\rho(a,b),x+y,x\parallel y,x\mid y, x\between y$ are the same as usual, and the definition of $a.x$ is the prefix composition.

\begin{definition}[Language semantics of prefix-expressions]
We define the interpretation of prefix-expressions $\sembrack{-}_{PRE}:\mathcal{T}_{PRE}\rightarrow 2^{\mathsf{SCP}}$ inductively as Table \ref{SPREE} shows.
\end{definition}

\begin{center}
    \begin{table}
        $$\sembrack{0}_{PRE}=\emptyset \quad \sembrack{a}_{PRE}=\{a\} \quad \sembrack{a.x}_{PRE}=\{a\}\cdot \sembrack{x}_{PRE}$$
        $$\sembrack{1}_{PRE}=\{1\} \quad \sembrack{x+y}_{PRE}=\sembrack{x}_{PRE}+\sembrack{y}_{PRE}$$
        $$\sembrack{x\parallel y}_{PRE}=\sembrack{x}_{PRE}\parallel\sembrack{y}_{PRE} \quad\sembrack{x\mid y}_{PRE}=\sembrack{x}_{PRE}\mid\sembrack{y}_{PRE}$$
        $$\sembrack{x\between y}_{PRE}=\sembrack{x}_{PRE}\between\sembrack{y}_{PRE}$$
        \caption{Language semantics of prefix-expressions}
        \label{SPREE}
    \end{table}
\end{center}

We define a prefix algebra as a tuple $(\Sigma,+,.,\parallel,\between,\mid,0,1)$, where $\Sigma$ is an alphabet, $+$, $.$, $\parallel$, $\between$ and $\mid$ are binary operators, and $0$ and $1$ are constants, which satisfies the axioms in Table \ref{AxiomsForPrefixL} for all $x,y\in \mathcal{T}_{PRE}$ and $a,b\in\Sigma$

\begin{center}
    \begin{table}
        \begin{tabular}{@{}ll@{}}
            \hline No. &Axiom\\
            $A1$ & $x+y=y+z$\\
            $A2$ & $x+(y+z)=(x+y)+z$\\
            $A3$ & $x+x=x$\\
            $A4$ & $a.(x+y)=a.x+a.y$\\
            $A5$ & $x+0=x$\\
            $A6$ & $a.0=0$\\
            $A7$ & $a.1=a$\\
            $P1$ & $x\between y=x\parallel y+x\mid y$\\
            $P2$ & $x\parallel y=y\parallel x$\\
            $P3$ & $x\parallel(y\parallel z)=(x\parallel y)\parallel z$\\
            $P4$ & $(x+y)\parallel z=x\parallel z+y\parallel z$\\
            $P5$ & $x\parallel(y+z)=x\parallel y+x\parallel z$\\
            $P6$ & $x\parallel 0=0$\\
            $P7$ & $0\parallel x=0$\\
            $P8$ & $x\parallel 1=x$\\
            $P9$ & $1\parallel x=x$\\
            $C1$ & $x\mid y=y\mid x$\\
            $C2$ & $(x+y)\mid z=x\mid z+y\mid z$\\
            $C3$ & $x\mid(y+z)=x\mid y+x\mid z$\\
            $C4$ & $x\mid 0=0$\\
            $C5$ & $0\mid x=0$\\
            $C6$ & $x\mid 1=0$\\
            $C7$ & $1\mid x=0$\\
        \end{tabular}
        \caption{Axioms of prefix algebra modulo language equivalence}
        \label{AxiomsForPrefixL}
    \end{table}
\end{center}

Since language equivalence is a congruence w.r.t. operators of the prefix algebra, we can only check the soundness of each axiom according to the definition of semantics of prefix-expressions. And also by use of communication merge, the pre-expressions are been transformed into the so-called series-parallel ones \cite{CKA3} \cite{CKA4} \cite{CKA7} free of N-shapes. Then we can get the following soundness and completeness theorem modulo language equivalence.

\begin{theorem}[Soundness and completeness of prefix algebra modulo language equivalence]
For all $x,y\in\mathcal{T}_{PRE}$, $x= y$ if and only if $\sembrack{x}_{PRE}=\sembrack{y}_{PRE}$.
\end{theorem}

\begin{definition}[Operational semantics of the prefix algebra modulo pomset, step and hp-bisimilarities]
Let the symbol $\downarrow$ denote the successful termination predicate. Then we give the TSS of prefix algebra as Table \ref{TRAMB}, where $a,b,c,\cdots\in \Sigma$, $x,y,x',y'\in\mathcal{T}_{PRE}$.
\end{definition}

\begin{center}
    \begin{table}
        $$\frac{}{1\downarrow}\quad\frac{}{a\xrightarrow{a}1}$$
        $$\frac{x\downarrow}{(x+y)\downarrow}\quad\frac{y\downarrow}{(x+y)\downarrow}\quad\frac{x\xrightarrow{a}x'}{x+y\xrightarrow{a}x'}\quad\frac{y\xrightarrow{b}y'}{x+y\xrightarrow{b}y'}$$
        $$\frac{}{a.x\xrightarrow{a}x}$$
        $$\frac{x\downarrow\quad y\downarrow}{(x\parallel y)\downarrow} \quad\frac{x\xrightarrow{a}x'\quad y\xrightarrow{b}y'}{x\parallel y\xrightarrow{\{a,b\}}x'\between y'}$$
        $$\frac{x\downarrow\quad y\downarrow}{(x\mid y)\downarrow} \quad\frac{x\xrightarrow{a}x'\quad y\xrightarrow{b}y'}{x\mid y\xrightarrow{\rho(a,b)}x'\between y'}$$
        \caption{Operational semantics of algebra modulo pomset, step and hp-bisimilarities}
        \label{TRAMB}
    \end{table}
\end{center}

The axioms of prefix algebra modulo pomset, step and hp-bisimilarities in Table \ref{AxiomsForPrefixB} for all $x,y\in \mathcal{T}_{PRE}$ and $a,b\in\Sigma$.

\begin{center}
    \begin{table}
        \begin{tabular}{@{}ll@{}}
            \hline No. &Axiom\\
            $A1$ & $x+y=y+z$\\
            $A2$ & $x+(y+z)=(x+y)+z$\\
            $A3$ & $x+x=x$\\
            $A4$ & $x+0=x$\\
            $A5$ & $a.1=a$\\
            $P1$ & $x\between y=x\parallel y+x\mid y$\\
            $P2$ & $x\parallel y=y\parallel x$\\
            $P3$ & $x\parallel(y\parallel z)=(x\parallel y)\parallel z$\\
            $P4$ & $(x+y)\parallel z=x\parallel z+y\parallel z$\\
            $P5$ & $x\parallel(y+z)=x\parallel y+x\parallel z$\\
            $P6$ & $x\parallel 0=0$\\
            $P7$ & $0\parallel x=0$\\
            $P8$ & $x\parallel 1=x$\\
            $P9$ & $1\parallel x=x$\\
            $C1$ & $x\mid y=y\mid x$\\
            $C2$ & $(x+y)\mid z=x\mid z+y\mid z$\\
            $C3$ & $x\mid(y+z)=x\mid y+x\mid z$\\
            $C4$ & $x\mid 0=0$\\
            $C5$ & $0\mid x=0$\\
            $C6$ & $x\mid 1=0$\\
            $C7$ & $1\mid x=0$\\
        \end{tabular}
        \caption{Axioms of prefix algebra modulo pomset, step and hp-bisimilarities}
        \label{AxiomsForPrefixB}
    \end{table}
\end{center}

Since pomset, step and hp-bisimilarities are all congruences w.r.t. operators of the prefix algebra, we can only check the soundness of each axiom according to the definition of semantics of the prefix algebra. And also by use of communication merge, the pre-expressions are been transformed into the so-called series-parallel ones \cite{CKA3} \cite{CKA4} \cite{CKA7} free of N-shapes. Then we can get the following soundness and completeness theorem.

\begin{theorem}[Soundness and completeness of the prefix algebra modulo pomset, step and hp-bisimilarities]
The prefix algebra is sound and complete modulo pomset, step and hp-bisimilarities.
\end{theorem}

\begin{definition}[Operational semantics of the prefix algebra modulo hhp-bisimilarity]
Let the symbol $\downarrow$ denote the successful termination predicate. Then we give the TSS of prefix algebra as Table \ref{TRAMB2}, where $a,b,c,\cdots\in \Sigma$, $x,y,x',y'\in\mathcal{T}_{PRE}$.
\end{definition}

\begin{center}
    \begin{table}
        $$\frac{}{1\downarrow}\quad\frac{}{a\xrightarrow{a}1}$$
        $$\frac{x\downarrow}{(x+y)\downarrow}\quad\frac{y\downarrow}{(x+y)\downarrow}\quad\frac{x\xrightarrow{a}x'}{x+y\xrightarrow{a}x'}\quad\frac{y\xrightarrow{b}y'}{x+y\xrightarrow{b}y'}$$
        $$\frac{}{a.x\xrightarrow{a}x}$$
        $$\frac{x\downarrow\quad y\downarrow}{(x\parallel y)\downarrow} \quad\frac{x\xrightarrow{a}x'\quad y\xrightarrow{b}y'}{x\parallel y\xrightarrow{\{a,b\}}x'\between y'}$$
        $$\frac{x\downarrow\quad y\downarrow}{(x\leftmerge y)\downarrow} \quad\frac{x\xrightarrow{a}x'\quad y\xrightarrow{b}y'\quad a\leq b}{x\leftmerge y\xrightarrow{\mset{a,b}}x'\between y'}$$
        $$\frac{x\downarrow\quad y\downarrow}{(x\mid y)\downarrow} \quad\frac{x\xrightarrow{a}x'\quad y\xrightarrow{b}y'}{x\mid y\xrightarrow{\rho(a,b)}x'\between y'}$$
        \caption{Operational semantics of algebra modulo hhp-bisimilarity}
        \label{TRAMB2}
    \end{table}
\end{center}

The axioms of prefix algebra modulo hhp-bisimilarity in Table \ref{AxiomsForPrefixB2} for all $x,y\in \mathcal{T}_{PRE}$ and $a,b\in\Sigma$.

\begin{center}
    \begin{table}
        \begin{tabular}{@{}ll@{}}
            \hline No. &Axiom\\
            $A1$ & $x+y=y+z$\\
            $A2$ & $x+(y+z)=(x+y)+z$\\
            $A3$ & $x+x=x$\\
            $A4$ & $x+0=x$\\
            $A5$ & $a.1=a$\\
            $P1$ & $x\between y=x\parallel y+x\mid y$\\
            $P2$ & $x\parallel y=y\parallel x$\\
            $P3$ & $x\parallel(y\parallel z)=(x\parallel y)\parallel z$\\
            $P4$ & $x\parallel y = x\leftmerge y+y\leftmerge x$\\
            $P5$ & $(x+y)\leftmerge z=x\leftmerge z+y\leftmerge z$\\
            $P6$ & $0\leftmerge x=0$\\
            $P7$ & $x\leftmerge 1=x$\\
            $P8$ & $1\leftmerge x=x$\\
            $C1$ & $x\mid y=y\mid x$\\
            $C2$ & $(x+y)\mid z=x\mid z+y\mid z$\\
            $C3$ & $x\mid(y+z)=x\mid y+x\mid z$\\
            $C4$ & $x\mid 0=0$\\
            $C5$ & $0\mid x=0$\\
            $C6$ & $x\mid 1=0$\\
            $C7$ & $1\mid x=0$\\
        \end{tabular}
        \caption{Axioms of prefix algebra modulo hhp-bisimilarity}
        \label{AxiomsForPrefixB2}
    \end{table}
\end{center}

Since hhp-bisimilarity is a congruences w.r.t. operators of the prefix algebra, we can only check the soundness of each axiom according to the definition of semantics of the prefix algebra. And also by use of communication merge, the pre-expressions are been transformed into the so-called series-parallel ones \cite{CKA3} \cite{CKA4} \cite{CKA7} free of N-shapes. Then we can get the following soundness and completeness theorem.

\begin{theorem}[Soundness and completeness of the prefix algebra modulo hhp-bisimilarity]
The prefix algebra is sound and complete modulo hhp-bisimilarity.
\end{theorem}

\subsubsubsection{Recursion}

We discuss recursion over the prefix algebra.

\begin{definition}[Recursion specification]
Let $\mathcal{N}$ be a finite set of names or variables called recursive variables. A recursive specification over $\mathcal{N}$ is a set of equations called recursive equations of the form $S=t_{S}$, exactly one equation for each $S\in\mathcal{N}$, where the right-hand side $t_{S}$ is an expression over the prefix algebra and elements of $\mathcal{N}$.
\end{definition}

\begin{definition}[Linear recursive specification]
A recursive specification over $\mathcal{N}$ is called linear if each right-hand $t_{S}$ of each recursive equation is a linear expression, which is defined recursively as follows.

\begin{enumerate}
  \item Expressions $1$, $0$, of the form $a.T$ with $a\in\Sigma$ and $T\in\mathcal{N}$, or of the form $(a_1\parallel\cdots\parallel a_n).T$ with $a_1,\cdots,a_n\in\Sigma$ and $T\in\mathcal{N}$ are linear expressions.
  \item An alternative composition $+$ of linear expressions is a linear expression.
\end{enumerate}
\end{definition}

\begin{corollary}[Soundness of recursion modulo language equivalence]
Let a recursion specification contain an equation $S=t$, then, $S$ is language equivalent to $t$.
\end{corollary}

\begin{corollary}[Elimination of recursion modulo language equivalence]
Let a recursive specification be over prefix algebra and recursive variables $\mathcal{N}$, then, it is language equivalent to a linear specification.
\end{corollary}

\begin{definition}[Operational semantics of prefix algebra with recursion]
Let the symbol $\downarrow$ denote the successful termination predicate. Then we give the TSS of prefix algebra with recursion as Table \ref{TRRecursion}, where $a,b,c,\cdots\in \Sigma$, $x\in\mathcal{T}_{PRE}$.
\end{definition}

\begin{center}
    \begin{table}
        $$\frac{t_{S}\downarrow\quad S=t_{S}}{S\downarrow}\quad\frac{t_{S}\xrightarrow{\mset{a_1,\cdots,a_n}}x\quad S=t_{S}}{S\xrightarrow{\mset{a_1,\cdots,a_n}}x}$$
        \caption{Operational semantics of recursion modulo bisimilarities}
        \label{TRRecursion}
    \end{table}
\end{center}

\begin{theorem}[Soundness of recursion modulo bisimilarities]
Let a recursion specification contain an equation $S=t$, then, 

\begin{enumerate}
  \item $S\sim_{p}t$.
  \item $S\sim_{s}t$.
  \item $S\sim_{hp}t$.
  \item $S\sim_{hhp}t$.
\end{enumerate}
\end{theorem}

\begin{theorem}[Elimination of recursion modulo bisimilarities]
Let a recursive specification be over prefix algebra and recursive variables $\mathcal{N}$, then,

\begin{enumerate}
  \item It is pomset bisimilar to a linear specification.
  \item It is step bisimilar to a linear specification.
  \item It is hp-bisimilar to a linear specification.
  \item It is hhp-bisimilar to a linear specification.
\end{enumerate}
\end{theorem}

\subsubsubsection{Encapsulation}

The following algebra is based on the prefix algebra.

\begin{definition}[Syntax of prefix algebra with encapsulation]
The expressions (terms) set $\mathcal{T}_{ENC}$ is defined inductively by the following grammar.

$$\mathcal{T}_{ENC}\ni x,y::=0|1|a,b\in\Sigma|\rho(a,b)|x+y|a.x|(a\parallel b).x|x\parallel y|x\mid y|x\between y|\partial_H(x)$$
\end{definition}

\begin{definition}[Language semantics of prefix-expressions with encapsulation]
We define the interpretation of prefix-expressions with encapsulation $\sembrack{-}_{ENC}:\mathcal{T}_{ENC}\rightarrow 2^{\mathsf{SCP}}$ inductively as Table \ref{LSPREE} shows.
\end{definition}

\begin{center}
    \begin{table}
        $$\sembrack{0}_{ENC}=\emptyset \quad \sembrack{a}_{ENC}=\{a\} \quad \sembrack{a.x}_{ENC}=\{a\}\cdot \sembrack{x}_{ENC}$$
        $$\sembrack{1}_{ENC}=\{1\} \quad \sembrack{x+y}_{ENC}=\sembrack{x}_{ENC}+\sembrack{y}_{ENC}$$
        $$\sembrack{x\parallel y}_{ENC}=\sembrack{x}_{ENC}\parallel\sembrack{y}_{ENC} \quad\sembrack{x\mid y}_{ENC}=\sembrack{x}_{ENC}\mid\sembrack{y}_{ENC}$$
        $$\sembrack{x\between y}_{ENC}=\sembrack{x}_{ENC}\between\sembrack{y}_{ENC}\quad \sembrack{\partial_H(x)}_{ENC}=\partial_H(\sembrack{x}_{ENC})$$
        \caption{Language semantics of prefix-expressions with encapsulation}
        \label{LSPREE}
    \end{table}
\end{center}

\begin{definition}[Operational semantics of the prefix algebra with encapsulation]
We give the TSS of prefix algebra with encapsulation as Table \ref{TRENC}, where $a,b,c,\cdots\in \Sigma$, $x,x'\in\mathcal{T}_{ENC}$.
\end{definition}

\begin{center}
    \begin{table}
        $$\frac{x\downarrow}{\partial_H(x)\downarrow}\quad\frac{x\xrightarrow{a}x'\quad a\notin H}{\partial_H(x)\xrightarrow{a}\partial_H(x')}$$
        \caption{Operational semantics of prefix-expressions with encapsulation modulo bisimilarities}
        \label{TRENC}
    \end{table}
\end{center}

The axioms of the prefix algebra with encapsulation are shown in Table \ref{AxiomsForENC} for all $x,y\in \mathcal{T}_{ENC}$ and $a,b\in\Sigma$.

\begin{center}
    \begin{table}
        \begin{tabular}{@{}ll@{}}
            \hline No. &Axiom\\
            $ENC1$ & $\partial_H(0)=0$\\
            $ENC2$ & $\partial_H(1)=1$\\
            $ENC3$ & $\partial_H(a.x)=0\quad a\in H$\\
            $ENC4$ & $\partial_H(a.x)=a.\partial_H(x)\quad a\notin H$\\
            $ENC5$ & $\partial_H(x+y)=\partial_H(x)+\partial_H(y)$\\
            $ENC6$ & $\partial_H(x\parallel y)=\partial_H(x)\parallel\partial_H(y)$\\
            $ENC7$ & $\partial_H(\partial_H(x))=\partial_H(x)$\\
            $ENC8$ & $\partial_{H_1}(\partial_{H_2}(x))=\partial_{H_2}(\partial_{H_1}(x))$\\
        \end{tabular}
        \caption{Axioms of encapsulation modulo bisimilarities}
        \label{AxiomsForENC}
    \end{table}
\end{center}

\begin{theorem}[Soundness of the prefix algebra with encapsulation modulo bisimilarities]
The axioms of the prefix algebra with encapsulation shown in Table \ref{AxiomsForENC} are sound modulo bisimilarities.
\end{theorem}

\begin{theorem}[Soundness of the prefix algebra with encapsulation modulo language equivalence]
The axioms of the prefix algebra with encapsulation shown in Table \ref{AxiomsForENC} are sound modulo language equivalence.
\end{theorem}

\subsubsubsection{Silent Step}

We use $\tau$ to denote silent step. The following algebra is based on the prefix algebra.

\begin{definition}[Syntax of the prefix algebra with silent step]
The expressions (terms) set $\mathcal{T}_{\tau}$ is defined inductively by the following grammar.

$$\mathcal{T}_{\tau}\ni x,y::=0|1|a,b\in\Sigma|\rho(a,b)|x+y|a.x|\tau.x|(a\parallel b).x|x\parallel y|x\mid y|x\between y$$
\end{definition}

\begin{definition}[Language semantics of prefix-expressions with silent step]
We define the interpretation of prefix-expressions with silent step $\sembrack{-}_{\tau}:\mathcal{T}_{\tau}\rightarrow 2^{\mathsf{SCP}}$ inductively as Table \ref{LSPRET} shows.
\end{definition}

\begin{center}
    \begin{table}
        $$\sembrack{0}_{\tau}=\emptyset \quad \sembrack{a}_{\tau}=\{a\} \quad \sembrack{\tau}_{\tau}=\{\tau\}\quad \sembrack{a.x}_{\tau}=\{a\}\cdot \sembrack{x}_{\tau}\quad \sembrack{\tau.x}_{\tau}=\{\tau\}\cdot \sembrack{x}_{\tau}$$
        $$\sembrack{1}_{\tau}=\{1\} \quad \sembrack{x+y}_{\tau}=\sembrack{x}_{\tau}+\sembrack{y}_{\tau}$$
        $$\sembrack{x\parallel y}_{\tau}=\sembrack{x}_{\tau}\parallel\sembrack{y}_{\tau} \quad\sembrack{x\mid y}_{\tau}=\sembrack{x}_{\tau}\mid\sembrack{y}_{\tau}$$
        $$\sembrack{x\between y}_{\tau}=\sembrack{x}_{\tau}\between\sembrack{y}_{\tau}$$
        \caption{Language semantics of prefix-expressions with silent step}
        \label{LSPRET}
    \end{table}
\end{center}

We show the $\tau$ laws modulo language equivalence in Table \ref{AxiomsForTauL}.

\begin{center}
    \begin{table}
        \begin{tabular}{@{}ll@{}}
            \hline No. &Axiom\\
            $T1$ & $\tau.x=x$\\
            $T2$ & $\tau\parallel x=x$\\
            $T3$ & $x\parallel\tau=x$\\
        \end{tabular}
        \caption{Axioms of prefix algebra with silent step modulo language equivalence}
        \label{AxiomsForTauL}
    \end{table}
\end{center}

\begin{theorem}[Soundness of the algebra with silent step modulo language equivalence]
The axioms of the prefix algebra with silent step shown in Table \ref{AxiomsForTauL} are sound modulo language equivalence.
\end{theorem} 

\begin{definition}[Branching pomset, step bisimulation]\label{BPSB}
Assume a special termination predicate $\downarrow$, and let $\surd$ represent a state with $\surd\downarrow$. Let $x,y\in\mathcal{T}_{\tau}$ be expressions. A branching pomset bisimulation is a relation $R\subseteq\mathcal{C}(x)\times\mathcal{C}(y)$, such that:
 \begin{enumerate}
   \item If $(\mathbf{C}_1,\mathbf{C}_2)\in R$, and $\mathbf{C}_1\xrightarrow{X}\mathbf{C}_1'$ then
   \begin{itemize}
     \item either $X\equiv \tau^*$, and $(\mathbf{C}_1',\mathbf{C}_2)\in R$;
     \item or there is a sequence of (zero or more) $\tau$-transitions $\mathbf{C}_2\xrightarrow{\tau^*} \mathbf{C}_2^0$, such that $(\mathbf{C}_1,\mathbf{C}_2^0)\in R$ and $\mathbf{C}_2^0\xRightarrow{X}\mathbf{C}_2'$ with $(\mathbf{C}_1',\mathbf{C}_2')\in R$.
   \end{itemize}
   \item If $(\mathbf{C}_1,\mathbf{C}_2)\in R$, and $\mathbf{C}_2\xrightarrow{X}\mathbf{C}_2'$ then
   \begin{itemize}
     \item either $X\equiv \tau^*$, and $(\mathbf{C}_1,\mathbf{C}_2')\in R$;
     \item or there is a sequence of (zero or more) $\tau$-transitions $\mathbf{C}_1\xrightarrow{\tau^*} \mathbf{C}_1^0$, such that $(\mathbf{C}_1^0,\mathbf{C}_2)\in R$ and $\mathbf{C}_1^0\xRightarrow{X}\mathbf{C}_1'$ with $(\mathbf{C}_1',\mathbf{C}_2')\in R$.
   \end{itemize}
   \item If $(\mathbf{C}_1,\mathbf{C}_2)\in R$ and $\mathbf{C}_1\downarrow$, then there is a sequence of (zero or more) $\tau$-transitions $\mathbf{C}_2\xrightarrow{\tau^*}\mathbf{C}_2^0$ such that $(\mathbf{C}_1,\mathbf{C}_2^0)\in R$ and $\mathbf{C}_2^0\downarrow$.
   \item If $(\mathbf{C}_1,\mathbf{C}_2)\in R$ and $\mathbf{C}_2\downarrow$, then there is a sequence of (zero or more) $\tau$-transitions $\mathbf{C}_1\xrightarrow{\tau^*}\mathbf{C}_1^0$ such that $(\mathbf{C}_1^0,\mathbf{C}_2)\in R$ and $\mathbf{C}_1^0\downarrow$.
 \end{enumerate}

We say that $x$, $y$ are branching pomset bisimilar, written $x\approx_{bp}y$, if there exists a branching pomset bisimulation $R$, such that $(\emptyset,\emptyset)\in R$.

By replacing pomset transitions with steps, we can get the definition of branching step bisimulation. When $x$ and $y$ are branching step bisimilar, we write $x\approx_{bs}y$.
\end{definition}

\begin{definition}[Rooted branching pomset, step bisimulation]\label{RBPSB}
Assume a special termination predicate $\downarrow$, and let $\surd$ represent a state with $\surd\downarrow$. Let $x,y\in\mathcal{T}_{\tau}$ be expressions. A rooted branching pomset bisimulation is a relation $R\subseteq\mathcal{C}(x)\times\mathcal{C}(y)$, such that:
 \begin{enumerate}
   \item If $(\mathbf{C}_1,\mathbf{C}_2)\in R$, and $\mathbf{C}_1\xrightarrow{X}\mathbf{C}_1'$ then $\mathbf{C}_2\xrightarrow{X}\mathbf{C}_2'$ with $\mathbf{C}_1'\approx_{bp}\mathbf{C}_2'$.
   \item If $(\mathbf{C}_1,\mathbf{C}_2)\in R$, and $\mathbf{C}_2\xrightarrow{X}\mathbf{C}_2'$ then $\mathbf{C}_1\xrightarrow{X}\mathbf{C}_1'$ with $\mathbf{C}_1'\approx_{bp}\mathbf{C}_2'$.
   \item If $(\mathbf{C}_1,\mathbf{C}_2)\in R$ and $\mathbf{C}_1\downarrow$, then $\mathbf{C}_2\downarrow$.
   \item If $(\mathbf{C}_1,\mathbf{C}_2)\in R$ and $\mathbf{C}_2\downarrow$, then $\mathbf{C}_1\downarrow$.
 \end{enumerate}

We say that $x$, $y$ are rooted branching pomset bisimilar, written $x\approx_{rbp}y$, if there exists a rooted branching pomset bisimulation $R$, such that $(\emptyset,\emptyset)\in R$.

By replacing pomset transitions with steps, we can get the definition of rooted branching step bisimulation. When $x$ and $y$ are rooted branching step bisimilar, we write $x\approx_{rbs}y$.
\end{definition}

\begin{definition}[Branching (hereditary) history-preserving bisimulation]\label{BHHPB}
Assume a special termination predicate $\downarrow$, and let $\surd$ represent a state with $\surd\downarrow$. A branching history-preserving (hp-) bisimulation is a posetal relation $R\subseteq\mathcal{C}(x)\overline{\times}\mathcal{C}(y)$ such that:

 \begin{enumerate}
   \item If $(\mathbf{C}_1,f,\mathbf{C}_2)\in R$, and $\mathbf{C}_1\xrightarrow{a_1}\mathbf{C}_1'$ then
   \begin{itemize}
     \item either $a_1\equiv \tau$, and $(\mathbf{C}_1',f[a_1\mapsto \tau],\mathbf{C}_2)\in R$;
     \item or there is a sequence of (zero or more) $\tau$-transitions $\mathbf{C}_2\xrightarrow{\tau^*} \mathbf{C}_2^0$, such that $(\mathbf{C}_1,f,\mathbf{C}_2^0)\in R$ and $\mathbf{C}_2^0\xrightarrow{a_2}\mathbf{C}_2'$ with $(\mathbf{C}_1',f[a_1\mapsto a_2],\mathbf{C}_2')\in R$.
   \end{itemize}
   \item If $(\mathbf{C}_1,f,\mathbf{C}_2)\in R$, and $\mathbf{C}_2\xrightarrow{a_2}\mathbf{C}_2'$ then
   \begin{itemize}
     \item either $a_2\equiv \tau$, and $(\mathbf{C}_1,f[a_2\mapsto \tau],\mathbf{C}_2')\in R$;
     \item or there is a sequence of (zero or more) $\tau$-transitions $\mathbf{C}_1\xrightarrow{\tau^*} \mathbf{C}_1^0$, such that $(\mathbf{C}_1^0,f,\mathbf{C}_2)\in R$ and $\mathbf{C}_1^0\xrightarrow{a_1}\mathbf{C}_1'$ with $(\mathbf{C}_1',f[a_2\mapsto a_1],\mathbf{C}_2')\in R$.
   \end{itemize}
   \item If $(\mathbf{C}_1,f,\mathbf{C}_2)\in R$ and $\mathbf{C}_1\downarrow$, then there is a sequence of (zero or more) $\tau$-transitions $\mathbf{C}_2\xrightarrow{\tau^*}\mathbf{C}_2^0$ such that $(\mathbf{C}_1,f,\mathbf{C}_2^0)\in R$ and $\mathbf{C}_2^0\downarrow$.
   \item If $(\mathbf{C}_1,f,\mathbf{C}_2)\in R$ and $\mathbf{C}_2\downarrow$, then there is a sequence of (zero or more) $\tau$-transitions $\mathbf{C}_1\xrightarrow{\tau^*}\mathbf{C}_1^0$ such that $(\mathbf{C}_1^0,f,\mathbf{C}_2)\in R$ and $\mathbf{C}_1^0\downarrow$.
 \end{enumerate}

$x,y$ are branching history-preserving (hp-)bisimilar and are written $x\approx_{bhp}y$ if there exists a branching hp-bisimulation $R$ such that $(\emptyset,\emptyset,\emptyset)\in R$.

A branching hereditary history-preserving (hhp-)bisimulation is a downward closed branching hp-bisimulation. $x,y$ are branching hereditary history-preserving (hhp-)bisimilar and are written $x\approx_{bhhp}y$.
\end{definition}

\begin{definition}[Rooted branching (hereditary) history-preserving bisimulation]\label{RBHHPB}
Assume a special termination predicate $\downarrow$, and let $\surd$ represent a state with $\surd\downarrow$. A rooted branching history-preserving (hp-) bisimulation is a weakly posetal relation $R\subseteq\mathcal{C}(x)\overline{\times}\mathcal{C}(y)$ such that:

 \begin{enumerate}
   \item If $(\mathbf{C}_1,f,\mathbf{C}_2)\in R$, and $\mathbf{C}_1\xrightarrow{a_1}\mathbf{C}_1'$, then $\mathbf{C}_2\xrightarrow{a_2}\mathbf{C}_2'$ with $\mathbf{C}_1'\approx_{bhp}\mathbf{C}_2'$.
   \item If $(\mathbf{C}_1,f,\mathbf{C}_2)\in R$, and $\mathbf{C}_2\xrightarrow{a_2}\mathbf{C}_2'$, then $\mathbf{C}_1\xrightarrow{a_1}\mathbf{C}_1'$ with $\mathbf{C}_1'\approx_{bhp}\mathbf{C}_2'$.
   \item If $(\mathbf{C}_1,f,\mathbf{C}_2)\in R$ and $\mathbf{C}_1\downarrow$, then $\mathbf{C}_2\downarrow$.
   \item If $(\mathbf{C}_1,f,\mathbf{C}_2)\in R$ and $\mathbf{C}_2\downarrow$, then $\mathbf{C}_1\downarrow$.
 \end{enumerate}

$x,y$ are rooted branching history-preserving (hp-)bisimilar and are written $x\approx_{rbhp}y$ if there exists a rooted branching hp-bisimulation $R$ such that $(\emptyset,\emptyset,\emptyset)\in R$.

A rooted branching hereditary history-preserving (hhp-)bisimulation is a downward closed rooted branching hp-bisimulation. $x,y$ are rooted branching hereditary history-preserving (hhp-)bisimilar and are written $x\approx_{rbhhp}y$.
\end{definition}

\begin{theorem}[Branching pomset bisimilarity implying language equivalence]
Given two expressions $x$ and $y$, if $x\approx_{bp}y$, then $x$ is language equivalent to $y$.
\end{theorem}

\begin{theorem}[Branching step bisimilarity implying language equivalence]
Given two expressions $x$ and $y$, if $x\approx_{bs}y$, then $x$ is language equivalent to $y$.
\end{theorem}

\begin{theorem}[Branching hp-bisimilarity implying language equivalence]
Given two expressions $x$ and $y$, if $x\approx_{bhp}y$, then $x$ is language equivalent to $y$.
\end{theorem}

\begin{theorem}[Branching hhp-bisimilarity implying language equivalence]
Given two expressions $x$ and $y$, if $x\approx_{bhhp}y$, then $x$ is language equivalent to $y$.
\end{theorem}

Note that the above branching pomset, step, hp-, hhp-bisimilarities preserve deadlocks. 

\begin{definition}[Operational semantics of the prefix algebra with silent step]
We give the TSS of the prefix algebra with silent step as Table \ref{TRTAU}, where $a,b,c,\cdots\in \Sigma$, $x,y,x',y'\in\mathcal{T}_{\tau}$.
\end{definition}

\begin{center}
    \begin{table}
        $$\frac{}{\tau.x\xrightarrow{\tau}x}$$
        $$\frac{x\xrightarrow{\tau}x'\quad y\xrightarrow{b}y'}{x\parallel y\xrightarrow{\{\tau,b\}}x'\between y'}$$
        \caption{Operational semantics of silent step modulo branching bisimilarities}
        \label{TRTAU}
    \end{table}
\end{center}

We show the $\tau$ laws modulo branching pomset, branching step and branching hp-bisimilarities in Table \ref{AxiomsForTauB}.

\begin{center}
    \begin{table}
        \begin{tabular}{@{}ll@{}}
            \hline No. &Axiom\\
            $T1$ & $a.\tau=a$\\
            $T2$ & $\tau\parallel x=x$\\
            $T3$ & $x\parallel\tau=x$\\
        \end{tabular}
        \caption{Axioms of prefix algebra with silent step modulo branching pomset, branching step and branching hp-bisimilarities}
        \label{AxiomsForTauB}
    \end{table}
\end{center}

\begin{theorem}[Soundness of the prefix algebra with silent step modulo branching pomset, branching step and branching hp-bisimilarities]
The axioms of the prefix algebra with silent step shown in Table \ref{AxiomsForTauB} are sound modulo branching pomset, branching step, branching hp-bisimilarities.
\end{theorem}

We show the $\tau$ laws modulo branching hhp-bisimilarity in Table \ref{AxiomsForTauB2}.

\begin{center}
    \begin{table}
        \begin{tabular}{@{}ll@{}}
            \hline No. &Axiom\\
            $T1$ & $a.\tau=a$\\
            $T2$ & $x\leftmerge\tau=x$\\
        \end{tabular}
        \caption{Axioms of prefix algebra with silent step modulo branching hhp-bisimilarity}
        \label{AxiomsForTauB2}
    \end{table}
\end{center}

\begin{theorem}[Soundness of the prefix algebra with silent step modulo branching hhp-bisimilarity]
The axioms of the prefix algebra with silent step shown in Table \ref{AxiomsForTauB2} are sound modulo branching hhp-bisimilarity.
\end{theorem}

\subsubsubsection{Abstraction}

The following algebra is based on the prefix algebra with silent step.

\begin{definition}[Syntax of the prefix algebra with silent step and abstraction]
The expressions (terms) set $\mathcal{T}_{TI}$ is defined inductively by the following grammar.

$$\mathcal{T}_{TI}\ni x,y::=0|1|a,b\in\Sigma|\rho(a,b)|x+y|a.x|\tau.x|(a\parallel b).x|x\parallel y|x\mid y|x\between y|\tau_I(x)$$
\end{definition}

\begin{definition}[Language semantics of prefix-expressions with silent step and abstraction]
We define the interpretation of prefix-expressions with silent step and abstraction $\sembrack{-}_{TI}:\mathcal{T}_{TI}\rightarrow 2^{\mathsf{SCP}}$ inductively as Table \ref{LSPRETI} shows.
\end{definition}

\begin{center}
    \begin{table}
        $$\sembrack{0}_{TI}=\emptyset \quad \sembrack{a}_{TI}=\{a\} \quad \sembrack{\tau}_{TI}=\{\tau\}\quad \sembrack{a.x}_{TI}=\{a\}\cdot \sembrack{x}_{TI}\quad \sembrack{\tau.x}_{TI}=\{\tau\}\cdot \sembrack{x}_{TI}$$
        $$\sembrack{1}_{TI}=\{1\} \quad \sembrack{x+y}_{TI}=\sembrack{x}_{TI}+\sembrack{y}_{TI}$$
        $$\sembrack{x\parallel y}_{TI}=\sembrack{x}_{TI}\parallel\sembrack{y}_{TI} \quad\sembrack{x\mid y}_{TI}=\sembrack{x}_{TI}\mid\sembrack{y}_{TI}$$
        $$\sembrack{x\between y}_{TI}=\sembrack{x}_{TI}\between\sembrack{y}_{TI}\quad \sembrack{\tau_I(x)}_{TI}=\tau_I(\sembrack{x}_{TI})$$
        \caption{Language semantics of prefix-expressions with silent step and abstraction}
        \label{LSPRETI}
    \end{table}
\end{center}

\begin{definition}[Operational semantics of the prefix algebra with silent step and abstraction]
We give the TSS of the prefix algebra with silent step and abstraction as Table \ref{TRABS}, where $a,b,c,\cdots\in \Sigma$, $x,x'\in\mathcal{T}_{TI}$.
\end{definition}

\begin{center}
    \begin{table}
        $$\frac{x\downarrow}{\tau_I(x)\downarrow}\quad\frac{x\xrightarrow{a}x'\quad a\notin I}{\tau_I(x)\xrightarrow{a}\tau_I(x')}\quad\frac{x\xrightarrow{a}x'\quad a\in I}{\tau_I(x)\xrightarrow{\tau}\tau_I(x')}$$
        \caption{Operational semantics of prefix-expressions with silent step and abstraction modulo branching bisimilarities}
        \label{TRABS}
    \end{table}
\end{center}

The axioms of the prefix algebra with silent step and abstraction are shown in Table \ref{AxiomsForABS} for all $x,y\in \mathcal{T}_{TI}$ and $a,b\in\Sigma$.

\begin{center}
    \begin{table}
        \begin{tabular}{@{}ll@{}}
            \hline No. &Axiom\\
            $TI1$ & $\tau_I(0)=0$\\
            $TI2$ & $\tau_I(1)=1$\\
            $TI3$ & $\tau_I(a.x)=\tau.\tau_I(x)\quad a\in I$\\
            $TI4$ & $\tau_I(a.x)=a.\tau_I(x)\quad a\notin I$\\
            $TI5$ & $\tau_I(x+y)=\tau_I(x)+\tau_I(y)$\\
            $TI6$ & $\tau_I(x\parallel y)=\tau_I(x)\parallel\tau_I(y)$\\
            $TI7$ & $\tau_I(\tau_I(x))=\tau_I(x)$\\
            $TI8$ & $\tau_{I_1}(\tau_{I_2}(x))=\tau_{I_2}(\tau_{I_1}(x))$\\
        \end{tabular}
        \caption{Axioms of prefix-expressions with silent step and abstraction modulo branching bisimilarities}
        \label{AxiomsForABS}
    \end{table}
\end{center}

\begin{theorem}[Soundness of the prefix algebra with silent step and abstraction modulo language equivalence]
The axioms of the prefix algebra with silent step and abstraction shown in Table \ref{AxiomsForABS} are sound modulo language equivalence.
\end{theorem} 

\begin{theorem}[Soundness of the prefix algebra with silent step and abstraction modulo branching bisimilarities]
The axioms of the prefix algebra with silent step and abstraction shown in Table \ref{AxiomsForABS} are sound modulo branching pomset, step, hp-, hhp-bisimilarities.
\end{theorem} 
\newpage\section{Hypotheses for Concurrency}\label{hfc} 

In this chapter, we introduce hypotheses for concurrency based the five equivalences: language equivalence, pomset bisimilarity, step bisimilarity, hp-bisimilarity and hhp-bisimilarity. In section \ref{soundnessH}, we introduce the soundness theorem. We introduce the completeness-related conclusions by use of the concept of reduction in section \ref{reductionH}.

\begin{definition}[Hypotheses]
A hypothesis is an inequation $x\leq y$ where $x,y\in\mathcal{T}_{SCR}$. When $H$ is a set of hypotheses, we write $=^H$ for the smallest BKAC congruence on $\mathcal{T}_{SCR}$ that satisfies the containments in $H$, i.e., whenever $x\leq y\in H$, also $x\leqq^H y$ ($x+y=^H y$). Correspondingly, we have defined five kinds of congruences over $\mathcal{T}_{SCR}$ in chapter \ref{cp}, and we denote $=^H_1$ and $\leqq^H_1$, $=^H_2$ and $\leqq^H_2$, $=^H_3$ and $\leqq^H_3$, $=^H_4$ and $\leqq^H_4$, and $=^H_5$ and $\leqq^H_5$ respectively.
\end{definition} 

It is no hard to show that $\leqq^H$ and its five kinds of variants are all preorders and further partial orders on $\mathcal{T}_{SCR}$, and all operators of $\mathcal{T}_{SCR}$ are monotone w.r.t. $\leqq^H$ and its five variants.

\subsection{Soundness}\label{soundnessH}

The augmented congruence $\leqq^H$ and its five variants on $\mathcal{T}_{SCR}$ should be sound modulo the five corresponding equivalences. To make the soundness, we also need the following definitions.

\begin{definition}[Pomsetc contexts]\label{pc}
Let $\square\notin\Sigma$. The set of series-communication-parallel pomsetc contexts, denoted $\mathsf{PC}^{\mathsf{SCP}}$, is the smallest subset of $\mathsf{Pomc}(\Sigma\cup\{\square\})$ satisfying the following rules of inference:

$$\frac{}{\square\in \mathsf{PC}^{\mathsf{SCP}}} \quad\frac{V\in\mathsf{SCP}\quad C\in\mathsf{PC}^{\mathsf{SCP}}}{V\cdot C\in \mathsf{PC}^{\mathsf{SCP}}} \quad\frac{C\in\mathsf{PC}^{\mathsf{SCP}}\quad V\in\mathsf{SCP}}{C\cdot V\in \mathsf{PC}^{\mathsf{SCP}}}$$

$$\frac{V\in\mathsf{SCP}\quad C\in\mathsf{PC}^{\mathsf{SCP}}}{V\between C\in \mathsf{PC}^{\mathsf{SCP}}}\quad\frac{V\in\mathsf{SCP}\quad C\in\mathsf{PC}^{\mathsf{SCP}}}{V\parallel C\in \mathsf{PC}^{\mathsf{SCP}}}$$

$$\frac{V\in\mathsf{SCP}\quad C\in\mathsf{PC}^{\mathsf{SCP}}}{V\mid C\in \mathsf{PC}^{\mathsf{SCP}}}\quad \frac{V\in\mathsf{SCP}\quad C\in\mathsf{PC}^{\mathsf{SCP}}}{V\leftmerge C\in \mathsf{PC}^{\mathsf{SCP}}}$$
\end{definition} 

\begin{definition}[Context plugging]
Let $C\in \mathsf{PC}^{\mathsf{SCP}}$ and $U\in\mathsf{Pomc}$, we write $C[U]$ for the pomsetc defined by induction on the structure of $C$:

$$\square[U]=U\quad(V\cdot C)[U]=V\cdot C[U]\quad(C\cdot V)[U]=C[U]\cdot V$$
$$(V\between C)[U]=V\between C[U]\quad (V\parallel C)[U]=V\parallel C[U]$$
$$(V\mid C)[U]=V\mid C[U]\quad (V\leftmerge C)[U]=V\leftmerge C[U]$$

We denote $C[L]$ as the pomsetc language $C[L]=\{C[U]:U\in L\}$, where $L\subseteq\mathsf{Pomc}$ and $C\in\mathsf{PC}^{\mathsf{SCP}}$.
\end{definition}

\begin{definition}[1-Closure]
Let $H$ be a set of hypotheses, and $L\subseteq\mathsf{Pomc}$, we define the 1-$H$-closure of $L$, written $L^{H^1}$, as the smallest language containing $L$ and satisfying:

$$\frac{x\leq y\in H\quad C\in\mathsf{PC}^{\mathsf{SCP}}\quad C[\sembrack{y}^{H^1}_{SCR}]\subseteq L^{H^1}}{C[\sembrack{x}^{H^1}_{SCR}]\subseteq L^{H^1}}$$
\end{definition}

\begin{definition}[2-Closure]
Let $H$ be a set of hypotheses, and $L\subseteq\mathsf{Pomc}$, we define the 2-$H$-closure of $L$, written $L^{H^2}$, as the smallest language containing $L$ and satisfying:

$$\frac{x\leq y\in H\quad C\in\mathsf{PC}^{\mathsf{SCP}}\quad C[y^{H^2}]\subseteq L^{H^2}\quad C[x]\downarrow}{C[x^{H^2}]\subseteq L^{H^2}\quad C[y]\downarrow}$$
$$\frac{x\leq y\in H\quad C\in\mathsf{PC}^{\mathsf{SCP}}\quad C[y^{H^2}]\subseteq L^{H^2}\quad \mathbf{C}(C[x])\xrightarrow{X}\mathbf{C}(C[x])'}{C[x^{H^2}]\subseteq L^{H^2}\quad \mathbf{C}(C[y])\xrightarrow{Y}\mathbf{C}(C[y])'}$$

Where $X\subseteq x$, $Y\subseteq y$ and $X\sim Y$.
\end{definition}

\begin{definition}[3-Closure]
Let $H$ be a set of hypotheses, and $L\subseteq\mathsf{Pomc}$, we define the 3-$H$-closure of $L$, written $L^{H^3}$, as the smallest language containing $L$ and satisfying:

$$\frac{x\leq y\in H\quad C\in\mathsf{PC}^{\mathsf{SCP}}\quad C[y^{H^3}_{SCR}]\subseteq L^{H^3}\quad C[x]\downarrow}{C[x^{H^3}]\subseteq L^{H^3}\quad C[y]\downarrow}$$
$$\frac{x\leq y\in H\quad C\in\mathsf{PC}^{\mathsf{SCP}}\quad C[y^{H^3}]\subseteq L^{H^3}\quad \mathbf{C}(C[x])\xrightarrow{X}\mathbf{C}(C[x])'}{C[x^{H^3}]\subseteq L^{H^3}\quad \mathbf{C}(C[y])\xrightarrow{Y}\mathbf{C}(C[y])'}$$

Where $X\subseteq x$, $Y\subseteq y$, $X\sim Y$, and all events in $X$ and $Y$ are without execution orders $\leq^e$ and communication orders $\leq^c$.
\end{definition}

\begin{definition}[4-Closure and 5-closure]
Let $H$ be a set of hypotheses, and $L\subseteq\mathsf{Pomc}$, we define the 4-$H$-closure of $L$, written $L^{H^4}$, as the smallest language containing $L$ and satisfying:

$$\frac{x\leq y\in H\quad C\in\mathsf{PC}^{\mathsf{SCP}}\quad C[y^{H^4}]\subseteq L^{H^4}\quad C[x]\downarrow}{C[x^{H^4}]\subseteq L^{H^4}\quad C[y]\downarrow}$$
$$\frac{x\leq y\in H\quad C\in\mathsf{PC}^{\mathsf{SCP}}\quad C[y^{H^4}]\subseteq L^{H^4}\quad \mathbf{C}(C[x])\xrightarrow{a_1}\mathbf{C}(C[x])'}{C[x^{H^4}]\subseteq L^{H^4}\quad \mathbf{C}(C[y])\xrightarrow{a_2}\mathbf{C}(C[y])'}$$

Where $R\subseteq\mathcal{C}(C[x])\overline{\times}\mathcal{C}(C[y])$, if $(\mathbf{C}(C[x]),f,\mathbf{C}(C[y]))\in R$, then $(\mathbf{C}(C[x])',f[a_1\mapsto a_2],\mathbf{C}(C[y])')\in R$.

A 5-closure is a downward closed 4-closure.
\end{definition}

Similarly to the reference \cite{CKA7}, we can prove the following two lemmas.

\begin{lemma}
The following conclusions hold:

\begin{enumerate}
  \item Let $L,K\subseteq\mathsf{Pomc}$, then $L\subseteq K^{H^1}$ if and only if $L^{H^1}\subseteq K^{H^1}$.
  \item Let $L,K\subseteq\mathsf{Pomc}$, then $L\subseteq K^{H^2}$ if and only if $L^{H^2}\subseteq K^{H^2}$.
  \item Let $L,K\subseteq\mathsf{Pomc}$, then $L\subseteq K^{H^3}$ if and only if $L^{H^3}\subseteq K^{H^3}$.
  \item Let $L,K\subseteq\mathsf{Pomc}$, then $L\subseteq K^{H^4}$ if and only if $L^{H^4}\subseteq K^{H^4}$.
  \item Let $L,K\subseteq\mathsf{Pomc}$, then $L\subseteq K^{H^5}$ if and only if $L^{H^5}\subseteq K^{H^5}$.
\end{enumerate}
\end{lemma}

\begin{lemma}
Let $L,K\subseteq\mathsf{Pomc}$, then the following hold:

$$(L\cup K)^{H^1}=(L^{H^1}\cup K^{H^1})^{H^1}\quad (L\cdot K)^{H^1}=(L^{H^1}\cdot K^{H^1})^{H^1}$$
$$(L\between K)^{H^1}=(L^{H^1}\between K^{H^1})^{H^1}\quad (L\parallel K)^{H^1}=(L^{H^1}\parallel K^{H^1})^{H^1}$$
$$(L\mid K)^{H^1}=(L^{H^1}\mid K^{H^1})^{H^1}\quad (L\leftmerge K)^{H^1}=(L^{H^1}\leftmerge K^{H^1})^{H^1}$$
$$(L^*)^{H^1}=((L^{H^1})^*)^{H^1}$$

$$(L\cup K)^{H^2}=(L^{H^2}\cup K^{H^2})^{H^2}\quad (L\cdot K)^{H^2}=(L^{H^2}\cdot K^{H^2})^{H^2}$$
$$(L\between K)^{H^2}=(L^{H^2}\between K^{H^2})^{H^2}\quad (L\parallel K)^{H^2}=(L^{H^2}\parallel K^{H^2})^{H^2}$$
$$(L\mid K)^{H^2}=(L^{H^2}\mid K^{H^2})^{H^2}\quad (L\leftmerge K)^{H^2}=(L^{H^2}\leftmerge K^{H^2})^{H^2}$$
$$(L^*)^{H^2}=((L^{H^2})^*)^{H^2}$$

$$(L\cup K)^{H^3}=(L^{H^3}\cup K^{H^3})^{H^3}\quad (L\cdot K)^{H^3}=(L^{H^3}\cdot K^{H^3})^{H^3}$$
$$(L\between K)^{H^3}=(L^{H^3}\between K^{H^3})^{H^3}\quad (L\parallel K)^{H^3}=(L^{H^3}\parallel K^{H^3})^{H^3}$$
$$(L\mid K)^{H^3}=(L^{H^3}\mid K^{H^3})^{H^3}\quad (L\leftmerge K)^{H^3}=(L^{H^3}\leftmerge K^{H^3})^{H^3}$$
$$(L^*)^{H^3}=((L^{H^3})^*)^{H^3}$$

$$(L\cup K)^{H^4}=(L^{H^4}\cup K^{H^4})^{H^4}\quad (L\cdot K)^{H^4}=(L^{H^4}\cdot K^{H^4})^{H^4}$$
$$(L\between K)^{H^4}=(L^{H^4}\between K^{H^4})^{H^4}\quad (L\parallel K)^{H^4}=(L^{H^4}\parallel K^{H^4})^{H^4}$$
$$(L\mid K)^{H^4}=(L^{H^4}\mid K^{H^4})^{H^4}\quad (L\leftmerge K)^{H^4}=(L^{H^4}\leftmerge K^{H^4})^{H^4}$$
$$(L^*)^{H^4}=((L^{H^4})^*)^{H^4}$$

$$(L\cup K)^{H^5}=(L^{H^5}\cup K^{H^5})^{H^5}\quad (L\cdot K)^{H^5}=(L^{H^5}\cdot K^{H^5})^{H^5}$$
$$(L\between K)^{H^5}=(L^{H^5}\between K^{H^5})^{H^5}\quad (L\parallel K)^{H^5}=(L^{H^5}\parallel K^{H^5})^{H^5}$$
$$(L\mid K)^{H^5}=(L^{H^5}\mid K^{H^5})^{H^5}\quad (L\leftmerge K)^{H^5}=(L^{H^5}\leftmerge K^{H^5})^{H^5}$$
$$(L^*)^{H^5}=((L^{H^5})^*)^{H^5}$$
\end{lemma}

Then, we can get the following five soundness theorems.

\begin{theorem}
If $x=^H_1 y$, then $\sembrack{x}^{H^1}_{SCR}=\sembrack{y}^{H^1}_{SCR}$.
\end{theorem}

\begin{theorem}
If $x=^H_2 y$, then $x^{H^2} \sim_p y^{H^2}$.
\end{theorem}

\begin{theorem}
If $x=^H_3 y$, then $x^{H^3} \sim_s y^{H^3}$.
\end{theorem}

\begin{theorem}
If $x=^H_4 y$, then $x^{H^4} \sim_{hp}y^{H^4}$.
\end{theorem}

\begin{theorem}
If $x=^H_5 y$, then $x^{H^5} \sim_{hhp}y^{H^5}$.
\end{theorem}

\subsection{Reduction}\label{reductionH}

After discussion of the soundness, we discuss the completeness and the decidability related to hypotheses.

The completeness problem means that: (1) does $\sembrack{x}^{H^1}_{SCR}=\sembrack{y}^{H^1}_{SCR}$ imply $x=^H_1 y$? (2) does $x^{H^2}\sim_p y^{H^2}$ imply $x=^H_2 y$? (3) does $x^{H^3}\sim_s y^{H^3}$ imply $x=^H_3 y$? (4) does $x^{H^4}\sim_{hp} y^{H^4}$ imply $x=^H_4 y$? (5) does $x^{H^5}\sim_{hhp} y^{H^5}$ imply $x=^H_5 y$?

And the decidability problem means that: (1) can we decide whether $\sembrack{x}^{H^1}_{SCR}=\sembrack{y}^{H^1}_{SCR}$? (2) can we decide whether $x^{H^2}\sim_p y^{H^2}$? (3) can we decide whether $x^{H^3}\sim_s y^{H^3}$? (4) can we decide whether $x^{H^4}\sim_{hp} y^{H^4}$? (5) can we decide whether $x^{H^5}\sim_{hhp} y^{H^5}$? 

Unfortunately, either of these properties may not hold, we need some intermediate solutions.

\begin{definition}[Decidability and completeness]
Let $H$ be a set of hypotheses and $x,y\in\mathcal{T}_{SCR}$. For decidability,

\begin{enumerate}
  \item We call $H$ 1-decidable if $\sembrack{x}^{H^1}_{SCR}=\sembrack{y}^{H^1}_{SCR}$.
  \item We call $H$ 2-decidable if $x^{H^2}\sim_p y^{H^2}$.
  \item We call $H$ 3-decidable if $x^{H^3}\sim_s y^{H^3}$.
  \item We call $H$ 4-decidable if $x^{H^4}\sim_{hp} y^{H^4}$.
  \item We call $H$ 5-decidable if $x^{H^5}\sim_{hhp} y^{H^5}$.
\end{enumerate}

For completeness,

\begin{enumerate}
  \item We call $H$ 1-complete if $\sembrack{x}^{H^1}_{SCR}=\sembrack{y}^{H^1}_{SCR}$ implies $x=^H_1 y$.
  \item We call $H$ 2-complete if $x^{H^2}\sim_p y^{H^2}$ implies $x=^H_2 y$.
  \item We call $H$ 3-complete if $x^{H^3}\sim_s y^{H^3}$ implies $x=^H_3 y$.
  \item We call $H$ 4-complete if $x^{H^4}\sim_{hp} y^{H^4}$ implies $x=^H_4 y$.
  \item We call $H$ 5-complete if $x^{H^5}\sim_{hhp} y^{H^5}$ implies $x=^H_5 y$.
\end{enumerate}
\end{definition}

\begin{definition}[Implication]
We say that $H$ implies $H'$ if for every hypothesis $x\leq y\in H'$, it holds that $x\leq y\in H$, and there are five variants:

\begin{enumerate}
  \item 1-implication: if $x\leqq^{H'}_1 y$, then $x\leqq^{H}_1 y$.
  \item 2-implication: if $x\leqq^{H'}_2 y$, then $x\leqq^{H}_2 y$.
  \item 3-implication: if $x\leqq^{H'}_3 y$, then $x\leqq^{H}_3 y$.
  \item 4-implication: if $x\leqq^{H'}_4 y$, then $x\leqq^{H}_4 y$.
  \item 5-implication: if $x\leqq^{H'}_5 y$, then $x\leqq^{H}_5 y$.
\end{enumerate}
\end{definition}

\begin{lemma}
Let $H$ and $H'$ be sets of hypotheses, then the following hold:

\begin{enumerate}
  \item For $x,y\in\mathcal{T}_{SCR}$,
  \begin{enumerate}
    \item If $H$ 1-implies $H'$ and $x=^{H'}_1 y$, then $x=^H_1 y$.
    \item If $H$ 2-implies $H'$ and $x=^{H'}_2 y$, then $x=^H_2 y$.
    \item If $H$ 3-implies $H'$ and $x=^{H'}_3 y$, then $x=^H_3 y$.
    \item If $H$ 4-implies $H'$ and $x=^{H'}_4 y$, then $x=^H_4 y$.
    \item If $H$ 5-implies $H'$ and $x=^{H'}_5 y$, then $x=^H_5 y$.
  \end{enumerate}
  \item For $L\subseteq\mathsf{Pomc}$,
  \begin{enumerate}
    \item If $H$ 1-implies $H'$, then $L^{H'^1}\subseteq L^{H^1}$.
    \item If $H$ 2-implies $H'$, then $L^{H'^2}\subseteq L^{H^2}$.
    \item If $H$ 3-implies $H'$, then $L^{H'^3}\subseteq L^{H^3}$.
    \item If $H$ 4-implies $H'$, then $L^{H'^4}\subseteq L^{H^4}$.
    \item If $H$ 5-implies $H'$, then $L^{H'^5}\subseteq L^{H^5}$.
  \end{enumerate}
  \item
  \begin{enumerate}
    \item If $H$ 1-implies $H'$ and $H'$ also 1-implies $H$, then $H$ is 1-decidable (resp. complete) if and only if $H'$ is too.
    \item If $H$ 2-implies $H'$ and $H'$ also 2-implies $H$, then $H$ is 2-decidable (resp. complete) if and only if $H'$ is too.
    \item If $H$ 3-implies $H'$ and $H'$ also 3-implies $H$, then $H$ is 3-decidable (resp. complete) if and only if $H'$ is too.
    \item If $H$ 4-implies $H'$ and $H'$ also 4-implies $H$, then $H$ is 4-decidable (resp. complete) if and only if $H'$ is too.
    \item If $H$ 5-implies $H'$ and $H'$ also 5-implies $H$, then $H$ is 5-decidable (resp. complete) if and only if $H'$ is too.
  \end{enumerate}
\end{enumerate}
\end{lemma}

\begin{definition}[1-Reduction]
Let $H$ and $H'$ be sets of hypotheses such that $H$ 1-implies $H'$. A computable function $r:\mathcal{T}_{SCR}\rightarrow\mathcal{T}_{SCR}$ is an 1-reduction from $H$ to $H'$ such that:

\begin{enumerate}
  \item For $x\in\mathcal{T}_{SCR}$, it holds that $x=^H_1 r(x)$.
  \item For $x,y\in\mathcal{T}_{SCR}$, if $\sembrack{x}^{H^1}_{SCR}=\sembrack{y}^{H^1}_{SCR}$, then $\sembrack{r(x)}^{H'^1}_{SCR}=\sembrack{r(y)}^{H'^1}_{SCR}$.
\end{enumerate}

If the above two conditions are replaced by the two following ones, $r$ is called a strong 1-reduction:

\begin{enumerate}
  \item For $x\in\mathcal{T}_{SCR}$, it holds that $r(x)\leqq^H_1 x$ and $x\leqq^{H'}_1 r(x)$.
  \item For $x\in\mathcal{T}_{SCR}$, it holds that $\sembrack{x}^{H^1}_{SCR}=\sembrack{r(x)}^{H'^1}_{SCR}$.
\end{enumerate}
\end{definition}

\begin{definition}[2-Reduction]
Let $H$ and $H'$ be sets of hypotheses such that $H$ 2-implies $H'$. A computable function $r:\mathcal{T}_{SCR}\rightarrow\mathcal{T}_{SCR}$ is a 2-reduction from $H$ to $H'$ such that:

\begin{enumerate}
  \item For $x\in\mathcal{T}_{SCR}$, it holds that $x=^H_2 r(x)$.
  \item For $x,y\in\mathcal{T}_{SCR}$, if $x^{H^2}\sim_p y^{H^2}$, then $r(x)^{H'^2}\sim_p r(y)^{H'^2}$.
\end{enumerate}

If the above two conditions are replaced by the two following ones, $r$ is called a strong 2-reduction:

\begin{enumerate}
  \item For $x\in\mathcal{T}_{SCR}$, it holds that $r(x)\leqq^H_2 x$ and $x\leqq^{H'}_2 r(x)$.
  \item For $x\in\mathcal{T}_{SCR}$, it holds that $x^{H^2}\sim_p r(x)^{H'^2}$.
\end{enumerate}
\end{definition}

\begin{definition}[3-Reduction]
Let $H$ and $H'$ be sets of hypotheses such that $H$ 3-implies $H'$. A computable function $r:\mathcal{T}_{SCR}\rightarrow\mathcal{T}_{SCR}$ is a 3-reduction from $H$ to $H'$ such that:

\begin{enumerate}
  \item For $x\in\mathcal{T}_{SCR}$, it holds that $x=^H_3 r(x)$.
  \item For $x,y\in\mathcal{T}_{SCR}$, if $x^{H^3}\sim_s y^{H^3}$, then $r(x)^{H'^3}\sim_s r(y)^{H'^3}$.
\end{enumerate}

If the above two conditions are replaced by the two following ones, $r$ is called a strong 3-reduction:

\begin{enumerate}
  \item For $x\in\mathcal{T}_{SCR}$, it holds that $r(x)\leqq^H_3 x$ and $x\leqq^{H'}_3 r(x)$.
  \item For $x\in\mathcal{T}_{SCR}$, it holds that $x^{H^3}\sim_s r(x)^{H'^3}$.
\end{enumerate}
\end{definition}

\begin{definition}[4-Reduction]
Let $H$ and $H'$ be sets of hypotheses such that $H$ 4-implies $H'$. A computable function $r:\mathcal{T}_{SCR}\rightarrow\mathcal{T}_{SCR}$ is a 4-reduction from $H$ to $H'$ such that:

\begin{enumerate}
  \item For $x\in\mathcal{T}_{SCR}$, it holds that $x=^H_4 r(x)$.
  \item For $x,y\in\mathcal{T}_{SCR}$, if $x^{H^4}\sim_{hp} y^{H^4}$, then $r(x)^{H'^4}\sim_{hp} r(y)^{H'^4}$.
\end{enumerate}

If the above two conditions are replaced by the two following ones, $r$ is called a strong 4-reduction:

\begin{enumerate}
  \item For $x\in\mathcal{T}_{SCR}$, it holds that $r(x)\leqq^H_4 x$ and $x\leqq^{H'}_4 r(x)$.
  \item For $x\in\mathcal{T}_{SCR}$, it holds that $x^{H^4}\sim_{hp} r(x)^{H'^4}$.
\end{enumerate}
\end{definition}

\begin{definition}[5-Reduction]
Let $H$ and $H'$ be sets of hypotheses such that $H$ 5-implies $H'$. A computable function $r:\mathcal{T}_{SCR}\rightarrow\mathcal{T}_{SCR}$ is a 5-reduction from $H$ to $H'$ such that:

\begin{enumerate}
  \item For $x\in\mathcal{T}_{SCR}$, it holds that $x=^H_5 r(x)$.
  \item For $x,y\in\mathcal{T}_{SCR}$, if $x^{H^5}\sim_{hhp} y^{H^5}$, then $r(x)^{H'^5}\sim_{hhp}r(y)^{H'^5}$.
\end{enumerate}

If the above two conditions are replaced by the two following ones, $r$ is called a strong 5-reduction:

\begin{enumerate}
  \item For $x\in\mathcal{T}_{SCR}$, it holds that $r(x)\leqq^H_5 x$ and $x\leqq^{H'}_5 r(x)$.
  \item For $x\in\mathcal{T}_{SCR}$, it holds that $x^{H^5}\sim_{hhp}r(x)^{H'^5}$.
\end{enumerate}
\end{definition}

\begin{lemma}
If $H$ is 1-reducible to $H'$ and $H'$ is 1-decidable (resp. complete), then so is $H$.
\end{lemma}

\begin{lemma}
If $H$ is 2-reducible to $H'$ and $H'$ is 2-decidable (resp. complete), then so is $H$.
\end{lemma}

\begin{lemma}
If $H$ is 3-reducible to $H'$ and $H'$ is 3-decidable (resp. complete), then so is $H$.
\end{lemma}

\begin{lemma}
If $H$ is 4-reducible to $H'$ and $H'$ is 4-decidable (resp. complete), then so is $H$.
\end{lemma}

\begin{lemma}
If $H$ is 5-reducible to $H'$ and $H'$ is 5-decidable (resp. complete), then so is $H$.
\end{lemma}

Finding a reduction $r$ sometime is difficult, the following provide two special kinds of reductions.

\subsubsection{Reification}

\begin{definition}[1-Reification]
Let $\Gamma\subseteq\Sigma$ be a fixed subalphabet, $H$ and $H'$ be sets of hypotheses such that $H$ 1-implies $H'$, and $r:\Sigma\rightarrow\mathcal{T}_{SCR}(\Gamma)$ be computable. We call $r$ an 1-reification when the following hold:

\begin{enumerate}
  \item For all $a\in\Sigma$, it holds that $r(a)=^H_1 a$.
  \item For all $x,y\in\mathcal{T}_{SCR}$ and $x\leq y\in H$, it holds that $r(x)\leqq^H_1 r(y)$.
\end{enumerate}
\end{definition}

\begin{definition}[2-Reification]
Let $\Gamma\subseteq\Sigma$ be a fixed subalphabet, $H$ and $H'$ be sets of hypotheses such that $H$ 2-implies $H'$, and $r:\Sigma\rightarrow\mathcal{T}_{SCR}(\Gamma)$ be computable. We call $r$ a 2-reification when the following hold:

\begin{enumerate}
  \item For all $a\in\Sigma$, it holds that $r(a)=^H_2 a$.
  \item For all $x,y\in\mathcal{T}_{SCR}$ and $x\leq y\in H$, it holds that $r(x)\leqq^H_2 r(y)$.
\end{enumerate}
\end{definition}

\begin{definition}[3-Reification]
Let $\Gamma\subseteq\Sigma$ be a fixed subalphabet, $H$ and $H'$ be sets of hypotheses such that $H$ 3-implies $H'$, and $r:\Sigma\rightarrow\mathcal{T}_{SCR}(\Gamma)$ be computable. We call $r$ a 3-reification when the following hold:

\begin{enumerate}
  \item For all $a\in\Sigma$, it holds that $r(a)=^H_3 a$.
  \item For all $x,y\in\mathcal{T}_{SCR}$ and $x\leq y\in H$, it holds that $r(x)\leqq^H_3 r(y)$.
\end{enumerate}
\end{definition}

\begin{definition}[4-Reification]
Let $\Gamma\subseteq\Sigma$ be a fixed subalphabet, $H$ and $H'$ be sets of hypotheses such that $H$ 4-implies $H'$, and $r:\Sigma\rightarrow\mathcal{T}_{SCR}(\Gamma)$ be computable. We call $r$ a 4-reification when the following hold:

\begin{enumerate}
  \item For all $a\in\Sigma$, it holds that $r(a)=^H_4 a$.
  \item For all $x,y\in\mathcal{T}_{SCR}$ and $x\leq y\in H$, it holds that $r(x)\leqq^H_4 r(y)$.
\end{enumerate}
\end{definition}

\begin{definition}[5-Reification]
Let $\Gamma\subseteq\Sigma$ be a fixed subalphabet, $H$ and $H'$ be sets of hypotheses such that $H$ 5-implies $H'$, and $r:\Sigma\rightarrow\mathcal{T}_{SCR}(\Gamma)$ be computable. We call $r$ a 5-reification when the following hold:

\begin{enumerate}
  \item For all $a\in\Sigma$, it holds that $r(a)=^H_5 a$.
  \item For all $x,y\in\mathcal{T}_{SCR}$ and $x\leq y\in H$, it holds that $r(x)\leqq^H_5 r(y)$.
\end{enumerate}
\end{definition}

\begin{lemma}
Let $\Gamma\subseteq\Sigma$ be a fixed subalphabet, $H$ and $H'$ be sets of hypotheses such that $H$ 1-implies $H'$, and $r:\Sigma\rightarrow\mathcal{T}_{SCR}(\Gamma)$ be an 1-reification, the following hold:

\begin{enumerate}
  \item For all $x\in\mathcal{T}_{SCR}$, it holds that $r(\sembrack{x}_{SCR})=^H_1\sembrack{r(x)}_{SCR}$.
  \item For all $L\subseteq\mathsf{SCP}(\Sigma)$, it holds that $r(L^{H^1})^{H'^1}=r(L)^{H'^1}$.
\end{enumerate}
\end{lemma}

\begin{lemma}
Let $\Gamma\subseteq\Sigma$ be a fixed subalphabet, $H$ and $H'$ be sets of hypotheses such that $H$ 2-implies $H'$, and $r:\Sigma\rightarrow\mathcal{T}_{SCR}(\Gamma)$ be a 2-reification, the following hold:

\begin{enumerate}
  \item For all $x\in\mathcal{T}_{SCR}$, it holds that $r(x^{H^2})=^H_2 r(x)^{H^2}$.
  \item For all $L\subseteq\mathsf{SCP}(\Sigma)$, it holds that $r(L^{H^2})^{H'^2}=r(L)^{H'^2}$.
\end{enumerate}
\end{lemma}

\begin{lemma}
Let $\Gamma\subseteq\Sigma$ be a fixed subalphabet, $H$ and $H'$ be sets of hypotheses such that $H$ 3-implies $H'$, and $r:\Sigma\rightarrow\mathcal{T}_{SCR}(\Gamma)$ be a 3-reification, the following hold:

\begin{enumerate}
  \item For all $x\in\mathcal{T}_{SCR}$, it holds that $r(x^{H^3})=^H_3 r(x)^{H^3}$.
  \item For all $L\subseteq\mathsf{SCP}(\Sigma)$, it holds that $r(L^{H^3})^{H'^3}=r(L)^{H'^3}$.
\end{enumerate}
\end{lemma}

\begin{lemma}
Let $\Gamma\subseteq\Sigma$ be a fixed subalphabet, $H$ and $H'$ be sets of hypotheses such that $H$ 4-implies $H'$, and $r:\Sigma\rightarrow\mathcal{T}_{SCR}(\Gamma)$ be a 4-reification, the following hold:

\begin{enumerate}
  \item For all $x\in\mathcal{T}_{SCR}$, it holds that $r(x^{H^4})=^H_4 r(x)^{H^4}$.
  \item For all $L\subseteq\mathsf{SCP}(\Sigma)$, it holds that $r(L^{H^4})^{H'^4}=r(L)^{H'^4}$.
\end{enumerate}
\end{lemma}

\begin{lemma}
Let $\Gamma\subseteq\Sigma$ be a fixed subalphabet, $H$ and $H'$ be sets of hypotheses such that $H$ 5-implies $H'$, and $r:\Sigma\rightarrow\mathcal{T}_{SCR}(\Gamma)$ be a 5-reification, the following hold:

\begin{enumerate}
  \item For all $x\in\mathcal{T}_{SCR}$, it holds that $r(x^{H^5})=^H_5 r(x)^{H^5}$.
  \item For all $L\subseteq\mathsf{SCP}(\Sigma)$, it holds that $r(L^{H^5})^{H'^5}=r(L)^{H'^5}$.
\end{enumerate}
\end{lemma}

\begin{lemma}
If $H$ and $H'$ be sets of hypotheses such that $H$ 1-implies $H'$, then any 1-reification $r$ from $H$ to $H'$ is an 1-reduction from $H$ to $H'$.
\end{lemma}

\begin{lemma}
If $H$ and $H'$ be sets of hypotheses such that $H$ 2-implies $H'$, then any 2-reification $r$ from $H$ to $H'$ is a 2-reduction from $H$ to $H'$.
\end{lemma}

\begin{lemma}
If $H$ and $H'$ be sets of hypotheses such that $H$ 3-implies $H'$, then any 3-reification $r$ from $H$ to $H'$ is a 3-reduction from $H$ to $H'$.
\end{lemma}

\begin{lemma}
If $H$ and $H'$ be sets of hypotheses such that $H$ 4-implies $H'$, then any 4-reification $r$ from $H$ to $H'$ is a 4-reduction from $H$ to $H'$.
\end{lemma}

\begin{lemma}
If $H$ and $H'$ be sets of hypotheses such that $H$ 5-implies $H'$, then any 5-reification $r$ from $H$ to $H'$ is a 5-reduction from $H$ to $H'$.
\end{lemma}

\subsubsection{Lifting}

We can generalize those reductional procedures from rational expressions in section \ref{rla} and lift them to scr-expressions.

We recall the definition of rational expressions $\mathcal{T}_{R}$ \cref{rationalexpressions}.

\rationalexpressions*

The corresponding algebra modulo language equivalence is called Kleene Algebra (KA), and the axioms of KA is shown in Table \ref{AxiomsForKA}, we use $=_{R,1}$ to denote the smallest congruence induced by KA's axioms, and $\leqq_{R,1}$ to stand for $x+y=_{R,1} y$ for $x,y\in\mathcal{T}_R$; and the corresponding algebra modulo bisimilarity is called Mil, and the axioms of Mil is shown in Table \ref{AxiomsForMil}, we use $=_{R,2}$ to denote the smallest congruence induced by Mil's axioms, and $\leqq_{R,2}$ to stand for $x+y=_{R,2} y$ for $x,y\in\mathcal{T}_R$. We write $=_R$ and $\leqq_R$ which stand both for $=_{R,1}$ and $\leqq_{R,1}$, and $=_{R,2}$ and $\leqq_{R,2}$.

\begin{definition}[Sequential hypotheses]
A sequential hypothesis is a hypothesis $x\leq y$ for $x,y\in\mathcal{T}_R$. Similarly, we write $=^H_R$ for the smallest KA congruence on $\mathcal{T}_{R}$ that satisfies the containments in $H$, i.e., whenever $x\leq y\in H$, also $x\leqq^H_R y$ ($x+y=^H_R y$).
\end{definition}

\begin{definition}[Sequential 1-closure]
Let $H$ be a set of hypotheses, and $L\subseteq\Sigma^*$, we define the sequential 1-$H$-closure of $L$, written $L^{\langle H^1\rangle}$, as the smallest language containing $L$ and satisfying:

$$\frac{x\leq y\in H\quad h,z\in\Sigma^*\quad h\cdot\sembrack{y}^{\langle H^1\rangle}_{R}\cdot z\subseteq L^{\langle H^1\rangle}}{h\cdot\sembrack{x}^{\langle H^1\rangle}_{R}\cdot z\subseteq L^{\langle H^1\rangle}}$$
\end{definition}

\begin{definition}[Sequential 2-closure]
Let $H$ be a set of sequential hypotheses, and $L\subseteq\Sigma^*$, we define the sequential 2-$H$-closure of $L$, written $L^{\langle H^2\rangle}$, as the smallest language containing $L$ and satisfying:

$$\frac{x\leq y\in H\quad h,z\in\Sigma^*\quad h\cdot y^{\langle H^2\rangle}\cdot z\subseteq L^{\langle H^2\rangle}\quad x\downarrow}{h\cdot x^{\langle H^2\rangle}\cdot z\subseteq L^{\langle H^2\rangle}\quad y\downarrow}$$
$$\frac{x\leq y\in H\quad h,z\in\Sigma^*\quad h\cdot y^{\langle H^2\rangle}\cdot z\subseteq L^{\langle H^2\rangle}\quad x\xrightarrow{a}x'}{h\cdot x^{\langle H^2\rangle}\cdot z\subseteq L^{\langle H^2\rangle}\quad y\xrightarrow{a}y'}$$
\end{definition}

\begin{definition}[Sequential implication]
We say that $H$ sequentially implies $H'$ if for every hypothesis $x\leq y\in H'$, it holds that $x\leq y\in H$, and there are two variants:

\begin{enumerate}
  \item Sequential 1-implication: if $x\leqq^{H'}_{R,1} y$, then $x\leqq^{H}_{R,1} y$.
  \item Sequential 2-implication: if $x\leqq^{H'}_{R,2} y$, then $x\leqq^{H}_{R,2} y$.
\end{enumerate}
\end{definition}

\begin{lemma}
Let $H$ and $H'$ be sets of hypotheses, then the following hold:

\begin{enumerate}
  \item For $x,y\in\mathcal{T}_{R}$,
  \begin{enumerate}
    \item If $H$ sequentially 1-implies $H'$ and $x=^{H'}_{R,1} y$, then $x=^H_{R,1} y$.
    \item If $H$ sequentially 2-implies $H'$ and $x=^{H'}_{R,2} y$, then $x=^H_{R,2} y$.
  \end{enumerate}
  \item For $L\subseteq\Sigma^*$,
  \begin{enumerate}
    \item If $H$ sequentially 1-implies $H'$, then $L^{\langle H'^1\rangle}\subseteq L^{\langle H^1\rangle}$.
    \item If $H$ sequentially 2-implies $H'$, then $L^{\langle H'^2\rangle}\subseteq L^{\langle H^2\rangle}$.
  \end{enumerate}
  \item
  \begin{enumerate}
    \item If $H$ sequentially 1-implies $H'$ and $H'$ also sequentially 1-implies $H$, then $H$ is decidable (resp. complete) if and only if $H'$ is too.
    \item If $H$ sequentially 2-implies $H'$ and $H'$ also sequentially 2-implies $H$, then $H$ is decidable (resp. complete) if and only if $H'$ is too.
  \end{enumerate}
\end{enumerate}
\end{lemma}

\begin{definition}[Sequential 1-reduction]
Let $H$ and $H'$ be sets of hypotheses such that $H$ sequentially 1-implies $H'$. A computable function $r:\mathcal{T}_{R}\rightarrow\mathcal{T}_{R}$ is a sequential 1-reduction from $H$ to $H'$ such that:

\begin{enumerate}
  \item For $x\in\mathcal{T}_{R}$, it holds that $r(x)\leqq^H_{R,1} x$ and $x\leqq^{H'}_{R,1} r(x)$.
  \item For $x,y\in\mathcal{T}_{R}$, it holds that $\sembrack{x}_{R}^{\langle H^1\rangle}=\sembrack{r(x)}_{R}^{\langle H'^1\rangle}$.
\end{enumerate}
\end{definition}

\begin{definition}[Sequential 2-reduction]
Let $H$ and $H'$ be sets of hypotheses such that $H$ sequentially 2-implies $H'$. A computable function $r:\mathcal{T}_{R}\rightarrow\mathcal{T}_{R}$ is a sequential 2-reduction from $H$ to $H'$ such that:

\begin{enumerate}
  \item For $x\in\mathcal{T}_{R}$, it holds that $r(x)\leqq^H_{R,2} x$ and $x\leqq^{H'}_{R,2} r(x)$.
  \item For $x,y\in\mathcal{T}_{R}$, it holds that $x^{\langle H^2\rangle}\sim_{HM}r(x)^{\langle H'^2\rangle}$.
\end{enumerate}
\end{definition}

\begin{definition}[Grounded hypotheses]
A sequential hypothesis $x\leq y$ is grounded if $y=a_1\cdots a_n$ for some $a_1,\cdots,a_n\in\Sigma$ with $n\geq 1$. A set of hypotheses $H$ is grounded if all its members are.
\end{definition}

\begin{lemma}
Let $H$ be grounded. 

\begin{enumerate}
  \item If $L\subseteq\Sigma^*$, then $L^{H^1}=L^{\langle H^1\rangle}$. Furthermore, for $L,L'\subseteq\mathsf{SCP}$, we have that $(L\between L')^{H^1}=L^{H^1}\between L'^{H^1}$, $(L\parallel L')^{H^1}=L^{H^1}\parallel L'^{H^1}$ and $(L\mid L')^{H^1}=L^{H^1}\mid L'^{H^1}$.
  \item If $L\subseteq\Sigma^*$, then $L^{H^2}=L^{\langle H^2\rangle}$. Furthermore, for $L,L'\subseteq\mathsf{SCP}$, we have that $(L\between L')^{H^2}=L^{H^2}\between L'^{H^2}$, $(L\parallel L')^{H^2}=L^{H^2}\parallel L'^{H^2}$ and $(L\mid L')^{H^2}=L^{H^2}\mid L'^{H^2}$.
  \item If $L\subseteq\Sigma^*$, then $L^{H^3}=L^{\langle H^2\rangle}$. Furthermore, for $L,L'\subseteq\mathsf{SCP}$, we have that $(L\between L')^{H^3}=L^{H^3}\between L'^{H^3}$, $(L\parallel L')^{H^3}=L^{H^3}\parallel L'^{H^3}$ and $(L\mid L')^{H^3}=L^{H^3}\mid L'^{H^3}$.
  \item If $L\subseteq\Sigma^*$, then $L^{H^4}=L^{\langle H^2\rangle}$. Furthermore, for $L,L'\subseteq\mathsf{SCP}$, we have that $(L\between L')^{H^4}=L^{H^4}\between L'^{H^4}$, $(L\parallel L')^{H^4}=L^{H^4}\parallel L'^{H^4}$ and $(L\mid L')^{H^4}=L^{H^4}\mid L'^{H^4}$.
  \item If $L\subseteq\Sigma^*$, then $L^{H^5}=L^{\langle H^2\rangle}$. Furthermore, for $L,L'\subseteq\mathsf{SCP}$, we have that $(L\between L')^{H^5}=L^{H^5}\between L'^{H^5}$, $(L\parallel L')^{H^5}=L^{H^5}\parallel L'^{H^5}$, $(L\mid L')^{H^5}=L^{H^5}\mid L'^{H^5}$ and $(L\leftmerge L')^{H^5}=L^{H^5}\leftmerge L'^{H^5}$.
\end{enumerate}

\end{lemma}

\begin{lemma}

Let $H$ and $H'$ be grounded, The following conclusions hold:

\begin{enumerate}
  \item $r$ be a sequential 1-reduction from $H$ to $H'$. If we extend $r$ to $r:\mathcal{T}_{SCR}\rightarrow\mathcal{T}_{SCR}$ by setting $r(x\between y)=r(x)\between r(y)$, $r(x\parallel y)=r(x)\parallel r(y)$ and $r(x\mid y)=r(x)\mid r(y)$, then, $r$ is a strong 1-reduction from $H$ to $H'$.
  \item $r$ be a sequential 2-reduction from $H$ to $H'$. If we extend $r$ to $r:\mathcal{T}_{SCR}\rightarrow\mathcal{T}_{SCR}$ by setting $r(x\between y)=r(x)\between r(y)$, $r(x\parallel y)=r(x)\parallel r(y)$ and $r(x\mid y)=r(x)\mid r(y)$, then, $r$ is a strong 2-reduction from $H$ to $H'$.
  \item $r$ be a sequential 2-reduction from $H$ to $H'$. If we extend $r$ to $r:\mathcal{T}_{SCR}\rightarrow\mathcal{T}_{SCR}$ by setting $r(x\between y)=r(x)\between r(y)$, $r(x\parallel y)=r(x)\parallel r(y)$ and $r(x\mid y)=r(x)\mid r(y)$, then, $r$ is a strong 3-reduction from $H$ to $H'$.
  \item $r$ be a sequential 2-reduction from $H$ to $H'$. If we extend $r$ to $r:\mathcal{T}_{SCR}\rightarrow\mathcal{T}_{SCR}$ by setting $r(x\between y)=r(x)\between r(y)$, $r(x\parallel y)=r(x)\parallel r(y)$ and $r(x\mid y)=r(x)\mid r(y)$, then, $r$ is a strong 4-reduction from $H$ to $H'$.
  \item $r$ be a sequential 2-reduction from $H$ to $H'$. If we extend $r$ to $r:\mathcal{T}_{SCR}\rightarrow\mathcal{T}_{SCR}$ by setting $r(x\between y)=r(x)\between r(y)$, $r(x\parallel y)=r(x)\parallel r(y)$, $r(x\mid y)=r(x)\mid r(y)$ and $r(x\leftmerge y)=r(x)\leftmerge r(y)$ then, $r$ is a strong 5-reduction from $H$ to $H'$.
\end{enumerate}
\end{lemma}

For such sequential reductions that can be acted as a strong reduction from $H$ to $H'$, we can specialize the results about series-communication rational systems to the so-called rational systems for rational expressions.

\begin{definition}[Rational system modulo language equivalence]
Let $Q$ be a finite set. A rational system modulo language equivalence on $Q$, is a pair $\mathcal{S}=\langle M,b\rangle$, where $M:Q^2\rightarrow\mathcal{T}_{R}$ and $b:Q\rightarrow\mathcal{T}_{R}$. Let $=_{R,1}$ be a KA language equivalence on $\mathcal{T}_{R}(\Delta)$ with $\Sigma\subseteq\Delta$ and $x\in\mathcal{T}_{R}$. We call $s:Q\rightarrow\mathcal{T}_{R}(\Delta)$ a $\langle =_{R,1},x\rangle$-solution to $\mathcal{S}$ if for $q\in Q$:

$$b(q)\cdot x+\sum_{q'\in Q}M(q,q')\cdot s(q')\leqq_{R,1} s(q)$$

Lastly, $s$ is the least $\langle =_{R,1},x\rangle$-solution, if for every such solution $s'$ and every $q\in Q$, we have $s(q)\leqq_{R,1} s'(q)$.
\end{definition}

\begin{definition}[Rational system modulo bisimilarity]
Let $Q$ be a finite set. A rational system modulo bisimilarity on $Q$, is a pair $\mathcal{S}=\langle M,b\rangle$, where $M:Q^2\rightarrow\mathcal{T}_{R}$ and $b:Q\rightarrow\mathcal{T}_{R}$. Let $=_{R,2}$ be an Mil bisimilarity on $\mathcal{T}_{R}(\Delta)$ with $\Sigma\subseteq\Delta$ and $x\in\mathcal{T}_{R}$. We call $s:Q\rightarrow\mathcal{T}_{R}(\Delta)$ a $\langle =_{R,2},x\rangle$-solution to $\mathcal{S}$ if for $q\in Q$:

$$b(q)\cdot x+\sum_{q'\in Q}M(q,q')\cdot s(q')\leqq_{R,2} s(q)$$

Lastly, $s$ is the least $\langle =_{R,2},x\rangle$-solution, if for every such solution $s'$ and every $q\in Q$, we have $s(q)\leqq_{R,2} s'(q)$.
\end{definition}

\begin{theorem}
Let $\mathcal{S}=\langle M,b\rangle$ be an rational system modulo language equivalence on $Q$. We can construct an $s:Q\rightarrow \mathcal{T}_{R}$ such that, for any KA language equivalence $=_{R,1}$ on $\mathcal{T}_{R}(\Delta)$ with $\Sigma\subseteq\Delta$ and any $x\in\mathcal{T}_{R}$, the $Q$-vector $s^x:Q\rightarrow\mathcal{T}_{R}$ is the least $\langle =_{R,1},x\rangle$-solution to $\mathcal{S}$; we call such an $s$ the least solution to $\mathcal{S}$.
\end{theorem}

\begin{theorem}
Let $\mathcal{S}=\langle M,b\rangle$ be an rational system modulo bisimilarity on $Q$. We can construct an $s:Q\rightarrow \mathcal{T}_{R}$ such that, for any Mil bisimilarity $=_{R,2}$ on $\mathcal{T}_{R}(\Delta)$ with $\Sigma\subseteq\Delta$ and any $x\in\mathcal{T}_{R}$, the $Q$-vector $s^x:Q\rightarrow\mathcal{T}_{R}$ is the least $\langle =_{R,2},x\rangle$-solution to $\mathcal{S}$; we call such an $s$ the least solution to $\mathcal{S}$.
\end{theorem}

\subsubsection{Decomposition}

\begin{lemma}
Let $H_0$, $H_1$ and $H'$ be sets of hypotheses. If $r$ is an 1-reduction from $H_0$ to $H'$, and $r'$ is an 1-reduction from $H'$ to $H_1$, then $r'\circ r$ is an 1-reduction from $H_0$ to $H_1$.
\end{lemma}

\begin{lemma}
Let $H_0$, $H_1$ and $H'$ be sets of hypotheses. If $r$ is a 2-reduction from $H_0$ to $H'$, and $r'$ is a 2-reduction from $H'$ to $H_1$, then $r'\circ r$ is a 2-reduction from $H_0$ to $H_1$.
\end{lemma}

\begin{lemma}
Let $H_0$, $H_1$ and $H'$ be sets of hypotheses. If $r$ is a 3-reduction from $H_0$ to $H'$, and $r'$ is a 3-reduction from $H'$ to $H_1$, then $r'\circ r$ is a 3-reduction from $H_0$ to $H_1$.
\end{lemma}

\begin{lemma}
Let $H_0$, $H_1$ and $H'$ be sets of hypotheses. If $r$ is a 4-reduction from $H_0$ to $H'$, and $r'$ is a 4-reduction from $H'$ to $H_1$, then $r'\circ r$ is a 4-reduction from $H_0$ to $H_1$.
\end{lemma}

\begin{lemma}
Let $H_0$, $H_1$ and $H'$ be sets of hypotheses. If $r$ is a 5-reduction from $H_0$ to $H'$, and $r'$ is a 5-reduction from $H'$ to $H_1$, then $r'\circ r$ is a 5-reduction from $H_0$ to $H_1$.
\end{lemma}

\begin{definition}[Factorization]
\begin{enumerate}
  \item $H$ 1-factorizes into $H_1,\cdots,H_n$ if for every $L\subseteq\mathsf{SCP}$ the following holds: $L^{H^1}=((L^{H_1^1})\cdots)^{H_n^1}$.
  \item $H$ 2-factorizes into $H_1,\cdots,H_n$ if for every $L\subseteq\mathsf{SCP}$ the following holds: $L^{H^2}=((L^{H_1^2})\cdots)^{H_n^2}$.
  \item $H$ 3-factorizes into $H_1,\cdots,H_n$ if for every $L\subseteq\mathsf{SCP}$ the following holds: $L^{H^3}=((L^{H_1^3})\cdots)^{H_n^3}$.
  \item $H$ 4-factorizes into $H_1,\cdots,H_n$ if for every $L\subseteq\mathsf{SCP}$ the following holds: $L^{H^4}=((L^{H_1^4})\cdots)^{H_n^4}$.
  \item $H$ 5-factorizes into $H_1,\cdots,H_n$ if for every $L\subseteq\mathsf{SCP}$ the following holds: $L^{H^5}=((L^{H_1^5})\cdots)^{H_n^5}$.
\end{enumerate}
\end{definition}

\begin{lemma}
Let $H$ be a set of hypotheses that 1-factorizes into $H_1,\cdots,H_n$, for $1\leq i\leq n$, $H$ 1-implies $H_i$ and $H_i$ strongly 1-reduces to $H_n$, then $H$ strongly 1-reduces to $H_n$.
\end{lemma}

\begin{lemma}
Let $H$ be a set of hypotheses that 2-factorizes into $H_1,\cdots,H_n$, for $1\leq i\leq n$, $H$ 2-implies $H_i$ and $H_i$ strongly 2-reduces to $H_n$, then $H$ strongly 2-reduces to $H_n$.
\end{lemma}

\begin{lemma}
Let $H$ be a set of hypotheses that 3-factorizes into $H_1,\cdots,H_n$, for $1\leq i\leq n$, $H$ 3-implies $H_i$ and $H_i$ strongly 3-reduces to $H_n$, then $H$ strongly 3-reduces to $H_n$.
\end{lemma}

\begin{lemma}
Let $H$ be a set of hypotheses that 4-factorizes into $H_1,\cdots,H_n$, for $1\leq i\leq n$, $H$ 4-implies $H_i$ and $H_i$ strongly 4-reduces to $H_n$, then $H$ strongly 4-reduces to $H_n$.
\end{lemma}

\begin{lemma}
Let $H$ be a set of hypotheses that 5-factorizes into $H_1,\cdots,H_n$, for $1\leq i\leq n$, $H$ 5-implies $H_i$ and $H_i$ strongly 5-reduces to $H_n$, then $H$ strongly 5-reduces to $H_n$.
\end{lemma}

\begin{lemma}
Let $H_1,\cdots,H_n$ be sets of hypotheses, if $H_i\cup H_j$ 1-factorizes into $H_i$, $H_j$, then $H_1\cup\cdots\cup H_n$ 1-factorizes into $H_1,\cdots,H_n$ where $1\leq i\leq j\leq n$.
\end{lemma} 

\begin{lemma}
Let $H_1,\cdots,H_n$ be sets of hypotheses, if $H_i\cup H_j$ 2-factorizes into $H_i$, $H_j$, then $H_1\cup\cdots\cup H_n$ 2-factorizes into $H_1,\cdots,H_n$ where $1\leq i\leq j\leq n$.
\end{lemma} 

\begin{lemma}
Let $H_1,\cdots,H_n$ be sets of hypotheses, if $H_i\cup H_j$ 3-factorizes into $H_i$, $H_j$, then $H_1\cup\cdots\cup H_n$ 3-factorizes into $H_1,\cdots,H_n$ where $1\leq i\leq j\leq n$.
\end{lemma} 

\begin{lemma}
Let $H_1,\cdots,H_n$ be sets of hypotheses, if $H_i\cup H_j$ 4-factorizes into $H_i$, $H_j$, then $H_1\cup\cdots\cup H_n$ 4-factorizes into $H_1,\cdots,H_n$ where $1\leq i\leq j\leq n$.
\end{lemma} 

\begin{lemma}
Let $H_1,\cdots,H_n$ be sets of hypotheses, if $H_i\cup H_j$ 5-factorizes into $H_i$, $H_j$, then $H_1\cup\cdots\cup H_n$ 5-factorizes into $H_1,\cdots,H_n$ where $1\leq i\leq j\leq n$.
\end{lemma} 
\newpage\section{The Exchange Laws}\label{tel} 

The so-called exchange laws describes the laws of two branches executing in parallel. As we discussed in \cref{conc}, the executions of different branches include the asynchronous executions in different branches and the communications between events in different branches, which are modeled by concurrent composition $\between$, parallel composition $\parallel$, communication merge $\mid$ and left parallel composition $\leftmerge$ in $\mathcal{T}_{SCR}$. BKAC with the exchange laws is also called Concurrent Kleene Algebra with Communication, abbreviated CKAC. Firstly, we give the exchange laws as sets $\mathsf{exchs}$ and $\mathsf{exchs}'$ of hypotheses. Then, we conclude that both $\mathsf{exchs}$ and $\mathsf{exchs}'$ reduce to the empty set of hypotheses in section \ref{reductionE}. Finally, we introduce some decomposition-related conclusions in section \ref{decompositionE}.

\begin{definition}[The exchange laws as hypotheses]
The set of hypotheses $\mathsf{exchs}$ is give by:

\begin{center}
$$\mathsf{exchb}=\{(x\between y)\cdot(z\between h)\leq(x\cdot z)\between(y\cdot h):x,y,z,h\in\mathcal{T}_{SCR}\}$$
$$\mathsf{exchp}=\{(x\parallel y)\cdot(z\between h)\leq(x\cdot z)\parallel(y\cdot h):x,y,z,h\in\mathcal{T}_{SCR}\}$$
$$\mathsf{exchl}=\{(a\leftmerge b)\cdot(x\between y)\leq(a\cdot x)\leftmerge(b\cdot y):a\leq^e b;a,b\in\Sigma;x,y\in\mathcal{T}_{SCR}\}$$
$$\mathsf{exchc}=\{\rho(a,b)\cdot(x\between y)\leq(a\cdot x)\mid(b\cdot y):a\leq^c b;a,b\in\Sigma;x,y\in\mathcal{T}_{SCR}\}$$
$$\mathsf{exchs}=\mathsf{exchb}\cup\mathsf{exchp}\cup\mathsf{exchc}$$
$$\mathsf{exchs}'=\mathsf{exchb}\cup\mathsf{exchl}\cup\mathsf{exchc}$$
\end{center}
\end{definition} 

The following lemma says that the exchange laws also hold in the level of pomsets and subsumption.

\begin{lemma}
Let $\sqsubseteq^{\mathsf{SCP}}$ be $\sqsubseteq$ restricted to $\mathsf{SCP}$, then $\sqsubseteq^{\mathsf{SCP}}$ is the smallest precongruence satisfying the exchange laws, i.e., for all $a,b\in\Sigma$ and $U,V,W,X\in\mathsf{SCP}$, these hold that:

$$(U\between V)\cdot(W\between X)\sqsubseteq^{\mathsf{SCP}}(U\cdot W)\between(V\cdot X)$$
$$(U\parallel V)\cdot(W\between X)\sqsubseteq^{\mathsf{SCP}}(U\cdot W)\parallel(V\cdot X)$$
$$(a\leftmerge b)\cdot(U\between V)\sqsubseteq^{\mathsf{SCP}}(a\cdot U)\leftmerge(b\cdot V):a\leq^e b$$
$$\rho(a,b)\cdot(U\between V)\sqsubseteq^{\mathsf{SCP}}(a\cdot U)\mid(b\cdot V):a\leq^c b$$
\end{lemma}

\begin{corollary}
Let $L\subseteq\mathsf{SCP}$, the following hold that:

\begin{enumerate}
  \item $U\in L^{\mathsf{exchs}^1}$ if and only if there exists a $V\in L$ such that $U\sqsubseteq V$.
  \item $U\in L^{\mathsf{exchs}^2}$ if and only if there exists a $V\in L$ such that $U\sqsubseteq V$.
  \item $U\in L^{\mathsf{exchs}^3}$ if and only if there exists a $V\in L$ such that $U\sqsubseteq V$.
  \item $U\in L^{\mathsf{exchs}^4}$ if and only if there exists a $V\in L$ such that $U\sqsubseteq V$.
  \item $U\in L^{\mathsf{exchs}'^5}$ if and only if there exists a $V\in L$ such that $U\sqsubseteq V$.
\end{enumerate}
\end{corollary}

\subsection{Reduction}\label{reductionE}

We will show that $\mathsf{exchs}$ and $\mathsf{exchs}'$ strongly reduce to the empty set of hypotheses.

\begin{definition}[1-Closure]
Let $x\in\mathcal{T}_{SCR}$, then $x\downarrow^1\in\mathcal{T}_{SCR}$ is an 1-closure of $x$ if $x\downarrow^1\leqq^{\mathsf{exchs}}_1 x\leqq_1 x\downarrow^1$ and $\sembrack{x\downarrow^1}^{\mathsf{exch}^1}_{SCR}=\sembrack{x}^{\mathsf{exch}^1}_{SCR}\downarrow^1$.
\end{definition}

\begin{definition}[2-Closure]
Let $x\in\mathcal{T}_{SCR}$, then $x\downarrow^2\in\mathcal{T}_{SCR}$ is a 2-closure of $x$ if $x\downarrow^2\leqq^{\mathsf{exchs}}_2 x\leqq_2 x\downarrow^2$ and $(x\downarrow^2)^{\mathsf{exch}^2}\sim_p x^{\mathsf{exch}^2}\downarrow^2$.
\end{definition}

\begin{definition}[3-Closure]
Let $x\in\mathcal{T}_{SCR}$, then $x\downarrow^3\in\mathcal{T}_{SCR}$ is a 3-closure of $x$ if $x\downarrow^3\leqq^{\mathsf{exchs}}_3 x\leqq_3 x\downarrow^3$ and $(x\downarrow^3)^{\mathsf{exch}^3}\sim_s x^{\mathsf{exch}^3}\downarrow^3$.
\end{definition}

\begin{definition}[4-Closure]
Let $x\in\mathcal{T}_{SCR}$, then $x\downarrow^4\in\mathcal{T}_{SCR}$ is a 4-closure of $x$ if $x\downarrow^4\leqq^{\mathsf{exchs}}_4 x\leqq_4 x\downarrow^4$ and $(x\downarrow^4)^{\mathsf{exch}^4}\sim_{hp}x^{\mathsf{exch}^4}\downarrow^4$.
\end{definition}

\begin{definition}[5-Closure]
Let $x\in\mathcal{T}_{SCR}$, then $x\downarrow^5\in\mathcal{T}_{SCR}$ is a 5-closure of $x$ if $x\downarrow^5\leqq^{\mathsf{exchs}'}_5 x\leqq_5 x\downarrow^5$ and $(x\downarrow^5)^{\mathsf{exch}'^5}\sim_{hhp}x^{\mathsf{exch}'^5}\downarrow^5$.
\end{definition}

The following is related to how to exactly compute such closures.

\begin{definition}[$\between$-depth]
The $\between$-depth of $x\in\mathcal{T}_{SCR}$ denoted $d_{\between}(x)$ is defined inductively on the structure of $x$ as follows.

$$d_{\between}(0)=0\quad d_{\between}(1)=0\quad d_{\between}(a)=0$$
$$d_{\between}(x\cdot y)=\max(d_{\between}(x),d_{\between}(y))\quad d_{\between}(x+ y)=\max(d_{\between}(x),d_{\between}(y))\quad d_{\between}(x^*)=d_{\between}(x)$$ 
$$d_{\between}(x\between y)=\max(d_{\between}(x\parallel y),d_{\between}(x\mid y))\quad d_{\between}(x\parallel y)=\max(d_{\between}(x),d_{\between}(y))+1$$
$$d_{\between}(x\leftmerge y)=\max(d_{\between}(x),d_{\between}(y))+1\quad d_{\between}(x\mid y)=\max(d_{\between}(x),d_{\between}(y))+1$$
\end{definition}

\begin{definition}[$\parallel$-depth]
The $\parallel$-depth of $x\in\mathcal{T}_{SCR}$ denoted $d_{\parallel}(x)$ is defined inductively on the structure of $x$ as follows.

$$d_{\parallel}(0)=0\quad d_{\parallel}(1)=0\quad d_{\parallel}(a)=0$$
$$d_{\parallel}(x\cdot y)=\max(d_{\parallel}(x),d_{\parallel}(y))\quad d_{\parallel}(x+ y)=\max(d_{\parallel}(x),d_{\parallel}(y))\quad d_{\parallel}(x^*)=d_{\parallel}(x)$$ 
$$d_{\parallel}(x\between y)=\max(d_{\parallel}(x\parallel y),d_{\parallel}(x\mid y))\quad d_{\parallel}(x\parallel y)=\max(d_{\parallel}(x),d_{\parallel}(y))+1$$
$$d_{\parallel}(x\leftmerge y)=\max(d_{\parallel}(x),d_{\parallel}(y))+1\quad d_{\parallel}(x\mid y)=\max(d_{\parallel}(x),d_{\parallel}(y))+1$$
\end{definition}

\begin{definition}[$\leftmerge$-depth]
The $\leftmerge$-depth of $x\in\mathcal{T}_{SCR}$ denoted $d_{\leftmerge}(x)$ is defined inductively on the structure of $x$ as follows.

$$d_{\leftmerge}(0)=0\quad d_{\leftmerge}(1)=0\quad d_{\leftmerge}(a)=0$$
$$d_{\leftmerge}(x\cdot y)=\max(d_{\leftmerge}(x),d_{\leftmerge}(y))\quad d_{\leftmerge}(x+ y)=\max(d_{\leftmerge}(x),d_{\leftmerge}(y))\quad d_{\leftmerge}(x^*)=d_{\leftmerge}(x)$$ 
$$d_{\leftmerge}(x\between y)=\max(d_{\leftmerge}(x\parallel y),d_{\leftmerge}(x\mid y))\quad d_{\leftmerge}(x\parallel y)=\max(d_{\leftmerge}(x\leftmerge y),d_{\leftmerge}(y\leftmerge x))$$
$$d_{\leftmerge}(x\leftmerge y)=\max(d_{\leftmerge}(x),d_{\leftmerge}(y))+1\quad d_{\leftmerge}(x\mid y)=\max(d_{\leftmerge}(x),d_{\leftmerge}(y))+1$$
\end{definition}

\begin{definition}[$\mid$-depth]
The $\mid$-depth of $x\in\mathcal{T}_{SCR}$ denoted $d_{\mid}(x)$ is defined inductively on the structure of $x$ as follows.

$$d_{\mid}(0)=0\quad d_{\mid}(1)=0\quad d_{\mid}(a)=0$$
$$d_{\mid}(x\cdot y)=\max(d_{\mid}(x),d_{\mid}(y))\quad d_{\mid}(x+ y)=\max(d_{\mid}(x),d_{\mid}(y))\quad d_{\mid}(x^*)=d_{\mid}(x)$$ 
$$d_{\mid}(x\between y)=\max(d_{\mid}(x\parallel y),d_{\mid}(x\mid y))\quad d_{\mid}(x\parallel y)=\max(d_{\mid}(x),d_{\mid}(y))+1$$
$$d_{\mid}(x\leftmerge y)=\max(d_{\mid}(x),d_{\mid}(y))+1\quad d_{\mid}(x\mid y)=\max(d_{\mid}(x),d_{\mid}(y))+1$$
\end{definition}

\begin{definition}[The main induction hypothesis]\label{mih}
The main induction hypothesis in computing a closure of $x\in\mathcal{T}_{SCR}$: 

\begin{enumerate}
  \item if $y\in\mathcal{T}_{SCR}$ and $d_{\between}(y)<d_{\between}(x)$, then we can compute the 1-closure of $y$, $y\downarrow^1$; the 2-closure of $y$, $y\downarrow^2$; the 3-closure of $y$, $y\downarrow^3$; the 4-closure of $y$, $y\downarrow^4$; and the 5-closure of $y$, $y\downarrow^5$.
  \item if $y\in\mathcal{T}_{SCR}$ and $d_{\parallel}(y)<d_{\parallel}(x)$, then we can compute the 1-closure of $y$, $y\downarrow^1$; the 2-closure of $y$, $y\downarrow^2$; the 3-closure of $y$, $y\downarrow^3$; and the 4-closure of $y$, $y\downarrow^4$.
  \item if $y\in\mathcal{T}_{SCR}$ and $d_{\mid}(y)<d_{\mid}(x)$, then we can compute the 1-closure of $y$, $y\downarrow^1$; the 2-closure of $y$, $y\downarrow^2$; the 3-closure of $y$, $y\downarrow^3$; the 4-closure of $y$, $y\downarrow^4$; and the 5-closure of $y$, $y\downarrow^5$.
  \item if $y\in\mathcal{T}_{SCR}$ and $d_{\leftmerge}(y)<d_{\leftmerge}(x)$, then we can compute the 5-closure of $y$, $y\downarrow^5$.
\end{enumerate}
\end{definition}

The closures of $x\in\mathcal{T}_{SCR}$ except concurrent composition $\between$, parallel composition $\parallel$, left parallel composition $\leftmerge$ and communication merge $\mid$ can be computed according to the following lemma.

\begin{lemma}
Let $L,K\subseteq\mathsf{SCP}$ and $a\in\Sigma$, the following hold:

$$\{1\}\downarrow^1=\{1\}\quad\{a\}\downarrow^1=\{a\}\quad (L\cup K)\downarrow^1=L\downarrow^1\cup K\downarrow^1\quad (L\cdot K)\downarrow^1=L\downarrow^1\cdot K\downarrow^1\quad L^*\downarrow^1=(L\downarrow^1)^*$$
$$\{1\}\downarrow^2=\{1\}\quad\{a\}\downarrow^2=\{a\}\quad (L\cup K)\downarrow^2=L\downarrow^2\cup K\downarrow^2\quad (L\cdot K)\downarrow^2=L\downarrow^2\cdot K\downarrow^2\quad L^*\downarrow^2=(L\downarrow^2)^*$$
$$\{1\}\downarrow^3=\{1\}\quad\{a\}\downarrow^3=\{a\}\quad (L\cup K)\downarrow^3=L\downarrow^3\cup K\downarrow^3\quad (L\cdot K)\downarrow^3=L\downarrow^3\cdot K\downarrow^3\quad L^*\downarrow^3=(L\downarrow^3)^*$$
$$\{1\}\downarrow^4=\{1\}\quad\{a\}\downarrow^4=\{a\}\quad (L\cup K)\downarrow^4=L\downarrow^4\cup K\downarrow^4\quad (L\cdot K)\downarrow^4=L\downarrow^4\cdot K\downarrow^4\quad L^*\downarrow^4=(L\downarrow^4)^*$$
$$\{1\}\downarrow^5=\{1\}\quad\{a\}\downarrow^5=\{a\}\quad (L\cup K)\downarrow^5=L\downarrow^5\cup K\downarrow^5\quad (L\cdot K)\downarrow^5=L\downarrow^5\cdot K\downarrow^5\quad L^*\downarrow^5=(L\downarrow^5)^*$$
\end{lemma}

The computation of closures of scr-expressions with the forms of $x\between y$, $x\parallel y$, $x\leftmerge y$ and $x\mid y$ where $x,y\in\mathcal{T}_{SCR}$ can be done by use of the exchange laws $\mathsf{exchs}$ and $\mathsf{exchs}'$.

\subsubsection{Preclosure}

For concurrent composition $\between$, parallel composition $\parallel$, left parallel composition $\leftmerge$ and communication merge $\mid$, we do not need a full closure and give the definition of preclosure as follows.

\begin{definition}[1-Preclosure]
Let $x\in\mathcal{T}_{SCR}$, an 1-preclosure of $x$ is an scr-expressions $\tilde{x}\in\mathcal{T}_{SCR}$ with $\tilde{x}\leqq^{\mathsf{exchs}}_1 x$ and $x\leqq_1\tilde{x}$. And if $U\in\sembrack{x}^{\mathsf{exchs}^1}_{SCR}\downarrow^1$ is a sequential prime, then $U\in\sembrack{\tilde{x}}^{\mathsf{exchs}^1}_{SCR}$.
\end{definition}

\begin{definition}[2-Preclosure]
Let $x\in\mathcal{T}_{SCR}$, a 2-preclosure of $x$ is an scr-expressions $\tilde{x}\in\mathcal{T}_{SCR}$ with $\tilde{x}\leqq^{\mathsf{exchs}}_2 x$ and $x\leqq_2\tilde{x}$. And if $U\in x^{\mathsf{exchs}^2}\downarrow^2$ is a sequential prime, then $U\in \tilde{x}^{\mathsf{exchs}^2}$.
\end{definition}

\begin{definition}[3-Preclosure]
Let $x\in\mathcal{T}_{SCR}$, a 3-preclosure of $x$ is an scr-expressions $\tilde{x}\in\mathcal{T}_{SCR}$ with $\tilde{x}\leqq^{\mathsf{exchs}}_3 x$ and $x\leqq_3\tilde{x}$. And if $U\in x^{\mathsf{exchs}^3}\downarrow^3$ is a sequential prime, then $U\in\tilde{x}^{\mathsf{exchs}^3}$.
\end{definition}

\begin{definition}[4-Preclosure]
Let $x\in\mathcal{T}_{SCR}$, a 4-preclosure of $x$ is an scr-expressions $\tilde{x}\in\mathcal{T}_{SCR}$ with $\tilde{x}\leqq^{\mathsf{exchs}}_4 x$ and $x\leqq_4\tilde{x}$. And if $U\in x^{\mathsf{exchs}^4}\downarrow^4$ is a sequential prime, then $U\in\tilde{x}^{\mathsf{exchs}^4}$.
\end{definition}

\begin{definition}[5-Preclosure]
Let $x\in\mathcal{T}_{SCR}$, a 5-preclosure of $x$ is an scr-expressions $\tilde{x}\in\mathcal{T}_{SCR}$ with $\tilde{x}\leqq^{\mathsf{exchs}'}_5 x$ and $x\leqq_5\tilde{x}$. And if $U\in x^{\mathsf{exchs}'^5}\downarrow^5$ is a sequential prime, then $U\in\tilde{x}^{\mathsf{exchs}'^5}$.
\end{definition}

\begin{definition}[Concurrent splitting]
We define $\triangle$ as the smallest subset of $\mathcal{T}_{SCR}\times\binom{\mathcal{T}_{SCR}}{2}$ that satisfying the following rules, where $x,y\in\mathcal{T}_{SCR}$ and $\langle x,\mset{\ell,r}\rangle\in\triangle$ is denoted by $\ell\triangle_x r$.

$$\frac{}{x\triangle_{x\between y} y}\quad \frac{\ell\triangle_{x}r\quad \ell'\triangle_{y}r'}{\ell\between \ell'\triangle_{x\between y}r\between r'} \quad\frac{\ell\triangle_x r}{\ell\triangle_{x\between y}r\between y} \quad\frac{\ell\triangle_y r}{x\between\ell\triangle_{x\between y}r}$$
$$\frac{}{x\triangle_{x\parallel y} y}\quad \frac{\ell\triangle_{x}r\quad \ell'\triangle_{y}r'}{\ell\parallel \ell'\triangle_{x\parallel y}r\parallel r'} \quad\frac{\ell\triangle_x r}{\ell\triangle_{x\parallel y}r\parallel y} \quad\frac{\ell\triangle_y r}{x\parallel\ell\triangle_{x\parallel y}r}$$
$$\frac{}{x\triangle_{x\leftmerge y} y}\quad \frac{\ell\triangle_{x}r\quad \ell'\triangle_{y}r'}{\ell\leftmerge \ell'\triangle_{x\leftmerge y}r\leftmerge r'} \quad\frac{\ell\triangle_x r}{\ell\triangle_{x\leftmerge y}r\leftmerge y} \quad\frac{\ell\triangle_y r}{x\leftmerge\ell\triangle_{x\leftmerge y}r}$$
$$\frac{}{x\triangle_{x\mid y} y}\quad \frac{\ell\triangle_{x}r\quad \ell'\triangle_{y}r'}{\ell\mid \ell'\triangle_{x\mid y}r\mid r'} \quad\frac{\ell\triangle_x r}{\ell\triangle_{x\mid y}r\mid y} \quad\frac{\ell\triangle_y r}{x\mid\ell\triangle_{x\mid y}r}$$
$$\frac{\ell\triangle_x r}{\ell\triangle_{x+y} r} \quad\frac{\ell\triangle_y r}{\ell\triangle_{x+y} r} \quad\frac{\ell\triangle_x r}{\ell\triangle_{x^*} r} \quad\frac{\ell\triangle_x r\quad y\in\mathcal{F}_{SCR}}{\ell\triangle_{x\cdot y} r} \quad\frac{\ell\triangle_y r\quad x\in\mathcal{F}_{SCR}}{\ell\triangle_{x\cdot y} r}$$

$\ell$ and $r$ from a concurrent splitting of $x$ when $\ell\triangle_x r$.
\end{definition}

\begin{lemma}
Let $x\in\mathcal{T}_{SCR}$, and $U,V\in\mathsf{SCP}$ are non-empty and $U\between V\in x$, then there exist $\ell,r\in\mathcal{T}_{SCR}$ with $\ell\triangle_x r$ such that $U\in \ell$ and $V\in r$.
\end{lemma}

\begin{lemma}
Let $x\in\mathcal{T}_{SCR}$, and $U,V\in\mathsf{SCP}$ are non-empty and $U\parallel V\in x$, then there exist $\ell,r\in\mathcal{T}_{SCR}$ with $\ell\triangle_x r$ such that $U\in \ell$ and $V\in r$.
\end{lemma}

\begin{lemma}
Let $x\in\mathcal{T}_{SCR}$, and $U,V\in\mathsf{SCP}$ are non-empty and $U\leftmerge V\in x$, then there exist $\ell,r\in\mathcal{T}_{SCR}$ with $\ell\triangle_x r$ such that $U\in \ell$ and $V\in r$.
\end{lemma}

\begin{lemma}
Let $x\in\mathcal{T}_{SCR}$, and $U,V\in\mathsf{SCP}$ are non-empty and $U\mid V\in x$, then there exist $\ell,r\in\mathcal{T}_{SCR}$ with $\ell\triangle_x r$ such that $U\in \ell$ and $V\in r$.
\end{lemma}

\begin{lemma}
Let $x\in\mathcal{T}_{SCR}$, then the following hold:

\begin{enumerate}
  \item There are finitely many $\ell,r\in\mathcal{T}_{SCR}$ such that $\ell\triangle_x r$.
  \item
  \begin{enumerate}
    \item if $\ell\triangle_x r$, then $\ell\between r\leqq_1 x$, $\ell\parallel r\leqq_1 x$ and $\ell\mid r\leqq_1 x$;
    \item if $\ell\triangle_x r$, then $\ell\between r\leqq_2 x$, $\ell\parallel r\leqq_2 x$ and $\ell\mid r\leqq_2 x$;
    \item if $\ell\triangle_x r$, then $\ell\between r\leqq_3 x$, $\ell\parallel r\leqq_3 x$ and $\ell\mid r\leqq_3 x$;
    \item if $\ell\triangle_x r$, then $\ell\between r\leqq_4 x$, $\ell\parallel r\leqq_4 x$ and $\ell\mid r\leqq_4 x$;
    \item if $\ell\triangle_x r$, then $\ell\between r\leqq_5 x$, $\ell\leftmerge r\leqq_5 x$ and $\ell\mid r\leqq_5 x$;
  \end{enumerate}
  \item If $\ell\triangle_x r$, then 
  \begin{enumerate}
    \item $d_{\between}(\ell),d_{\between}(r)<d_{\between}(x)$;
    \item $d_{\parallel}(\ell),d_{\parallel}(r)<d_{\parallel}(x)$;
    \item $d_{\leftmerge}(\ell),d_{\leftmerge}(r)<d_{\leftmerge}(x)$;
    \item $d_{\mid}(\ell),d_{\mid}(r)<d_{\mid}(x)$.
  \end{enumerate}
\end{enumerate}
\end{lemma}

\begin{definition}[Syntactic construction of preclosures]
Let $x,y\in\mathcal{T}_{SCR}$, and the main induction hypothesis in \cref{mih} applies to $x\between y$, $x\parallel y$, $x\leftmerge y$ and $x\mid y$. The scr-expressions $x\odot_{\between}^1 y$, $x\odot_{\between}^2 y$, $x\odot_{\between}^3 y$, $x\odot_{\between}^4 y$ and $x\odot_{\between}^5 y$ are defined as follows:

$$x\odot_{\between}^1 y=\sum_{\ell\triangle_{x\between y}r}\ell\downarrow^1\between r\downarrow^1$$
$$x\odot_{\between}^2 y=\sum_{\ell\triangle_{x\between y}r}\ell\downarrow^2\between r\downarrow^2$$
$$x\odot_{\between}^3 y=\sum_{\ell\triangle_{x\between y}r}\ell\downarrow^3\between r\downarrow^3$$
$$x\odot_{\between}^4 y=\sum_{\ell\triangle_{x\between y}r}\ell\downarrow^4\between r\downarrow^4$$
$$x\odot_{\between}^5 y=\sum_{\ell\triangle_{x\between y}r}\ell\downarrow^5\between r\downarrow^5$$

The scr-expressions $x\odot_{\parallel}^1 y$, $x\odot_{\parallel}^2 y$, $x\odot_{\parallel}^3 y$ and $x\odot_{\parallel}^4 y$ are defined as follows:

$$x\odot_{\parallel}^1 y=\sum_{\ell\triangle_{x\parallel y}r}\ell\downarrow^1\parallel r\downarrow^1$$
$$x\odot_{\parallel}^2 y=\sum_{\ell\triangle_{x\parallel y}r}\ell\downarrow^2\parallel r\downarrow^2$$
$$x\odot_{\parallel}^3 y=\sum_{\ell\triangle_{x\parallel y}r}\ell\downarrow^3\parallel r\downarrow^3$$
$$x\odot_{\parallel}^4 y=\sum_{\ell\triangle_{x\parallel y}r}\ell\downarrow^4\parallel r\downarrow^4$$

The scr-expressions $x\odot_{\mid}^1 y$, $x\odot_{\mid}^2 y$, $x\odot_{\mid}^3 y$, $x\odot_{\mid}^4 y$ and $x\odot_{\mid}^5 y$ are defined as follows:

$$x\odot_{\mid}^1 y=\sum_{\ell\triangle_{x\mid y}r}\ell\downarrow^1\mid r\downarrow^1$$
$$x\odot_{\mid}^2 y=\sum_{\ell\triangle_{x\mid y}r}\ell\downarrow^2\mid r\downarrow^2$$
$$x\odot_{\mid}^3 y=\sum_{\ell\triangle_{x\mid y}r}\ell\downarrow^3\mid r\downarrow^3$$
$$x\odot_{\mid}^4 y=\sum_{\ell\triangle_{x\mid y}r}\ell\downarrow^4\mid r\downarrow^4$$
$$x\odot_{\mid}^5 y=\sum_{\ell\triangle_{x\mid y}r}\ell\downarrow^5\mid r\downarrow^5$$

The scr-expression $x\odot_{\leftmerge}^5 y$ are defined as follows:

$$x\odot_{\leftmerge}^5 y=\sum_{\ell\triangle_{x\leftmerge y}r}\ell\downarrow^5\leftmerge r\downarrow^5$$
\end{definition}

\begin{lemma}
Let $x,y\in\mathcal{T}_{SCR}$, and the main induction hypothesis in \cref{mih} applies to $x\between y$, $x\parallel y$, $x\leftmerge y$ and $x\mid y$, then,

\begin{enumerate}
  \item 
  \begin{enumerate}
    \item $x\odot_{\between}^1 y$ is the 1-preclosure of $x\between y$;
    \item $x\odot_{\between}^2 y$ is the 2-preclosure of $x\between y$;
    \item $x\odot_{\between}^3 y$ is the 3-preclosure of $x\between y$;
    \item $x\odot_{\between}^4 y$ is the 4-preclosure of $x\between y$;
    \item $x\odot_{\between}^5 y$ is the 5-preclosure of $x\between y$.
  \end{enumerate}
  \item 
  \begin{enumerate}
    \item $x\odot_{\parallel}^1 y$ is the 1-preclosure of $x\parallel y$;
    \item $x\odot_{\parallel}^2 y$ is the 2-preclosure of $x\parallel y$;
    \item $x\odot_{\parallel}^3 y$ is the 3-preclosure of $x\parallel y$;
    \item $x\odot_{\parallel}^4 y$ is the 4-preclosure of $x\parallel y$.
  \end{enumerate}
  \item 
  \begin{enumerate}
    \item $x\odot_{\mid}^1 y$ is the 1-preclosure of $x\mid y$;
    \item $x\odot_{\mid}^2 y$ is the 2-preclosure of $x\mid y$;
    \item $x\odot_{\mid}^3 y$ is the 3-preclosure of $x\mid y$;
    \item $x\odot_{\mid}^4 y$ is the 4-preclosure of $x\mid y$;
    \item $x\odot_{\mid}^5 y$ is the 5-preclosure of $x\mid y$.
  \end{enumerate}
  \item $x\odot_{\leftmerge}^5 y$ is the 5-preclosure of $x\leftmerge y$.
\end{enumerate}
\end{lemma}

\subsubsection{Closure}

\begin{definition}[Sequential splitting]
We define $\nabla$ as the smallest subset of $\mathcal{T}_{SCR}^3$ that satisfying the following rules, where $a\in\Sigma$, $x,y\in\mathcal{T}_{SCR}$ and $\langle x,\ell,r\rangle\in\nabla$ is denoted by $\ell\nabla_x r$.

$$\frac{}{1\nabla_1 1} \quad\frac{}{a\nabla_a 1} \quad\frac{}{1\nabla_a a} \quad\frac{}{1\nabla_{x^*} 1} \quad\frac{\ell\nabla_x r}{\ell\nabla_{x+y} r} \quad\frac{\ell\nabla_y r}{\ell\nabla_{x+y} r}$$
$$\quad\frac{\ell\nabla_x r}{\ell\nabla_{x\cdot y} r\cdot y} \quad\frac{\ell\nabla_y r}{x\cdot\ell\nabla_{x\cdot y} r} \quad\frac{\ell\nabla_x r}{x^*\cdot\ell\nabla_{x^*} r\cdot x^*}$$
$$\frac{\ell\nabla_{x}r\quad \ell'\nabla_{y}r'}{\ell\between \ell'\nabla_{x\between y}r\between r'} \quad\frac{\ell\nabla_{x}r\quad \ell'\nabla_{y}r'}{\ell\parallel \ell'\nabla_{x\parallel y}r\between r'}\quad \frac{\ell\nabla_{x}r\quad \ell'\nabla_{y}r'}{\ell\leftmerge \ell'\nabla_{x\leftmerge y}r\between r'} \quad\frac{\ell\nabla_{x}r\quad \ell'\nabla_{y}r'}{\ell\mid \ell'\nabla_{x\mid y}r\between r'}$$
$\ell$ and $r$ from a sequential splitting of $x$ when $\ell\nabla_x r$.
\end{definition}

\begin{lemma}
Let $x\in\mathcal{T}_{SCR}$, and $U,V\in\mathsf{SCP}$, then the following conclusions hold: 

\begin{enumerate}
  \item If $U\cdot V\in x\downarrow^1$, then there exist $\ell,r\in\mathcal{T}_{SCR}$ with $\ell\nabla_x r$ such that $U\in\ell\downarrow^1$ and $V\in r\downarrow^1$.
  \item If $U\cdot V\in x\downarrow^2$, then there exist $\ell,r\in\mathcal{T}_{SCR}$ with $\ell\nabla_x r$ such that $U\in\ell\downarrow^2$ and $V\in r\downarrow^2$.
  \item If $U\cdot V\in x\downarrow^3$, then there exist $\ell,r\in\mathcal{T}_{SCR}$ with $\ell\nabla_x r$ such that $U\in\ell\downarrow^3$ and $V\in r\downarrow^3$.
  \item If $U\cdot V\in x\downarrow^4$, then there exist $\ell,r\in\mathcal{T}_{SCR}$ with $\ell\nabla_x r$ such that $U\in\ell\downarrow^4$ and $V\in r\downarrow^4$.
  \item If $U\cdot V\in x\downarrow^5$, then there exist $\ell,r\in\mathcal{T}_{SCR}$ with $\ell\nabla_x r$ such that $U\in\ell\downarrow^5$ and $V\in r\downarrow^5$.
\end{enumerate}
\end{lemma}

\begin{lemma}
Let $x\in\mathcal{T}_{SCR}$, there exist $\ell_1,\cdots,\ell_n\in\mathcal{T}_{SCR}$ and $r_1,\cdots,r_n\in\mathcal{F}_{SCR}$ such that for $1\leq i\leq n$ it hold that $\ell_i\nabla_x r_i$ and $x=_1 \ell_1+\cdots+\ell_n$.
\end{lemma}

\begin{lemma}
Let $x\in\mathcal{T}_{SCR}$, there exist $\ell_1,\cdots,\ell_n\in\mathcal{T}_{SCR}$ and $r_1,\cdots,r_n\in\mathcal{F}_{SCR}$ such that for $1\leq i\leq n$ it hold that $\ell_i\nabla_x r_i$ and $x=_2 \ell_1+\cdots+\ell_n$.
\end{lemma}

\begin{lemma}
Let $x\in\mathcal{T}_{SCR}$, there exist $\ell_1,\cdots,\ell_n\in\mathcal{T}_{SCR}$ and $r_1,\cdots,r_n\in\mathcal{F}_{SCR}$ such that for $1\leq i\leq n$ it hold that $\ell_i\nabla_x r_i$ and $x=_3 \ell_1+\cdots+\ell_n$.
\end{lemma}

\begin{lemma}
Let $x\in\mathcal{T}_{SCR}$, there exist $\ell_1,\cdots,\ell_n\in\mathcal{T}_{SCR}$ and $r_1,\cdots,r_n\in\mathcal{F}_{SCR}$ such that for $1\leq i\leq n$ it hold that $\ell_i\nabla_x r_i$ and $x=_4 \ell_1+\cdots+\ell_n$.
\end{lemma}

\begin{lemma}
Let $x\in\mathcal{T}_{SCR}$, there exist $\ell_1,\cdots,\ell_n\in\mathcal{T}_{SCR}$ and $r_1,\cdots,r_n\in\mathcal{F}_{SCR}$ such that for $1\leq i\leq n$ it hold that $\ell_i\nabla_x r_i$ and $x=_5 \ell_1+\cdots+\ell_n$.
\end{lemma}

\begin{lemma}
Let $x\in\mathcal{T}_{SCR}$, then the following hold:

\begin{enumerate}
  \item There are finitely many $\ell,r\in\mathcal{T}_{SCR}$ such that $\ell\nabla_x r$.
  \item
  \begin{enumerate}
    \item if $\ell\nabla_x r$, then $\ell\cdot r\leqq^{\mathsf{exchs}}_1 x$;
    \item if $\ell\nabla_x r$, then $\ell\cdot r\leqq^{\mathsf{exchs}}_2 x$;
    \item if $\ell\nabla_x r$, then $\ell\cdot r\leqq^{\mathsf{exchs}}_3 x$;
    \item if $\ell\nabla_x r$, then $\ell\cdot r\leqq^{\mathsf{exchs}}_4 x$;
    \item if $\ell\nabla_x r$, then $\ell\cdot r\leqq^{\mathsf{exchs}'}_5 x$;
  \end{enumerate}
  \item If $\ell\nabla_x r$, then 
  \begin{enumerate}
    \item $d_{\between}(\ell),d_{\between}(r)\leq d_{\between}(x)$;
    \item $d_{\parallel}(\ell),d_{\parallel}(r)\leq d_{\parallel}(x)$;
    \item $d_{\leftmerge}(\ell),d_{\leftmerge}(r)\leq d_{\leftmerge}(x)$;
    \item $d_{\mid}(\ell),d_{\mid}(r)\leq d_{\mid}(x)$.
  \end{enumerate}
\end{enumerate}
\end{lemma}

\begin{definition}
Let $x\in\mathcal{T}_{SCR}$, the set of (right-hand) remainders of $x$ written $R(x)$, is the smallest set containing $x$ such that if $y\in R(x)$ and $\ell\nabla_y r$, then $r\in R(x)$.
\end{definition}

\begin{lemma}
Let $x\in\mathcal{T}_{SCR}$, $R(x)$ is finite. Furthermore, if $x'\in R(x)$, then

\begin{enumerate}
  \item $d_{\between}(x')\leq d_{\between}(x)$.
  \item $d_{\parallel}(x')\leq d_{\parallel}(x)$.
  \item $d_{\leftmerge}(x')\leq d_{\leftmerge}(x)$.
  \item $d_{\mid}(x')\leq d_{\mid}(x)$.
\end{enumerate}
\end{lemma}

\begin{definition}
Let $x,y\in\mathcal{T}_{SCR}$, and suppose that the main induction hypothesis in \cref{mih} applies to $x\between y$. We define a system $\mathcal{S}=\langle M,b\rangle$ on $Q=\{z\between h: z\in R(x), h\in R(y)\}$, with components given by

$$M(z\between h, z'\between h')=\sum_{z_0\triangle_z z',h_0\triangle_h h'}z_0\odot h_0$$
$$b(z\between h)=[z\between h\in\mathcal{F}_{SCR}]$$

Let $s_1$, $s_2$, $s_3$, $s_4$ and $s_5$ be the least solutions to $\mathcal{S}$ obtained through \cref{ls1}, \cref{ls2}, \cref{ls3}, \cref{ls4} and \cref{ls5} respectively, we write $x\otimes_{\between}^1 y$ for $s_1(x\between y)$, $x\otimes_{\between}^2 y$ for $s_2(x\between y)$, $x\otimes_{\between}^3 y$ for $s_3(x\between y)$, $x\otimes_{\between}^4 y$ for $s_4(x\between y)$ and $x\otimes_{\between}^5 y$ for $s_5(x\between y)$.
\end{definition}

\begin{definition}
Let $x,y\in\mathcal{T}_{SCR}$, and suppose that the main induction hypothesis in \cref{mih} applies to $x\parallel y$. We define a system $\mathcal{S}=\langle M,b\rangle$ on $Q=\{z\parallel h: z\in R(x), h\in R(y)\}$, with components given by

$$M(z\parallel h, z'\parallel h')=\sum_{z_0\triangle_z z',h_0\triangle_h h'}z_0\odot h_0$$
$$b(z\parallel h)=[z\parallel h\in\mathcal{F}_{SCR}]$$

Let $s_1$, $s_2$, $s_3$ and $s_4$ be the least solutions to $\mathcal{S}$ obtained through \cref{ls1}, \cref{ls2}, \cref{ls3} and \cref{ls4} respectively, we write $x\otimes_{\parallel}^1 y$ for $s_1(x\parallel y)$, $x\otimes_{\parallel}^2 y$ for $s_2(x\parallel y)$, $x\otimes_{\parallel}^3 y$ for $s_3(x\parallel y)$ and $x\otimes_{\parallel}^4 y$ for $s_4(x\parallel y)$.
\end{definition}

\begin{definition}
Let $x,y\in\mathcal{T}_{SCR}$, and suppose that the main induction hypothesis in \cref{mih} applies to $x\mid y$. We define a system $\mathcal{S}=\langle M,b\rangle$ on $Q=\{z\mid h: z\in R(x), h\in R(y)\}$, with components given by

$$M(z\mid h, z'\mid h')=\sum_{z_0\triangle_z z',h_0\triangle_h h'}z_0\odot h_0$$
$$b(z\mid h)=[z\mid h\in\mathcal{F}_{SCR}]$$

Let $s_1$, $s_2$, $s_3$, $s_4$ and $s_5$ be the least solutions to $\mathcal{S}$ obtained through \cref{ls1}, \cref{ls2}, \cref{ls3}, \cref{ls4} and \cref{ls5} respectively, we write $x\otimes_{\mid}^1 y$ for $s_1(x\mid y)$, $x\otimes_{\mid}^2 y$ for $s_2(x\mid y)$, $x\otimes_{\mid}^3 y$ for $s_3(x\mid y)$, $x\otimes_{\mid}^4 y$ for $s_4(x\mid y)$ and $x\otimes_{\mid}^5 y$ for $s_5(x\mid y)$.
\end{definition}

\begin{definition}
Let $x,y\in\mathcal{T}_{SCR}$, and suppose that the main induction hypothesis in \cref{mih} applies to $x\leftmerge y$. We define a system $\mathcal{S}=\langle M,b\rangle$ on $Q=\{z\leftmerge h: z\in R(x), h\in R(y)\}$, with components given by

$$M(z\leftmerge h, z'\leftmerge h')=\sum_{z_0\triangle_z z',h_0\triangle_h h'}z_0\odot h_0$$
$$b(z\leftmerge h)=[z\leftmerge h\in\mathcal{F}_{SCR}]$$

Let $s_5$ be the least solution to $\mathcal{S}$ obtained through \cref{ls5}, we write $x\otimes_{\leftmerge}^5 y$ for $s_5(x\leftmerge y)$.
\end{definition}

\begin{lemma}\label{closure}
Let $x,y\in\mathcal{T}_{SCR}$, and suppose that the main induction hypothesis in \cref{mih} applies to $x\between y$, $x\parallel y$, $x\mid y$ and $x\leftmerge y$, then,

\begin{enumerate}
  \item 
  \begin{enumerate}
    \item $x\otimes_{\between}^1 y$ is an 1-closure of $x\between y$;
    \item $x\otimes_{\between}^2 y$ is a 2-closure of $x\between y$;
    \item $x\otimes_{\between}^3 y$ is a 3-closure of $x\between y$;
    \item $x\otimes_{\between}^4 y$ is a 4-closure of $x\between y$;
    \item $x\otimes_{\between}^5 y$ is a 5-closure of $x\between y$.
  \end{enumerate}
  \item 
  \begin{enumerate}
    \item $x\otimes_{\parallel}^1 y$ is an 1-closure of $x\parallel y$;
    \item $x\otimes_{\parallel}^2 y$ is a 2-closure of $x\parallel y$;
    \item $x\otimes_{\parallel}^3 y$ is a 3-closure of $x\parallel y$;
    \item $x\otimes_{\parallel}^4 y$ is a 4-closure of $x\parallel y$.
  \end{enumerate}
  \item 
  \begin{enumerate}
    \item $x\otimes_{\mid}^1 y$ is an 1-closure of $x\mid y$;
    \item $x\otimes_{\mid}^2 y$ is a 2-closure of $x\mid y$;
    \item $x\otimes_{\mid}^3 y$ is a 3-closure of $x\mid y$;
    \item $x\otimes_{\mid}^4 y$ is a 4-closure of $x\mid y$;
    \item $x\otimes_{\mid}^5 y$ is a 5-closure of $x\mid y$.
  \end{enumerate}
  \item $x\otimes_{\leftmerge}^5 y$ is a 5-closure of $x\leftmerge y$.
\end{enumerate}
\end{lemma}

\begin{theorem}
Let $x\in\mathcal{T}_{SCR}$, we can compute an 1-closure $x\downarrow^1$ of $x$ according to \cref{closure}. Hence, $\mathsf{exchs}$ strongly 1-reduces to $\emptyset$.
\end{theorem}

\begin{theorem}
Let $x\in\mathcal{T}_{SCR}$, we can compute a 2-closure $x\downarrow^2$ of $x$ according to \cref{closure}. Hence, $\mathsf{exchs}$ strongly 2-reduces to $\emptyset$.
\end{theorem}

\begin{theorem}
Let $x\in\mathcal{T}_{SCR}$, we can compute a 3-closure $x\downarrow^3$ of $x$ according to \cref{closure}. Hence, $\mathsf{exchs}$ strongly 3-reduces to $\emptyset$.
\end{theorem}

\begin{theorem}
Let $x\in\mathcal{T}_{SCR}$, we can compute a 4-closure $x\downarrow^4$ of $x$ according to \cref{closure}. Hence, $\mathsf{exchs}$ strongly 4-reduces to $\emptyset$.
\end{theorem}

\begin{theorem}
Let $x\in\mathcal{T}_{SCR}$, we can compute a 5-closure $x\downarrow^5$ of $x$ according to \cref{closure}. Hence, $\mathsf{exchs}'$ strongly 5-reduces to $\emptyset$.
\end{theorem}

\begin{corollary}
Let $x,y\in\mathcal{T}_{SCR}$, the following hold:
\begin{enumerate}
  \item It is decidable whether
  \begin{enumerate}
    \item $\sembrack{x}_{SCR}\downarrow^1=\sembrack{y}_{SCR}\downarrow^1$;
    \item $x\downarrow^2\sim_p y\downarrow^2$;
    \item $x\downarrow^3\sim_s y\downarrow^3$;
    \item $x\downarrow^4\sim_{hp} y\downarrow^4$;
    \item $x\downarrow^5\sim_{hhp} y\downarrow^5$.
  \end{enumerate}
  \item 
  \begin{enumerate}
    \item $\sembrack{x}_{SCR}\downarrow^1=\sembrack{y}_{SCR}\downarrow^1$ if and only if $x=^{\mathsf{exchs}}_1 y$;
    \item $x\downarrow^2\sim_p y\downarrow^2$ if and only if $x=^{\mathsf{exchs}}_2 y$;
    \item $x\downarrow^3\sim_s y\downarrow^3$ if and only if $x=^{\mathsf{exchs}}_3 y$;
    \item $x\downarrow^4\sim_{hp} y\downarrow^4$ if and only if $x=^{\mathsf{exchs}}_4 y$;
    \item $x\downarrow^5\sim_{hhp} y\downarrow^5$ if and only if $x=^{\mathsf{exchs}'}_5 y$.
  \end{enumerate}
\end{enumerate}
\end{corollary}

\subsection{Decomposition}\label{decompositionE}

\begin{lemma}
Let $H$ be a set of hypotheses, then $H\cup\mathsf{exchs}$ strongly 1-reduces to $\emptyset$, they hold that:

\begin{enumerate}
  \item $H$ strongly 1-reduces to $\emptyset$.
  \item $H\cup\mathsf{exchs}$ 1-factorizes into $H,\mathsf{exchs}$ or $\mathsf{exchs},H$.
\end{enumerate}
\end{lemma}

\begin{lemma}
Let $H$ be a set of hypotheses, then $H\cup\mathsf{exchs}$ strongly 2-reduces to $\emptyset$, they hold that:

\begin{enumerate}
  \item $H$ strongly 2-reduces to $\emptyset$.
  \item $H\cup\mathsf{exchs}$ 2-factorizes into $H,\mathsf{exchs}$ or $\mathsf{exchs},H$.
\end{enumerate}
\end{lemma}

\begin{lemma}
Let $H$ be a set of hypotheses, then $H\cup\mathsf{exchs}$ strongly 3-reduces to $\emptyset$, they hold that:

\begin{enumerate}
  \item $H$ strongly 3-reduces to $\emptyset$.
  \item $H\cup\mathsf{exchs}$ 3-factorizes into $H,\mathsf{exchs}$ or $\mathsf{exchs},H$.
\end{enumerate}
\end{lemma}

\begin{lemma}
Let $H$ be a set of hypotheses, then $H\cup\mathsf{exchs}$ strongly 4-reduces to $\emptyset$, they hold that:

\begin{enumerate}
  \item $H$ strongly 4-reduces to $\emptyset$.
  \item $H\cup\mathsf{exchs}$ 4-factorizes into $H,\mathsf{exchs}$ or $\mathsf{exchs},H$.
\end{enumerate}
\end{lemma}

\begin{lemma}
Let $H$ be a set of hypotheses, then $H\cup\mathsf{exchs}$ strongly 5-reduces to $\emptyset$, they hold that:

\begin{enumerate}
  \item $H$ strongly 5-reduces to $\emptyset$.
  \item $H\cup\mathsf{exchs}'$ 5-factorizes into $H,\mathsf{exchs}'$ or $\mathsf{exchs}',H$.
\end{enumerate}
\end{lemma}

\begin{definition}[Simple hypotheses]
A hypotheses $x\leq y$ is called left-simple if $x=1$ or $x=a$ with $a\in\mathcal{T}_{SCR}$; similarly, a hypotheses $x\leq y$ is called right-simple if $y=1$ or $y=a$ with $a\in\mathcal{T}_{SCR}$. A set of hypotheses is called left-simple (resp. right-simple) if each of its element is.
\end{definition}

We have defined the concept of pomsetc context $\mathsf{PC}$ in \cref{pc}, next, we give the concept of $\mathsf{SCP}$-pomsetc context denoted $\mathsf{PC}^{\mathsf{SCP}}$.

\begin{definition}[$\mathsf{SCP}$-pomsetc contexts]
For $C=[\mathbf{c}]\in\mathsf{PC}$ with the node labelled by $\square$, $s_{\square}\in S_{\mathbf{c}}$, and a labelled poset $\mathbf{u}$,  we write the plugging $\mathbf{c}[\mathbf{u}]$, where:

\begin{enumerate}
  \item $S_{\mathbf{c}[\mathbf{u}]}=S_{\mathbf{c}}\cup S_{\mathbf{c}}\setminus\{s_{\square}\}$.
  \item $\lambda_{\mathbf{c}[\mathbf{u}]}(s)=\lambda_{\mathbf{c}}(s)$ if $s\in S_{\mathbf{c}}\setminus\{s_{\square}\}$; $\lambda_{\mathbf{c}[\mathbf{u}]}(s)=\lambda_{\mathbf{u}}(s)$ if $s\in S_{\mathbf{u}}$.
  \item $\leq^e_{\mathbf{c}[\mathbf{u}]}$ is the smallest relation on $S_{\mathbf{c}[\mathbf{u}]}$ satisfying the following rules:
  
  $$\frac{s\leq^e_{\mathbf{u}} s'}{s \leq^e_{\mathbf{c}[\mathbf{u}]} s'}\quad \frac{s \leq^e_{\mathbf{c}} s'}{s \leq^e_{\mathbf{c}[\mathbf{u}]} s'} \quad\frac{s_{\square \leq^e_{\mathbf{c}}}s\quad s'\in S_{\mathbf{u}}}{s'\leq^e_{\mathbf{c}[\mathbf{u}]} s} \quad \frac{s'\in S_{\mathbf{u}}\quad s\leq^e_{\mathbf{c}}s_{\square}}{s\leq^e_{\mathbf{c}[\mathbf{u}]} s'}$$
  \item $\leq^c_{\mathbf{c}[\mathbf{u}]}$ is the smallest relation on $S_{\mathbf{c}[\mathbf{u}]}$ satisfying the following rules:
  
  $$\frac{s\leq^c_{\mathbf{u}} s'}{s \leq^c_{\mathbf{c}[\mathbf{u}]} s'}\quad \frac{s \leq^c_{\mathbf{c}} s'}{s \leq^c_{\mathbf{c}[\mathbf{u}]} s'} \quad\frac{s_{\square \leq^c_{\mathbf{c}}}s\quad s'\in S_{\mathbf{u}}}{s'\leq^c_{\mathbf{c}[\mathbf{u}]} s} \quad \frac{s'\in S_{\mathbf{u}}\quad s\leq^c_{\mathbf{c}}s_{\square}}{s\leq^c_{\mathbf{c}[\mathbf{u}]} s'}$$
\end{enumerate} 
\end{definition}

\begin{lemma}
Let $C\in\mathsf{PC}$, the following hold:

\begin{enumerate}
  \item $C\in\mathsf{PC}^{\mathsf{SCP}}$ if and only if $C$ is series-communication-parallel.
  \item if $C=[\mathbf{c}]\in\mathsf{PC}^{\mathsf{SCP}}$ and $U=[\mathbf{u}]\in\mathsf{Pomc}$, then $C[U]=[\mathbf{c}[\mathbf{u}]]$.
\end{enumerate}
\end{lemma}

\begin{lemma}
Let $C\in\mathsf{PC}$, $U\in\mathsf{Pomc}$ and $a\in\Sigma$, the following hold:

\begin{enumerate}
  \item If $C[a]\sqsubseteq U$, then we can construct $C'\in\mathsf{PC}$ with $C\sqsubseteq C'$ and $C'[a]=U$.
  \item If $U\sqsubseteq C[a]$, then we can construct $C'\in\mathsf{PC}$ with $C'\sqsubseteq C$ and $C'[a]=U$.
\end{enumerate}

Moreover, if $U\in\mathsf{SCP}$, then $C'\in\mathsf{SCP}$.
\end{lemma}

\begin{lemma}
Let $C\in\mathsf{PC}$ and $U\mathsf{SCP}$ with $C[1]=U$, the following hold:

\begin{enumerate}
  \item We can construct a $C'\in\mathsf{PC}^{\mathsf{SCP}}$ with $C'[1]=U$ and $C\sqsubseteq C'$.
  \item We can construct a $C'\in\mathsf{PC}^{\mathsf{SCP}}$ with $C'[1]=U$ and $C'\sqsubseteq C$.
\end{enumerate}
\end{lemma}

\begin{lemma}
Let $C\in\mathsf{PC}$ and $U\in\mathsf{Pomc}$, the following hold:

\begin{enumerate}
  \item If $C[1]\sqsubseteq U$, then we can construct a $C'\in\mathsf{PC}$ with $C'[1]=U$ and $C\sqsubseteq C'$.
  \item If $U\sqsubseteq C[1]$, then we can construct a $C'\in\mathsf{PC}^{\mathsf{SCP}}$ with $C'[1]=U$ and $C'\sqsubseteq C$.
\end{enumerate}

Moreover, if $U\in\mathsf{SCP}$, then $C'\in\mathsf{SCP}$.
\end{lemma}

\begin{lemma}
Let $H$ be a set of hypotheses, $x,y\in\mathcal{T}_{SCR}$, $L\subseteq\mathsf{SCP}$ and $C\in\mathsf{PC}^{\mathsf{SCP}}$, the following hold:

\begin{enumerate}
  \item 
  \begin{enumerate}
    \item if $H$ is right-simple and $x\leq y\in H$ with $C[\sembrack{y}_{SCR}]\subseteq(L^{H^1})^{\mathsf{exchs}^1}$, then $C[\sembrack{x}_{SCR}]\subseteq(L^{H^1})^{\mathsf{exchs}^1}$;
    \item if $H$ is right-simple and $x\leq y\in H$ with $C[y]\subseteq(L^{H^2})^{\mathsf{exchs}^2}$, then $C[x]\subseteq(L^{H^2})^{\mathsf{exchs}^2}$;
    \item if $H$ is right-simple and $x\leq y\in H$ with $C[y]\subseteq(L^{H^3})^{\mathsf{exchs}^3}$, then $C[x]\subseteq(L^{H^3})^{\mathsf{exchs}^3}$;
    \item if $H$ is right-simple and $x\leq y\in H$ with $C[y]\subseteq(L^{H^4})^{\mathsf{exchs}^4}$, then $C[x]\subseteq(L^{H^4})^{\mathsf{exchs}^4}$;
    \item if $H$ is right-simple and $x\leq y\in H$ with $C[y]\subseteq(L^{H^5})^{\mathsf{exchs}'^5}$, then $C[x]\subseteq(L^{H^5})^{\mathsf{exchs}'^5}$.
  \end{enumerate}
  
  \item 
  \begin{enumerate}
    \item if $H$ is left-simple and $x\leq y\in H$ with $C[\sembrack{y}_{SCR}]\subseteq(L^{\mathsf{exchs}^1})^{H^1}$, then $C[\sembrack{x}_{SCR}]\subseteq(L^{\mathsf{exchs}^1})^{H^1}$;
    \item if $H$ is left-simple and $x\leq y\in H$ with $C[y]\subseteq(L^{\mathsf{exchs}^2})^{H^2}$, then $C[x]\subseteq(L^{\mathsf{exchs}^2})^{H^2}$;
    \item if $H$ is left-simple and $x\leq y\in H$ with $C[y]\subseteq(L^{\mathsf{exchs}^3})^{H^3}$, then $C[x]\subseteq(L^{\mathsf{exchs}^3})^{H^3}$;
    \item if $H$ is left-simple and $x\leq y\in H$ with $C[y]\subseteq(L^{\mathsf{exchs}^4})^{H^4}$, then $C[x]\subseteq(L^{\mathsf{exchs}^4})^{H^4}$;
    \item if $H$ is left-simple and $x\leq y\in H$ with $C[y]\subseteq(L^{\mathsf{exchs}'^5})^{H^5}$, then $C[x]\subseteq(L^{\mathsf{exchs}'^5})^{H^5}$.
  \end{enumerate}
\end{enumerate}
\end{lemma}

\begin{theorem}
Let $H$ be a set of hypotheses, the following hold:

\begin{enumerate}
  \item 
  \begin{enumerate}
    \item if $H$ is right-simple, then $H\cup\mathsf{exchs}$ 1-factorizes into $H,\mathsf{exchs}$;
    \item if $H$ is right-simple, then $H\cup\mathsf{exchs}$ 2-factorizes into $H,\mathsf{exchs}$;
    \item if $H$ is right-simple, then $H\cup\mathsf{exchs}$ 3-factorizes into $H,\mathsf{exchs}$;
    \item if $H$ is right-simple, then $H\cup\mathsf{exchs}$ 4-factorizes into $H,\mathsf{exchs}$;
    \item if $H$ is right-simple, then $H\cup\mathsf{exchs}'$ 5-factorizes into $H,\mathsf{exchs}'$.
  \end{enumerate}
  \item 
  \begin{enumerate}
    \item if $H$ is left-simple, then $H\cup\mathsf{exchs}$ 1-factorizes into $\mathsf{exchs},H$;
    \item if $H$ is left-simple, then $H\cup\mathsf{exchs}$ 2-factorizes into $\mathsf{exchs},H$;
    \item if $H$ is left-simple, then $H\cup\mathsf{exchs}$ 3-factorizes into $\mathsf{exchs},H$;
    \item if $H$ is left-simple, then $H\cup\mathsf{exchs}$ 4-factorizes into $\mathsf{exchs},H$;
    \item if $H$ is left-simple, then $H\cup\mathsf{exchs}'$ 5-factorizes into $\mathsf{exchs}',H$.
  \end{enumerate}
\end{enumerate}
\end{theorem}

\begin{corollary}
Let $H$ be a set of hypotheses, the following hold:

\begin{enumerate}
  \item If $H$ is left-simple or right-simple and $H$ strongly 1-reduces to $\emptyset$, then $H\cup\mathsf{exchs}$ strongly 1-reduces to $\emptyset$, too.
  \item If $H$ is left-simple or right-simple and $H$ strongly 2-reduces to $\emptyset$, then $H\cup\mathsf{exchs}$ strongly 2-reduces to $\emptyset$, too.
  \item If $H$ is left-simple or right-simple and $H$ strongly 3-reduces to $\emptyset$, then $H\cup\mathsf{exchs}$ strongly 3-reduces to $\emptyset$, too.
  \item If $H$ is left-simple or right-simple and $H$ strongly 4-reduces to $\emptyset$, then $H\cup\mathsf{exchs}$ strongly 4-reduces to $\emptyset$, too.
  \item If $H$ is left-simple or right-simple and $H$ strongly 5-reduces to $\emptyset$, then $H\cup\mathsf{exchs}'$ strongly 5-reduces to $\emptyset$, too.
\end{enumerate}
\end{corollary}

\begin{corollary}
Let $H_{\ell}$ be a set of hypotheses and left-simple, $H_r$ be a set of hypotheses and right-simple, then the following hold:

\begin{enumerate}
  \item if $H_{\ell}\cup H_r$ 1-factorizes into $H_{\ell}, H_r$, then $H_{\ell}\cup\mathsf{exchs}\cup H_r$ 1-factorizes into $H_{\ell},\mathsf{exchs}, H_r$;
  \item if $H_{\ell}\cup H_r$ 2-factorizes into $H_{\ell}, H_r$, then $H_{\ell}\cup\mathsf{exchs}\cup H_r$ 2-factorizes into $H_{\ell},\mathsf{exchs}, H_r$;
  \item if $H_{\ell}\cup H_r$ 3-factorizes into $H_{\ell}, H_r$, then $H_{\ell}\cup\mathsf{exchs}\cup H_r$ 3-factorizes into $H_{\ell},\mathsf{exchs}, H_r$;
  \item if $H_{\ell}\cup H_r$ 4-factorizes into $H_{\ell}, H_r$, then $H_{\ell}\cup\mathsf{exchs}\cup H_r$ 4-factorizes into $H_{\ell},\mathsf{exchs}, H_r$;
  \item if $H_{\ell}\cup H_r$ 5-factorizes into $H_{\ell}, H_r$, then $H_{\ell}\cup\mathsf{exchs}'\cup H_r$ 5-factorizes into $H_{\ell},\mathsf{exchs}', H_r$.
\end{enumerate}
\end{corollary}

\begin{corollary}
Let $H_{\ell}$ be a set of hypotheses and left-simple, $H_r$ be a set of hypotheses and right-simple, the following hold:

\begin{enumerate}
  \item If both of $H_{\ell}$ and $H_r$ strongly 1-reduce to $\emptyset$, and $H_{\ell}\cup H_r$ 1-factorizes into $H_{\ell}, H_r$, then $H_{\ell}\cup\mathsf{exchs}\cup H_r$ strongly 1-reduces to $\emptyset$.
  \item If both of $H_{\ell}$ and $H_r$ strongly 2-reduce to $\emptyset$, and $H_{\ell}\cup H_r$ 2-factorizes into $H_{\ell}, H_r$, then $H_{\ell}\cup\mathsf{exchs}\cup H_r$ strongly 2-reduces to $\emptyset$.
  \item If both of $H_{\ell}$ and $H_r$ strongly 3-reduce to $\emptyset$, and $H_{\ell}\cup H_r$ 3-factorizes into $H_{\ell}, H_r$, then $H_{\ell}\cup\mathsf{exchs}\cup H_r$ strongly 3-reduces to $\emptyset$.
  \item If both of $H_{\ell}$ and $H_r$ strongly 4-reduce to $\emptyset$, and $H_{\ell}\cup H_r$ 4-factorizes into $H_{\ell}, H_r$, then $H_{\ell}\cup\mathsf{exchs}\cup H_r$ strongly 4-reduces to $\emptyset$.
  \item If both of $H_{\ell}$ and $H_r$ strongly 5-reduce to $\emptyset$, and $H_{\ell}\cup H_r$ 5-factorizes into $H_{\ell}, H_r$, then $H_{\ell}\cup\mathsf{exchs}'\cup H_r$ strongly 5-reduces to $\emptyset$.
\end{enumerate}
\end{corollary} 
\newpage\section{Control Flow}\label{controlflow}

Standard control flow can model programming constructs such as conditionals and while-loops, in this chapter, we extend CKAC with Observations, abbreviated CKACO, which is also an extension of CKAO (Concurrent Kleene Algebra with Observations) \cite{CKA7} that origins from the study of Kleene Algebra with Test (KAT) \cite{KA6} \cite{KA7}.

Firstly, we give the definition of the set of Boolean observations.

\begin{definition}[Boolean observations]
We fix a finite set of primitive observations denoted $\Omega$. The set of propositional terms, denoted $\mathcal{T}_{BA}$, is inductively defined by the following grammars:

$$\mathcal{T}_{BA}\ni p,q::=\bot|\top|o\in\Omega|p\vee q|p\wedge q|\bar{p}$$
\end{definition}

For $p,q\in\mathcal{T}_{BA}$, we have the axioms of Boolean observations as Table \ref{AxiomsForBA} shows, where $p\leqq q$ if $p\vee q=q$. Sometimes, for unambiguity, we denote the $=$ and $\leqq$ relations defined by the axioms in Table \ref{AxiomsForBA} as $=_{BA}$ and $\leqq_{BA}$.

\begin{center}
    \begin{table}
        \begin{tabular}{@{}ll@{}}
            \hline No. &Axiom\\
            $BA1$ & $p\vee\bot=p$\\
            $BA2$ & $p\vee q=q\vee p$\\
            $BA3$ & $p\vee\bar{p}=\top$\\
            $BA4$ & $p\vee(q\vee r)=(p\vee q)\vee r$\\
            $BA5$ & $p\wedge\top=p$\\
            $BA6$ & $p\wedge q=q\wedge p$\\
            $BA7$ & $p\wedge\bar{p}=\bot$\\
            $BA8$ & $p\wedge(q\wedge r)=(p\wedge q)\wedge r$\\
            $BA9$ & $p\vee(q\wedge r)=(p\vee q)\wedge(p\vee r)$\\
            $BA10$ & $p\wedge(q\vee r)=(p\wedge q)\vee(p\wedge r)$\\
        \end{tabular}
        \caption{Axioms of Boolean observations}
        \label{AxiomsForBA}
    \end{table}
\end{center}

We denote the set of atoms of the above Boolean algebra as $\mathsf{At}$ for $2^{\Omega}$. For every $\alpha\in\mathsf{At}$, there exists a canonical corresponding propositional term $\pi_{\alpha}$, and every $p\in\mathcal{T}_{BA}$ is equivalent to the disjunction of all $\pi_{\alpha}$ with $\pi_{\alpha}\leqq p$.

We define the terms of CKACO as $\mathcal{T}(\Sigma\cup\mathcal{T}_{BA})$, the CKAC terms over $\Sigma\cup\mathcal{T}_{BA}$, denoted $\mathcal{T}_{CKACO}$. 

\begin{definition}[CKACO]
We define the following set of hypotheses over $\mathcal{T}_{CKACO}$:

$$\mathsf{bool}=\{p=q:p=_{BA}q\textrm{ for }p,q\in\mathcal{T}_{BA}\}$$
$$\mathsf{contr}=\{p\wedge q\leq p\cdot q:p,q\in\mathcal{T}_{BA}\}$$
$$\mathsf{glue}=\{0=\bot\}\cup\{p+q=p\vee q:p,q\in\mathcal{T}_{BA}\}$$
$$\mathsf{obs}=\mathsf{bool}\cup\mathsf{contr}\cup\mathsf{glue}\cup\mathsf{exchs}$$
$$\mathsf{obs}'=\mathsf{bool}\cup\mathsf{contr}\cup\mathsf{glue}\cup\mathsf{exchs}'$$

The semantics of CKACO is given by $\sembrack{-}_{SCR}\downarrow^{\mathsf{obs}}$ and $\sembrack{-}_{SCR}\downarrow^{\mathsf{obs}'}$.
\end{definition}

We define the following set of hypotheses $\mathsf{contr}'$:

$$\mathsf{contr}'=\{\alpha\leq\alpha\cdot\alpha:\alpha\in\mathsf{At}\}$$

And let $\Gamma=\mathsf{At}\cup\Sigma\subseteq\mathcal{T}_{BA}\cup\Sigma$ and $r:\Sigma\cup\mathcal{T}_{BA}\rightarrow\mathcal{T}(\Gamma)$ be the following reification:

\begin{equation}
r(a)=\begin{cases}
       \sum_{\alpha\leqq_{BA}p~\alpha}, & \mbox{if } a=p\in\mathcal{T}_{BA} \\
       a, & \mbox{if }a\in\Sigma.
     \end{cases}\nonumber
\end{equation}

\begin{lemma}
The hypotheses $\mathsf{obs}$ 1-reduce to $\mathsf{exchs}\cup\mathsf{contr}'$.
\end{lemma}

\begin{lemma}
The hypotheses $\mathsf{obs}$ 2-reduce to $\mathsf{exchs}\cup\mathsf{contr}'$.
\end{lemma}

\begin{lemma}
The hypotheses $\mathsf{obs}$ 3-reduce to $\mathsf{exchs}\cup\mathsf{contr}'$.
\end{lemma}

\begin{lemma}
The hypotheses $\mathsf{obs}$ 4-reduce to $\mathsf{exchs}\cup\mathsf{contr}'$.
\end{lemma}

\begin{lemma}
The hypotheses $\mathsf{obs}'$ 5-reduce to $\mathsf{exchs}'\cup\mathsf{contr}'$.
\end{lemma}

\begin{lemma}
The hypotheses $\mathsf{exchs}\cup\mathsf{contr}'$ 1-factorize into $\mathsf{exchs}$ and $\mathsf{contr}'$.
\end{lemma}

\begin{lemma}
The hypotheses $\mathsf{exchs}\cup\mathsf{contr}'$ 2-factorize into $\mathsf{exchs}$ and $\mathsf{contr}'$.
\end{lemma}

\begin{lemma}
The hypotheses $\mathsf{exchs}\cup\mathsf{contr}'$ 3-factorize into $\mathsf{exchs}$ and $\mathsf{contr}'$.
\end{lemma}

\begin{lemma}
The hypotheses $\mathsf{exchs}\cup\mathsf{contr}'$ 4-factorize into $\mathsf{exchs}$ and $\mathsf{contr}'$.
\end{lemma}

\begin{lemma}
The hypotheses $\mathsf{exchs}'\cup\mathsf{contr}'$ 5-factorize into $\mathsf{exchs}'$ and $\mathsf{contr}'$.
\end{lemma}

\begin{lemma}
The hypotheses $\mathsf{contr}'$ strongly 1-reduce to $\emptyset$.
\end{lemma}

\begin{lemma}
The hypotheses $\mathsf{contr}'$ strongly 2-reduce to $\emptyset$.
\end{lemma}

\begin{lemma}
The hypotheses $\mathsf{contr}'$ strongly 3-reduce to $\emptyset$.
\end{lemma}

\begin{lemma}
The hypotheses $\mathsf{contr}'$ strongly 4-reduce to $\emptyset$.
\end{lemma}

\begin{lemma}
The hypotheses $\mathsf{contr}'$ strongly 5-reduce to $\emptyset$.
\end{lemma}

\begin{theorem}[Soundness and Completeness of CKACO]
For $x,y\in\mathcal{T}_{CKACO}$, 

\begin{enumerate}
  \item 
  \begin{enumerate}
    \item $x=^{\mathsf{obs}}_1 y$ if and only if $(\sembrack{x}_{SCR}\downarrow^1)^{\mathsf{obs}^1}=(\sembrack{y}_{SCR}\downarrow^1)^{\mathsf{obs}^1}$;
    \item $x=^{\mathsf{obs}}_2 y$ if and only if $(x\downarrow^2)^{\mathsf{obs}^2}\sim_p (y\downarrow^2)^{\mathsf{obs}^2}$;
    \item $x=^{\mathsf{obs}}_3 y$ if and only if $(x\downarrow^3)^{\mathsf{obs}^3}\sim_s (y\downarrow^3)^{\mathsf{obs}^3}$;
    \item $x=^{\mathsf{obs}}_4 y$ if and only if $(x\downarrow^4)^{\mathsf{obs}^4}\sim_{hp}(y\downarrow^4)^{\mathsf{obs}^4}$;
    \item $x=^{\mathsf{obs}}_2 y$ if and only if $(x\downarrow^5)^{\mathsf{obs}'^5}\sim_{hhp}(y\downarrow^5)^{\mathsf{obs}'^5}$.
  \end{enumerate}
  \item 
  \begin{enumerate}
    \item it is decidable whether $(\sembrack{x}_{SCR}\downarrow^1)^{\mathsf{obs}^1}=(\sembrack{y}_{SCR}\downarrow^1)^{\mathsf{obs}^1}$;
    \item it is decidable whether $(x\downarrow^2)^{\mathsf{obs}^2}\sim_p (y\downarrow^2)^{\mathsf{obs}^2}$;
    \item it is decidable whether $(x\downarrow^3)^{\mathsf{obs}^3}\sim_s (y\downarrow^3)^{\mathsf{obs}^3}$;
    \item it is decidable whether $(x\downarrow^4)^{\mathsf{obs}^4}\sim_{hp}(y\downarrow^4)^{\mathsf{obs}^4}$;
    \item it is decidable whether $(x\downarrow^5)^{\mathsf{obs}'^5}\sim_{hhp}(y\downarrow^5)^{\mathsf{obs}'^5}$.
  \end{enumerate}
\end{enumerate}
\end{theorem} 
\newpage\section{Parallel Star}\label{ps}

The so-called series-communication-parallel rational language is the series-communication rational language with parallel star. In this chapter, we introduce series-communication-parallel rational language with parallel star in section \ref{scprl}. Then we introduce the algebra modulo language equivalence in section \ref{aml} and the algebra modulo bisimilarities in section \ref{amb}.

\subsection{Series-Communication-Parallel Rational Language}\label{scprl}

We define the syntax and language semantics of the series-communication-parallel rational (scpr-) expressions.

\begin{definition}[Syntax of scpr-expressions]
We define the set of scpr-expressions $\mathcal{T}_{SCPR}$ as follows.

$$\mathcal{T}_{SCPR}\ni x,y::=0|1|a,b\in\Sigma|\rho(a,b)|x+y|x\cdot y|x^*|x\parallel y|x^{\langle *\rangle}|x\mid y|x\between y$$
\end{definition}

In the definition of scpr-expressions, the atomic actions include actions in $a,b\in\Sigma$, the constant $0$ denoted inaction without any behaviour, the constant $1$ denoted empty action which terminates immediately and successfully, and also the communication actions $\rho(a,b)$. The operator $+$ is the alternative composition, i.e., the program $x+y$ either executes $x$ or $y$ alternatively. The operator $\cdot$ is the sequential composition, i.e., the program $x\cdot y$ firstly executes $x$ followed $y$. The Kleene star $x^*$ can execute $x$ for some number of times sequentially (maybe zero). The operator $\parallel$ is the parallel composition, i.e., the program $x\parallel y$ executes $x$ and $y$ in parallel. The parallel star $x^{\langle *\rangle}$ can execute $x$ for some number of times in parallel (maybe zero). The program $x\mid y$ executes with synchronous communications. The program $x\between y$ means $x$ and $y$ execute concurrently, i.e., in parallel but maybe with unstructured communications.

\begin{definition}[Language semantics of scpr-expressions]
We define the interpretation of scpr-expressions $\sembrack{-}_{SCPR}:\mathcal{T}_{SCPR}\rightarrow 2^{\mathsf{SCP}}$ inductively as Table \ref{LSEBKAC} shows.
\end{definition}

\begin{center}
    \begin{table}
        $$\sembrack{0}_{SCPR}=\emptyset \quad \sembrack{a}_{SCPR}=\{a\} \quad \sembrack{x\cdot y}_{SCPR}=\sembrack{x}_{SCPR}\cdot \sembrack{y}_{SCPR}$$
        $$\sembrack{1}_{SCPR}=\{1\} \quad \sembrack{x+y}_{SCPR}=\sembrack{x}_{SCPR}+\sembrack{y}_{SCPR} \quad\sembrack{x^*}_{SCPR}=\sembrack{x}^*_{SCPR}$$
        $$\sembrack{x\parallel y}_{SCPR}=\sembrack{x}_{SCPR}\parallel\sembrack{y}_{SCPR} \quad\sembrack{x^{\langle *\rangle}}_{SCPR}=\sembrack{x}^{\langle *\rangle}_{SCPR}$$
        $$\sembrack{x\mid y}_{SCPR}=\sembrack{x}_{SCPR}\mid\sembrack{y}_{SCPR}\quad \sembrack{x\between y}_{SCPR}=\sembrack{x}_{SCPR}\between\sembrack{y}_{SCPR}$$
        \caption{Language semantics of scpr-expressions}
        \label{LSEBKAC}
    \end{table}
\end{center}

\subsection{Algebra Modulo Language Equivalence}\label{aml} 

We define an extended Bi-Kleene algebra with communication (EBKAC) as a tuple $(\Sigma,+,\cdot,^*,\parallel,^{\langle *\rangle},\between,\mid,0,1)$, where $\Sigma$ is an alphabet, $^*$ and $^{\langle *\rangle}$ are unary, $+$, $\cdot$, $\parallel$, $\between$ and $\mid$ are binary operators, and $0$ and $1$ are constants, which satisfies the axioms in Table \ref{AxiomsForEBKACL} for all $x,y,z\in \mathcal{T}_{SCPR}$, where $x\leqq y$ means $x+y=y$.

\begin{center}
    \begin{table}
        \begin{tabular}{@{}ll@{}}
            \hline No. &Axiom\\
            $A1$ & $x+y=y+z$\\
            $A2$ & $x+(y+z)=(x+y)+z$\\
            $A3$ & $x+x=x$\\
            $A4$ & $(x+y)\cdot z=x\cdot z+y\cdot z$\\
            $A5$ & $x\cdot(y+z)=x\cdot y+x\cdot z$\\
            $A6$ & $x\cdot(y\cdot z)=(x\cdot y)\cdot z$\\
            $A7$ & $x+0=x$\\
            $A8$ & $0\cdot x=0$\\
            $A9$ & $x\cdot 0=0$\\
            $A10$ & $x\cdot 1=x$\\
            $A11$ & $1\cdot x=x$\\
            $P1$ & $x\between y=x\parallel y+x\mid y$\\
            $P2$ & $x\parallel y=y\parallel x$\\
            $P3$ & $x\parallel(y\parallel z)=(x\parallel y)\parallel z$\\
            $P4$ & $(x+y)\parallel z=x\parallel z+y\parallel z$\\
            $P5$ & $x\parallel(y+z)=x\parallel y+x\parallel z$\\
            $P6$ & $x\parallel 0=0$\\
            $P7$ & $0\parallel x=0$\\
            $P8$ & $x\parallel 1=x$\\
            $P9$ & $1\parallel x=x$\\
            $C1$ & $x\mid y=y\mid x$\\
            $C2$ & $(x+y)\mid z=x\mid z+y\mid z$\\
            $C3$ & $x\mid(y+z)=x\mid y+x\mid z$\\
            $C4$ & $x\mid 0=0$\\
            $C5$ & $0\mid x=0$\\
            $C6$ & $x\mid 1=0$\\
            $C7$ & $1\mid x=0$\\
            $A12$ & $1+x\cdot x^*=x^*$\\
            $A13$ & $1+x^*\cdot x=x^*$\\
            $A14$ & $x+y\cdot z\leqq z\Rightarrow y^*\cdot x\leqq z$\\
            $A15$ & $x+y\cdot z\leqq y\Rightarrow x\cdot z^*\leqq y$\\
            $P10$ & $1+x\parallel x^{\langle *\rangle}=x^{\langle *\rangle}$\\
            $P11$ & $1+x^{\langle *\rangle}\parallel x=x^{\langle *\rangle}$\\
            $P12$ & $x+y\parallel z\leqq z\Rightarrow y^{\langle *\rangle}\parallel x\leqq z$\\
            $P13$ & $x+y\parallel z\leqq y\Rightarrow x\parallel z^{\langle *\rangle}\leqq y$\\
        \end{tabular}
        \caption{Axioms of EBKAC modulo language equivalence}
        \label{AxiomsForEBKACL}
    \end{table}
\end{center}

Since language equivalence is a congruence w.r.t. the operators of EBKAC, we can only check the soundness of each axiom in Table \ref{AxiomsForEBKACL} according to the definition of semantics of scpr-expressions. And also by use of communication merge, the scpr-expressions are been transformed into the so-called series-parallel ones \cite{CKA3} \cite{CKA4} \cite{CKA7} free of N-shapes. Then we can get the following soundness and completeness theorem \cite{CKA3} \cite{CKA4} \cite{CKA7}.

\begin{theorem}[Soundness and completeness of EBKAC]
For all $x,y\in\mathcal{T}_{SCPR}$, $x= y$ if and only if $\sembrack{x}_{SCPR}=\sembrack{y}_{SCPR}$.
\end{theorem} 

\begin{theorem}
Let $x, y\in\mathcal{T}_{SCPR}$. It is decidable whether $\sembrack{x}_{SCPR}=\sembrack{y}_{SCPR}$.
\end{theorem}

\begin{definition}
We define $\mathcal{F}_{SCPR}$ as smallest subset of $\mathcal{T}_{SCPR}$ satisfying the following rules:

$$\frac{}{1\in\mathcal{F}_{SCPR}}\quad \frac{x\in\mathcal{F}_{SCPR}\quad y\in\mathcal{T}_{SCPR}}{x+y\in\mathcal{F}_{SCPR}\quad y+x\in\mathcal{F}_{SCPR}} \quad\frac{x\in\mathcal{T}_{SCPR}}{x^*\in\mathcal{F}_{SCPR}\quad x^{\langle *\rangle}\in\mathcal{F}_{SCPR}}$$
$$\frac{x\in\mathcal{F}_{SCPR}\quad y\in\mathcal{F}_{SCPR}}{x\cdot y\in\mathcal{F}_{SCPR}\quad x\between y\in\mathcal{F}_{SCPR}\quad x\parallel y\in\mathcal{F}_{SCPR}\quad x\mid y\in\mathcal{F}_{SCPR}}$$
\end{definition}

\subsection{Algebra Modulo Bisimilarities}\label{amb}

The signature of EBKAC is defined as a tuple $(\Sigma,+,\cdot,^*,\parallel,^{\langle *\rangle},\between,\mid,0,1)$ includes a set of atomic actions $\Sigma$ and $a,b,c,\cdots\in \Sigma$, two special constants with inaction or deadlock denoted $0$ and empty action denoted $1$, six binary functions with sequential composition denoted $\cdot$, alternative composition denoted $+$, parallel composition denoted $\parallel$, concurrent composition $\between$ and communication merge $\mid$, and also three unary functions with sequential iteration denoted $^*$ and parallel iteration denoted $^{\langle *\rangle}$. 

\begin{definition}[Operational semantics of EBKAC modulo pomset, step and hp-bisimilarities]
Let the symbol $\downarrow$ denote the successful termination predicate. Then we give the TSS of EBKAC modulo pomset, step and hp-bisimilarities as Table \ref{TREBKAC}, where $a,b,c,\cdots\in \Sigma$, $x,y,x',y'\in\mathcal{T}_{SCPR}$.
\end{definition}

\begin{center}
    \begin{table}
        $$\frac{}{1\downarrow}\quad\frac{}{a\xrightarrow{a}1}$$
        $$\frac{x\downarrow}{(x+y)\downarrow}\quad\frac{y\downarrow}{(x+y)\downarrow}\quad\frac{x\xrightarrow{a}x'}{x+y\xrightarrow{a}x'}\quad\frac{y\xrightarrow{b}y'}{x+y\xrightarrow{b}y'}$$
        $$\frac{x\downarrow\quad y\downarrow}{(x\cdot y)\downarrow} \quad\frac{x\xrightarrow{a}x'}{x\cdot y\xrightarrow{a}x'\cdot y} \quad\frac{x\downarrow\quad y\xrightarrow{b}y'}{x\cdot y\xrightarrow{b}y'}$$
        $$\frac{x\downarrow\quad y\downarrow}{(x\parallel y)\downarrow} \quad\frac{x\xrightarrow{a}x'\quad y\xrightarrow{b}y'}{x\parallel y\xrightarrow{\mset{a,b}}x'\between y'}$$
        $$\frac{x\downarrow\quad y\downarrow}{(x\mid y)\downarrow} \quad\frac{x\xrightarrow{a}x'\quad y\xrightarrow{b}y'}{x\mid y\xrightarrow{\rho(a,b)}x'\between y'}$$
        $$\frac{x\downarrow}{(x^*)\downarrow} \quad\frac{x\xrightarrow{a}x'}{x^*\xrightarrow{a}x'\cdot x^*}$$
        $$\frac{x\downarrow}{(x^{\langle *\rangle})\downarrow} \quad\frac{x\xrightarrow{a}x'}{x^{\langle *\rangle}\xrightarrow{a}x'\parallel x^*}$$
        \caption{Operational semantics of algebra modulo pomset, step and hp-bisimilarities}
        \label{TREBKAC}
    \end{table}
\end{center}

Note that there is no any transition rules related to the constant $0$. Then the axiomatic system of EBKAC modulo pomset, step and hp-bisimilarities is shown in Table \ref{AxiomsForEBKACB}.

\begin{center}
    \begin{table}
        \begin{tabular}{@{}ll@{}}
            \hline No. &Axiom\\
            $A1$ & $x+y=y+z$\\
            $A2$ & $x+(y+z)=(x+y)+z$\\
            $A3$ & $x+x=x$\\
            $A4$ & $(x+y)\cdot z=x\cdot z+y\cdot z$\\
            $A5$ & $x\cdot(y\cdot z)=(x\cdot y)\cdot z$\\
            $A6$ & $x+0=x$\\
            $A7$ & $0\cdot x=0$\\
            $A8$ & $x\cdot 1=x$\\
            $A9$ & $1\cdot x=x$\\
            $P1$ & $x\between y=x\parallel y+x\mid y$\\
            $P2$ & $x\parallel y=y\parallel x$\\
            $P3$ & $x\parallel(y\parallel z)=(x\parallel y)\parallel z$\\
            $P4$ & $(x+y)\parallel z=x\parallel z+y\parallel z$\\
            $P5$ & $x\parallel(y+z)=x\parallel y+x\parallel z$\\
            $P6$ & $x\parallel 0=0$\\
            $P7$ & $0\parallel x=0$\\
            $P8$ & $x\parallel 1=x$\\
            $P9$ & $1\parallel x=x$\\
            $C1$ & $x\mid y=y\mid x$\\
            $C2$ & $(x+y)\mid z=x\mid z+y\mid z$\\
            $C3$ & $x\mid(y+z)=x\mid y+x\mid z$\\
            $C5$ & $x\mid 0=0$\\
            $C6$ & $0\mid x=0$\\
            $C7$ & $x\mid 1=0$\\
            $C8$ & $1\mid x=0$\\
            $A10$ & $1+x\cdot x^*=x^*$\\
            $A11$ & $(1+x)^*=x^*$\\
            $A12$ & $x+y\cdot z\leqq z\Rightarrow y^*\cdot x\leqq z$\\
            $A13$ & $x+y\cdot z\leqq y\Rightarrow x\cdot z^*\leqq y$\\
            $P10$ & $1+x\parallel x^{\langle *\rangle}=x^{\langle *\rangle}$\\
            $P11$ & $(1+x)^{\langle *\rangle}=x^{\langle *\rangle}$\\
            $P12$ & $x+y\parallel z\leqq z\Rightarrow y^{\langle *\rangle}\parallel x\leqq z$\\
            $P13$ & $x+y\parallel z\leqq y\Rightarrow x\parallel z^{\langle *\rangle}\leqq y$\\
        \end{tabular}
        \caption{Axioms of EBKAC modulo pomset, step and hp-bisimilarities}
        \label{AxiomsForEBKACB}
    \end{table}
\end{center}

Note that there are two significant differences between the axiomatic systems of EBKAC modulo bisimilarites and language equivalence, the axioms $x\cdot 0=0$ and $x\cdot(y+z)=x\cdot y+x\cdot z$ of EBKAC do not hold modulo bisimilarities.

Since pomset, step and hp-bisimilarities are all congruences w.r.t. the operators $\cdot$, $+$, $^*$, $^{\langle *\rangle}$, $\between$, $\parallel$ and $\mid$, pomset, step and hp-similarities are all precongruences w.r.t. the operators $\cdot$, $+$, $^*$, $^{\langle *\rangle}$, $\between$, $\parallel$ and $\mid$, we can only check the soundness of each axiom in Table \ref{AxiomsForEBKACB} according to the definition of TSS of scpr-expressions in Table \ref{TREBKAC}.

\begin{theorem}[Soundness of EBKAC modulo pomset (bi)similarity]
EBKAC is sound modulo pomset (bi)similarity w.r.t. scpr-expressions.
\end{theorem}

\begin{theorem}[Soundness of EBKAC modulo step (bi)similarity]
EBKAC is sound modulo step (bi)similarity w.r.t. scpr-expressions.
\end{theorem}

\begin{theorem}[Soundness of EBKAC modulo hp-(bi)similarity]
EBKAC is sound modulo hp-(bi)similarity w.r.t. scpr-expressions.
\end{theorem}

For hhp-bisimilarity, an auxiliary binary operator called left-parallelism denoted $\leftmerge$ would be added into the syntax of $\mathcal{T}_{SCPR}$. The following transition rules of $\leftmerge$ should be added into the operational semantics of scpr-expressions.

$$\frac{x\downarrow\quad y\downarrow}{(x\leftmerge y)\downarrow} \quad\frac{x\xrightarrow{a}x'\quad y\xrightarrow{b}y'\quad a\leq b}{x\leftmerge y\xrightarrow{\mset{a,b}}x'\between y'}$$

Then the axiomatic system of EBKAC modulo hhp-bisimilarity is shown in Table \ref{AxiomsForEBKACB2}.

Note that, the left-parallelism operator $\leftmerge$ is unnecessary to be added into the language semantics, pomset bisimilarity, step bisimilarity and hp-bisimilarity semantics.

\begin{center}
    \begin{table}
        \begin{tabular}{@{}ll@{}}
            \hline No. &Axiom\\
            $A1$ & $x+y=y+z$\\
            $A2$ & $x+(y+z)=(x+y)+z$\\
            $A3$ & $x+x=x$\\
            $A4$ & $(x+y)\cdot z=x\cdot z+y\cdot z$\\
            $A5$ & $x\cdot(y\cdot z)=(x\cdot y)\cdot z$\\
            $A6$ & $x+0=x$\\
            $A7$ & $0\cdot x=0$\\
            $A8$ & $x\cdot 1=x$\\
            $A9$ & $1\cdot x=x$\\
            $P1$ & $x\between y=x\parallel y+x\mid y$\\
            $P2$ & $x\parallel y=y\parallel x$\\
            $P3$ & $x\parallel(y\parallel z)=(x\parallel y)\parallel z$\\
            $P4$ & $x\parallel y = x\leftmerge y+y\leftmerge x$\\
            $P5$ & $(x+y)\leftmerge z=x\leftmerge z+y\leftmerge z$\\
            $P6$ & $0\leftmerge x=0$\\
            $P7$ & $x\leftmerge 1=x$\\
            $P8$ & $1\leftmerge x=x$\\
            $C1$ & $x\mid y=y\mid x$\\
            $C2$ & $(x+y)\mid z=x\mid z+y\mid z$\\
            $C3$ & $x\mid(y+z)=x\mid y+x\mid z$\\
            $C4$ & $x\mid 0=0$\\
            $C5$ & $0\mid x=0$\\
            $C6$ & $x\mid 1=0$\\
            $C7$ & $1\mid x=0$\\
            $A10$ & $1+x\cdot x^*=x^*$\\
            $A11$ & $(1+x)^*=x^*$\\
            $A12$ & $x+y\cdot z\leqq z\Rightarrow y^*\cdot x\leqq z$\\
            $A13$ & $x+y\cdot z\leqq y\Rightarrow x\cdot z^*\leqq y$\\
            $P9$ & $1+x\parallel x^{\langle *\rangle}=x^{\langle *\rangle}$\\
            $P10$ & $(1+x)^{\langle *\rangle}=x^{\langle *\rangle}$\\
            $P11$ & $x+y\parallel z\leqq z\Rightarrow y^{\langle *\rangle}\parallel x\leqq z$\\
            $P12$ & $x+y\parallel z\leqq y\Rightarrow x\parallel z^{\langle *\rangle}\leqq y$\\
        \end{tabular}
        \caption{Axioms of EBKAC modulo hhp-bisimilarity}
        \label{AxiomsForEBKACB2}
    \end{table}
\end{center}

Since hhp-bisimilarity is a congruences w.r.t. the operators $\cdot$, $+$, $^*$, $^{\langle *\rangle}$, $\between$, $\parallel$, $\leftmerge$ and $\mid$, hhp-similarity is a precongruences w.r.t. the operators $\cdot$, $+$, $^*$, $^{\langle *\rangle}$, $\between$, $\parallel$, $\leftmerge$ and $\mid$, we can only check the soundness of each axiom in Table \ref{AxiomsForEBKACB2} according to the definition of TSS of scpr-expressions in Table \ref{TREBKAC} and the additional transition rules of $\leftmerge$.

\begin{theorem}[Soundness of EBKAC modulo hhp-(bi)similarity]
EBKAC is sound modulo hhp-(bi)similarity w.r.t. scpr-expressions.
\end{theorem}

Then there are two questions: (R) the problem of recognizing whether a given process graph is bisimilar to one in the image of the process interpretation of a $\mathcal{T}_{SCPR}$ expression, and (A) whether a natural adaptation of Salomaa’s complete proof system for language equivalence of $\mathcal{T}_{SCPR}$ expressions is complete for bisimilarities of the process interpretation of $\mathcal{T}_{SCPR}$ expressions. While (R) is decidable in principle, it is just a pomset extension to the problem of recognizing whether a given process graph is bisimilar to one in the image of the process interpretation of a star expression \cite{AP8}.

As mentioned in the section \ref{intro}, just very recently, Grabmayer \cite{MF6} claimed to have proven that Mil is complete w.r.t. a specific kind of process graphs called LLEE-1-charts which is equal to regular expressions, modulo the corresponding kind of bisimilarity called 1-bisimilarity. Based on this work, we believe that we can get the completeness conclusion based on the corresponding truly concurrent bisimilarities and let the proof of the completeness be open.

\begin{theorem}[Completeness of EBKAC modulo pomset (bi)similarity]
EBKAC is complete modulo pomset (bi)similarity w.r.t. scpr-expressions.
\end{theorem}

\begin{theorem}[Completeness of EBKAC modulo step (bi)similarity]
EBKAC is complete modulo step (bi)similarity w.r.t. scpr-expressions.
\end{theorem}

\begin{theorem}[Completeness of EBKAC modulo hp-(bi)similarity]
EBKAC is complete modulo hp-(bi)similarity w.r.t. scpr-expressions.
\end{theorem}

\begin{theorem}[Completeness of EBKAC modulo hhp-(bi)similarity]
EBKAC is complete modulo hhp-(bi)similarity w.r.t. scpr-expressions.
\end{theorem} 

\begin{theorem}
Let $x, y\in\mathcal{T}_{SCPR}$. It is decidable whether $x\sim_p y$.
\end{theorem}

\begin{theorem}
Let $x, y\in\mathcal{T}_{SCPR}$. It is decidable whether $x\sim_s y$.
\end{theorem}

\begin{theorem}
Let $x, y\in\mathcal{T}_{SCPR}$. It is decidable whether $x\sim_{hp}y$.
\end{theorem}

\begin{theorem}
Let $x, y\in\mathcal{T}_{SCPR}$. It is decidable whether $x\sim_{hhp}y$.
\end{theorem} 

\subsection{Series-Communication-Parallel Rational Systems}

We have already defined five kinds of $=$ relations of EBKAC modulo language equivalence, pomset bisimilarity, step bisimilarity, hp-bisimilarity, and hhp-bisimilarity and the corresponding preorders $\leqq$ in Tables \ref{AxiomsForEBKACL}, \ref{AxiomsForEBKACB} and \ref{AxiomsForEBKACB2}, we denote the corresponding $=$ and $\leqq$ as $=_1$ and $\leqq_1$, $=_2$ and $\leqq_2$, $=_3$ and $\leqq_3$, $=_4$ and $\leqq_4$, and $=_5$ and $\leqq_5$ respectively.

\begin{definition}[Series-communication-parallel rational system modulo language equivalence]
Let $Q$ be a finite set. A series-communication-parallel rational system modulo language equivalence on $Q$, or called scpr-system modulo language equivalence, is a pair $\mathcal{S}=\langle M,b\rangle$, where $M:Q^2\rightarrow\mathcal{T}_{SCPR}$ and $b:Q\rightarrow\mathcal{T}_{SCPR}$. Let $=_1$ be an EBKAC language equivalence on $\mathcal{T}_{SCPR}(\Delta)$ with $\Sigma\subseteq\Delta$. We call $s:Q\rightarrow\mathcal{T}_{SCPR}(\Delta)$ a $=_1$-solution to $\mathcal{S}$ if for $q\in Q$:

$$b(q)+\sum_{q'\in Q}M(q,q')\cdot s(q')\leqq_1 s(q)$$

Lastly, $s$ is the least $=_1$-solution, if for every such solution $s'$ and every $q\in Q$, we have $s(q)\leqq_1 s'(q)$.
\end{definition}

\begin{definition}[Series-communication-parallel rational system modulo pomset bisimilarity]
Let $Q$ be a finite set. A series-communication-parallel rational system modulo pomset bisimilarity on $Q$, or called scpr-system modulo posmet bisimilarity, is a pair $\mathcal{S}=\langle M,b\rangle$, where $M:Q^2\rightarrow\mathcal{T}_{SCPR}$ and $b:Q\rightarrow\mathcal{T}_{SCPR}$. Let $=_2$ be an EBKAC pomset bisimilarity on $\mathcal{T}_{SCPR}(\Delta)$ with $\Sigma\subseteq\Delta$. We call $s:Q\rightarrow\mathcal{T}_{SCPR}(\Delta)$ a $=_2$-solution to $\mathcal{S}$ if for $q\in Q$:

$$b(q)+\sum_{q'\in Q}M(q,q')\cdot s(q')\leqq_2 s(q)$$

Lastly, $s$ is the least $=_2$-solution, if for every such solution $s'$ and every $q\in Q$, we have $s(q)\leqq_2 s'(q)$.
\end{definition}

\begin{definition}[Series-communication-parallel rational system modulo step bisimilarity]
Let $Q$ be a finite set. A series-communication-parallel rational system modulo step bisimilarity on $Q$, or called scpr-system modulo step bisimilarity, is a pair $\mathcal{S}=\langle M,b\rangle$, where $M:Q^2\rightarrow\mathcal{T}_{SCPR}$ and $b:Q\rightarrow\mathcal{T}_{SCPR}$. Let $=_3$ be an EBKAC step bisimilarity on $\mathcal{T}_{SCPR}(\Delta)$ with $\Sigma\subseteq\Delta$. We call $s:Q\rightarrow\mathcal{T}_{SCPR}(\Delta)$ a $=_3$-solution to $\mathcal{S}$ if for $q\in Q$:

$$b(q)+\sum_{q'\in Q}M(q,q')\cdot s(q')\leqq_3 s(q)$$

Lastly, $s$ is the least $=_3$-solution, if for every such solution $s'$ and every $q\in Q$, we have $s(q)\leqq_3 s'(q)$.
\end{definition}

\begin{definition}[Series-communication-parallel rational system modulo hp-bisimilarity]
Let $Q$ be a finite set. A series-communication-parallel rational system modulo hp-bisimilarity on $Q$, or called scpr-system modulo hp-bisimilarity, is a pair $\mathcal{S}=\langle M,b\rangle$, where $M:Q^2\rightarrow\mathcal{T}_{SCPR}$ and $b:Q\rightarrow\mathcal{T}_{SCPR}$. Let $=_4$ be an EBKAC hp-bisimilarity on $\mathcal{T}_{SCPR}(\Delta)$ with $\Sigma\subseteq\Delta$. We call $s:Q\rightarrow\mathcal{T}_{SCPR}(\Delta)$ a $=_4$-solution to $\mathcal{S}$ if for $q\in Q$:

$$b(q)+\sum_{q'\in Q}M(q,q')\cdot s(q')\leqq_4 s(q)$$

Lastly, $s$ is the least $=_4$-solution, if for every such solution $s'$ and every $q\in Q$, we have $s(q)\leqq_4 s'(q)$.
\end{definition}

\begin{definition}[Series-communication-parallel rational system modulo hhp-bisimilarity]
Let $Q$ be a finite set. A series-communication-parallel rational system modulo hhp-bisimilarity on $Q$, or called scpr-system modulo hhp-bisimilarity, is a pair $\mathcal{S}=\langle M,b\rangle$, where $M:Q^2\rightarrow\mathcal{T}_{SCPR}$ and $b:Q\rightarrow\mathcal{T}_{SCPR}$. Let $=_5$ be an EBKAC hhp-bisimilarity on $\mathcal{T}_{SCPR}(\Delta)$ with $\Sigma\subseteq\Delta$. We call $s:Q\rightarrow\mathcal{T}_{SCPR}(\Delta)$ a $=_5$-solution to $\mathcal{S}$ if for $q\in Q$:

$$b(q)+\sum_{q'\in Q}M(q,q')\cdot s(q')\leqq_5 s(q)$$

Lastly, $s$ is the least $=_5$-solution, if for every such solution $s'$ and every $q\in Q$, we have $s(q)\leqq_5 s'(q)$.
\end{definition}

\begin{theorem}
Let $\mathcal{S}=\langle M,b\rangle$ be an scpr-system on $Q$ modulo language equivalence. We can construct an $s:Q\rightarrow \mathcal{T}_{SCPR}$ that, for any EBKAC equivalence $=_1$ on $\mathcal{T}_{SCPR}(\Delta)$ with $\Sigma\subseteq\Delta$ and any $x\in\mathcal{T}_{SCPR}$, the $Q$-vector $s:Q\rightarrow\mathcal{T}_{SCPR}$ is the least $=_1$-solution to $\mathcal{S}$; we call such an $s$ the least solution to $\mathcal{S}$.
\end{theorem}

\begin{theorem}
Let $\mathcal{S}=\langle M,b\rangle$ be an scpr-system on $Q$ modulo pomset bisimilarity. We can construct an $s:Q\rightarrow \mathcal{T}_{SCPR}$ that, for any EBKAC equivalence $=_2$ on $\mathcal{T}_{SCPR}(\Delta)$ with $\Sigma\subseteq\Delta$ and any $x\in\mathcal{T}_{SCPR}$, the $Q$-vector $s:Q\rightarrow\mathcal{T}_{SCPR}$ is the least $=_2$-solution to $\mathcal{S}$; we call such an $s$ the least solution to $\mathcal{S}$.
\end{theorem}

\begin{theorem}
Let $\mathcal{S}=\langle M,b\rangle$ be an scpr-system on $Q$ modulo step bisimilarity. We can construct an $s:Q\rightarrow \mathcal{T}_{SCPR}$ that, for any EBKAC equivalence $=_3$ on $\mathcal{T}_{SCPR}(\Delta)$ with $\Sigma\subseteq\Delta$ and any $x\in\mathcal{T}_{SCPR}$, the $Q$-vector $s:Q\rightarrow\mathcal{T}_{SCPR}$ is the least $=_3$-solution to $\mathcal{S}$; we call such an $s$ the least solution to $\mathcal{S}$.
\end{theorem}

\begin{theorem}
Let $\mathcal{S}=\langle M,b\rangle$ be an scpr-system on $Q$ modulo hp-bisimilarity. We can construct an $s:Q\rightarrow \mathcal{T}_{SCPR}$ that, for any EBKAC equivalence $=_4$ on $\mathcal{T}_{SCPR}(\Delta)$ with $\Sigma\subseteq\Delta$ and any $x\in\mathcal{T}_{SCPR}$, the $Q$-vector $s:Q\rightarrow\mathcal{T}_{SCPR}$ is the least $=_4$-solution to $\mathcal{S}$; we call such an $s$ the least solution to $\mathcal{S}$.
\end{theorem}

\begin{theorem}
Let $\mathcal{S}=\langle M,b\rangle$ be an scpr-system on $Q$ modulo hhp-bisimilarity. We can construct an $s:Q\rightarrow \mathcal{T}_{SCPR}$ that, for any EBKAC equivalence $=_5$ on $\mathcal{T}_{SCPR}(\Delta)$ with $\Sigma\subseteq\Delta$ and any $x\in\mathcal{T}_{SCPR}$, the $Q$-vector $s:Q\rightarrow\mathcal{T}_{SCPR}$ is the least $=_5$-solution to $\mathcal{S}$; we call such an $s$ the least solution to $\mathcal{S}$.
\end{theorem} 
\newpage\section{Pomsetc Automata}\label{pa}

There exist two types of automata recognizing finite N-free pomsets: the branching automata \cite{BA1} \cite{BA2} \cite{BA3} \cite{BA4} and the pomset automata \cite{PA1} \cite{PA2}, and in general case, they are equivalent in terms of expressive power \cite{PA3}.

From \cite{CKA7}, with the assumptions in chapter \ref{cp}, i.e., the causalities among parallel branches are all communications, we know that a pomsetc with N-shape can be structured and then is transformed into a N-free (without N-shapes) pomset, so it is the so-called series-parallel\cite{CKA3} \cite{CKA4} \cite{CKA7}. 

From the background of (true) concurrency, as a bridge between truly concurrent process algebra \cite{APTC} \cite{APTC2} vs. concurrent Kleene algebra and automata theory, we also adopt naturally the pomset automata \cite{PA1} \cite{PA2} as the basic computational and concurrent model. And we add merge transition function into the definition of pomset automaton from \cite{PA1} \cite{PA2} below, for the existence of communication merge, and the extended pomset automaton is called pomsetc automaton.

Firstly, we introduce pomsetc automaton. Then in section \ref{tcbbpa}, we introduce truly concurrent bisimilarities based on pomsetc automata. Finally, we introduce the fork-acyclicity property of pomsetc automata with a correspondence to scr-expressions in section \ref{f-a}, and the well-nestedness property of pomsetc automata with a correspondence to scpr-expressions in section \ref{w-n}.

\begin{definition}[Pomsetc automaton]
A pomsetc automaton (PA) is a tuple $A=(Q,F,\delta,\gamma,\eta)$ where:

\begin{enumerate}
  \item $Q$ is a finite set of states.
  \item $F\subseteq Q$ is the set of accepting states.
  \item $\delta:Q\times\Sigma\rightarrow 2^Q$ is the sequential transition function which is the transition of traditional Kleene automata.
  \item $\gamma:Q\times\mathbb{M}(Q)\rightarrow 2^Q$ is the parallel transition function where $\mathbb{M}(Q)$ is the set of finite multisets with elements in $Q$, and there are only finite many $\phi\in\mathbb{M}(Q)$ with $\gamma(q,\phi)\neq\emptyset$ for all $q\in Q$.
  \item $\eta:\mathbb{M}(Q)\times Q\rightarrow 2^Q$ is the merge transition function where $\mathbb{M}(Q)$ is the same as in the parallel transition function.
\end{enumerate}
\end{definition}

The PA accepting $a\cdot (b\parallel c)\cdot d$ is illustrated in Figure \ref{figure:example11}, while Figure \ref{figure:example12} shows the PA accepting $a\cdot (b\mid c)\cdot d$. And we illustrate the PA accepting $a\cdot (b\between c)\cdot d$ in Figure \ref{figure:example13}, in which the splitting of states is denoted two arcs. We know that the combination of the PA in Figure \ref{figure:example11} and that in Figure \ref{figure:example12} is equivalent to the one in Figure \ref{figure:example13}, and we can draw the combination PA and leave it to the readers. Note that in these three figures, $q_1$ forks into $q_3$ and $q_4$, i.e., in these PAs, $\gamma(q_1,\mset{q_3,q_4})$ is a parallel transition. While in Figure \ref{figure:example12}, $\eta(\mset{q_3,q_4},q_5)$ is a merge transition.

\begin{figure}
  \centering
  \begin{tikzpicture}[every node/.style={transform shape}]
    \node[state] (q0) {$q_0$};
    \node[state,right=15mm of q0] (q1) {$q_1$};
    \node[state,above right=7mm of q1] (q3) {$q_3$};
    \node[state,below right=7mm of q1] (q4) {$q_4$};
    \node[state,right=15mm of q1] (q2) {$q_2$};
    \node[state,accepting,right=15mm of q2] (q5) {$q_5$};

    \draw (q0) edge[->] node[above] {$a$} (q1);
    \draw[-] (q1) edge (q3);
    \draw[-] (q1) edge (q4);
    \draw[dashed] (q1) edge[->] (q2);
    \draw[rounded corners=5pt,->] (q3) -| node[above,xshift=-4em] {$b$} (q5);
    \draw[rounded corners=5pt,->] (q4) -| node[below,xshift=-4em] {$c$} (q5);
    \draw (q1) + (-45:7mm) arc (-45:45:7mm);
    \draw (q2) edge[->] node[above] {$d$} (q5);
  \end{tikzpicture}
  \caption{PA accepting $a \cdot (b \parallel c) \cdot d$.}\label{figure:example11}
\end{figure}

\begin{figure}
  \centering
  \begin{tikzpicture}[every node/.style={transform shape}]
    \node[state] (q0) {$q_0$};
    \node[state,right=15mm of q0] (q1) {$q_1$};
    \node[state,above right=7mm of q1] (q3) {$q_3$};
    \node[state,below right=7mm of q1] (q4) {$q_4$};
    \node[state,right=15mm of q1] (q2) {$q_2$};
    \node[state,accepting,right=15mm of q2] (q6) {$q_6$};
    \node[state,right=15mm of q6] (q5) {$q_5$};

    \draw (q0) edge[->] node[above] {$a$} (q1);
    \draw[-] (q1) edge (q3);
    \draw[-] (q1) edge (q4);
    \draw[dashed] (q1) edge[->] (q2);
    \draw (q5) edge[->] node[above] {$a\mid b$} (q6);
    \draw[-][rounded corners=5pt] (q3) -| (q5);
    \draw[-][rounded corners=5pt] (q4) -| (q5);
    \draw (q1) + (-45:7mm) arc (-45:45:7mm);
    \draw (q2) edge[->] node[above] {$d$} (q6);
  \end{tikzpicture}
  \caption{PA accepting $a \cdot (b \mid c) \cdot d$.}\label{figure:example12}
\end{figure}

\begin{figure}
  \centering
  \begin{tikzpicture}[every node/.style={transform shape}]
    \node[state] (q0) {$q_0$};
    \node[state,right=15mm of q0] (q1) {$q_1$};
    \node[state,above right=7mm of q1] (q3) {$q_3$};
    \node[state,below right=7mm of q1] (q4) {$q_4$};
    \node[state,right=15mm of q1] (q2) {$q_2$};
    \node[state,accepting,right=15mm of q2] (q5) {$q_5$};

    \draw (q0) edge[->] node[above] {$a$} (q1);
    \draw[-] (q1) edge (q3);
    \draw[-] (q1) edge (q4);
    \draw[dashed] (q1) edge[->] (q2);
    \draw[rounded corners=5pt,->] (q3) -| node[above,xshift=-4em] {$b$} (q5);
    \draw[rounded corners=5pt,->] (q4) -| node[below,xshift=-4em] {$c$} (q5);
    \draw (q1) + (-45:7mm) arc (-45:45:7mm);
    \draw (q1) + (-45:8mm) arc (-45:45:8mm);
    \draw (q2) edge[->] node[above] {$d$} (q5);
  \end{tikzpicture}
  \caption{PA accepting $a \cdot (b \between c) \cdot d$.}\label{figure:example13}
\end{figure}

In Figure \ref{figure:example21}, we illustrate the PA accepting $(a\cdot b\cdot c)\between (d\cdot e\cdot f)$ with $\rho(b,e)$ defined, while in Figure \ref{figure:example22} we show an equivalent PA accepting the same language. In Figure \ref{figure:example22}, $\gamma(q_0,\mset{q_1,q_3})$ and $\gamma(q_6,\mset{q_7,q_8})$ are parallel transitions, while $\eta(\mset{q_2,q_4},q_5)$ is a merge transition.

\begin{figure}
  \centering
  \begin{tikzpicture}[every node/.style={transform shape}]
    \node[state] (q0) {$q_0$};
    \node[state,above right=7mm of q0] (q1) {$q_1$};
    \node[state,right=7mm of q1] (q2) {$q_2$};
    \node[state,right=7mm of q2] (q3) {$q_3$};
    \node[state,below right=7mm of q0] (q4) {$q_4$};
    \node[state,right=7mm of q4] (q5) {$q_5$};
    \node[state,right=7mm of q5] (q6) {$q_6$};
    \node[state,accepting,right=50mm of q0] (q7) {$q_7$};

    \draw[-] (q0) edge (q1);
    \draw[-] (q0) edge (q4);
    \draw (q1) edge[->] node[above] {$a$} (q2);
    \draw (q2) edge[->] node[above] {$b$} (q3);
    \draw (q3) edge[->] node[above] {$c$} (q7);
    \draw (q4) edge[->] node[below] {$d$} (q5);
    \draw (q5) edge[->] node[below,sloped] {$e$} (q6);
    \draw (q6) edge[->] node[below,sloped] {$f$} (q7);
    \draw[dotted] (q2) edge (q5);
    \draw (q0) + (-45:7mm) arc (-45:45:7mm);
    \draw (q0) + (-45:8mm) arc (-45:45:8mm);
  \end{tikzpicture}
  \caption{PA accepting $(a \cdot b\cdot c) \between (d \cdot e\cdot f)$ with $\rho(b,e)$ defined.}\label{figure:example21}
\end{figure}

\begin{figure}
  \centering
  \begin{tikzpicture}[every node/.style={transform shape}]
    \node[state] (q0) {$q_0$};
    \node[state,above right=7mm of q0] (q1) {$q_1$};
    \node[state,right=7mm of q1] (q2) {$q_2$};
    \node[state,below right=7mm of q0] (q3) {$q_3$};
    \node[state,right=7mm of q3] (q4) {$q_4$};
    \node[state,right=30mm of q0] (q5) {$q_5$};
    \node[state,right=14mm of q5] (q6) {$q_6$};
    \node[state,above right=7mm of q6] (q7) {$q_7$};
    \node[state,below right=7mm of q6] (q8) {$q_8$};
    \node[state,accepting,right=14mm of q6] (q9) {$q_9$};

    \draw[-] (q0) edge (q1);
    \draw[-] (q0) edge (q3);
    \draw (q1) edge[->] node[above] {$a$} (q2);
    \draw (q3) edge[->] node[below] {$d$} (q4);
    \draw[-] (q2) edge (q5);
    \draw[-] (q4) edge (q5);
    \draw (q5) edge[->] node[above] {$\rho(b,e)$} (q6);
    \draw[-] (q6) edge (q7);
    \draw[-] (q6) edge (q8);
    \draw (q7) edge[->] node[above,sloped] {$c$} (q9);
    \draw (q8) edge[->] node[below,sloped] {$f$} (q9);
    \draw (q0) + (-45:7mm) arc (-45:45:7mm);
    \draw (q6) + (-45:7mm) arc (-45:45:7mm);
  \end{tikzpicture}
  \caption{Equivalent PA accepting $(a \cdot b\cdot c) \between (d \cdot e\cdot f)$ with $\rho(b,e)$ defined.}\label{figure:example22}
\end{figure}

\begin{definition}[Run relation]
Let $\mathsf{SCP}(\Sigma)$ be the series-communication-parallel pomsetc of $\Sigma$, $a\in\Sigma$, $q,q',q'',q_i\in Q$ and $U,U_1,U_2,U_i\subseteq \mathsf{SCP}(\Sigma)$. We define the run relation $\xrightarrow[A]{}\subseteq Q\times \mathsf{SCP}(\Sigma)\times Q$ on a PA $A$ as the smallest relation satisfying:

\begin{enumerate}
  \item $q\xrightarrow[A]{1}q$ for all $q\in Q$.
  \item $q\xrightarrow[A]{a}q'$ if and only if $q'\in\delta(q,a)$.
  \item If $q\xrightarrow[A]{U_1}q''$ and $q''\xrightarrow[A]{U_2}q'$, then $q\xrightarrow[A]{U_1\cdot U_2}q'$.
  \item For all $n>1$, if $q_i\xrightarrow[A]{U_i}q_i'$ for $i\in\{1,\cdots,n\}$, $q''\in\gamma(q,\{q_1,\cdots,q_n\})$, $q'\in\eta(\{q_1',\cdots,q_n'\},q'')$, then $q\xrightarrow[A]{\parallel U_i}q'$.
  \item For all $n>1$, if $q_i\xrightarrow[A]{U_i}q_i'\in F$ for $i\in\{1,\cdots,n\}$, $q'\in\gamma(q,\{q_1,\cdots,q_n\})$, then $q\xrightarrow[A]{\parallel U_i}q'$.
\end{enumerate}

For a run relation $q\xrightarrow[A]{U}q'$: (1) if it is applied to the first rule, then it is called a trivial run; (2) if the second rule is applied last, then it is called a sequential unit run; (3) if it is applied to the fourth or fifth rule, it is called a parallel unit run, the sequential unit run and the parallel unit run are all called unit runs; (4) if the third rule is applied, then it is called a composite run.
\end{definition}

\begin{lemma}[Run composition]
Let $q\xrightarrow[A]{U}q'$ be a run relation, then there exist $q=q_0,\cdots,q_{\ell}=q'\in Q$ and $U_1,\cdots, U_{\ell}\in\mathsf{SCP}$ with $U=U_1\cdots U_{\ell}$ and each $q_{i-1}\xrightarrow[A]{U_i}q_i$ is a unit run, for all $1\leq i\leq\ell$.
\end{lemma}

\begin{definition}[Language of PA]
The PA $A=(Q,A,\delta,\gamma,\eta)$ accepts the language by $q\in Q$, is the set $L_A(q)=\{U\in\mathsf{SCP}(\Sigma)$:$q\xrightarrow[A]{U}q'\in F\}$. 
\end{definition}

\begin{definition}[Deadlock]
A state $q\in Q$ a PA $A$ is a deadlock state if and only if it does not have any outgoing transitions and it does not allow successful termination, i.e., for all $U\subseteq\mathsf{SCP}(\Sigma)$, $q\xnrightarrow{U}$ and $q\notin F$. A transition system has a deadlock if and only if it has a reachable deadlock state; it is deadlock free if and only if it does not have a deadlock.
\end{definition}

\subsection{Truly Concurrent Bisimilarities Based on Pomsetc Automata}\label{tcbbpa}

\begin{definition}[Pomset, step bisimulation]
Let $A=(Q,F,\delta,\gamma,\eta)$ and $A'=(Q',F',\delta',\gamma',\eta')$ be two pomsetc automata with the same alphabet $\Sigma$. The automata $A$ and $A'$ are pomset bisimilar, denoted $A\sim_p A'$, if and only if there is a relation $R$ between their reachable states that preserves transitions and termination:

\begin{enumerate}
  \item $R$ relate reachable states, i.e., every reachable state of $A$ is related to a reachable state of $A'$ and every reachable state of $A'$ is related to a reachable state of $A$.
  \item For $p,q\in Q$, whenever $p$ is related to $p'\in Q'$, $pRp'$ and $p\xrightarrow[A]{U}q$ with $U\subseteq \mathsf{SCP}(\Sigma)$, then there is state $q'\in Q'$ with $p'\xrightarrow[A']{U}q'$ and $qRq'$.
  \item For $p,q\in Q$, whenever $p$ is related to $p'\in Q'$, $pRp'$ and $p'\xrightarrow[A']{U}q'$ with $U\subseteq \mathsf{SCP}(\Sigma)$, then there is state $q\in Q$ in $A$ with $p\xrightarrow[A]{U}q$ and $qRq'$.
  \item Whenever $pRp'$, then $p\in F$ if and only if $p'\in F'$.
\end{enumerate}

When the events in $U$ are pairwise concurrent (without causalities), we get the definition of step bisimulation, the automata $A$ and $A'$ are step bisimilar, denoted $A\sim_s A'$.
\end{definition}

\begin{definition}[Configuration]
A (finite) configuration in $A=(Q,F,\delta,\gamma,\eta)$ is a (finite) consistent subset of events (without alternative composition +) $\mathbf{C}\subseteq \mathsf{SCP}(\Sigma)$, closed with respect to causality (i.e. $\lceil \mathbf{C}\rceil=\mathbf{C}$). The set of finite configurations of $A=(Q,F,\delta,\gamma,\eta)$ is denoted by $\mathcal{C}(A)$.
\end{definition}

\begin{definition}[Posetal product]
Given two pomsetc automata $A_1=(Q_1,F_1,\delta_1,\gamma_1,\eta_1)$, $A_2=(Q_2,F_2,\delta_2,\gamma_2,\eta_2)$, the posetal product of their configurations, denoted $\mathcal{C}(A_1)\overline{\times}\mathcal{C}(A_2)$, is defined as

$$\{(\mathbf{C}_1,f,\mathbf{C}_2)|\mathbf{C}_1\in\mathcal{C}(A_1),\mathbf{C}_2\in\mathcal{C}(A_2),f:\mathbf{C}_1\rightarrow \mathbf{C}_2 \textrm{ isomorphism}\}$$

A subset $R\subseteq\mathcal{C}(A_1)\overline{\times}\mathcal{C}(A_2)$ is called a posetal relation. We say that $R$ is downward closed when for any $(\mathbf{C}_1,f,\mathbf{C}_2),(\mathbf{C}_1',f',\mathbf{C}_2')\in \mathcal{C}(A_1)\overline{\times}\mathcal{C}(A_2)$, if $(\mathbf{C}_1,f,\mathbf{C}_2)\subseteq (\mathbf{C}_1',f',\mathbf{C}_2')$ pointwise and $(\mathbf{C}_1',f',\mathbf{C}_2')\in R$, then $(\mathbf{C}_1,f,\mathbf{C}_2)\in R$.

For $f:X_1\rightarrow X_2$, we define $f[a_1\mapsto a_2]:X_1\cup\{a_1\}\rightarrow X_2\cup\{a_2\}$, $z\in X_1\cup\{a_1\}$,(1)$f[a_1\mapsto a_2](z)=
a_2$,if $z=a_1$;(2)$f[a_1\mapsto a_2](z)=f(z)$, otherwise. Where $X_1\subseteq \mathsf{SCP}(\Sigma_1)$, $X_2\subseteq \mathsf{SCP}(\Sigma_2)$, $a_1\in \mathsf{SCP}(\Sigma_1)$, $a_2\in \mathsf{SCP}(\Sigma_2)$.
\end{definition}

\begin{definition}[(Hereditary) history-preserving bisimulation]
Let $A_1=(Q_1,F_1,\delta_1,\gamma_1,\eta_1)$ and $A_2=(Q_2,F_2,\delta_2,\gamma_2,\eta_2)$ be two pomsetc automata. A history-preserving (hp-) bisimulation is a posetal relation $R\subseteq\mathcal{C}(A_1)\overline{\times}\mathcal{C}(A_2)$ such that if $(\mathbf{C}_1,f,\mathbf{C}_2)\in R$, and $\mathbf{C}_1\xrightarrow[A_1]{a_1} \mathbf{C}_1'$, then $\mathbf{C}_2\xrightarrow[A_2]{a_2} \mathbf{C}_2'$, with $(\mathbf{C}_1',f[a_1\mapsto a_2],\mathbf{C}_2')\in R$, and vice-versa. $A_1$ and $A_2$ are history-preserving (hp-)bisimilar and are written $A_1\sim_{hp}A_2$ if there exists a hp-bisimulation $R$ such that $(\emptyset,\emptyset,\emptyset)\in R$.

A hereditary history-preserving (hhp-)bisimulation is a downward closed hp-bisimulation. $A_1,A_2$ are hereditary history-preserving (hhp-)bisimilar and are written $A_1\sim_{hhp}A_2$.
\end{definition}

Note that the above pomset, step, hp-, hhp-bisimilarities preserve deadlocks.

\subsection{Fork-acyclicity}\label{f-a}

It has already been proven that the so-called fork-acyclic pomset automaton just exactly accepts series rational (sr) language. In the following, we extend the related concepts and conclusions from \cite{CKA7} and prove that fork-acyclic PA with merge transitions exactly accepts series-communication rational (scr) languages. And also, the laws of scr-expressions are sound and maybe complete modulo truly concurrent bisimilarities based on pomsetc automata.

\begin{definition}[Support relation]
The support relation $\preceq$ of $A$ is the smallest preorder on $Q$ and for $q\in Q$:

$$\frac{a\in\Sigma\quad q'\in \delta(q,a)}{q'\preceq_{A}q}$$
$$\frac{\phi\in\mathbb{M}(Q)\quad q'\in \gamma(q,\phi)}{q'\preceq_{A} q}\quad\frac{r\in\phi\in\mathbb{M}(Q)\quad \gamma(q,\phi)\neq\emptyset}{r\preceq_{A}q}$$
$$\frac{\phi\in\mathbb{M}(Q)\quad q'\in \eta(\phi,q)}{q'\preceq_{A} q}\quad\frac{r\in\phi\in\mathbb{M}(Q)\quad \eta(\phi,q)\neq\emptyset}{q\preceq_{A}r}$$

We call the strict support relation $\prec_A$ if $q'\preceq_A q$ and $q\npreceq_A q'$ then $q'\prec_A q$ holds.
\end{definition}

\begin{definition}[Fork-acyclicity]
A PA $A$ is called fork-cyclic if for some $q,r\in Q$ such that $r$ is a fork target of $q$, we have that $q\preceq_{A}r$; $A$ is fork-acyclic if it is not fork-cyclic.
\end{definition}

\begin{definition}[Depth of pomsetc automaton]
If the pomsetc automaton $A$ is finite and fork-acyclic, the depth of $q\in Q$ in $A$ denoted $D_A(q)$ is the maximum $n$ such that there exist $q_1,\cdots,q_n\in Q$ with $q_1\prec_A q_1\prec_A\cdots\prec_A q_n =q$. The depth of A denoted $D_A$ is the maximum of $D_A(q)$ for all $q\in Q$ in $A$.
\end{definition}

\begin{definition}[Support]
$Q'\subseteq Q$ is support-closed if for all $q\in Q'$ with $q'\preceq_A q$ then $q'\in Q'$. The support of $q\in Q$ denoted $\pi_A(q)$ is the smallest support-closed set containing $q$.
\end{definition}

\begin{definition}[Bounded]
If $\pi_A(q)$ is finite for all $q\in Q$ in $A$, then $A$ is called bounded.
\end{definition}

\begin{definition}[Implementation]
$A=(Q,F,\delta,\gamma,\eta)$ and $A'=(Q',F',\delta',\gamma',\eta')$ are two pomsetc automata, then $A'$ implements $A$ if the following hold:

\begin{enumerate}
  \item $Q\subseteq Q'$ such that if $q\in Q$, then $L_A(q)=L_{A'}(q)$.
  \item If $A$ is fork-acyclic, then so is $A'$.
\end{enumerate}
\end{definition}

\begin{definition}[Support-closed restricted pomsetc automaton]
Let $Q'\subseteq Q$ be support-closed, the support-closed restricted PA of the PA $A=(Q,F,\delta,\gamma,\eta)$, denoted $A[Q']=(Q',F\cap Q',\delta',\gamma',\eta')$, where $\delta':Q'\times\Sigma\rightarrow 2^{Q'}$, $\gamma':Q'\times\mathbb{M}(Q')\rightarrow 2^{Q'}$ and $\eta':\mathbb{M}(Q')\times Q'\rightarrow 2^{Q'}$ with:

$$\delta'(q,a)=\delta(q,a)\quad \gamma'(q,\phi)=\gamma(q,\phi)\quad \eta'(\phi,q)=\eta(\phi,q)$$

where $q\in Q'$, $a\in\Sigma$ and $\phi\in\mathbb{M}(Q')$.
\end{definition}

\begin{lemma}
Let PAs $A=(Q,F,\delta,\gamma,\eta)$ and $A[Q']=(Q',F\cap Q',\delta',\gamma',\eta')$, if $Q'$ is support-closed, then $A[Q']$ implements $A$, and if $A$ is bounded, then $A[Q']$ is bounded.
\end{lemma}

\subsubsection{Expressions to Pomsetc Automata}

Given an scr-expression $x$, we show that how to obtain a fork-acyclic and finite PA with some state accepting $\sembrack{x}_{SCR}$. Similarly to the process of sr-expression, we firstly construct the so-called series-communication rational syntactic PA.

\begin{definition}[Series-communication rational syntactic pomsetc automaton]
Let $x\in\mathcal{T}_{SCR}$ and $S\subseteq\mathcal{T}_{SCR}$: (1)$x\star S=S$, if $x\in\mathcal{F}_{SCR}$; (2) $x\star S=\emptyset$, otherwise. We define the series-communication rational syntactic PA as $A_{SCR}=(\mathcal{T}_{SCR},\mathcal{F}_{SCR},\delta_{SCR},\gamma_{SCR},\eta_{SCR})$, where $\delta_{SCR}:\mathcal{T}_{SCR}\times\Sigma\rightarrow 2^{\mathcal{T}_{SCR}}$ is defined inductively as follows.

$$\delta_{SCR}(0,a)=\emptyset \quad \delta_{SCR}(1,a)=\emptyset \quad \delta_{SCR}(b,a)=\{1:a=b\}$$
$$\delta_{SCR}(x+y,a)=\delta_{SCR}(x,a)\cup\delta_{SCR}(y,a) \quad \delta_{SCR}(x\cdot y,a)=\delta_{SCR}(x,a)\fatsemi y\;\cup\; x\star\delta_{SCR}(y,a)$$
$$\delta_{SCR}(x^*,a)=\delta_{SCR}(x,a)\fatsemi x^* \quad \delta_{SCR}(x\between y,a)=\emptyset$$
$$\delta_{SCR}(x\parallel y,a)=\emptyset \quad \delta_{SCR}(x\mid y,a)=\emptyset$$

$\gamma_{SCR}:\mathcal{T}_{SCR}\times\mathbb{M}(\mathcal{T}_{SCR})\rightarrow 2^{\mathcal{T}_{SCR}}$ is defined inductively as follows.

$$\gamma_{SCR}(0,\phi)=\emptyset \quad \gamma_{SCR}(1,\phi)=\emptyset \quad \gamma_{SCR}(b,\phi)=\emptyset$$
$$\gamma_{SCR}(x+y,\phi)=\gamma_{SCR}(x,\phi)\cup\gamma_{SCR}(y,\phi) \quad \gamma_{SCR}(x\cdot y,\phi)=\gamma_{SCR}(x,\phi)\fatsemi y\;\cup\; x\star\gamma_{SCR}(y,\phi)$$
$$\gamma_{SCR}(x^*,\phi)=\gamma_{SCR}(x,\phi)\fatsemi x^* \quad \gamma_{SCR}(x\between y,\phi)=\gamma_{SCR}(x\parallel y,\phi)\;\cup\;\gamma_{SCR}(x\mid y,\phi)$$
$$\gamma_{SCR}(x\parallel y,\phi)=\{1:\phi=\mset{x,y}\} \quad \gamma_{SCR}(x\mid y,\phi)=\{\rho(a,b)\}\fatsemi\gamma_{SCR}(x'\between y',\phi)\;\textrm{with}\;x\mid y=(a\cdot x')\mid(b\cdot y')$$

$\eta_{SCR}:\mathbb{M}(\mathcal{T}_{SCR})\times\mathcal{T}_{SCR}\rightarrow 2^{\mathcal{T}_{SCR}}$ is defined inductively as follows.

$$\eta_{SCR}(\phi, 0)=\emptyset \quad \eta_{SCR}(\phi, 1)=\{1\} \quad \eta_{SCR}(\phi, b)=\{b\}$$
$$\eta_{SCR}(\phi, x+y)=\eta_{SCR}(\phi, x)\cup\eta_{SCR}(\phi, y) \quad \eta_{SCR}(\phi, x\cdot y)=\eta_{SCR}(\phi, x)\fatsemi y\;\cup\; x\star\eta_{SCR}(\phi, y)$$
$$\eta_{SCR}(\phi, x^*)=\eta_{SCR}(\phi, x)\fatsemi x^* \quad \eta_{SCR}(\phi, x\between y)=\eta_{SCR}(\phi, x\parallel y)\;\cup\;\eta_{SCR}(\phi, x\mid y)$$
$$\eta_{SCR}(\phi, x\parallel y)=\{1\}\fatsemi\gamma_{SCR}(x\parallel y,\phi') \quad \eta_{SCR}(\phi, x\mid y)=\{1\}\fatsemi\gamma_{SCR}(x\mid y,\phi')$$
\end{definition}

\begin{lemma}
Let $x_1,x_2\in\mathcal{T}_{SCR}$ and $U\in\mathsf{SCP}$. The following two conclusions are equivalent:

\begin{enumerate}
  \item There exists a $y\in\mathcal{F}_{SCR}$ such that $x_1+x_2\xrightarrow[A_{SCR}]{U} y$.
  \item There exists a $y\in\mathcal{F}_{SCR}$ such that $x_1\xrightarrow[A_{SCR}]{U} y$ or $x_2\xrightarrow[A_{SCR}]{U} y$.
\end{enumerate}
\end{lemma}

\begin{lemma}
Let $x_1,x_2\in\mathcal{T}_{SCR}$, $U\in\mathsf{SCP}$ and $\ell\in\mathbb{N}$. The following two conclusions are equivalent:

\begin{enumerate}
  \item There exists a $y\in\mathcal{F}_{SCR}$ such that $x_1\cdot x_2\xrightarrow[A_{SCR}]{U} y$ of length $\ell$.
  \item $U=U_1\cdot U_2$, then there exist $y_1,y_2\in\mathcal{F}_{SCR}$ such that $x_1\xrightarrow[A_{SCR}]{U_1} y_1$ or $x_2\xrightarrow[A_{SCR}]{U_2} y_2$ of length at most $\ell$.
\end{enumerate}
\end{lemma}

\begin{lemma}
Let $x_1,x_2\in\mathcal{T}_{SCR}$, $U\in\mathsf{SCP}$. The following two conclusions are equivalent:

\begin{enumerate}
  \item There exists a $y\in\mathcal{F}_{SCR}$ such that $x_1\parallel x_2\xrightarrow[A_{SCR}]{U} y$.
  \item $U=U_1\parallel U_2$, then there exist $y_1,y_2\in\mathcal{F}_{SCR}$ such that $x_1\xrightarrow[A_{SCR}]{U_1} y_1$ or $x_2\xrightarrow[A_{SCR}]{U_2} y_2$.
\end{enumerate}
\end{lemma}

\begin{lemma}
Let $x_1,x_2\in\mathcal{T}_{SCR}$, $U\in\mathsf{SCP}$. The following two conclusions are equivalent:

\begin{enumerate}
  \item There exists a $y\in\mathcal{F}_{SCR}$ such that $x_1\mid x_2\xrightarrow[A_{SCR}]{U} y$.
  \item $U=U_1\mid U_2$, then there exist $y_1,y_2\in\mathcal{F}_{SCR}$ such that $x_1\xrightarrow[A_{SCR}]{U_1} y_1$ or $x_2\xrightarrow[A_{SCR}]{U_2} y_2$.
\end{enumerate}
\end{lemma}

\begin{lemma}
Let $x_1,x_2\in\mathcal{T}_{SCR}$, $U\in\mathsf{SCP}$. The following two conclusions are equivalent:

\begin{enumerate}
  \item There exists a $y\in\mathcal{F}_{SCR}$ such that $x_1\leftmerge x_2\xrightarrow[A_{SCR}]{U} y$.
  \item $U=U_1\leftmerge U_2$, then there exist $y_1,y_2\in\mathcal{F}_{SCR}$ such that $x_1\xrightarrow[A_{SCR}]{U_1} y_1$ or $x_2\xrightarrow[A_{SCR}]{U_2} y_2$.
\end{enumerate}
\end{lemma}

\begin{lemma}
Let $x_1,x_2\in\mathcal{T}_{SCR}$, $U\in\mathsf{SCP}$. The following two conclusions are equivalent:

\begin{enumerate}
  \item There exists a $y\in\mathcal{F}_{SCR}$ such that $x_1\between x_2\xrightarrow[A_{SCR}]{U} y$.
  \item $U=U_1\between U_2$, then there exist $y_1,y_2\in\mathcal{F}_{SCR}$ such that $x_1\xrightarrow[A_{SCR}]{U_1} y_1$ or $x_2\xrightarrow[A_{SCR}]{U_2} y_2$.
\end{enumerate}
\end{lemma}

\begin{lemma}
Let $x\in\mathcal{T}_{SCR}$, $U\in\mathsf{SCP}$. The following two conclusions are equivalent:

\begin{enumerate}
  \item There exists a $y\in\mathcal{F}_{SCR}$ such that $x^*\xrightarrow[A_{SCR}]{U} y$.
  \item $U=U_1\cdots U_n$, then there exist $y_i\in\mathcal{F}_{SCR}$ such that $x\xrightarrow[A_{SCR}]{U_i} y_i$ for $1\leq i\leq n$.
\end{enumerate}
\end{lemma}

\begin{lemma}
Let $x,y\in\mathcal{T}_{SCR}$, then the following hold:

\begin{enumerate}
  \item $L_{SCR}(x+y)=L_{SCR}(x)+L_{SCR}(y)$.
  \item $L_{SCR}(x\cdot y)=L_{SCR}(x)\cdot L_{SCR}(y)$.
  \item $L_{SCR}(x^*)=L_{SCR}(x)^*$.
  \item $L_{SCR}(x\between y)=L_{SCR}(x)\between L_{SCR}(y)$.
  \item $L_{SCR}(x\parallel y)=L_{SCR}(x)\parallel L_{SCR}(y)$.
  \item $L_{SCR}(x\mid y)=L_{SCR}(x)\mid L_{SCR}(y)$.
  \item $L_{SCR}(x\leftmerge y)=L_{SCR}(x)\leftmerge L_{SCR}(y)$.
\end{enumerate}
\end{lemma}

\begin{lemma}
For all $x\in\mathcal{T}_{SCR}$, it holds that $L_{SCR}(x)=\sembrack{x}_{SCR}$.
\end{lemma}

\begin{lemma}
Let $x,y\in\mathcal{T}_{SCR}$, if $x\preceq_{SCR} y$, then $d_{\between}(x)\leq d_{\between}(y)$, $d_{\parallel}(x)\leq d_{\parallel}(y)$, $d_{\leftmerge}(x)\leq d_{\leftmerge}(y)$ and $d_{\mid}(x)\leq d_{\mid}(y)$.
\end{lemma}

\begin{lemma}
Let $x,y\in\mathcal{T}_{SCR}$, if $y$ is a fork target of $x$ in the syntactic PA, then $d_{\between}(y)\leq d_{\between}(x)$, $d_{\parallel}(y)\leq d_{\parallel}(x)$, $d_{\leftmerge}(y)\leq d_{\leftmerge}(x)$ and $d_{\mid}(y)\leq d_{\mid}(x)$, and the syntactic PA is fork-acyclic.
\end{lemma}

\begin{definition}
Let $x_1,x_2\in\mathcal{T}_{SCR}$, $R:\mathcal{T}_{SCR}\rightarrow 2^{\mathcal{T}_{SCR}}$ is defined inductively as follows:

$$R(0)=\{0\}\quad R(1)=\{1\} \quad R(a)=\{a,1\}$$
$$R(x_1+x_2)=R(x_1)\cup R(x_2)\quad R(x_1\cdot x_2)=R(x_1)\fatsemi x_2\;\cup\; R(x_1)\cup R(x_2)$$
$$R(x_1^*)=R(x_1)\fatsemi x_1^*\;\cup\; R(x_1)\cup\{x_1^*\}\quad R(x_1\between x_2)=R(x_1)\cup R(x_2)\cup\{x_1\between x_2,1\}$$
$$R(x_1\parallel x_2)=R(x_1)\cup R(x_2)\cup\{x_1\parallel x_2,1\}\quad R(x_1\mid x_2)=R(x_1)\cup R(x_2)\cup\{x_1\mid x_2,1\}$$
$$R(x_1\leftmerge x_2)=R(x_1)\cup R(x_2)\cup\{x_1\leftmerge x_2,1\}$$
\end{definition}

\begin{lemma}
For every $x\in\mathcal{T}_{SCR}$, they hold that:

\begin{enumerate}
  \item $x\in R(x)$.
  \item $R(x)$ is support-closed.
  \item The syntactic PA is bounded.
\end{enumerate}
\end{lemma}

\begin{theorem}[Expressions to pomsetc automata]\label{etoa}
For every $x\in\mathcal{T}_{SCR}$, we can obtain a fork-acyclic and finite PA $A$ with a state $q$ such that $L_{A}(q)=\sembrack{x}_{SCR}$.
\end{theorem}

\begin{theorem}
For $x,y\in\mathcal{T}_{SCR}$, according to \cref{etoa}, we obtain two corresponding PA $A_x$ and $A_y$. If $x\sim_p y$, then $A_x\sim_p A_y$.
\end{theorem}

\begin{theorem}
For $x,y\in\mathcal{T}_{SCR}$, according to \cref{etoa}, we obtain two corresponding PA $A_x$ and $A_y$. If $x\sim_s y$, then $A_x\sim_s A_y$.
\end{theorem}

\begin{theorem}
For $x,y\in\mathcal{T}_{SCR}$, according to \cref{etoa}, we obtain two corresponding PA $A_x$ and $A_y$. If $x\sim_{hp} y$, then $A_x\sim_{hp} A_y$.
\end{theorem}

\begin{theorem}
For $x,y\in\mathcal{T}_{SCR}$, according to \cref{etoa}, we obtain two corresponding PA $A_x$ and $A_y$. If $x\sim_{hhp} y$, then $A_x\sim_{hhp} A_y$.
\end{theorem}

\subsubsection{Pomsetc Automata to Expressions}

In this section, we show that the language accepted by a state in any fork-acyclic and finite automaton can be implemented by a series-communication rational expression.

\begin{lemma}
If $A=(Q,F,\delta,\gamma,\eta)$ be a pomsetc automaton, then $L_A:Q\rightarrow 2^{SCP}$ is the least function $t:Q\rightarrow 2^{SCP}$ (w.r.t. the pointwise inclusion order) such that for all $q\in Q$, the following hold:

$$\frac{q\in\mathcal{F}_{SCR}}{1\in t(q)} \quad \frac{a\in\Sigma\quad q'\in\delta(q,a)}{a\cdot t(q')}\subseteq t(q)$$
$$\frac{q'\in\gamma(q,\mset{r_1,\cdots,r_n})}{(t(r_1)\between \cdots\between t(r_n))\cdot t(q')\subseteq t(q)} \quad \frac{\gamma(q,\mset{r_1,\cdots,r_n})\quad \eta(\mset{r_1,\cdots,r_n},q')}{(t(r_1)\between \cdots\between t(r_n))\cdot t(q')\subseteq t(q)}$$
\end{lemma}

\begin{definition}[1-Solution of a PA]
Let $A=(Q,F,\delta,\gamma,\eta)$ be a PA, and let $=_1$ be a BKAC language congruence on $\mathcal{T}_{SCR}(\Delta)$ with $\Sigma\subseteq\Delta$. We say that $s:Q\rightarrow\mathcal{T}_{SCR}(\Delta)$ is an $=_1$-solution to $A$, if for every $q\in Q$:

\begin{center}
$[q\in \mathcal{F}_{SCR}]+\sum_{q'\in\delta(q,a)}a\cdot s(q')+\sum_{q'\in\gamma(q,\mset{r_1,\cdots,r_n})}(s(r_1)\between\cdots\between s(r_n))\cdot s(q')\linebreak
+\sum_{\gamma(q,\mset{r_1,\cdots,r_n}),\eta(\mset{r_1,\cdots,r_n},q')}(s(r_1)\between\cdots\between s(r_n))\cdot s(q')\leqq_1 s(q)$
\end{center}

Also, $s$ is a least $=_1$-solution to $A$ if for every $=_1$-solution $s'$ it holds that $s(q)\leqq_1 s'(q)$ for all $q\in Q$. We call $s:Q\rightarrow\mathcal{T}_{SCR}$ the least 1-solution to $A$ if it is the least $=_1$-solution for any BKAC language congruence $=_1$.
\end{definition}

\begin{lemma}
Let $A=(Q,F,\delta,\gamma,\eta)$ be a pomsetc automaton. If $s:Q\rightarrow\mathcal{T}_{SCR}$ is the least 1-solution to $A$, then it holds that $L_A(q)=\sembrack{s(q)}_{SCR}$ for $q\in Q$.
\end{lemma}

\begin{lemma}
Let $A$ be a fork-acyclic and finite PA, then we can construct the least 1-solution to $A$.
\end{lemma}

\begin{theorem}[Pomsetc automata to expressions]
If $A=(Q,F,\delta,\gamma,\eta)$ is a fork-acyclic and finite PA, then we can construct for every $q\in Q$ a series-communication rational expression $x\in\mathcal{T}_{SCR}$ such that $L_{A}(q)=\sembrack{x}_{SCR}$.
\end{theorem}

\begin{corollary}[Kleene theorem for series-communication rational language]
Let $L\subseteq\mathsf{SCP}$, then $L$ is series-communication rational if and only if it is accepted by a finite and fork-acyclic pomsetc automaton.
\end{corollary}

\begin{definition}[2-Solution of a PA]
Let $A=(Q,F,\delta,\gamma,\eta)$ be a PA, and let $=_2$ be a BKAC pomset bisimilar congruence on $\mathcal{T}_{SCR}(\Delta)$ with $\Sigma\subseteq\Delta$. We say that $s:Q\rightarrow\mathcal{T}_{SCR}(\Delta)$ is an $=_2$-solution to $A$, if for every $q\in Q$:

\begin{center}
$[q\in \mathcal{F}_{SCR}]+\sum_{q'\in\delta(q,a)}a\cdot s(q')+\sum_{q'\in\gamma(q,\mset{r_1,\cdots,r_n})}(s(r_1)\between\cdots\between s(r_n))\cdot s(q')\linebreak
+\sum_{\gamma(q,\mset{r_1,\cdots,r_n}),\eta(\mset{r_1,\cdots,r_n},q')}(s(r_1)\between\cdots\between s(r_n))\cdot s(q')\leqq_2 s(q)$
\end{center}

Also, $s$ is a least $=_2$-solution to $A$ if for every $=_2$-solution $s'$ it holds that $s(q)\leqq_2 s'(q)$ for all $q\in Q$. We call $s:Q\rightarrow\mathcal{T}_{SCR}$ the least 2-solution to $A$ if it is the least $=_2$-solution for any BKAC pomset bisimilar congruence $=_2$.
\end{definition}

\begin{lemma}
Let $A$ be a fork-acyclic and finite PA, then we can construct the least 2-solution to $A$.
\end{lemma}

\begin{theorem}[Pomsetc automata to expressions modulo pomset bisimilarity]
If $A=(Q,F,\delta,\gamma,\eta)$ and $A=(Q',F',\delta',\gamma',\eta')$  are fork-acyclic and finite PA, then we can construct for each $q\in Q$ a series-communication rational expression $x\in\mathcal{T}_{SCR}$ and for each $q'\in Q$ a series-communication rational expression $x'\in\mathcal{T}_{SCR}$, such that if $A\sim_p A'$ then $x\sim_p x'$.
\end{theorem}

\begin{corollary}[Kleene theorem for series-communication rational language modulo pomset bisimilarity]
Let $L\subseteq\mathsf{SCP}$, then $L$ is series-communication rational and $x,y\subseteq L$ with $x\sim_p y$, if and only if there exist finite and fork-acyclic pomsetc automata $A_x$ and $A_y$ such that $A_x\sim_p A_y$.
\end{corollary}

\begin{definition}[3-Solution of a PA]
Let $A=(Q,F,\delta,\gamma,\eta)$ be a PA, and let $=_3$ be a BKAC step bisimilar congruence on $\mathcal{T}_{SCR}(\Delta)$ with $\Sigma\subseteq\Delta$. We say that $s:Q\rightarrow\mathcal{T}_{SCR}(\Delta)$ is an $=_3$-solution to $A$, if for every $q\in Q$:

\begin{center}
$[q\in \mathcal{F}_{SCR}]+\sum_{q'\in\delta(q,a)}a\cdot s(q')+\sum_{q'\in\gamma(q,\mset{r_1,\cdots,r_n})}(s(r_1)\between\cdots\between s(r_n))\cdot s(q')\linebreak
+\sum_{\gamma(q,\mset{r_1,\cdots,r_n}),\eta(\mset{r_1,\cdots,r_n},q')}(s(r_1)\between\cdots\between s(r_n))\cdot s(q')\leqq_3 s(q)$
\end{center}

Also, $s$ is a least $=_3$-solution to $A$ if for every $=_3$-solution $s'$ it holds that $s(q)\leqq_3 s'(q)$ for all $q\in Q$. We call $s:Q\rightarrow\mathcal{T}_{SCR}$ the least 3-solution to $A$ if it is the least $=_3$-solution for any BKAC step bisimilar congruence $=_3$.
\end{definition}

\begin{lemma}
Let $A$ be a fork-acyclic and finite PA, then we can construct the least 3-solution to $A$.
\end{lemma}

\begin{theorem}[Pomsetc automata to expressions modulo step bisimilarity]
If $A=(Q,F,\delta,\gamma,\eta)$ and $A=(Q',F',\delta',\gamma',\eta')$  are fork-acyclic and finite PA, then we can construct for each $q\in Q$ a series-communication rational expression $x\in\mathcal{T}_{SCR}$ and for each $q'\in Q$ a series-communication rational expression $x'\in\mathcal{T}_{SCR}$, such that if $A\sim_p A'$ then $x\sim_s x'$.
\end{theorem}

\begin{corollary}[Kleene theorem for series-communication rational language modulo step bisimilarity]
Let $L\subseteq\mathsf{SCP}$, then $L$ is series-communication rational and $x,y\subseteq L$ with $x\sim_s y$, if and only if there exist finite and fork-acyclic pomsetc automata $A_x$ and $A_y$ such that $A_x\sim_s A_y$.
\end{corollary}

\begin{definition}[4-Solution of a PA]
Let $A=(Q,F,\delta,\gamma,\eta)$ be a PA, and let $=_4$ be a BKAC hp-bisimilar congruence on $\mathcal{T}_{SCR}(\Delta)$ with $\Sigma\subseteq\Delta$. We say that $s:Q\rightarrow\mathcal{T}_{SCR}(\Delta)$ is an $=_4$-solution to $A$, if for every $q\in Q$:

\begin{center}
$[q\in \mathcal{F}_{SCR}]+\sum_{q'\in\delta(q,a)}a\cdot s(q')+\sum_{q'\in\gamma(q,\mset{r_1,\cdots,r_n})}(s(r_1)\between\cdots\between s(r_n))\cdot s(q')\linebreak
+\sum_{\gamma(q,\mset{r_1,\cdots,r_n}),\eta(\mset{r_1,\cdots,r_n},q')}(s(r_1)\between\cdots\between s(r_n))\cdot s(q')\leqq_4 s(q)$
\end{center}

Also, $s$ is a least $=_4$-solution to $A$ if for every $=_4$-solution $s'$ it holds that $s(q)\leqq_4 s'(q)$ for all $q\in Q$. We call $s:Q\rightarrow\mathcal{T}_{SCR}$ the least 4-solution to $A$ if it is the least $=_4$-solution for any BKAC hp-bisimilar congruence $=_4$.
\end{definition}

\begin{lemma}
Let $A$ be a fork-acyclic and finite PA, then we can construct the least 4-solution to $A$.
\end{lemma}

\begin{theorem}[Pomsetc automata to expressions modulo hp-bisimilarity]
If $A=(Q,F,\delta,\gamma,\eta)$ and $A=(Q',F',\delta',\gamma',\eta')$  are fork-acyclic and finite PA, then we can construct for each $q\in Q$ a series-communication rational expression $x\in\mathcal{T}_{SCR}$ and for each $q'\in Q$ a series-communication rational expression $x'\in\mathcal{T}_{SCR}$, such that if $A\sim_{hp} A'$ then $x\sim_{hp} x'$.
\end{theorem}

\begin{corollary}[Kleene theorem for series-communication rational language modulo hp-bisimilarity]
Let $L\subseteq\mathsf{SCP}$, then $L$ is series-communication rational and $x,y\subseteq L$ with $x\sim_{hp} y$, if and only if there exist finite and fork-acyclic pomsetc automata $A_x$ and $A_y$ such that $A_x\sim_{hp} A_y$.
\end{corollary}

\begin{definition}[5-Solution of a PA]
Let $A=(Q,F,\delta,\gamma,\eta)$ be a PA, and let $=_5$ be a BKAC hhp-bisimilar congruence on $\mathcal{T}_{SCR}(\Delta)$ with $\Sigma\subseteq\Delta$. We say that $s:Q\rightarrow\mathcal{T}_{SCR}(\Delta)$ is an $=_5$-solution to $A$, if for every $q\in Q$:

\begin{center}
$[q\in \mathcal{F}_{SCR}]+\sum_{q'\in\delta(q,a)}a\cdot s(q')+\sum_{q'\in\gamma(q,\mset{r_1,\cdots,r_n})}(s(r_1)\between\cdots\between s(r_n))\cdot s(q')\linebreak
+\sum_{\gamma(q,\mset{r_1,\cdots,r_n}),\eta(\mset{r_1,\cdots,r_n},q')}(s(r_1)\between\cdots\between s(r_n))\cdot s(q')\leqq_5 s(q)$
\end{center}

Also, $s$ is a least $=_5$-solution to $A$ if for every $=_5$-solution $s'$ it holds that $s(q)\leqq_5 s'(q)$ for all $q\in Q$. We call $s:Q\rightarrow\mathcal{T}_{SCR}$ the least 5-solution to $A$ if it is the least $=_5$-solution for any BKAC hhp-bisimilar congruence $=_5$.
\end{definition}

\begin{lemma}
Let $A$ be a fork-acyclic and finite PA, then we can construct the least 5-solution to $A$.
\end{lemma}

\begin{theorem}[Pomsetc automata to expressions modulo hhp-bisimilarity]
If $A=(Q,F,\delta,\gamma,\eta)$ and $A=(Q',F',\delta',\gamma',\eta')$  are fork-acyclic and finite PA, then we can construct for each $q\in Q$ a series-communication rational expression $x\in\mathcal{T}_{SCR}$ and for each $q'\in Q$ a series-communication rational expression $x'\in\mathcal{T}_{SCR}$, such that if $A\sim_{hhp} A'$ then $x\sim_{hhp} x'$.
\end{theorem}

\begin{corollary}[Kleene theorem for series-communication rational language modulo hhp-bisimilarity]
Let $L\subseteq\mathsf{SCP}$, then $L$ is series-communication rational and $x,y\subseteq L$ with $x\sim_{hhp} y$, if and only if there exist finite and fork-acyclic pomsetc automata $A_x$ and $A_y$ such that $A_x\sim_{hhp} A_y$.
\end{corollary}

\subsection{Well-nestedness}\label{w-n}

It has already been proven that the so-called well-nested pomset automaton just exactly accepts series-parallel rational (spr) language. In the following, we extend the related concepts and conclusions from \cite{CKA7} and prove that well-nested PA with merge transitions exactly accepts series-communication-parallel rational (scpr) languages. And also, the laws of scpr-expressions are sound and maybe complete modulo truly concurrent bisimilarities based on pomsetc automata.

Parallel star allows an unbounded number of events to occur in parallel, we need the following concepts.

\begin{definition}[Pomsetc width]
The width of a finite pomsetc $U=[\mathbf{u}]\in\mathsf{Pomc}$ is the size of maximum of the largest $\leq^e_{\mathbf{u}}$-antichain and the largest $\leq^c_{\mathbf{u}}$-antichain.
\end{definition}

\begin{definition}[Pomsetc depth]
The depth of $U\in \mathsf{SCP}$ denoted $d(U)$ is defined inductively as follows:

\begin{enumerate}
  \item $d(U)=0$ if $U$ is empty or primitive.
  \item $d(U)=1+\max_{1\leq i\leq n}d(U_i)$ if $U$ is sequential with sequential factorization $U_1,\cdots,U_n$.
  \item $d(U)=1+\max_{1\leq i\leq n}d(U_i)$ if $U$ is parallel with parallel factorization $\mset{U_1,\cdots,U_n}$.
\end{enumerate}
\end{definition}

\begin{definition}[Recursive states]
Let $A=(Q,F,\delta,\gamma,\eta)$ be a PA, $q\in Q$ is recursive if:

\begin{enumerate}
  \item For all $a\in\Sigma$, $q'\in \delta(q,a)$, then $q'\prec_{A}q$.
  \item For all $\phi\in\mathbb{M}(Q)$, $q'\in \gamma(q,\phi)$, then $q'\prec_{A}q$.
  \item If $\phi\in\mathbb{M}(Q)$ with $q'\in \gamma(q,\phi)$, then either (a) $\phi=\mset{q}\sqcup\psi$ with for all $r\in\psi$ we have $r\prec_{A}q$, and $q'$ does not have any outgoing transitions, or (b) for all $r\in\phi$ we have $r\prec_{A}q$.
  \item For all $\phi\in\mathbb{M}(Q)$, $\gamma(\phi,q)$ and $r\in\phi$, then $q\prec_{A}r$.
\end{enumerate}
\end{definition}

\begin{definition}[Progressive states]
Let $A=(Q,F,\delta,\gamma,\eta)$ be a PA, $q\in Q$ is progressive if, whenever $\phi\in\mathbb{M}(Q)$ with $\gamma(q,\phi)\neq\emptyset$, $r\prec_{A}q$ for all $r\in\phi$.
\end{definition}

\begin{definition}[Well-nestedness]
Let $A=(Q,F,\delta,\gamma,\eta)$ be a PA, $A$ is well-nested if every state is either recursive or progressive.
\end{definition}

\begin{lemma}
Let PAs $A=(Q,F,\delta,\gamma,\eta)$ and $A[Q']=(Q',F\cap Q',\delta',\gamma',\eta')$, if $Q'$ is support-closed and $A$ is well-nested, then $A[Q']$ is well-nested.
\end{lemma}

\subsubsection{Expressions to Pomsetc Automata}

Given an scpr-expression $x$, we show that how to obtain a well-nested and finite PA with some state accepting $\sembrack{x}_{SCPR}$. Similarly to the process of sr-expression, we firstly construct the so-called series-communication-parallel rational syntactic PA.

\begin{definition}[Series-communication-parallel rational syntactic pomsetc automaton]
Let $x\in\mathcal{T}_{SCPR}$ and $S\subseteq\mathcal{T}_{SCPR}$: (1)$x\star S=S$, if $x\in\mathcal{F}_{SCPR}$; (2) $x\star S=\emptyset$, otherwise. We define the series-communication-parallel rational syntactic PA as $A_{SCPR}=(\mathcal{T}_{SCPR},\mathcal{F}_{SCPR},\delta_{SCPR},\gamma_{SCPR},\eta_{SCPR})$, where $\delta_{SCPR}:\mathcal{T}_{SCPR}\times\Sigma\rightarrow 2^{\mathcal{T}_{SCPR}}$ is defined inductively as follows.

$$\delta_{SCPR}(0,a)=\emptyset \quad \delta_{SCPR}(1,a)=\emptyset \quad \delta_{SCPR}(b,a)=\{1:a=b\}$$
$$\delta_{SCPR}(x+y,a)=\delta_{SCPR}(x,a)\cup\delta_{SCPR}(y,a) \quad \delta_{SCPR}(x\cdot y,a)=\delta_{SCPR}(x,a)\fatsemi y\;\cup\; x\star\delta_{SCPR}(y,a)$$
$$\delta_{SCPR}(x^*,a)=\delta_{SCPR}(x,a)\fatsemi x^* \quad \delta_{SCPR}(x^{\langle *\rangle},a)=\emptyset \quad \delta_{SCPR}(x\between y,a)=\emptyset$$
$$\delta_{SCPR}(x\parallel y,a)=\emptyset \quad \delta_{SCPR}(x\mid y,a)=\emptyset$$

$\gamma_{SCPR}:\mathcal{T}_{SCPR}\times\mathbb{M}(\mathcal{T}_{SCPR})\rightarrow 2^{\mathcal{T}_{SCPR}}$ is defined inductively as follows.

$$\gamma_{SCPR}(0,\phi)=\emptyset \quad \gamma_{SCPR}(1,\phi)=\emptyset \quad \gamma_{SCPR}(b,\phi)=\emptyset$$
$$\gamma_{SCPR}(x+y,\phi)=\gamma_{SCPR}(x,\phi)\cup\gamma_{SCPR}(y,\phi) \quad \gamma_{SCPR}(x\cdot y,\phi)=\gamma_{SCPR}(x,\phi)\fatsemi y\;\cup\; x\star\gamma_{SCPR}(y,\phi)$$
$$\gamma_{SCPR}(x^*,\phi)=\gamma_{SCPR}(x,\phi)\fatsemi x^* \quad \gamma_{SCPR}(x^{\langle *\rangle},\phi)=\{1:\phi=\mset{x,x^{\langle *\rangle}}\}\quad \gamma_{SCPR}(x\between y,\phi)=\gamma_{SCPR}(x\parallel y,\phi)\;\cup\;\gamma_{SCPR}(x\mid y,\phi)$$
$$\gamma_{SCPR}(x\parallel y,\phi)=\{1:\phi=\mset{x,y}\} \quad \gamma_{SCPR}(x\mid y,\phi)=\{\rho(a,b)\}\fatsemi\gamma_{SCPR}(x'\between y',\phi)\;\textrm{with}\;x\mid y=(a\cdot x')\mid(b\cdot y')$$

$\eta_{SCPR}:\mathbb{M}(\mathcal{T}_{SCPR})\times\mathcal{T}_{SCPR}\rightarrow 2^{\mathcal{T}_{SCPR}}$ is defined inductively as follows.

$$\eta_{SCPR}(\phi, 0)=\emptyset \quad \eta_{SCPR}(\phi, 1)=\{1\} \quad \eta_{SCPR}(\phi, b)=\{b\}$$
$$\eta_{SCPR}(\phi, x+y)=\eta_{SCPR}(\phi, x)\cup\eta_{SCPR}(\phi, y) \quad \eta_{SCPR}(\phi, x\cdot y)=\eta_{SCPR}(\phi, x)\fatsemi y\;\cup\; x\star\eta_{SCPR}(\phi, y)$$
$$\eta_{SCPR}(\phi, x^*)=\eta_{SCPR}(\phi, x)\fatsemi x^* \quad \eta_{SCPR}(\phi,x^{\langle *\rangle})=\{1\}\fatsemi\gamma_{SCPR}(x^{\langle *\rangle},\phi')$$
$$\eta_{SCPR}(\phi, x\between y)=\eta_{SCPR}(\phi, x\parallel y)\;\cup\;\eta_{SCPR}(\phi, x\mid y)\quad\eta_{SCPR}(\phi, x\parallel y)=\{1\}\fatsemi\gamma_{SCPR}(x\parallel y,\phi')$$ 
$$\eta_{SCPR}(\phi, x\mid y)=\{1\}\fatsemi\gamma_{SCPR}(x\mid y,\phi')$$
\end{definition}

\begin{lemma}
Let $x_1,x_2\in\mathcal{T}_{SCPR}$ and $U\in\mathsf{SCP}$. The following two conclusions are equivalent:

\begin{enumerate}
  \item There exists a $y\in\mathcal{F}_{SCPR}$ such that $x_1+x_2\xrightarrow[A_{SCPR}]{U} y$.
  \item There exists a $y\in\mathcal{F}_{SCPR}$ such that $x_1\xrightarrow[A_{SCPR}]{U} y$ or $x_2\xrightarrow[A_{SCPR}]{U} y$.
\end{enumerate}
\end{lemma}

\begin{lemma}
Let $x_1,x_2\in\mathcal{T}_{SCPR}$, $U\in\mathsf{SCP}$ and $\ell\in\mathbb{N}$. The following two conclusions are equivalent:

\begin{enumerate}
  \item There exists a $y\in\mathcal{F}_{SCPR}$ such that $x_1\cdot x_2\xrightarrow[A_{SCPR}]{U} y$ of length $\ell$.
  \item $U=U_1\cdot U_2$, then there exist $y_1,y_2\in\mathcal{F}_{SCPR}$ such that $x_1\xrightarrow[A_{SCPR}]{U_1} y_1$ or $x_2\xrightarrow[A_{SCPR}]{U_2} y_2$ of length at most $\ell$.
\end{enumerate}
\end{lemma}

\begin{lemma}
Let $x_1,x_2\in\mathcal{T}_{SCPR}$, $U\in\mathsf{SCP}$. The following two conclusions are equivalent:

\begin{enumerate}
  \item There exists a $y\in\mathcal{F}_{SCPR}$ such that $x_1\parallel x_2\xrightarrow[A_{SCPR}]{U} y$.
  \item $U=U_1\parallel U_2$, then there exist $y_1,y_2\in\mathcal{F}_{SCPR}$ such that $x_1\xrightarrow[A_{SCPR}]{U_1} y_1$ or $x_2\xrightarrow[A_{SCPR}]{U_2} y_2$.
\end{enumerate}
\end{lemma}

\begin{lemma}
Let $x_1,x_2\in\mathcal{T}_{SCPR}$, $U\in\mathsf{SCP}$. The following two conclusions are equivalent:

\begin{enumerate}
  \item There exists a $y\in\mathcal{F}_{SCPR}$ such that $x_1\mid x_2\xrightarrow[A_{SCPR}]{U} y$.
  \item $U=U_1\mid U_2$, then there exist $y_1,y_2\in\mathcal{F}_{SCPR}$ such that $x_1\xrightarrow[A_{SCPR}]{U_1} y_1$ or $x_2\xrightarrow[A_{SCPR}]{U_2} y_2$.
\end{enumerate}
\end{lemma}

\begin{lemma}
Let $x_1,x_2\in\mathcal{T}_{SCPR}$, $U\in\mathsf{SCP}$. The following two conclusions are equivalent:

\begin{enumerate}
  \item There exists a $y\in\mathcal{F}_{SCPR}$ such that $x_1\leftmerge x_2\xrightarrow[A_{SCPR}]{U} y$.
  \item $U=U_1\leftmerge U_2$, then there exist $y_1,y_2\in\mathcal{F}_{SCPR}$ such that $x_1\xrightarrow[A_{SCPR}]{U_1} y_1$ or $x_2\xrightarrow[A_{SCPR}]{U_2} y_2$.
\end{enumerate}
\end{lemma}

\begin{lemma}
Let $x_1,x_2\in\mathcal{T}_{SCPR}$, $U\in\mathsf{SCP}$. The following two conclusions are equivalent:

\begin{enumerate}
  \item There exists a $y\in\mathcal{F}_{SCPR}$ such that $x_1\between x_2\xrightarrow[A_{SCPR}]{U} y$.
  \item $U=U_1\between U_2$, then there exist $y_1,y_2\in\mathcal{F}_{SCPR}$ such that $x_1\xrightarrow[A_{SCPR}]{U_1} y_1$ or $x_2\xrightarrow[A_{SCPR}]{U_2} y_2$.
\end{enumerate}
\end{lemma}

\begin{lemma}
Let $x\in\mathcal{T}_{SCPR}$, $U\in\mathsf{SCP}$. The following two conclusions are equivalent:

\begin{enumerate}
  \item There exists a $y\in\mathcal{F}_{SCPR}$ such that $x^*\xrightarrow[A_{SCPR}]{U} y$.
  \item $U=U_1\cdots U_n$, then there exist $y_i\in\mathcal{F}_{SCPR}$ such that $x\xrightarrow[A_{SCPR}]{U_i} y_i$ for $1\leq i\leq n$.
\end{enumerate}
\end{lemma}

\begin{lemma}
Let $x\in\mathcal{T}_{SCPR}$, $U\in\mathsf{SCP}$. The following two conclusions are equivalent:

\begin{enumerate}
  \item There exists a $y\in\mathcal{F}_{SCPR}$ such that $x^{\langle *\rangle}\xrightarrow[A_{SCPR}]{U} y$.
  \item $U=U_1\parallel\cdots\parallel U_n$, then there exist $y_i\in\mathcal{F}_{SCPR}$ such that $x\xrightarrow[A_{SCPR}]{U_i} y_i$ for $1\leq i\leq n$.
\end{enumerate}
\end{lemma}

\begin{lemma}
Let $x,y\in\mathcal{T}_{SCPR}$, then the following hold:

\begin{enumerate}
  \item $L_{SCPR}(x+y)=L_{SCPR}(x)+L_{SCPR}(y)$.
  \item $L_{SCPR}(x\cdot y)=L_{SCPR}(x)\cdot L_{SCPR}(y)$.
  \item $L_{SCPR}(x^*)=L_{SCPR}(x)^*$.
  \item $L_{SCPR}(x^{\langle *\rangle})=L_{SCPR}(x)^{\langle *\rangle}$.
  \item $L_{SCPR}(x\between y)=L_{SCPR}(x)\between L_{SCPR}(y)$.
  \item $L_{SCPR}(x\parallel y)=L_{SCPR}(x)\parallel L_{SCPR}(y)$.
  \item $L_{SCPR}(x\mid y)=L_{SCPR}(x)\mid L_{SCPR}(y)$.
  \item $L_{SCPR}(x\leftmerge y)=L_{SCPR}(x)\leftmerge L_{SCPR}(y)$.
\end{enumerate}
\end{lemma}

\begin{lemma}
For all $x\in\mathcal{T}_{SCPR}$, it holds that $L_{SCPR}(x)=\sembrack{x}_{SCPR}$.
\end{lemma}

\begin{definition}[$\langle *\rangle$-depth]
We extend the domain of $d_{\between}$, $d_{\parallel}$ and $d_{\mid}$ to $\mathcal{T}_{SCPR}$ by defining $d_{\between}(x^{\langle *\rangle})=d_{\between}(x)$, $d_{\parallel}(x^{\langle *\rangle})=d_{\parallel}(x)$ and $d_{\mid}(x^{\langle *\rangle})=d_{\mid}(x)$. And we define the $\langle *\rangle$-depth of $x\in\mathcal{T}_{SCPR}$ denoted $d_{\langle *\rangle}(x)$ is defined inductively on the structure of $x$ as follows.

$$d_{\langle *\rangle}(0)=0\quad d_{\langle *\rangle}(1)=0\quad d_{\langle *\rangle}(a)=0$$
$$d_{\langle *\rangle}(x\cdot y)=\max(d_{\langle *\rangle}(x),d_{\langle *\rangle}(y))\quad d_{\langle *\rangle}(x+ y)=\max(d_{\langle *\rangle}(x),d_{\langle *\rangle}(y))\quad d_{\langle *\rangle}(x^*)=d_{\langle *\rangle}(x)$$ 
$$d_{\langle *\rangle}(x\between y)=\max(d_{\langle *\rangle}(x\parallel y),d_{\langle *\rangle}(x\mid y))\quad d_{\langle *\rangle}(x\parallel y)=\max(d_{\langle *\rangle}(x),d_{\langle *\rangle}(y))$$
$$d_{\langle *\rangle}(x\mid y)=\max(d_{\langle *\rangle}(x),d_{\langle *\rangle}(y))\quad d_{\langle *\rangle}(x\leftmerge y)=\max(d_{\langle *\rangle}(x),d_{\langle *\rangle}(y))\quad d_{\langle *\rangle}(x^{\langle *\rangle})=d_{\langle *\rangle}(x)+1$$
\end{definition}

\begin{lemma}
Let $x,y\in\mathcal{T}_{SCPR}$, if $x\preceq_{SCPR} y$, then $d_{\between}(x)\leq d_{\between}(y)$, $d_{\parallel}(x)\leq d_{\parallel}(y)$, $d_{\leftmerge}(x)\leq d_{\leftmerge}(y)$, $d_{\mid}(x)\leq d_{\mid}(y)$ and $d_{\langle *\rangle}(x)\leq d_{\langle *\rangle}(y)$.
\end{lemma}

\begin{lemma}
Let $x,y\in\mathcal{T}_{SCPR}$, if $x\preceq_{SCPR} y^{\langle *\rangle}$ and $d_{\langle *\rangle}(x)=d_{\langle *\rangle}(y^{\langle *\rangle})$, then $x=y^{\langle *\rangle}$.
\end{lemma}

\begin{lemma}
Let $x,y,z,h\in\mathcal{T}_{SCPR}$ and $\phi\in\mathbb{M}(\mathcal{T}_{SCPR})$ with $\gamma(x,\phi)\neq\emptyset$, then $\phi=\mset{y,z}$ with either (1) $y\prec_{SCPR}x$ and $g\prec_{SCPR}x$, or (2) $y\prec_{SCPR} x$ and $z=h^{\langle *\rangle}$ for some $h\in\mathcal{T}_{SCPR}$.
\end{lemma}

\begin{lemma}
Every $x\in\mathcal{T}_{SCPR}$ is either recursive or progressive in $A_{SCPR}$, and the syntactic PA is well-nested.
\end{lemma}

\begin{definition}
Let $x_1,x_2\in\mathcal{T}_{SCPR}$, $R:\mathcal{T}_{SCPR}\rightarrow 2^{\mathcal{T}_{SCPR}}$ is defined inductively as follows:

$$R(0)=\{0\}\quad R(1)=\{1\} \quad R(a)=\{a,1\}$$
$$R(x_1+x_2)=R(x_1)\cup R(x_2)\quad R(x_1\cdot x_2)=R(x_1)\fatsemi x_2\;\cup\; R(x_1)\cup R(x_2)$$
$$R(x_1^*)=R(x_1)\fatsemi x_1^*\;\cup\; R(x_1)\cup\{x_1^*\}\quad R(x_1^{\langle *\rangle})=R(x_1)\cup\{x_1^{\langle *\rangle},1\} \quad R(x_1\between x_2)=R(x_1)\cup R(x_2)\cup\{x_1\between x_2,1\}$$
$$R(x_1\parallel x_2)=R(x_1)\cup R(x_2)\cup\{x_1\parallel x_2,1\}\quad R(x_1\mid x_2)=R(x_1)\cup R(x_2)\cup\{x_1\mid x_2,1\}$$
$$R(x_1\leftmerge x_2)=R(x_1)\cup R(x_2)\cup\{x_1\leftmerge x_2,1\}$$
\end{definition}

\begin{lemma}
For every $x\in\mathcal{T}_{SCPR}$, they hold that:

\begin{enumerate}
  \item $x\in R(x)$.
  \item $R(x)$ is support-closed.
  \item The syntactic PA is bounded.
\end{enumerate}
\end{lemma}

\begin{theorem}[Expressions to pomsetc automata]\label{etoa2}
For every $x\in\mathcal{T}_{SCPR}$, we can obtain a well-nested and finite PA $A$ with a state $q$ such that $L_{A}(q)=\sembrack{x}_{SCPR}$.
\end{theorem}

\begin{theorem}
For $x,y\in\mathcal{T}_{SCPR}$, according to \cref{etoa2}, we obtain two corresponding PA $A_x$ and $A_y$. If $x\sim_p y$, then $A_x\sim_p A_y$.
\end{theorem}

\begin{theorem}
For $x,y\in\mathcal{T}_{SCPR}$, according to \cref{etoa2}, we obtain two corresponding PA $A_x$ and $A_y$. If $x\sim_s y$, then $A_x\sim_s A_y$.
\end{theorem}

\begin{theorem}
For $x,y\in\mathcal{T}_{SCPR}$, according to \cref{etoa2}, we obtain two corresponding PA $A_x$ and $A_y$. If $x\sim_{hp} y$, then $A_x\sim_{hp} A_y$.
\end{theorem}

\begin{theorem}
For $x,y\in\mathcal{T}_{SCPR}$, according to \cref{etoa2}, we obtain two corresponding PA $A_x$ and $A_y$. If $x\sim_{hhp} y$, then $A_x\sim_{hhp} A_y$.
\end{theorem}

\subsubsection{Pomsetc Automata to Expressions}

In this section, we show that the language accepted by a state in any well-nested and finite automaton can be implemented by a series-communication-parallel rational expression.

\begin{definition}[1-Solution of a PA]
Let $A=(Q,F,\delta,\gamma,\eta)$ be a PA, and let $=_1$ be an EBKAC language congruence on $\mathcal{T}_{SCPR}(\Delta)$ with $\Sigma\subseteq\Delta$. We say that $s:Q\rightarrow\mathcal{T}_{SCPR}(\Delta)$ is an $=_1$-solution to $A$, if for every $q\in Q$:

\begin{center}
$[q\in \mathcal{F}_{SCPR}]+\sum_{q'\in\delta(q,a)}a\cdot s(q')+\sum_{q'\in\gamma(q,\mset{r_1,\cdots,r_n})}(s(r_1)\between\cdots\between s(r_n))\cdot s(q')\linebreak
+\sum_{\gamma(q,\mset{r_1,\cdots,r_n}),\eta(\mset{r_1,\cdots,r_n},q')}(s(r_1)\between\cdots\between s(r_n))\cdot s(q')\leqq_1 s(q)$
\end{center}

Also, $s$ is a least $=_1$-solution to $A$ if for every $=_1$-solution $s'$ it holds that $s(q)\leqq_1 s'(q)$ for all $q\in Q$. We call $s:Q\rightarrow\mathcal{T}_{SCPR}$ the least 1-solution to $A$ if it is the least $=_1$-solution for any EBKAC language congruence $=_1$.
\end{definition}

\begin{lemma}
Let $A=(Q,F,\delta,\gamma,\eta)$ be a pomsetc automaton. If $s:Q\rightarrow\mathcal{T}_{SCPR}$ is the least 1-solution to $A$, then it holds that $L_A(q)=\sembrack{s(q)}_{SCPR}$ for $q\in Q$.
\end{lemma}

\begin{lemma}
Let $A$ be a well-nested and finite PA, then we can construct the least 1-solution to $A$.
\end{lemma}

\begin{theorem}[Pomsetc automata to expressions]
If $A=(Q,F,\delta,\gamma,\eta)$ is a well-nested and finite PA, then we can construct for every $q\in Q$ a series-communication-parallel rational expression $x\in\mathcal{T}_{SCPR}$ such that $L_{A}(q)=\sembrack{x}_{SCPR}$.
\end{theorem}

\begin{corollary}[Kleene theorem for series-communication-parallel rational language]
Let $L\subseteq\mathsf{SCP}$, then $L$ is series-communication-parallel rational if and only if it is accepted by a finite and well-nested pomsetc automaton.
\end{corollary}

\begin{definition}[2-Solution of a PA]
Let $A=(Q,F,\delta,\gamma,\eta)$ be a PA, and let $=_2$ be an EBKAC pomset bisimilar congruence on $\mathcal{T}_{SCPR}(\Delta)$ with $\Sigma\subseteq\Delta$. We say that $s:Q\rightarrow\mathcal{T}_{SCPR}(\Delta)$ is an $=_2$-solution to $A$, if for every $q\in Q$:

\begin{center}
$[q\in \mathcal{F}_{SCPR}]+\sum_{q'\in\delta(q,a)}a\cdot s(q')+\sum_{q'\in\gamma(q,\mset{r_1,\cdots,r_n})}(s(r_1)\between\cdots\between s(r_n))\cdot s(q')\linebreak
+\sum_{\gamma(q,\mset{r_1,\cdots,r_n}),\eta(\mset{r_1,\cdots,r_n},q')}(s(r_1)\between\cdots\between s(r_n))\cdot s(q')\leqq_2 s(q)$
\end{center}

Also, $s$ is a least $=_2$-solution to $A$ if for every $=_2$-solution $s'$ it holds that $s(q)\leqq_2 s'(q)$ for all $q\in Q$. We call $s:Q\rightarrow\mathcal{T}_{SCPR}$ the least 2-solution to $A$ if it is the least $=_2$-solution for any EBKAC pomset bisimilar congruence $=_2$.
\end{definition}

\begin{lemma}
Let $A$ be a well-nested and finite PA, then we can construct the least 2-solution to $A$.
\end{lemma}

\begin{theorem}[Pomsetc automata to expressions modulo pomset bisimilarity]
If $A=(Q,F,\delta,\gamma,\eta)$ and $A=(Q',F',\delta',\gamma',\eta')$  are well-nested and finite PA, then we can construct for each $q\in Q$ a series-communication-parallel rational expression $x\in\mathcal{T}_{SCPR}$ and for each $q'\in Q$ a series-communication-parallel rational expression $x'\in\mathcal{T}_{SCPR}$, such that if $A\sim_p A'$ then $x\sim_p x'$.
\end{theorem}

\begin{corollary}[Kleene theorem for series-communication-parallel rational language modulo pomset bisimilarity]
Let $L\subseteq\mathsf{SCP}$, then $L$ is series-communication-parallel rational and $x,y\subseteq L$ with $x\sim_p y$, if and only if there exist finite and well-nested pomsetc automata $A_x$ and $A_y$ such that $A_x\sim_p A_y$.
\end{corollary}

\begin{definition}[3-Solution of a PA]
Let $A=(Q,F,\delta,\gamma,\eta)$ be a PA, and let $=_3$ be an EBKAC step bisimilar congruence on $\mathcal{T}_{SCPR}(\Delta)$ with $\Sigma\subseteq\Delta$. We say that $s:Q\rightarrow\mathcal{T}_{SCPR}(\Delta)$ is an $=_3$-solution to $A$, if for every $q\in Q$:

\begin{center}
$[q\in \mathcal{F}_{SCPR}]+\sum_{q'\in\delta(q,a)}a\cdot s(q')+\sum_{q'\in\gamma(q,\mset{r_1,\cdots,r_n})}(s(r_1)\between\cdots\between s(r_n))\cdot s(q')\linebreak
+\sum_{\gamma(q,\mset{r_1,\cdots,r_n}),\eta(\mset{r_1,\cdots,r_n},q')}(s(r_1)\between\cdots\between s(r_n))\cdot s(q')\leqq_3 s(q)$
\end{center}

Also, $s$ is a least $=_3$-solution to $A$ if for every $=_3$-solution $s'$ it holds that $s(q)\leqq_3 s'(q)$ for all $q\in Q$. We call $s:Q\rightarrow\mathcal{T}_{SCPR}$ the least 3-solution to $A$ if it is the least $=_3$-solution for any EBKAC step bisimilar congruence $=_3$.
\end{definition}

\begin{lemma}
Let $A$ be a well-nested and finite PA, then we can construct the least 3-solution to $A$.
\end{lemma}

\begin{theorem}[Pomsetc automata to expressions modulo step bisimilarity]
If $A=(Q,F,\delta,\gamma,\eta)$ and $A=(Q',F',\delta',\gamma',\eta')$  are well-nested and finite PA, then we can construct for each $q\in Q$ a series-communication-parallel rational expression $x\in\mathcal{T}_{SCPR}$ and for each $q'\in Q$ a series-communication-parallel rational expression $x'\in\mathcal{T}_{SCPR}$, such that if $A\sim_p A'$ then $x\sim_s x'$.
\end{theorem}

\begin{corollary}[Kleene theorem for series-communication-parallel rational language modulo step bisimilarity]
Let $L\subseteq\mathsf{SCP}$, then $L$ is series-communication-parallel rational and $x,y\subseteq L$ with $x\sim_s y$, if and only if there exist finite and well-nested pomsetc automata $A_x$ and $A_y$ such that $A_x\sim_s A_y$.
\end{corollary}

\begin{definition}[4-Solution of a PA]
Let $A=(Q,F,\delta,\gamma,\eta)$ be a PA, and let $=_4$ be an EBKAC hp-bisimilar congruence on $\mathcal{T}_{SCPR}(\Delta)$ with $\Sigma\subseteq\Delta$. We say that $s:Q\rightarrow\mathcal{T}_{SCPR}(\Delta)$ is an $=_4$-solution to $A$, if for every $q\in Q$:

\begin{center}
$[q\in \mathcal{F}_{SCPR}]+\sum_{q'\in\delta(q,a)}a\cdot s(q')+\sum_{q'\in\gamma(q,\mset{r_1,\cdots,r_n})}(s(r_1)\between\cdots\between s(r_n))\cdot s(q')\linebreak
+\sum_{\gamma(q,\mset{r_1,\cdots,r_n}),\eta(\mset{r_1,\cdots,r_n},q')}(s(r_1)\between\cdots\between s(r_n))\cdot s(q')\leqq_4 s(q)$
\end{center}

Also, $s$ is a least $=_4$-solution to $A$ if for every $=_4$-solution $s'$ it holds that $s(q)\leqq_4 s'(q)$ for all $q\in Q$. We call $s:Q\rightarrow\mathcal{T}_{SCPR}$ the least 4-solution to $A$ if it is the least $=_4$-solution for any EBKAC hp-bisimilar congruence $=_4$.
\end{definition}

\begin{lemma}
Let $A$ be a well-nested and finite PA, then we can construct the least 4-solution to $A$.
\end{lemma}

\begin{theorem}[Posmetc automata to expressions modulo hp-bisimilarity]
If $A=(Q,F,\delta,\gamma,\eta)$ and $A=(Q',F',\delta',\gamma',\eta')$  are well-nested and finite PA, then we can construct for each $q\in Q$ a series-communication-parallel rational expression $x\in\mathcal{T}_{SCPR}$ and for each $q'\in Q$ a series-communication-parallel rational expression $x'\in\mathcal{T}_{SCPR}$, such that if $A\sim_{hp} A'$ then $x\sim_{hp} x'$.
\end{theorem}

\begin{corollary}[Kleene theorem for series-communication-parallel rational language modulo hp-bisimilarity]
Let $L\subseteq\mathsf{SCP}$, then $L$ is series-communication-parallel rational and $x,y\subseteq L$ with $x\sim_{hp} y$, if and only if there exist finite and well-nested pomsetc automata $A_x$ and $A_y$ such that $A_x\sim_{hp} A_y$.
\end{corollary}

\begin{definition}[5-Solution of a PA]
Let $A=(Q,F,\delta,\gamma,\eta)$ be a PA, and let $=_5$ be an EBKAC hhp-bisimilar congruence on $\mathcal{T}_{SCPR}(\Delta)$ with $\Sigma\subseteq\Delta$. We say that $s:Q\rightarrow\mathcal{T}_{SCPR}(\Delta)$ is an $=_5$-solution to $A$, if for every $q\in Q$:

\begin{center}
$[q\in \mathcal{F}_{SCPR}]+\sum_{q'\in\delta(q,a)}a\cdot s(q')+\sum_{q'\in\gamma(q,\mset{r_1,\cdots,r_n})}(s(r_1)\between\cdots\between s(r_n))\cdot s(q')\linebreak
+\sum_{\gamma(q,\mset{r_1,\cdots,r_n}),\eta(\mset{r_1,\cdots,r_n},q')}(s(r_1)\between\cdots\between s(r_n))\cdot s(q')\leqq_5 s(q)$
\end{center}

Also, $s$ is a least $=_5$-solution to $A$ if for every $=_5$-solution $s'$ it holds that $s(q)\leqq_5 s'(q)$ for all $q\in Q$. We call $s:Q\rightarrow\mathcal{T}_{SCPR}$ the least 5-solution to $A$ if it is the least $=_5$-solution for any EBKAC hhp-bisimilar congruence $=_5$.
\end{definition}

\begin{lemma}
Let $A$ be a well-nested and finite PA, then we can construct the least 5-solution to $A$.
\end{lemma}

\begin{theorem}[Pomsetc automata to expressions modulo hhp-bisimilarity]
If $A=(Q,F,\delta,\gamma,\eta)$ and $A=(Q',F',\delta',\gamma',\eta')$  are well-nested and finite PA, then we can construct for each $q\in Q$ a series-communication-parallel rational expression $x\in\mathcal{T}_{SCPR}$ and for each $q'\in Q$ a series-communication-parallel rational expression $x'\in\mathcal{T}_{SCPR}$, such that if $A\sim_{hhp} A'$ then $x\sim_{hhp} x'$.
\end{theorem}

\begin{corollary}[Kleene theorem for series-communication-parallel rational language modulo hhp-bisimilarity]
Let $L\subseteq\mathsf{SCP}$, then $L$ is series-communication-parallel rational and $x,y\subseteq L$ with $x\sim_{hhp} y$, if and only if there exist finite and well-nested pomsetc automata $A_x$ and $A_y$ such that $A_x\sim_{hhp} A_y$.
\end{corollary} 
\newpage\section{Step Automata}\label{sa}

In this chapter, firstly, we introduce step automaton. Then in section \ref{tcbbsa}, we introduce truly concurrent bisimilarities based on step automata. Then, we introduce the fork-acyclicity property of step automata with a correspondence to scr-expressions in section \ref{f-a2}, the well-nestedness property of step automata with a correspondence to scpr-expressions in section \ref{w-n2}. Finally, we introduce the step Turing machine in \cref{stm}.

\begin{definition}[Step automaton]
A step automaton (SA) is a tuple $A=(Q,F,\delta,\gamma)$ where:

\begin{enumerate}
  \item $Q$ is a finite set of states.
  \item $F\subseteq Q$ is the set of accepting states.
  \item $\delta:Q\times\Sigma\rightarrow 2^Q$ is the sequential transition function which is the transition of traditional Kleene automata.
  \item $\gamma:Q\times U\rightarrow 2^Q$ is the step transition function where $U\in \mathsf{SCP}(\Sigma)$ and $\mathsf{SCP}(\Sigma)$ is the series-parallel step of $\Sigma$, which is a step of actions.
\end{enumerate}
\end{definition}

The SA accepting $a\cdot (b\parallel c)\cdot d$ is illustrated in Figure \ref{figure:example33}. 

\begin{figure}
  \centering
  \begin{tikzpicture}[every node/.style={transform shape}]
    \node[state] (q0) {$q_0$};
    \node[state,right=15mm of q0] (q1) {$q_1$};
    \node[state,right=15mm of q1] (q2) {$q_2$};
    \node[state,accepting,right=15mm of q2] (q3) {$q_3$};

    \draw (q0) edge[->] node[above] {$a$} (q1);
    \draw (q1) edge[->] node[above] {$\semangle{b,c}$} (q2);
    \draw (q2) edge[->] node[above] {$d$} (q3);
  \end{tikzpicture}
  \caption{SA accepting $a \cdot (b \parallel c) \cdot d$.}\label{figure:example33}
\end{figure}

\begin{definition}[Run relation]
Let $\mathsf{SCP}(\Sigma)$ be the series-parallel step of $\Sigma$, $a\in\Sigma$, $q,q'\in Q$ and $U\subseteq \mathsf{SCP}(\Sigma)$. We define the run relation $\xrightarrow[A]{}\subseteq Q\times \mathsf{SCP}(\Sigma)\times Q$ on a SA $A$ as the smallest relation satisfying:

\begin{enumerate}
  \item $q\xrightarrow[A]{1}q$ for all $q\in Q$.
  \item $q\xrightarrow[A]{a}q'$ if and only if $q'\in\delta(q,a)$.
  \item $q\xrightarrow[A]{U}q'$ if and only if $q'\in\gamma(q,U)$.
\end{enumerate}

Each $q\xrightarrow[A]{1}q$, $q\xrightarrow[A]{a}q'$, and $q\xrightarrow[A]{U}q'$ is called a unit run.
\end{definition}

\begin{definition}[Paths]
Let $A=(Q,F,\delta,\gamma)$ be an SA. We generalize the run relation $\rightarrow$ to paths $\xtworightarrow{}$, for $w\in \mathsf{SCP}(\Sigma)^*$ and $q_i,q_j\in Q$, then $q_i\xtworightarrow[A]{w}q_j$ denotes a path from $q_i$ to $q_j$ with label $w$, which can be derived as follows:

\begin{enumerate}
  \item For all $q\in Q$, it holds that $q\xtworightarrow[A]{1}q$.
  \item For all $q_i,q_j,q_k\in Q$, if $q_i\xrightarrow[A]{U}q_j$ and $q_j\xtworightarrow[A]{w}q_k$, then $q_i\xtworightarrow[A]{Uw}q_k$.
\end{enumerate}
\end{definition}

\begin{definition}[Language of SA]
The SA $A=(Q,A,\delta,\gamma)$ accepts the language by $q\in Q$, is the set $L_A(q)=\{w\in\mathsf{SCP}(\Sigma)^*:q\xtworightarrow[A]{w}q'\in F\}$. 
\end{definition}

\subsection{Truly Concurrent Bisimilarities Based on Step Automata}\label{tcbbsa}

\begin{definition}[Pomset, step bisimulation]
Let $A=(Q,F,\delta,\gamma)$ and $A'=(Q',F',\delta',\gamma')$ be two step automata with the same alphabet $\Sigma$. The automata $A$ and $A'$ are pomset bisimilar, denoted $A\sim_p A'$, if and only if there is a relation $R$ between their reachable states that preserves transitions and termination:

\begin{enumerate}
  \item $R$ relate reachable states, i.e., every reachable state of $A$ is related to a reachable state of $A'$ and every reachable state of $A'$ is related to a reachable state of $A$.
  \item For $p,q\in Q$, whenever $p$ is related to $p'\in Q'$, $pRp'$ and $p\xrightarrow[A]{U}q$ with $U\subseteq \mathsf{SCP}(\Sigma)$, then there is state $q'\in Q'$ with $p'\xrightarrow[A']{U}q'$ and $qRq'$.
  \item For $p,q\in Q$, whenever $p$ is related to $p'\in Q'$, $pRp'$ and $p'\xrightarrow[A']{U}q'$ with $U\subseteq \mathsf{SCP}(\Sigma)$, then there is state $q\in Q$ in $A$ with $p\xrightarrow[A]{U}q$ and $qRq'$.
  \item Whenever $pRp'$, then $p\in F$ if and only if $p'\in F'$.
\end{enumerate}

When the events in $U$ are pairwise concurrent (without causalities), we get the definition of step bisimulation, the automata $A$ and $A'$ are step bisimilar, denoted $A\sim_s A'$.
\end{definition}

\begin{definition}[Configuration]
A (finite) configuration in $A=(Q,F,\delta,\gamma)$ is a (finite) consistent subset of events (without alternative composition +) $\mathbf{C}\subseteq \mathsf{SCP}(\Sigma)$, closed with respect to causality (i.e. $\lceil \mathbf{C}\rceil=\mathbf{C}$). The set of finite configurations of $A=(Q,F,\delta,\gamma)$ is denoted by $\mathcal{C}(A)$.
\end{definition}

\begin{definition}[Posetal product]
Given two step automata $A_1=(Q_1,F_1,\delta_1,\gamma_1)$, $A_2=(Q_2,F_2,\delta_2,\gamma_2)$, the posetal product of their configurations, denoted $\mathcal{C}(A_1)\overline{\times}\mathcal{C}(A_2)$, is defined as

$$\{(\mathbf{C}_1,f,\mathbf{C}_2)|\mathbf{C}_1\in\mathcal{C}(A_1),\mathbf{C}_2\in\mathcal{C}(A_2),f:\mathbf{C}_1\rightarrow \mathbf{C}_2 \textrm{ isomorphism}\}$$

A subset $R\subseteq\mathcal{C}(A_1)\overline{\times}\mathcal{C}(A_2)$ is called a posetal relation. We say that $R$ is downward closed when for any $(\mathbf{C}_1,f,\mathbf{C}_2),(\mathbf{C}_1',f',\mathbf{C}_2')\in \mathcal{C}(A_1)\overline{\times}\mathcal{C}(A_2)$, if $(\mathbf{C}_1,f,\mathbf{C}_2)\subseteq (\mathbf{C}_1',f',\mathbf{C}_2')$ pointwise and $(\mathbf{C}_1',f',\mathbf{C}_2')\in R$, then $(\mathbf{C}_1,f,\mathbf{C}_2)\in R$.

For $f:X_1\rightarrow X_2$, we define $f[a_1\mapsto a_2]:X_1\cup\{a_1\}\rightarrow X_2\cup\{a_2\}$, $z\in X_1\cup\{a_1\}$,(1)$f[a_1\mapsto a_2](z)=
a_2$,if $z=a_1$;(2)$f[a_1\mapsto a_2](z)=f(z)$, otherwise. Where $X_1\subseteq \mathsf{SCP}(\Sigma_1)$, $X_2\subseteq \mathsf{SCP}(\Sigma_2)$, $a_1\in \mathsf{SCP}(\Sigma_1)$, $a_2\in \mathsf{SCP}(\Sigma_2)$.
\end{definition}

\begin{definition}[(Hereditary) history-preserving bisimulation]
Let $A_1=(Q_1,F_1,\delta_1,\gamma_1)$ and $A_2=(Q_2,F_2,\delta_2,\gamma_2)$ be two step automata. A history-preserving (hp-) bisimulation is a posetal relation $R\subseteq\mathcal{C}(A_1)\overline{\times}\mathcal{C}(A_2)$ such that if $(\mathbf{C}_1,f,\mathbf{C}_2)\in R$, and $\mathbf{C}_1\xrightarrow[A_1]{a_1} \mathbf{C}_1'$, then $\mathbf{C}_2\xrightarrow[A_2]{a_2} \mathbf{C}_2'$, with $(\mathbf{C}_1',f[a_1\mapsto a_2],\mathbf{C}_2')\in R$, and vice-versa. $A_1$ and $A_2$ are history-preserving (hp-)bisimilar and are written $A_1\sim_{hp}A_2$ if there exists a hp-bisimulation $R$ such that $(\emptyset,\emptyset,\emptyset)\in R$.

A hereditary history-preserving (hhp-)bisimulation is a downward closed hp-bisimulation. $A_1,A_2$ are hereditary history-preserving (hhp-)bisimilar and are written $A_1\sim_{hhp}A_2$.
\end{definition}

Note that the above pomset, step, hp-, hhp-bisimilarities preserve deadlocks.

\subsection{Fork-acyclicity}\label{f-a2}

It has already been proven that the so-called fork-acyclic pomset automaton just exactly accepts series rational (sr) language. In the following, we extend the related concepts and conclusions from \cite{CKA7} and prove that fork-acyclic SA with merge transitions exactly accepts series-communication rational (scr) languages. And also, the laws of scr-expressions are sound and maybe complete modulo truly concurrent bisimilarities based on step automata.

\begin{definition}[Support relation]
The support relation $\preceq$ of $A$ is the smallest preorder on $Q$ and for $q\in Q$:

$$\frac{a\in\Sigma\quad q'\in \delta(q,a)}{q'\preceq_{A}q}$$
$$\frac{U\in\mathsf{SCP}(\Sigma)\quad q'\in \gamma(q,U)}{q'\preceq_{A}q}$$

We call the strict support relation $\prec_A$ if $q'\preceq_A q$ and $q\npreceq_A q'$ then $q'\prec_A q$ holds.
\end{definition}

\begin{definition}[Fork-acyclicity]
A SA $A$ is called fork-cyclic if for some $q,r\in Q$ such that $r$ is a fork target of $q$, we have that $q\preceq_{A}r$; $A$ is fork-acyclic if it is not fork-cyclic.
\end{definition}

\begin{definition}[Depth of step automaton]
If the step automaton $A$ is finite and fork-acyclic, the depth of $q\in Q$ in $A$ denoted $D_A(q)$ is the maximum $n$ such that there exist $q_1,\cdots,q_n\in Q$ with $q_1\prec_A q_1\prec_A\cdots\prec_A q_n =q$. The depth of A denoted $D_A$ is the maximum of $D_A(q)$ for all $q\in Q$ in $A$.
\end{definition}

\begin{definition}[Support]
$Q'\subseteq Q$ is support-closed if for all $q\in Q'$ with $q'\preceq_A q$ then $q'\in Q'$. The support of $q\in Q$ denoted $\pi_A(q)$ is the smallest support-closed set containing $q$.
\end{definition}

\begin{definition}[Bounded]
If $\pi_A(q)$ is finite for all $q\in Q$ in $A$, then $A$ is called bounded.
\end{definition}

\begin{definition}[Implementation]
$A=(Q,F,\delta,\gamma)$ and $A'=(Q',F',\delta',\gamma')$ are two step automata, then $A'$ implements $A$ if the following hold:

\begin{enumerate}
  \item $Q\subseteq Q'$ such that if $q\in Q$, then $L_A(q)=L_{A'}(q)$.
  \item If $A$ is fork-acyclic, then so is $A'$.
\end{enumerate}
\end{definition}

\begin{definition}[Support-closed restricted step automaton]
Let $Q'\subseteq Q$ be support-closed, the support-closed restricted SA of the SA $A=(Q,F,\delta,\gamma)$, denoted $A[Q']=(Q',F\cap Q',\delta',\gamma')$, where $\delta':Q'\times\Sigma\rightarrow 2^{Q'}$ and $\gamma':Q'\times U\rightarrow 2^{Q'}$ with:

$$\delta'(q,a)=\delta(q,a)\quad \gamma'(q,U)=\gamma(q,U)$$

where $q\in Q'$, $a\in\Sigma$ and $U\in\mathsf{SCP}(\Sigma)$.
\end{definition}

\begin{lemma}
Let PAs $A=(Q,F,\delta,\gamma)$ and $A[Q']=(Q',F\cap Q',\delta',\gamma')$, if $Q'$ is support-closed, then $A[Q']$ implements $A$, and if $A$ is bounded, then $A[Q']$ is bounded.
\end{lemma}

\subsubsection{Expressions to Step Automata}

Given an scr-expression $x$, we show that how to obtain a fork-acyclic and finite SA with some state accepting $\sembrack{x}_{SCR}$. Similarly to the process of sr-expression, we firstly construct the so-called series-communication rational syntactic SA.

\begin{definition}[Series-communication rational syntactic step automaton]
Let $x\in\mathcal{T}_{SCR}$ and $S\subseteq\mathcal{T}_{SCR}$: (1)$x\star S=S$, if $x\in\mathcal{F}_{SCR}$; (2) $x\star S=\emptyset$, otherwise. We define the series-communication rational syntactic SA as $A_{SCR}=(\mathcal{T}_{SCR},\mathcal{F}_{SCR},\delta_{SCR},\gamma_{SCR})$, where $\delta_{SCR}:\mathcal{T}_{SCR}\times\Sigma\rightarrow 2^{\mathcal{T}_{SCR}}$ is defined inductively as follows.

$$\delta_{SCR}(0,a)=\emptyset \quad \delta_{SCR}(1,a)=\emptyset \quad \delta_{SCR}(b,a)=\{1:a=b\}$$
$$\delta_{SCR}(x+y,a)=\delta_{SCR}(x,a)\cup\delta_{SCR}(y,a) \quad \delta_{SCR}(x\cdot y,a)=\delta_{SCR}(x,a)\fatsemi y\;\cup\; x\star\delta_{SCR}(y,a)$$
$$\delta_{SCR}(x^*,a)=\delta_{SCR}(x,a)\fatsemi x^* \quad \delta_{SCR}(x\between y,a)=\emptyset$$
$$\delta_{SCR}(x\parallel y,a)=\emptyset \quad \delta_{SCR}(x\mid y,a)=\emptyset$$

$\gamma_{SCR}:\mathcal{T}_{SCR}\times U\rightarrow 2^{\mathcal{T}_{SCR}}$ is defined inductively as follows.

$$\gamma_{SCR}(0,U)=\emptyset \quad \gamma_{SCR}(1,U)=\emptyset \quad \gamma_{SCR}(V,U)=\{1:U=V\}$$
$$\gamma_{SCR}(x+y,U)=\gamma_{SCR}(x,U)\cup\delta_{SCR}(y,U) \quad \gamma_{SCR}(x\cdot y,U)=\gamma_{SCR}(x,U)\fatsemi y\;\cup\; x\star\gamma_{SCR}(y,U)$$
$$\gamma_{SCR}(x^*,U)=\gamma_{SCR}(x,U)\fatsemi x^* \quad \gamma_{SCR}(x\between y,U)=\gamma_{SCR}(x\parallel y,U)\;\cup\;\gamma_{SCR}(x\mid y,U)$$
$$\gamma_{SCR}(x\parallel y,U)=\{1:U=\semangle{x,y}\}\;\cup\;\{U\}\fatsemi\gamma_{SCR}(x'\between y',U')\;\textrm{with}\;x\parallel y=U\cdot(x'\between y')$$
$$\gamma_{SCR}(x\mid y,U)=\{1:U=\semangle{x,y}\}\;\cup\;\{U\}\fatsemi\gamma_{SCR}(x'\mid y',U')\;\textrm{with}\;x\mid y=U\cdot(x'\between y')$$
\end{definition}

\begin{lemma}
Let $x_1,x_2\in\mathcal{T}_{SCR}$ and $w\in\mathsf{SCP}(\Sigma)^*$. The following two conclusions are equivalent:

\begin{enumerate}
  \item There exists a $y\in\mathcal{F}_{SCR}$ such that $x_1+x_2\xtworightarrow[A_{SCR}]{w} y$.
  \item There exists a $y\in\mathcal{F}_{SCR}$ such that $x_1\xtworightarrow[A_{SCR}]{w} y$ or $x_2\xtworightarrow[A_{SCR}]{w} y$.
\end{enumerate}
\end{lemma}

\begin{lemma}
Let $x_1,x_2\in\mathcal{T}_{SCR}$, $w\in\mathsf{SCP}(\Sigma)^*$ and $\ell\in\mathbb{N}$. The following two conclusions are equivalent:

\begin{enumerate}
  \item There exists a $y\in\mathcal{F}_{SCR}$ such that $x_1\cdot x_2\xtworightarrow[A_{SCR}]{w} y$ of length $\ell$.
  \item $w=w_1\cdot w_2$, then there exist $y_1,y_2\in\mathcal{F}_{SCR}$ such that $x_1\xtworightarrow[A_{SCR}]{w_1} y_1$ or $x_2\xtworightarrow[A_{SCR}]{w_2} y_2$ of length at most $\ell$.
\end{enumerate}
\end{lemma}

\begin{lemma}
Let $x_1,x_2\in\mathcal{T}_{SCR}$, $w\in\mathsf{SCP}(\Sigma)^*$. The following two conclusions are equivalent:

\begin{enumerate}
  \item There exists a $y\in\mathcal{F}_{SCR}$ such that $x_1\parallel x_2\xtworightarrow[A_{SCR}]{w} y$.
  \item $w=w_1\parallel w_2$, then there exist $y_1,y_2\in\mathcal{F}_{SCR}$ such that $x_1\xtworightarrow[A_{SCR}]{w_1} y_1$ or $x_2\xtworightarrow[A_{SCR}]{w_2} y_2$.
\end{enumerate}
\end{lemma}

\begin{lemma}
Let $x_1,x_2\in\mathcal{T}_{SCR}$, $w\in\mathsf{SCP}(\Sigma)^*$. The following two conclusions are equivalent:

\begin{enumerate}
  \item There exists a $y\in\mathcal{F}_{SCR}$ such that $x_1\mid x_2\xtworightarrow[A_{SCR}]{w} y$.
  \item $w=w_1\mid w_2$, then there exist $y_1,y_2\in\mathcal{F}_{SCR}$ such that $x_1\xtworightarrow[A_{SCR}]{w_1} y_1$ or $x_2\xtworightarrow[A_{SCR}]{w_2} y_2$.
\end{enumerate}
\end{lemma}

\begin{lemma}
Let $x_1,x_2\in\mathcal{T}_{SCR}$, $w\in\mathsf{SCP}(\Sigma)^*$. The following two conclusions are equivalent:

\begin{enumerate}
  \item There exists a $y\in\mathcal{F}_{SCR}$ such that $x_1\leftmerge x_2\xtworightarrow[A_{SCR}]{w} y$.
  \item $w=w_1\leftmerge w_2$, then there exist $y_1,y_2\in\mathcal{F}_{SCR}$ such that $x_1\xtworightarrow[A_{SCR}]{w_1} y_1$ or $x_2\xtworightarrow[A_{SCR}]{w_2} y_2$.
\end{enumerate}
\end{lemma}

\begin{lemma}
Let $x_1,x_2\in\mathcal{T}_{SCR}$, $w\in\mathsf{SCP}(\Sigma)^*$. The following two conclusions are equivalent:

\begin{enumerate}
  \item There exists a $y\in\mathcal{F}_{SCR}$ such that $x_1\between x_2\xtworightarrow[A_{SCR}]{w} y$.
  \item $w=w_1\between w_2$, then there exist $y_1,y_2\in\mathcal{F}_{SCR}$ such that $x_1\xtworightarrow[A_{SCR}]{w_1} y_1$ or $x_2\xtworightarrow[A_{SCR}]{w_2} y_2$.
\end{enumerate}
\end{lemma}

\begin{lemma}
Let $x\in\mathcal{T}_{SCR}$, $w\in\mathsf{SCP}(\Sigma)^*$. The following two conclusions are equivalent:

\begin{enumerate}
  \item There exists a $y\in\mathcal{F}_{SCR}$ such that $x^*\xtworightarrow[A_{SCR}]{w} y$.
  \item $w=w_1\cdots w_n$, then there exist $y_i\in\mathcal{F}_{SCR}$ such that $x\xtworightarrow[A_{SCR}]{w_i} y_i$ for $1\leq i\leq n$.
\end{enumerate}
\end{lemma}

\begin{lemma}
Let $x,y\in\mathcal{T}_{SCR}$, then the following hold:

\begin{enumerate}
  \item $L_{SCR}(x+y)=L_{SCR}(x)+L_{SCR}(y)$.
  \item $L_{SCR}(x\cdot y)=L_{SCR}(x)\cdot L_{SCR}(y)$.
  \item $L_{SCR}(x^*)=L_{SCR}(x)^*$.
  \item $L_{SCR}(x\between y)=L_{SCR}(x)\between L_{SCR}(y)$.
  \item $L_{SCR}(x\parallel y)=L_{SCR}(x)\parallel L_{SCR}(y)$.
  \item $L_{SCR}(x\mid y)=L_{SCR}(x)\mid L_{SCR}(y)$.
  \item $L_{SCR}(x\leftmerge y)=L_{SCR}(x)\leftmerge L_{SCR}(y)$.
\end{enumerate}
\end{lemma}

\begin{lemma}
For all $x\in\mathcal{T}_{SCR}$, it holds that $L_{SCR}(x)=\sembrack{x}_{SCR}$.
\end{lemma}

\begin{lemma}
Let $x,y\in\mathcal{T}_{SCR}$, if $x\preceq_{SCR} y$, then $d_{\between}(x)\leq d_{\between}(y)$, $d_{\parallel}(x)\leq d_{\parallel}(y)$, $d_{\leftmerge}(x)\leq d_{\leftmerge}(y)$ and $d_{\mid}(x)\leq d_{\mid}(y)$.
\end{lemma}

\begin{lemma}
Let $x,y\in\mathcal{T}_{SCR}$, if $y$ is a fork target of $x$ in the syntactic SA, then $d_{\between}(y)\leq d_{\between}(x)$, $d_{\parallel}(y)\leq d_{\parallel}(x)$, $d_{\leftmerge}(y)\leq d_{\leftmerge}(x)$ and $d_{\mid}(y)\leq d_{\mid}(x)$, and the syntactic SA is fork-acyclic.
\end{lemma}

\begin{definition}
Let $x_1,x_2\in\mathcal{T}_{SCR}$, $R:\mathcal{T}_{SCR}\rightarrow 2^{\mathcal{T}_{SCR}}$ is defined inductively as follows:

$$R(0)=\{0\}\quad R(1)=\{1\} \quad R(a)=\{a,1\}$$
$$R(x_1+x_2)=R(x_1)\cup R(x_2)\quad R(x_1\cdot x_2)=R(x_1)\fatsemi x_2\;\cup\; R(x_1)\cup R(x_2)$$
$$R(x_1^*)=R(x_1)\fatsemi x_1^*\;\cup\; R(x_1)\cup\{x_1^*\}\quad R(x_1\between x_2)=R(x_1)\cup R(x_2)\cup\{x_1\between x_2,1\}$$
$$R(x_1\parallel x_2)=R(x_1)\cup R(x_2)\cup\{x_1\parallel x_2,1\}\quad R(x_1\mid x_2)=R(x_1)\cup R(x_2)\cup\{x_1\mid x_2,1\}$$
$$R(x_1\leftmerge x_2)=R(x_1)\cup R(x_2)\cup\{x_1\leftmerge x_2,1\}$$
\end{definition}

\begin{lemma}
For every $x\in\mathcal{T}_{SCR}$, they hold that:

\begin{enumerate}
  \item $x\in R(x)$.
  \item $R(x)$ is support-closed.
  \item The syntactic SA is bounded.
\end{enumerate}
\end{lemma}

\begin{theorem}[Expressions to step automata]\label{etoa3}
For every $x\in\mathcal{T}_{SCR}$, we can obtain a fork-acyclic and finite SA $A$ with a state $q$ such that $L_{A}(q)=\sembrack{x}_{SCR}$.
\end{theorem}

\begin{theorem}
For $x,y\in\mathcal{T}_{SCR}$, according to \cref{etoa3}, we obtain two corresponding SA $A_x$ and $A_y$. If $x\sim_p y$, then $A_x\sim_p A_y$.
\end{theorem}

\begin{theorem}
For $x,y\in\mathcal{T}_{SCR}$, according to \cref{etoa3}, we obtain two corresponding SA $A_x$ and $A_y$. If $x\sim_s y$, then $A_x\sim_s A_y$.
\end{theorem}

\begin{theorem}
For $x,y\in\mathcal{T}_{SCR}$, according to \cref{etoa3}, we obtain two corresponding SA $A_x$ and $A_y$. If $x\sim_{hp} y$, then $A_x\sim_{hp} A_y$.
\end{theorem}

\begin{theorem}
For $x,y\in\mathcal{T}_{SCR}$, according to \cref{etoa3}, we obtain two corresponding SA $A_x$ and $A_y$. If $x\sim_{hhp} y$, then $A_x\sim_{hhp} A_y$.
\end{theorem}

\subsubsection{Step Automata to Expressions}

In this section, we show that the language accepted by a state in any fork-acyclic and finite automaton can be implemented by a series-communication rational expression.

\begin{lemma}
If $A=(Q,F,\delta,\gamma)$ be a step automaton, then $L_A:Q\rightarrow 2^{SCP}$ is the least function $t:Q\rightarrow 2^{SCP}$ (w.r.t. the pointwise inclusion order) such that for all $q\in Q$, the following hold:

$$\frac{q\in\mathcal{F}_{SCR}}{1\in t(q)} \quad \frac{a\in\Sigma\quad q'\in\delta(q,a)}{a\cdot t(q')}\subseteq t(q)$$
$$\frac{U\in\mathsf{SCP}(\Sigma)\quad q'\in\gamma(q,U)}{U\cdot t(q')}\subseteq t(q)$$
\end{lemma}

\begin{definition}[1-Solution of a SA]
Let $A=(Q,F,\delta,\gamma)$ be a SA, and let $=_1$ be a BKAC language congruence on $\mathcal{T}_{SCR}(\Delta)$ with $\Sigma\subseteq\Delta$. We say that $s:Q\rightarrow\mathcal{T}_{SCR}(\Delta)$ is an $=_1$-solution to $A$, if for every $q\in Q$:

\begin{center}
$[q\in \mathcal{F}_{SCR}]+\sum_{q'\in\delta(q,a)}a\cdot s(q')+\sum_{q'\in\gamma(q,U)}U\cdot s(q')\leqq_1 s(q)$
\end{center}

Also, $s$ is a least $=_1$-solution to $A$ if for every $=_1$-solution $s'$ it holds that $s(q)\leqq_1 s'(q)$ for all $q\in Q$. We call $s:Q\rightarrow\mathcal{T}_{SCR}$ the least 1-solution to $A$ if it is the least $=_1$-solution for any BKAC language congruence $=_1$.
\end{definition}

\begin{lemma}
Let $A=(Q,F,\delta,\gamma)$ be a step automaton. If $s:Q\rightarrow\mathcal{T}_{SCR}$ is the least 1-solution to $A$, then it holds that $L_A(q)=\sembrack{s(q)}_{SCR}$ for $q\in Q$.
\end{lemma}

\begin{lemma}
Let $A$ be a fork-acyclic and finite SA, then we can construct the least 1-solution to $A$.
\end{lemma}

\begin{theorem}[Step automata to expressions]
If $A=(Q,F,\delta,\gamma)$ is a fork-acyclic and finite SA, then we can construct for every $q\in Q$ a series-communication rational expression $x\in\mathcal{T}_{SCR}$ such that $L_{A}(q)=\sembrack{x}_{SCR}$.
\end{theorem}

\begin{corollary}[Kleene theorem for series-communication rational language]
Let $L\subseteq\mathsf{SCP}$, then $L$ is series-communication rational if and only if it is accepted by a finite and fork-acyclic step automaton.
\end{corollary}

\begin{definition}[2-Solution of a SA]
Let $A=(Q,F,\delta,\gamma)$ be a SA, and let $=_2$ be a BKAC pomset bisimilar congruence on $\mathcal{T}_{SCR}(\Delta)$ with $\Sigma\subseteq\Delta$. We say that $s:Q\rightarrow\mathcal{T}_{SCR}(\Delta)$ is an $=_2$-solution to $A$, if for every $q\in Q$:

\begin{center}
$[q\in \mathcal{F}_{SCR}]+\sum_{q'\in\delta(q,a)}a\cdot s(q')+\sum_{q'\in\gamma(q,U)}U\cdot s(q')\leqq_2 s(q)$
\end{center}

Also, $s$ is a least $=_2$-solution to $A$ if for every $=_2$-solution $s'$ it holds that $s(q)\leqq_2 s'(q)$ for all $q\in Q$. We call $s:Q\rightarrow\mathcal{T}_{SCR}$ the least 2-solution to $A$ if it is the least $=_2$-solution for any BKAC pomset bisimilar congruence $=_2$.
\end{definition}

\begin{lemma}
Let $A$ be a fork-acyclic and finite SA, then we can construct the least 2-solution to $A$.
\end{lemma}

\begin{theorem}[Step automata to expressions modulo pomset bisimilarity]
If $A=(Q,F,\delta,\gamma)$ and $A=(Q',F',\delta',\gamma')$  are fork-acyclic and finite SA, then we can construct for each $q\in Q$ a series-communication rational expression $x\in\mathcal{T}_{SCR}$ and for each $q'\in Q$ a series-communication rational expression $x'\in\mathcal{T}_{SCR}$, such that if $A\sim_p A'$ then $x\sim_p x'$.
\end{theorem}

\begin{corollary}[Kleene theorem for series-communication rational language modulo pomset bisimilarity]
Let $L\subseteq\mathsf{SCP}$, then $L$ is series-communication rational and $x,y\subseteq L$ with $x\sim_p y$, if and only if there exist finite and fork-acyclic step automata $A_x$ and $A_y$ such that $A_x\sim_p A_y$.
\end{corollary}

\begin{definition}[3-Solution of a SA]
Let $A=(Q,F,\delta,\gamma)$ be a SA, and let $=_3$ be a BKAC step bisimilar congruence on $\mathcal{T}_{SCR}(\Delta)$ with $\Sigma\subseteq\Delta$. We say that $s:Q\rightarrow\mathcal{T}_{SCR}(\Delta)$ is an $=_3$-solution to $A$, if for every $q\in Q$:

\begin{center}
$[q\in \mathcal{F}_{SCR}]+\sum_{q'\in\delta(q,a)}a\cdot s(q')+\sum_{q'\in\gamma(q,U)}U\cdot s(q')\leqq_3 s(q)$
\end{center}

Also, $s$ is a least $=_3$-solution to $A$ if for every $=_3$-solution $s'$ it holds that $s(q)\leqq_3 s'(q)$ for all $q\in Q$. We call $s:Q\rightarrow\mathcal{T}_{SCR}$ the least 3-solution to $A$ if it is the least $=_3$-solution for any BKAC step bisimilar congruence $=_3$.
\end{definition}

\begin{lemma}
Let $A$ be a fork-acyclic and finite SA, then we can construct the least 3-solution to $A$.
\end{lemma}

\begin{theorem}[Step automata to expressions modulo step bisimilarity]
If $A=(Q,F,\delta,\gamma)$ and $A=(Q',F',\delta',\gamma')$  are fork-acyclic and finite SA, then we can construct for each $q\in Q$ a series-communication rational expression $x\in\mathcal{T}_{SCR}$ and for each $q'\in Q$ a series-communication rational expression $x'\in\mathcal{T}_{SCR}$, such that if $A\sim_p A'$ then $x\sim_s x'$.
\end{theorem}

\begin{corollary}[Kleene theorem for series-communication rational language modulo step bisimilarity]
Let $L\subseteq\mathsf{SCP}$, then $L$ is series-communication rational and $x,y\subseteq L$ with $x\sim_s y$, if and only if there exist finite and fork-acyclic step automata $A_x$ and $A_y$ such that $A_x\sim_s A_y$.
\end{corollary}

\begin{definition}[4-Solution of a SA]
Let $A=(Q,F,\delta,\gamma)$ be a SA, and let $=_4$ be a BKAC hp-bisimilar congruence on $\mathcal{T}_{SCR}(\Delta)$ with $\Sigma\subseteq\Delta$. We say that $s:Q\rightarrow\mathcal{T}_{SCR}(\Delta)$ is an $=_4$-solution to $A$, if for every $q\in Q$:

\begin{center}
$[q\in \mathcal{F}_{SCR}]+\sum_{q'\in\delta(q,a)}a\cdot s(q')+\sum_{q'\in\gamma(q,U)}U\cdot s(q')\leqq_4 s(q)$
\end{center}

Also, $s$ is a least $=_4$-solution to $A$ if for every $=_4$-solution $s'$ it holds that $s(q)\leqq_4 s'(q)$ for all $q\in Q$. We call $s:Q\rightarrow\mathcal{T}_{SCR}$ the least 4-solution to $A$ if it is the least $=_4$-solution for any BKAC hp-bisimilar congruence $=_4$.
\end{definition}

\begin{lemma}
Let $A$ be a fork-acyclic and finite SA, then we can construct the least 4-solution to $A$.
\end{lemma}

\begin{theorem}[Step automata to expressions modulo hp-bisimilarity]
If $A=(Q,F,\delta,\gamma)$ and $A=(Q',F',\delta',\gamma')$  are fork-acyclic and finite SA, then we can construct for each $q\in Q$ a series-communication rational expression $x\in\mathcal{T}_{SCR}$ and for each $q'\in Q$ a series-communication rational expression $x'\in\mathcal{T}_{SCR}$, such that if $A\sim_{hp} A'$ then $x\sim_{hp} x'$.
\end{theorem}

\begin{corollary}[Kleene theorem for series-communication rational language modulo hp-bisimilarity]
Let $L\subseteq\mathsf{SCP}$, then $L$ is series-communication rational and $x,y\subseteq L$ with $x\sim_{hp} y$, if and only if there exist finite and fork-acyclic step automata $A_x$ and $A_y$ such that $A_x\sim_{hp} A_y$.
\end{corollary}

\begin{definition}[5-Solution of a SA]
Let $A=(Q,F,\delta,\gamma)$ be a SA, and let $=_5$ be a BKAC hhp-bisimilar congruence on $\mathcal{T}_{SCR}(\Delta)$ with $\Sigma\subseteq\Delta$. We say that $s:Q\rightarrow\mathcal{T}_{SCR}(\Delta)$ is an $=_5$-solution to $A$, if for every $q\in Q$:

\begin{center}
$[q\in \mathcal{F}_{SCR}]+\sum_{q'\in\delta(q,a)}a\cdot s(q')+\sum_{q'\in\gamma(q,U)}U\cdot s(q')\leqq_5 s(q)$
\end{center}

Also, $s$ is a least $=_5$-solution to $A$ if for every $=_5$-solution $s'$ it holds that $s(q)\leqq_5 s'(q)$ for all $q\in Q$. We call $s:Q\rightarrow\mathcal{T}_{SCR}$ the least 5-solution to $A$ if it is the least $=_5$-solution for any BKAC hhp-bisimilar congruence $=_5$.
\end{definition}

\begin{lemma}
Let $A$ be a fork-acyclic and finite SA, then we can construct the least 5-solution to $A$.
\end{lemma}

\begin{theorem}[Step automata to expressions modulo hhp-bisimilarity]
If $A=(Q,F,\delta,\gamma)$ and $A=(Q',F',\delta',\gamma')$  are fork-acyclic and finite SA, then we can construct for each $q\in Q$ a series-communication rational expression $x\in\mathcal{T}_{SCR}$ and for each $q'\in Q$ a series-communication rational expression $x'\in\mathcal{T}_{SCR}$, such that if $A\sim_{hhp} A'$ then $x\sim_{hhp} x'$.
\end{theorem}

\begin{corollary}[Kleene theorem for series-communication rational language modulo hhp-bisimilarity]
Let $L\subseteq\mathsf{SCP}$, then $L$ is series-communication rational and $x,y\subseteq L$ with $x\sim_{hhp} y$, if and only if there exist finite and fork-acyclic step automata $A_x$ and $A_y$ such that $A_x\sim_{hhp} A_y$.
\end{corollary}

\subsection{Well-nestedness}\label{w-n2}

It has already been proven that the so-called well-nested pomset automaton just exactly accepts series-parallel rational (spr) language. In the following, we extend the related concepts and conclusions from \cite{CKA7} and prove that well-nested SA with merge transitions exactly accepts series-communication-parallel rational (scpr) languages. And also, the laws of scpr-expressions are sound and maybe complete modulo truly concurrent bisimilarities based on step automata.

Parallel star allows an unbounded number of events to occur in parallel, we need the following concepts.

\begin{definition}[Pomsetc width]
The width of a finite pomsetc $U=[\mathbf{u}]\in\mathsf{Pomc}$ is the size of maximum of the largest $\leq^e_{\mathbf{u}}$-antichain and the largest $\leq^c_{\mathbf{u}}$-antichain.
\end{definition}

\begin{definition}[Pomsetc depth]
The depth of $U\in \mathsf{SCP}$ denoted $d(U)$ is defined inductively as follows:

\begin{enumerate}
  \item $d(U)=0$ if $U$ is empty or primitive.
  \item $d(U)=1+\max_{1\leq i\leq n}d(U_i)$ if $U$ is sequential with sequential factorization $U_1,\cdots,U_n$.
  \item $d(U)=1+\max_{1\leq i\leq n}d(U_i)$ if $U$ is parallel with parallel factorization $\mset{U_1,\cdots,U_n}$.
\end{enumerate}
\end{definition}

\begin{definition}[Recursive states]
Let $A=(Q,F,\delta,\gamma)$ be a SA, $q\in Q$ is recursive if:

\begin{enumerate}
  \item For all $a\in\Sigma$, $q'\in \delta(q,a)$, then $q'\prec_{A}q$.
  \item For all $U\in\mathsf{SCP}(\Sigma)$, $q'\in \gamma(q,U)$, then $q'\prec_{A}q$.
\end{enumerate}
\end{definition}

\begin{definition}[Well-nestedness]
Let $A=(Q,F,\delta,\gamma)$ be a SA, $A$ is well-nested if every state is recursive.
\end{definition}

\begin{lemma}
Let PAs $A=(Q,F,\delta,\gamma)$ and $A[Q']=(Q',F\cap Q',\delta',\gamma')$, if $Q'$ is support-closed and $A$ is well-nested, then $A[Q']$ is well-nested.
\end{lemma}

\subsubsection{Expressions to Step Automata}

Given an scpr-expression $x$, we show that how to obtain a well-nested and finite SA with some state accepting $\sembrack{x}_{SCPR}$. Similarly to the process of sr-expression, we firstly construct the so-called series-communication-parallel rational syntactic SA.

\begin{definition}[Series-communication-parallel rational syntactic step automaton]
Let $x\in\mathcal{T}_{SCPR}$ and $S\subseteq\mathcal{T}_{SCPR}$: (1)$x\star S=S$, if $x\in\mathcal{F}_{SCPR}$; (2) $x\star S=\emptyset$, otherwise. We define the series-communication-parallel rational syntactic SA as $A_{SCPR}=(\mathcal{T}_{SCPR},\mathcal{F}_{SCPR},\delta_{SCPR},\gamma_{SCPR})$, where $\delta_{SCPR}:\mathcal{T}_{SCPR}\times\Sigma\rightarrow 2^{\mathcal{T}_{SCPR}}$ is defined inductively as follows.

$$\delta_{SCPR}(0,a)=\emptyset \quad \delta_{SCPR}(1,a)=\emptyset \quad \delta_{SCPR}(b,a)=\{1:a=b\}$$
$$\delta_{SCPR}(x+y,a)=\delta_{SCPR}(x,a)\cup\delta_{SCPR}(y,a) \quad \delta_{SCPR}(x\cdot y,a)=\delta_{SCPR}(x,a)\fatsemi y\;\cup\; x\star\delta_{SCPR}(y,a)$$
$$\delta_{SCPR}(x^*,a)=\delta_{SCPR}(x,a)\fatsemi x^* \quad \delta_{SCPR}(x^{\langle *\rangle},a)=\emptyset \quad \delta_{SCPR}(x\between y,a)=\emptyset$$
$$\delta_{SCPR}(x\parallel y,a)=\emptyset \quad \delta_{SCPR}(x\mid y,a)=\emptyset$$

$\gamma_{SCPR}:\mathcal{T}_{SCPR}\times U\rightarrow 2^{\mathcal{T}_{SCPR}}$ is defined inductively as follows.

$$\gamma_{SCPR}(0,U)=\emptyset \quad \gamma_{SCPR}(1,U)=\emptyset \quad \gamma_{SCPR}(V,U)=\{1:U=V\}$$
$$\gamma_{SCPR}(x+y,U)=\gamma_{SCPR}(x,U)\cup\gamma_{SCPR}(y,U) \quad \gamma_{SCPR}(x\cdot y,U)=\gamma_{SCPR}(x,U)\fatsemi y\;\cup\; x\star\gamma_{SCPR}(y,U)$$
$$\gamma_{SCPR}(x^*,U)=\gamma_{SCPR}(x,U)\fatsemi x^* \quad \gamma_{SCPR}(x^{\langle *\rangle},U)=\{1:U=\semangle{x,x^{\langle *\rangle}}\}\;\cup\;\{1\}\fatsemi\gamma_{SCPR}(x^{\langle *\rangle},U)$$
$$\gamma_{SCPR}(x\between y,U)=\gamma_{SCPR}(x\parallel y,U)\;\cup\;\gamma_{SCPR}(x\mid y,U)$$
$$\gamma_{SCPR}(x\parallel y,U)=\{1:U=\semangle{x,y}\}\;\cup\;\{U\}\fatsemi\gamma_{SCPR}(x'\between y',U')\;\textrm{with}\;x\parallel y=U\cdot(x'\between y')$$
$$\gamma_{SCPR}(x\mid y,U)=\{1:U=\semangle{x,y}\}\;\cup\;\{U\}\fatsemi\gamma_{SCPR}(x'\mid y',U')\;\textrm{with}\;x\mid y=U\cdot(x'\between y')$$
\end{definition}

\begin{lemma}
Let $x_1,x_2\in\mathcal{T}_{SCPR}$ and $w\in\mathsf{SCP}(\Sigma)^*$. The following two conclusions are equivalent:

\begin{enumerate}
  \item There exists a $y\in\mathcal{F}_{SCPR}$ such that $x_1+x_2\xtworightarrow[A_{SCPR}]{w} y$.
  \item There exists a $y\in\mathcal{F}_{SCPR}$ such that $x_1\xtworightarrow[A_{SCPR}]{w} y$ or $x_2\xtworightarrow[A_{SCPR}]{w} y$.
\end{enumerate}
\end{lemma}

\begin{lemma}
Let $x_1,x_2\in\mathcal{T}_{SCPR}$, $w\in\mathsf{SCP}(\Sigma)^*$ and $\ell\in\mathbb{N}$. The following two conclusions are equivalent:

\begin{enumerate}
  \item There exists a $y\in\mathcal{F}_{SCPR}$ such that $x_1\cdot x_2\xtworightarrow[A_{SCPR}]{w} y$ of length $\ell$.
  \item $w=w_1\cdot w_2$, then there exist $y_1,y_2\in\mathcal{F}_{SCPR}$ such that $x_1\xtworightarrow[A_{SCPR}]{w_1} y_1$ or $x_2\xtworightarrow[A_{SCPR}]{w_2} y_2$ of length at most $\ell$.
\end{enumerate}
\end{lemma}

\begin{lemma}
Let $x_1,x_2\in\mathcal{T}_{SCPR}$, $w\in\mathsf{SCP}(\Sigma)^*$. The following two conclusions are equivalent:

\begin{enumerate}
  \item There exists a $y\in\mathcal{F}_{SCPR}$ such that $x_1\parallel x_2\xtworightarrow[A_{SCPR}]{w} y$.
  \item $w=w_1\parallel w_2$, then there exist $y_1,y_2\in\mathcal{F}_{SCPR}$ such that $x_1\xtworightarrow[A_{SCPR}]{w_1} y_1$ or $x_2\xtworightarrow[A_{SCPR}]{w_2} y_2$.
\end{enumerate}
\end{lemma}

\begin{lemma}
Let $x_1,x_2\in\mathcal{T}_{SCPR}$, $w\in\mathsf{SCP}(\Sigma)^*$. The following two conclusions are equivalent:

\begin{enumerate}
  \item There exists a $y\in\mathcal{F}_{SCPR}$ such that $x_1\mid x_2\xtworightarrow[A_{SCPR}]{w} y$.
  \item $w=w_1\mid w_2$, then there exist $y_1,y_2\in\mathcal{F}_{SCPR}$ such that $x_1\xtworightarrow[A_{SCPR}]{w_1} y_1$ or $x_2\xtworightarrow[A_{SCPR}]{w_2} y_2$.
\end{enumerate}
\end{lemma}

\begin{lemma}
Let $x_1,x_2\in\mathcal{T}_{SCPR}$, $w\in\mathsf{SCP}(\Sigma)^*$. The following two conclusions are equivalent:

\begin{enumerate}
  \item There exists a $y\in\mathcal{F}_{SCPR}$ such that $x_1\leftmerge x_2\xtworightarrow[A_{SCPR}]{w} y$.
  \item $w=w_1\leftmerge w_2$, then there exist $y_1,y_2\in\mathcal{F}_{SCPR}$ such that $x_1\xtworightarrow[A_{SCPR}]{w_1} y_1$ or $x_2\xtworightarrow[A_{SCPR}]{w_2} y_2$.
\end{enumerate}
\end{lemma}

\begin{lemma}
Let $x_1,x_2\in\mathcal{T}_{SCPR}$, $w\in\mathsf{SCP}(\Sigma)^*$. The following two conclusions are equivalent:

\begin{enumerate}
  \item There exists a $y\in\mathcal{F}_{SCPR}$ such that $x_1\between x_2\xtworightarrow[A_{SCPR}]{w} y$.
  \item $w=w_1\between w_2$, then there exist $y_1,y_2\in\mathcal{F}_{SCPR}$ such that $x_1\xtworightarrow[A_{SCPR}]{w_1} y_1$ or $x_2\xtworightarrow[A_{SCPR}]{w_2} y_2$.
\end{enumerate}
\end{lemma}

\begin{lemma}
Let $x\in\mathcal{T}_{SCPR}$, $w\in\mathsf{SCP}(\Sigma)^*$. The following two conclusions are equivalent:

\begin{enumerate}
  \item There exists a $y\in\mathcal{F}_{SCPR}$ such that $x^*\xtworightarrow[A_{SCPR}]{w} y$.
  \item $w=w_1\cdots w_n$, then there exist $y_i\in\mathcal{F}_{SCPR}$ such that $x\xtworightarrow[A_{SCPR}]{w_i} y_i$ for $1\leq i\leq n$.
\end{enumerate}
\end{lemma}

\begin{lemma}
Let $x\in\mathcal{T}_{SCPR}$, $w\in\mathsf{SCP}(\Sigma)^*$. The following two conclusions are equivalent:

\begin{enumerate}
  \item There exists a $y\in\mathcal{F}_{SCPR}$ such that $x^{\langle *\rangle}\xtworightarrow[A_{SCPR}]{w} y$.
  \item $w=w_1\parallel\cdots\parallel w_n$, then there exist $y_i\in\mathcal{F}_{SCPR}$ such that $x\xtworightarrow[A_{SCPR}]{w_i} y_i$ for $1\leq i\leq n$.
\end{enumerate}
\end{lemma}

\begin{lemma}
Let $x,y\in\mathcal{T}_{SCPR}$, then the following hold:

\begin{enumerate}
  \item $L_{SCPR}(x+y)=L_{SCPR}(x)+L_{SCPR}(y)$.
  \item $L_{SCPR}(x\cdot y)=L_{SCPR}(x)\cdot L_{SCPR}(y)$.
  \item $L_{SCPR}(x^*)=L_{SCPR}(x)^*$.
  \item $L_{SCPR}(x^{\langle *\rangle})=L_{SCPR}(x)^{\langle *\rangle}$.
  \item $L_{SCPR}(x\between y)=L_{SCPR}(x)\between L_{SCPR}(y)$.
  \item $L_{SCPR}(x\parallel y)=L_{SCPR}(x)\parallel L_{SCPR}(y)$.
  \item $L_{SCPR}(x\mid y)=L_{SCPR}(x)\mid L_{SCPR}(y)$.
  \item $L_{SCPR}(x\leftmerge y)=L_{SCPR}(x)\leftmerge L_{SCPR}(y)$.
\end{enumerate}
\end{lemma}

\begin{lemma}
For all $x\in\mathcal{T}_{SCPR}$, it holds that $L_{SCPR}(x)=\sembrack{x}_{SCPR}$.
\end{lemma}

\begin{definition}[$\langle *\rangle$-depth]
We extend the domain of $d_{\between}$, $d_{\parallel}$ and $d_{\mid}$ to $\mathcal{T}_{SCPR}$ by defining $d_{\between}(x^{\langle *\rangle})=d_{\between}(x)$, $d_{\parallel}(x^{\langle *\rangle})=d_{\parallel}(x)$ and $d_{\mid}(x^{\langle *\rangle})=d_{\mid}(x)$. And we define the $\langle *\rangle$-depth of $x\in\mathcal{T}_{SCPR}$ denoted $d_{\langle *\rangle}(x)$ is defined inductively on the structure of $x$ as follows.

$$d_{\langle *\rangle}(0)=0\quad d_{\langle *\rangle}(1)=0\quad d_{\langle *\rangle}(a)=0$$
$$d_{\langle *\rangle}(x\cdot y)=\max(d_{\langle *\rangle}(x),d_{\langle *\rangle}(y))\quad d_{\langle *\rangle}(x+ y)=\max(d_{\langle *\rangle}(x),d_{\langle *\rangle}(y))\quad d_{\langle *\rangle}(x^*)=d_{\langle *\rangle}(x)$$ 
$$d_{\langle *\rangle}(x\between y)=\max(d_{\langle *\rangle}(x\parallel y),d_{\langle *\rangle}(x\mid y))\quad d_{\langle *\rangle}(x\parallel y)=\max(d_{\langle *\rangle}(x),d_{\langle *\rangle}(y))$$
$$d_{\langle *\rangle}(x\mid y)=\max(d_{\langle *\rangle}(x),d_{\langle *\rangle}(y))\quad d_{\langle *\rangle}(x\leftmerge y)=\max(d_{\langle *\rangle}(x),d_{\langle *\rangle}(y))\quad d_{\langle *\rangle}(x^{\langle *\rangle})=d_{\langle *\rangle}(x)+1$$
\end{definition}

\begin{lemma}
Let $x,y\in\mathcal{T}_{SCPR}$, if $x\preceq_{SCPR} y$, then $d_{\between}(x)\leq d_{\between}(y)$, $d_{\parallel}(x)\leq d_{\parallel}(y)$, $d_{\leftmerge}(x)\leq d_{\leftmerge}(y)$, $d_{\mid}(x)\leq d_{\mid}(y)$ and $d_{\langle *\rangle}(x)\leq d_{\langle *\rangle}(y)$.
\end{lemma}

\begin{lemma}
Let $x,y\in\mathcal{T}_{SCPR}$, if $x\preceq_{SCPR} y^{\langle *\rangle}$ and $d_{\langle *\rangle}(x)=d_{\langle *\rangle}(y^{\langle *\rangle})$, then $x=y^{\langle *\rangle}$.
\end{lemma}

\begin{lemma}
Let $x,y,z,h\in\mathcal{T}_{SCPR}$ and $U\in\mathsf{SCP}(\Sigma)$ with $\gamma(x,U)\neq\emptyset$, then $U=\semangle{y,z}$ with either (1) $y\prec_{SCPR}x$ and $g\prec_{SCPR}x$, or (2) $y\prec_{SCPR} x$ and $z=h^{\langle *\rangle}$ for some $h\in\mathcal{T}_{SCPR}$.
\end{lemma}

\begin{lemma}
Every $x\in\mathcal{T}_{SCPR}$ is recursive in $A_{SCPR}$, and the syntactic SA is well-nested.
\end{lemma}

\begin{definition}
Let $x_1,x_2\in\mathcal{T}_{SCPR}$, $R:\mathcal{T}_{SCPR}\rightarrow 2^{\mathcal{T}_{SCPR}}$ is defined inductively as follows:

$$R(0)=\{0\}\quad R(1)=\{1\} \quad R(a)=\{a,1\}$$
$$R(x_1+x_2)=R(x_1)\cup R(x_2)\quad R(x_1\cdot x_2)=R(x_1)\fatsemi x_2\;\cup\; R(x_1)\cup R(x_2)$$
$$R(x_1^*)=R(x_1)\fatsemi x_1^*\;\cup\; R(x_1)\cup\{x_1^*\}\quad R(x_1^{\langle *\rangle})=R(x_1)\cup\{x_1^{\langle *\rangle},1\} \quad R(x_1\between x_2)=R(x_1)\cup R(x_2)\cup\{x_1\between x_2,1\}$$
$$R(x_1\parallel x_2)=R(x_1)\cup R(x_2)\cup\{x_1\parallel x_2,1\}\quad R(x_1\mid x_2)=R(x_1)\cup R(x_2)\cup\{x_1\mid x_2,1\}$$
$$R(x_1\leftmerge x_2)=R(x_1)\cup R(x_2)\cup\{x_1\leftmerge x_2,1\}$$
\end{definition}

\begin{lemma}
For every $x\in\mathcal{T}_{SCPR}$, they hold that:

\begin{enumerate}
  \item $x\in R(x)$.
  \item $R(x)$ is support-closed.
  \item The syntactic SA is bounded.
\end{enumerate}
\end{lemma}

\begin{theorem}[Expressions to step automata]\label{etoa4}
For every $x\in\mathcal{T}_{SCPR}$, we can obtain a well-nested and finite SA $A$ with a state $q$ such that $L_{A}(q)=\sembrack{x}_{SCPR}$.
\end{theorem}

\begin{theorem}
For $x,y\in\mathcal{T}_{SCPR}$, according to \cref{etoa4}, we obtain two corresponding SA $A_x$ and $A_y$. If $x\sim_p y$, then $A_x\sim_p A_y$.
\end{theorem}

\begin{theorem}
For $x,y\in\mathcal{T}_{SCPR}$, according to \cref{etoa4}, we obtain two corresponding SA $A_x$ and $A_y$. If $x\sim_s y$, then $A_x\sim_s A_y$.
\end{theorem}

\begin{theorem}
For $x,y\in\mathcal{T}_{SCPR}$, according to \cref{etoa4}, we obtain two corresponding SA $A_x$ and $A_y$. If $x\sim_{hp} y$, then $A_x\sim_{hp} A_y$.
\end{theorem}

\begin{theorem}
For $x,y\in\mathcal{T}_{SCPR}$, according to \cref{etoa4}, we obtain two corresponding SA $A_x$ and $A_y$. If $x\sim_{hhp} y$, then $A_x\sim_{hhp} A_y$.
\end{theorem}

\subsubsection{Step Automata to Expressions}

In this section, we show that the language accepted by a state in any well-nested and finite automaton can be implemented by a series-communication-parallel rational expression.

\begin{definition}[1-Solution of a SA]
Let $A=(Q,F,\delta,\gamma)$ be a SA, and let $=_1$ be an EBKAC language congruence on $\mathcal{T}_{SCPR}(\Delta)$ with $\Sigma\subseteq\Delta$. We say that $s:Q\rightarrow\mathcal{T}_{SCPR}(\Delta)$ is an $=_1$-solution to $A$, if for every $q\in Q$:

\begin{center}
$[q\in \mathcal{F}_{SCPR}]+\sum_{q'\in\delta(q,a)}a\cdot s(q')+\sum_{q'\in\gamma(q,U)}U\cdot s(q')\leqq_1 s(q)$
\end{center}

Also, $s$ is a least $=_1$-solution to $A$ if for every $=_1$-solution $s'$ it holds that $s(q)\leqq_1 s'(q)$ for all $q\in Q$. We call $s:Q\rightarrow\mathcal{T}_{SCPR}$ the least 1-solution to $A$ if it is the least $=_1$-solution for any EBKAC language congruence $=_1$.
\end{definition}

\begin{lemma}
Let $A=(Q,F,\delta,\gamma)$ be a step automaton. If $s:Q\rightarrow\mathcal{T}_{SCPR}$ is the least 1-solution to $A$, then it holds that $L_A(q)=\sembrack{s(q)}_{SCPR}$ for $q\in Q$.
\end{lemma}

\begin{lemma}
Let $A$ be a well-nested and finite SA, then we can construct the least 1-solution to $A$.
\end{lemma}

\begin{theorem}[Step automata to expressions]
If $A=(Q,F,\delta,\gamma)$ is a well-nested and finite SA, then we can construct for every $q\in Q$ a series-communication-parallel rational expression $x\in\mathcal{T}_{SCPR}$ such that $L_{A}(q)=\sembrack{x}_{SCPR}$.
\end{theorem}

\begin{corollary}[Kleene theorem for series-communication-parallel rational language]
Let $L\subseteq\mathsf{SCP}$, then $L$ is series-communication-parallel rational if and only if it is accepted by a finite and well-nested step automaton.
\end{corollary}

\begin{definition}[2-Solution of a SA]
Let $A=(Q,F,\delta,\gamma)$ be a SA, and let $=_2$ be an EBKAC pomset bisimilar congruence on $\mathcal{T}_{SCPR}(\Delta)$ with $\Sigma\subseteq\Delta$. We say that $s:Q\rightarrow\mathcal{T}_{SCPR}(\Delta)$ is an $=_2$-solution to $A$, if for every $q\in Q$:

\begin{center}
$[q\in \mathcal{F}_{SCPR}]+\sum_{q'\in\delta(q,a)}a\cdot s(q')+\sum_{q'\in\gamma(q,U)}U\cdot s(q')\leqq_2 s(q)$
\end{center}

Also, $s$ is a least $=_2$-solution to $A$ if for every $=_2$-solution $s'$ it holds that $s(q)\leqq_2 s'(q)$ for all $q\in Q$. We call $s:Q\rightarrow\mathcal{T}_{SCPR}$ the least 2-solution to $A$ if it is the least $=_2$-solution for any EBKAC pomset bisimilar congruence $=_2$.
\end{definition}

\begin{lemma}
Let $A$ be a well-nested and finite SA, then we can construct the least 2-solution to $A$.
\end{lemma}

\begin{theorem}[Step automata to expressions modulo pomset bisimilarity]
If $A=(Q,F,\delta,\gamma)$ and $A=(Q',F',\delta',\gamma')$  are well-nested and finite SA, then we can construct for each $q\in Q$ a series-communication-parallel rational expression $x\in\mathcal{T}_{SCPR}$ and for each $q'\in Q$ a series-communication-parallel rational expression $x'\in\mathcal{T}_{SCPR}$, such that if $A\sim_p A'$ then $x\sim_p x'$.
\end{theorem}

\begin{corollary}[Kleene theorem for series-communication-parallel rational language modulo pomset bisimilarity]
Let $L\subseteq\mathsf{SCP}$, then $L$ is series-communication-parallel rational and $x,y\subseteq L$ with $x\sim_p y$, if and only if there exist finite and well-nested step automata $A_x$ and $A_y$ such that $A_x\sim_p A_y$.
\end{corollary}

\begin{definition}[3-Solution of a SA]
Let $A=(Q,F,\delta,\gamma)$ be a SA, and let $=_3$ be an EBKAC step bisimilar congruence on $\mathcal{T}_{SCPR}(\Delta)$ with $\Sigma\subseteq\Delta$. We say that $s:Q\rightarrow\mathcal{T}_{SCPR}(\Delta)$ is an $=_3$-solution to $A$, if for every $q\in Q$:

\begin{center}
$[q\in \mathcal{F}_{SCPR}]+\sum_{q'\in\delta(q,a)}a\cdot s(q')+\sum_{q'\in\gamma(q,U)}U\cdot s(q')\leqq_3 s(q)$
\end{center}

Also, $s$ is a least $=_3$-solution to $A$ if for every $=_3$-solution $s'$ it holds that $s(q)\leqq_3 s'(q)$ for all $q\in Q$. We call $s:Q\rightarrow\mathcal{T}_{SCPR}$ the least 3-solution to $A$ if it is the least $=_3$-solution for any EBKAC step bisimilar congruence $=_3$.
\end{definition}

\begin{lemma}
Let $A$ be a well-nested and finite SA, then we can construct the least 3-solution to $A$.
\end{lemma}

\begin{theorem}[Step automata to expressions modulo step bisimilarity]
If $A=(Q,F,\delta,\gamma)$ and $A=(Q',F',\delta',\gamma')$  are well-nested and finite SA, then we can construct for each $q\in Q$ a series-communication-parallel rational expression $x\in\mathcal{T}_{SCPR}$ and for each $q'\in Q$ a series-communication-parallel rational expression $x'\in\mathcal{T}_{SCPR}$, such that if $A\sim_p A'$ then $x\sim_s x'$.
\end{theorem}

\begin{corollary}[Kleene theorem for series-communication-parallel rational language modulo step bisimilarity]
Let $L\subseteq\mathsf{SCP}$, then $L$ is series-communication-parallel rational and $x,y\subseteq L$ with $x\sim_s y$, if and only if there exist finite and well-nested step automata $A_x$ and $A_y$ such that $A_x\sim_s A_y$.
\end{corollary}

\begin{definition}[4-Solution of a SA]
Let $A=(Q,F,\delta,\gamma)$ be a SA, and let $=_4$ be an EBKAC hp-bisimilar congruence on $\mathcal{T}_{SCPR}(\Delta)$ with $\Sigma\subseteq\Delta$. We say that $s:Q\rightarrow\mathcal{T}_{SCPR}(\Delta)$ is an $=_4$-solution to $A$, if for every $q\in Q$:

\begin{center}
$[q\in \mathcal{F}_{SCPR}]+\sum_{q'\in\delta(q,a)}a\cdot s(q')+\sum_{q'\in\gamma(q,U)}U\cdot s(q')\leqq_4 s(q)$
\end{center}

Also, $s$ is a least $=_4$-solution to $A$ if for every $=_4$-solution $s'$ it holds that $s(q)\leqq_4 s'(q)$ for all $q\in Q$. We call $s:Q\rightarrow\mathcal{T}_{SCPR}$ the least 4-solution to $A$ if it is the least $=_4$-solution for any EBKAC hp-bisimilar congruence $=_4$.
\end{definition}

\begin{lemma}
Let $A$ be a well-nested and finite SA, then we can construct the least 4-solution to $A$.
\end{lemma}

\begin{theorem}[Step automata to expressions modulo hp-bisimilarity]
If $A=(Q,F,\delta,\gamma)$ and $A=(Q',F',\delta',\gamma')$  are well-nested and finite SA, then we can construct for each $q\in Q$ a series-communication-parallel rational expression $x\in\mathcal{T}_{SCPR}$ and for each $q'\in Q$ a series-communication-parallel rational expression $x'\in\mathcal{T}_{SCPR}$, such that if $A\sim_{hp} A'$ then $x\sim_{hp} x'$.
\end{theorem}

\begin{corollary}[Kleene theorem for series-communication-parallel rational language modulo hp-bisimilarity]
Let $L\subseteq\mathsf{SCP}$, then $L$ is series-communication-parallel rational and $x,y\subseteq L$ with $x\sim_{hp} y$, if and only if there exist finite and well-nested step automata $A_x$ and $A_y$ such that $A_x\sim_{hp} A_y$.
\end{corollary}

\begin{definition}[5-Solution of a SA]
Let $A=(Q,F,\delta,\gamma)$ be a SA, and let $=_5$ be an EBKAC hhp-bisimilar congruence on $\mathcal{T}_{SCPR}(\Delta)$ with $\Sigma\subseteq\Delta$. We say that $s:Q\rightarrow\mathcal{T}_{SCPR}(\Delta)$ is an $=_5$-solution to $A$, if for every $q\in Q$:

\begin{center}
$[q\in \mathcal{F}_{SCPR}]+\sum_{q'\in\delta(q,a)}a\cdot s(q')+\sum_{q'\in\gamma(q,U)}U\cdot s(q')\leqq_5 s(q)$
\end{center}

Also, $s$ is a least $=_5$-solution to $A$ if for every $=_5$-solution $s'$ it holds that $s(q)\leqq_5 s'(q)$ for all $q\in Q$. We call $s:Q\rightarrow\mathcal{T}_{SCPR}$ the least 5-solution to $A$ if it is the least $=_5$-solution for any EBKAC hhp-bisimilar congruence $=_5$.
\end{definition}

\begin{lemma}
Let $A$ be a well-nested and finite SA, then we can construct the least 5-solution to $A$.
\end{lemma}

\begin{theorem}[Step automata to expressions modulo hhp-bisimilarity]
If $A=(Q,F,\delta,\gamma)$ and $A=(Q',F',\delta',\gamma')$  are well-nested and finite SA, then we can construct for each $q\in Q$ a series-communication-parallel rational expression $x\in\mathcal{T}_{SCPR}$ and for each $q'\in Q$ a series-communication-parallel rational expression $x'\in\mathcal{T}_{SCPR}$, such that if $A\sim_{hhp} A'$ then $x\sim_{hhp} x'$.
\end{theorem}

\begin{corollary}[Kleene theorem for series-communication-parallel rational language modulo hhp-bisimilarity]
Let $L\subseteq\mathsf{SCP}$, then $L$ is series-communication-parallel rational and $x,y\subseteq L$ with $x\sim_{hhp} y$, if and only if there exist finite and well-nested step automata $A_x$ and $A_y$ such that $A_x\sim_{hhp} A_y$.
\end{corollary} 

\subsection{Step Turing Machine}\label{stm} 

Firstly, we give the definition of step Turing machine.

\begin{figure}[!htbp]
 \centering
 \includegraphics{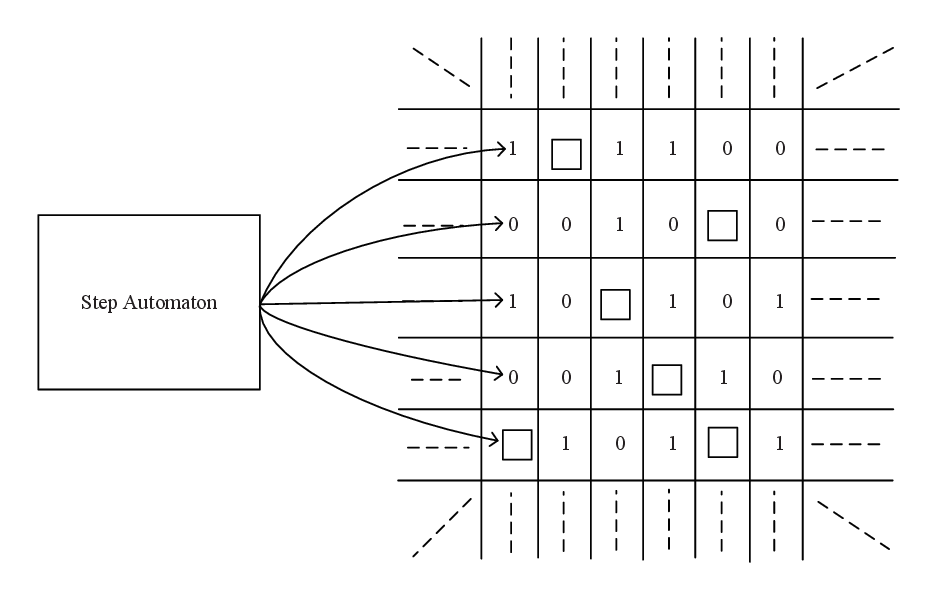}
 \caption{Step Turing machine}
 \label{steptm}
\end{figure} 

\begin{definition}[Step Turing machine]
A step Turing machine (STM) is defined as an octuple $M=(Q,q_0,F,\Sigma,\Gamma,\delta)$ where

\begin{itemize}
  \item $Q$ is a finite set of control states.
  \item $q_0\in Q$ is the initial state.
  \item $F\subseteq Q$ is the set of final states.
  \item $\Sigma$ is a finite alphabet including special actions 0, 1.
  \item $\Gamma$ is the set of tape symbols consisting of $0$, $1$ and the blank symbol $\square$.
  \item $\delta: (Q\setminus F)\times U\times (\Gamma\cup\{\epsilon\})^{\langle *\rangle}\rightarrow Q\times (\Gamma\cup\{\epsilon\})^{\langle *\rangle}\times (D=\{L,R\})$ is a finite set of step transitions, where $U\in \mathsf{SCP}(\Sigma)$ and $\mathsf{SCP}(\Sigma)$ is the series-parallel step of $\Sigma$, $D=\{L,R\}$ is the read/write heads' moving direction: left or right, and if $\epsilon$ is read, it means that we are looking at an empty part of the tape; if $\epsilon$ is written, then a symbol on the tape will be erased. For $(q_i,U,(\Gamma\cup\{\epsilon\})^{\langle *\rangle})\rightarrow (q_j, (\Gamma\cup\{\epsilon\})^{\langle *\rangle},D)$, we write it as $q_i\xrightarrow{U,(\Gamma\cup\{\epsilon\})^{\langle *\rangle},(\Gamma\cup\{\epsilon\})^{\langle *\rangle},D}q_j$, where $q_i,q_j\in Q$. It means that the STM is in state $q_i$ and reading symbols $(\Gamma\cup\{\epsilon\})^{\langle *\rangle}$ on the planar tape in parallel, will execute a step of actions $U$, change the symbols on the planar tape to new symbols $(\Gamma\cup\{\epsilon\})^{\langle *\rangle}$, move one step left if $D=L$ and right if $D=R$, and therefore evolve into state $q_j$. Note that, when $U=1$, the current contents of the planar step tape remain unchanged, and then the heads move left or right. 
\end{itemize}

As \cref{steptm} illustrates, an STM has a step automaton providing finite state control, and a planar tape within which exists maybe infinite classical tapes and each classical tape is with a read/write head and has infinite cells. Each cell may contain a tape symbol either being $0$, or $1$, or $\square$. The whole planar step tape move left/right, and read/write contents from/into the cells within a step in parallel.
\end{definition}

The STM is off-line, i.e., input to an STM is written on the planar step tape with an element of $\{0,1\}^{\langle*\rangle^*}$. 

The configurations of an STM are determined by the contents of its planar step tape and the locations of the corresponding read/write heads.

\begin{definition}[Configurations of an STM]
The current configuration of an STM consists of the contents and the current location of read/write heads of the planar step tape. We use a bar over a symbol of the contents to denote the current location of the tape head, for example, $\square1\bar{0}1\square$ is a configuration of a classical Turing machine.

For the planar step tape with intermediate contents:
  $$
  \left[\begin{array}{ccc}
     0 & 1 & 0\\
     1 & 1 & 0\\
     1 & 0 & 1\\
     0 &   & 0\\
       &   & 1
  \end{array}\right]
  $$
  the possible configurations are:
  \begin{small}
  $$
  \left[\begin{array}{ccccc}
     \bar{\square} & 0 & 1 & 0 & \square\\
     \bar{\square} & 1 & 1 & 0 & \square\\
     \bar{\square} & 1 & 0 & 1 & \square\\
     \bar{\square} & 0 &   & 0 & \square\\
     \bar{\square} &   &   & 1 & \square
  \end{array}\right]
  \left[\begin{array}{ccccc}
     \square & \bar{0} & 1 & 0 & \square\\
     \square & \bar{1} & 1 & 0 & \square\\
     \square & \bar{1} & 0 & 1 & \square\\
     \square & \bar{0} &   & 0 & \square\\
     \square &   &   & 1 & \square
  \end{array}\right]
  \left[\begin{array}{ccccc}
     \square & 0 & \bar{1} & 0 & \square\\
     \square & 1 & \bar{1} & 0 & \square\\
     \square & 1 & \bar{0} & 1 & \square\\
     \square & 0 &   & 0 & \square\\
     \square &   &   & 1 & \square
  \end{array}\right]
  \left[\begin{array}{ccccc}
     \square & 0 & 1 & \bar{0} & \square\\
     \square & 1 & 1 & \bar{0} & \square\\
     \square & 1 & 0 & \bar{1} & \square\\
     \square & 0 &   & \bar{0} & \square\\
     \square &   &   & \bar{1} & \square
  \end{array}\right]
  \left[\begin{array}{ccccc}
     \square & 0 & 1 & 0 & \bar{\square}\\
     \square & 1 & 1 & 0 & \bar{\square}\\
     \square & 1 & 0 & 1 & \bar{\square}\\
     \square & 0 &   & 0 & \bar{\square}\\
     \square &   &   & 1 & \bar{\square}
  \end{array}\right]
  $$
  \end{small}
\end{definition}

\begin{definition}[States of an STM]
Let $M=(Q,q_0,F,\Sigma,\Gamma,\delta)$ be an STM. The set of states is determined by the finite control states and the configurations, i.e., $\{(q,\square^{\langle *\rangle})|q\in Q,\square^{\langle *\rangle} \\\textrm{is the beginning of the planar step tape}\}\cup\{(q,\square^{\langle *\rangle}x\square^{\langle *\rangle},\bar{ })|q\in Q,x\in\Gamma^{\langle *\rangle^*}\}$, where $\bar{ }$ denotes the location of read/write head, on one of the elements of $\square x\square$ or $\square^{\langle *\rangle}x\square^{\langle *\rangle}$.

The initial state is $(q_0,\overline{\square^{\langle *\rangle}})$, and if $(q,\overline{\square^{\langle *\rangle}})\in F$, then $q\in F$.
\end{definition}

\begin{definition}[Transition system of an STM]
The transition system of an STM is following.

  $(q_i,\square^{\langle *\rangle}x\bar{d}\square^{\langle *\rangle})\xrightarrow{U}(q_j,\square^{\langle *\rangle}xe\overline{\square^{\langle *\rangle}})$ if and only if $q_i\xrightarrow{U,d,e,R}q_j$, where $d,e\in\Gamma^{\langle *\rangle},x\in\Gamma^{\langle *\rangle^*}$.
  
  $(q_i,\square^{\langle *\rangle}x\bar{d}fy\square^{\langle *\rangle})\xrightarrow{U}(q_j,\square^{\langle *\rangle}xe\bar{f}y\square^{\langle *\rangle})$ if and only if $q_i\xrightarrow{U,d,e,R}q_j$, where $d,e,f\in\Gamma^{\langle *\rangle},x,y\in\Gamma^{\langle *\rangle^*}$.
  
  $(q_i,\square^{\langle *\rangle}\bar{d}x\square^{\langle *\rangle})\xrightarrow{U}(q_j,\overline{\square^{\langle *\rangle}}ex\square^{\langle *\rangle})$ if and only if $q_i\xrightarrow{U,d,e,L}q_j$, where $d,e\in\Gamma^{\langle *\rangle},x\in\Gamma^{\langle *\rangle^*}$.
  
  $(q_i,\square^{\langle *\rangle}xf\bar{d}y\square^{\langle *\rangle})\xrightarrow{U}(q_j,\square^{\langle *\rangle}x\bar{f}ey\square^{\langle *\rangle})$ if and only if $q_i\xrightarrow{U,d,e,L}q_j$, where $d,e,f\in\Gamma^{\langle *\rangle},x,y\in\Gamma^{\langle *\rangle^*}$.
  
  $(q_i,\square^{\langle *\rangle}\bar{d}\square^{\langle *\rangle})\xrightarrow{U}(q_j,\square^{\langle *\rangle}\overline{\square^{\langle *\rangle}})$ if and only if $q_i\xrightarrow{U,d,\epsilon,R}q_j$, where $d\in\Gamma^{\langle *\rangle}$.
  
  $(q_i,\square^{\langle *\rangle}x\bar{d}\square^{\langle *\rangle})\xrightarrow{U}(q_j,\square^{\langle *\rangle}x\overline{\square^{\langle *\rangle}})$ if and only if $q_i\xrightarrow{U,d,\epsilon,R}q_j$, where $d\in\Gamma^{\langle *\rangle},x\in\Gamma^{\langle *\rangle^*}$.
  
  $(q_i,\square^{\langle *\rangle}\bar{d}fx\square^{\langle *\rangle})\xrightarrow{U}(q_j,\square^{\langle *\rangle}\bar{f}x\square^{\langle *\rangle})$ if and only if $q_i\xrightarrow{U,d,\epsilon,R}q_j$, where $d,f\in\Gamma^{\langle *\rangle},x\in\Gamma^{\langle *\rangle^*}$.
  
  $(q_i,\square^{\langle *\rangle}\bar{d}\square^{\langle *\rangle})\xrightarrow{U}(q_j,\overline{\square^{\langle *\rangle}})$ if and only if $q_i\xrightarrow{U,d,\epsilon,L}q_j$, where $d\in\Gamma^{\langle *\rangle}$.
  
  $(q_i,\square^{\langle *\rangle}\bar{d}x\square^{\langle *\rangle})\xrightarrow{U}(q_j,\overline{\square^{\langle *\rangle}}x\square^{\langle *\rangle})$ if and only if $q_i\xrightarrow{U,d,\epsilon,L}q_j$, where $d\in\Gamma^{\langle *\rangle},x\in\Gamma^{\langle *\rangle^*}$.
  
  $(q_i,\square^{\langle *\rangle}xf\bar{d}\square^{\langle *\rangle})\xrightarrow{U}(q_j,\square^{\langle *\rangle}x\bar{f}\square^{\langle *\rangle})$ if and only if $q_i\xrightarrow{U,d,\epsilon,L}q_j$, where $d,f\in\Gamma^{\langle *\rangle},x\in\Gamma^{\langle *\rangle^*}$.
  
  $(q_i,\overline{\square^{\langle *\rangle}})\xrightarrow{U}(q_j,\square^{\langle *\rangle}d\overline{\square^{\langle *\rangle}})$ if and only if $q_i\xrightarrow{U,\epsilon,d,R}q_j$, where $d\in\Gamma^{\langle *\rangle}$.
  
  $(q_i,\overline{\square^{\langle *\rangle}}fx\square^{\langle *\rangle})\xrightarrow{U}(q_j,\square^{\langle *\rangle}d\bar{f}x\square^{\langle *\rangle})$ if and only if $q_i\xrightarrow{U,\epsilon,d,R}q_j$, where $d\in\Gamma^{\langle *\rangle},x\in\Gamma^{\langle *\rangle^*}$.
  
  $(q_i,\overline{\square^{\langle *\rangle}})\xrightarrow{U}(q_j,\overline{\square^{\langle *\rangle}}d\square^{\langle *\rangle})$ if and only if $q_i\xrightarrow{U,\epsilon,d,L}q_j$, where $d\in\Gamma^{\langle *\rangle}$.
  
  $(q_i,\square^{\langle *\rangle}xf\overline{\square^{\langle *\rangle}})\xrightarrow{U}(q_j,\square^{\langle *\rangle}x\bar{f}d\square^{\langle *\rangle})$ if and only if $q_i\xrightarrow{U,\epsilon,d,L}q_j$, where $d,f\in\Gamma^{\langle *\rangle},x\in\Gamma^{\langle *\rangle^*}$.
  
  $(q_i,\overline{\square^{\langle *\rangle}})\xrightarrow{U}(q_j,\overline{\square^{\langle *\rangle}})$ if and only if $q_i\xrightarrow{U,\epsilon,\epsilon,R}q_j$.
  
  $(q_i,\overline{\square^{\langle *\rangle}}fx\square^{\langle *\rangle})\xrightarrow{U}(q_j,\square^{\langle *\rangle}\bar{f}x\square^{\langle *\rangle})$ if and only if $q_i\xrightarrow{U,\epsilon,\epsilon,R}q_j$, where $d\in\Gamma^{\langle *\rangle}$.
  
  $(q_i,\overline{\square^{\langle *\rangle}})\xrightarrow{U}(q_j,\overline{\square^{\langle *\rangle}})$ if and only if $q_i\xrightarrow{U,\epsilon,\epsilon,L}q_j$.
  
  $(q_i,\square^{\langle *\rangle}xf\overline{\square^{\langle *\rangle}})\xrightarrow{U}(q_j,\square^{\langle *\rangle}x\bar{f}\square^{\langle *\rangle})$ if and only if $q_i\xrightarrow{U,\epsilon,\epsilon,L}q_j$, where $f\in\Gamma^{\langle *\rangle}, x\in\Gamma^{\langle *\rangle^*}$.
\end{definition}

Similarly to the definition of path for step automata, we can define the path $\xtworightarrow{}$ of transition $\rightarrow$.

\begin{definition}[Language accepted by an STM]
Let $M=(Q,q_0,F,\Sigma,\Gamma,\delta)$ be an STM. The language accepted by $M$, is the set $L_M=\{w\in\mathsf{SCP}(\Sigma)^{\langle*\rangle^*}:(q_0,\overline{\square^{\langle *\rangle^*}})\xtworightarrow{w}(q',\overline{\square^{\langle *\rangle^*}}),q'\in F\}$. 
\end{definition}

\subsubsection{Church-Turing Thesis}

\begin{definition}[Computation of an STM]
An atomic action $a\in\Sigma$ is the one that an STM can compute individually, and correspondingly, a step contains some atomic actions without partial orders between them pairwise. We say the computation of an STM, if the STM do computations step by step from the initial state to one of the final states. And we say that the STM computes the function $f$ on a domain of data $\Gamma^{\langle*\rangle^*}$ if for all inputs $w\in\Gamma^{\langle*\rangle^*}$, it has the outputs $f(w)$.
\end{definition}

\begin{definition}[Computability]
 A problem is computable if it can be computed by a classical Turing machine. An algorithm for a function is a classical Turing machine computing this function.
\end{definition} 

\begin{theorem}
The class of computable languages and functions defined by an STM is the same as the class of computable languages and functions defined by a classical Turing machine.
\end{theorem} 
\bibliographystyle{elsarticle-num}
\newpage\bibliography{Refs-CC}

\end{document}